\documentclass[a4paper,11pt]{article}

\usepackage{tikz}
\usepackage{subcaption}
\usepackage{jheppub} 
\usepackage{braket}
\usepackage{soul}

\newcommand\Eqn[1]     {Eq.\,(\ref{#1})}
\newcommand\Eqns[2]    {Eqs.\,(\ref{#1}) and~(\ref{#2})}

\newcommand\eqn[1]     {eq.\,(\ref{#1})}
\newcommand\eqns[2]    {eqs.\,(\ref{#1}) and~(\ref{#2})}

\newcommand{\eps}{\epsilon}
\newcommand{\al}{\alpha}
\newcommand{\T}{\mathbf{T}}
\newcommand{\nn}{\nonumber}
\newcommand{\bea}{\begin{eqnarray}}
\newcommand{\eea}{\end{eqnarray}}
\newcommand{\beqa}{\begin{eqnarray}}
\newcommand{\eeqa}{\end{eqnarray}}
\newcommand{\be}{\begin{equation}}
\newcommand{\ee}{\end{equation}}
\newcommand{\beq}{\begin{equation}}
\newcommand{\eeq}{\end{equation}}
\newcommand{\dbar}{d\hspace*{-0.08em}\bar{}\hspace*{0.1em}}
\newcommand{\as}{\alpha_s}
\newcommand{\ord}{{\cal O}}
\newcommand{\CA}{C_A}
\newcommand{\nf}{n_f}

\allowdisplaybreaks

\title{\boldmath The Two-Loop Lipatov Vertex in QCD}

\author[a,b]{Samuel Abreu,}
\author[a]{Giuseppe De Laurentis,}
\author[c,d]{Giulio Falcioni,}
\author[a]{Einan Gardi,}
\author[e]{Calum Milloy,}
\author[e]{Leonardo Vernazza}

\affiliation[a]{Higgs Centre for Theoretical Physics, School of Physics and Astronomy,\\ The University of Edinburgh, Edinburgh EH9 3FD, Scotland, UK}
\affiliation[b]{Theoretical Physics Department, CERN, Geneva 1211, Switzerland}
\affiliation[c]{Physik-Institut, Universit\"{a}t Z\"{u}rich, Winterthurerstrasse 190, 8057 Z\"{u}rich, Switzerland}
\affiliation[d]{Dipartimento di Fisica, Universit\`{a} di Torino, Via Pietro Giuria 1, I-10125 Torino, Italy}
\affiliation[e]{INFN, Sezione di Torino, Via Pietro Giuria 1, I-10125 Torino, Italy}

\preprint{\begin{flushright}
    CERN-TH-2024-226, ZU-TH 68/24
\end{flushright}}

\emailAdd{Samuel.Abreu@ed.ac.uk}
\emailAdd{Giuseppe.DeLaurentis@ed.ac.uk}
\emailAdd{Giulio.Falcioni@physik.uzh.ch}
\emailAdd{Einan.Gardi@ed.ac.uk}
\emailAdd{calummilloy526@gmail.com}
\emailAdd{Leonardo.Vernazza@to.infn.it}

\abstract{High-energy factorization of $2\to 2$ amplitudes in QCD has been recently pushed to the next-to-next-to-leading logarithmic order by determining the three-loop gluon Regge trajectory. This was based on computing multi-Reggeon exchanges
using rapidity evolution in the shock-wave formalism, and disentangling between the Regge pole and Regge cut contributions. 
In the present paper we extend the relevant theoretical framework to $2\to 3$ processes, and compute all multi-Reggeon exchanges necessary for extracting the two-loop Reggeon-gluon-Reggeon Lipatov vertex from $2\to 3$ amplitudes.
Then, specializing general amplitude methods to multi-Regge kinematics, we derive analytic expressions for non-planar two-loop~$gg\to ggg$, $gq\to ggq$ and $qq\to qgq$ QCD amplitudes
in that limit. Matching these to the multi-Reggeon computation, we determine the QCD Lipatov vertex in dimensional regularization at two loops through finite terms. We also determine the one-loop vertex through~${\cal O}(\epsilon^4)$. All results are expressed in a compact form in terms of a basis of single-valued generalised polylogarithms, manifesting target-projectile symmetry and reality properties.
Furthermore, our basis of functions is explicitly finite in the soft limit, featuring delicate cancellation of spurious rational poles by transcendental functions.  
Agreement between all three partonic channels, as well agreement of the maximal weight contributions with the super Yang-Mills Lipatov vertex provide robust checks of the result.
}

\begin{document}
\noindent

\maketitle
\flushbottom


\section{Introduction}

The Regge limit of QCD scattering, $s\gg -t$, has long inspired physicists, leading to profound theoretical developments. In the early days, important observations were made based on general scattering theory~\cite{Collins:1977jy}, where the asymptotic behaviour of amplitudes was studied in terms of their singularities in the complex angular momentum plane,
corresponding to Regge poles and Regge cuts. It emerged that high-energy asymptotic properties are naturally attributed to amplitude components with a definite signature, namely ones that are either even (symmetric) or odd (antisymmetric) under kinematic $s\leftrightarrow u$ interchange. The leading power behaviour of an amplitude of a given signature is then governed by the $t$-channel exchange of the highest spin particle.

After the birth of QCD, it soon emerged~\cite{Lipatov:1976zz} that the effective degree of freedom exchanged in the $t$-channel is a \emph{Reggeized gluon}, also dubbed \emph{Reggeon}. This object inherits its basic properties of odd signature and octet colour charge from the gluon, and is understood to govern the power-like growth of partonic  amplitudes as $(s/(-t))^{1+\alpha_g(t)}$ where $\alpha_g(t)$ is the gluon Regge trajectory~\cite{Fadin:1995km,Fadin:1996tb,Fadin:1995xg,Fadin:1995dd,Fadin:1998sh,Fadin:2015zea,Fadin:2006bj}. 
The concept of a Reggeon has subsequently played a key role in studying the Regge limit over the past half a century. In particular, compound states of two Reggeons are understood to dominate the high-energy behaviour of even-signature amplitudes and cross sections~\cite{Forshaw:1997dc}. In the framework of BFKL~\cite{Fadin:1975cb,Kuraev:1976ge,Kuraev:1977fs,Balitsky:1978ic,Fadin:1998py}, where unitarity is used to relate the forward limit of elastic amplitudes to multi-parton inelastic ones, interactions between Reggeons generate rapidity evolution. The same equation governs the small Bjorken-$x$ limit of parton density functions (see e.g.~\cite{Kovchegov:2012mbw}).

While questions such as the asymptotic behaviour of hadronic scattering cross sections are fundamentally non-perturbative, the Regge limit is rich and fascinating already at the perturbative level. 
In particular, one may address a broad range of questions pertaining to the structure of partonic amplitudes, their factorization and exponentiation properties, as well as their Regge singularities. In this context, the past decade has seen exciting progress due to the development of effective methods which directly describe the Regge limit: the multi-Reggeon effective theory~\cite{Caron-Huot:2013fea,Caron-Huot:2017fxr,Caron-Huot:2017zfo,Caron-Huot:2020grv,Caron-Huot:2020vlo,Falcioni:2020lvv,Falcioni:2021buo,Falcioni:2021dgr,Abreu:2024mpk,Buccioni:2024gzo} based on the shock-wave formalism (see also~\cite{Fadin:2017nka,Fadin:2023aen,Fadin:2024hbe} for an alternative approach)  and Glauber SCET~\cite{Rothstein:2016bsq,Moult:2022lfy,Gao:2024qsg,Gao:2024fyz}. These effective methods are synergic with the computation of multi-loop partonic scattering amplitude in general kinematics, where much progress has been made in recent years. The synergy also motivates the present study.

At leading power in the Regge limit, $s\gg -t$, partonic scattering amplitudes feature major simplifications already at tree level, including the dominance of $t$-channel gluon exchange and flavour and helicity conservation of the scattered partons. 
More profoundly, partonic amplitudes of odd signature feature all-order factorization and exponentiation of high-energy logarithms.
At leading (LL) and next-to-leading (NLL) logarithmic accuracy, these all-order properties can be understood directly as a manifestation of the fact that the amplitude is governed by the exchange of a single Reggeon, and hence features a Regge pole.
Starting at next-to-next-to-leading logarithmic accuracy (NNLL), this is no more the case: at this logarithmic order factorization violating terms appear~\cite{DelDuca:2001gu,DelDuca:2011wkl,DelDuca:2013ara,DelDuca:2014cya}. This  has been first observed at two loops~\cite{DelDuca:2001gu} by noting that a pure Regge-pole factorization ansatz would be inconsistent with explicit amplitude results of the three partonic channels ($qq$, $qg$ and $gg$).  
It is now understood~\cite{Caron-Huot:2017fxr,Fadin:2017nka,Falcioni:2020lvv,Falcioni:2021buo,Falcioni:2021dgr} that these factorization violations are generated by the $t$-channel exchange of multiple Reggons -- specifically three Reggeons and their mixing under evolution with a single Reggeon --  giving rise to Regge cuts.

Key to pushing our understanding of amplitudes in the Regge limit to NNLL accuracy has been the development of a theoretical framework for direct computation of multi-Reggeon exchange~\cite{Caron-Huot:2013fea,Caron-Huot:2017fxr,Caron-Huot:2017zfo,Caron-Huot:2020grv,Caron-Huot:2020vlo,Falcioni:2020lvv,Falcioni:2021buo,Falcioni:2021dgr,Abreu:2024mpk,Buccioni:2024gzo}. This framework is based on 
the shock-wave formalism in which the scattered projectile is described by a set of infinite lightlike Wilson lines $U(z_{\perp})$ in the background of the target. In this formalism, correlators of products of Wilson lines admit the Balitsky-JIMWLK ~\cite{Balitsky:1995ub,
Jalilian-Marian:1996mkd,JalilianMarian:1996xn,JalilianMarian:1997gr} rapidity evolution equation, which provides a non-linear generalization of BFKL. While the original goal of setting up this formalism was to describe high gluon density saturation effects at cross-section level (see e.g.~\cite{Kovchegov:2012mbw}), Simon Caron-Huot~\cite{Caron-Huot:2013fea} has demonstrated its applicability to partonic amplitudes in the dilute, perturbative regime. Essential to this was to trade the use of correlators of Wilson lines $U(z_{\perp}) = e^{ig_s {\bf T}^a W^a(z_{\perp})}$, where $U(z_{\perp})\simeq 1$ in said regime,  for those of their exponent $W^a(z_{\perp})$, and to interpret the $W$ fields as individual Reggeons. By construction, $W$ fields have odd signature and carry octet colour charge, consistent with Reggeons. More profoundly, their interaction in the $2-2\epsilon$ transverse space, which generates rapidity evolution according to the Balitsky-JIMWLK equation (solved order by order in perturbation theory), matches partonic QCD amplitudes in $4-2\epsilon$ dimensions in the high-energy limit.  

The interpretation of $W$ as the Reggeon field has far-reaching consequences~\cite{Caron-Huot:2013fea,Caron-Huot:2017fxr,Caron-Huot:2017zfo,Caron-Huot:2020grv,Caron-Huot:2020vlo,Falcioni:2020lvv,Falcioni:2021buo,Falcioni:2021dgr,Abreu:2024mpk,Buccioni:2024gzo}. First, conceptually, it decouples the two-dimensional transverse space dynamics from that involving the collision energy, realising Lipatov's ambitious goal of~\cite{Lipatov:1995pn}, at least in part. Second, it provides an efficient\footnote{These computations are done directly in the $2-2\epsilon$ transverse space, and hence require no rapidity cutoffs. Furthermore, they typically involve planar integrals, even when the colour factors are non-planar.} and practical framework to perform computations of the multi-Reggeon components of partonic QCD amplitudes -- precisely these components which break Regge-pole factorization and give rise to Regge cuts.       

With these multi-Reggeon predictions at hand, a precise separation between Regge pole and Regge cut contributions to amplitudes is within reach. It requires, however, one more essential element~\cite{Falcioni:2021dgr}, namely the special nature of the planar theory. Specifically, one uses the fact that in the planar limit, four and five-point amplitudes only have Regge poles.\footnote{Focusing here on $2\to 3$ amplitudes, Regge cuts do not feature in the planar limit. For higher-point amplitudes, $n\geq 6$,
Regge cuts appear in certain kinematic regions even in the planar limit, see e.g.~\cite{Bartels:2008ce,Bartels:2009vkz,Lipatov:2009nt,Dixon:2014voa,DelDuca:2019tur,Bartels:2021thv}. For the same reason, the Bern-Dixon-Smirnov ansatz~\cite{Bern:2005iz} describing the exponentiation of planar amplitudes in super Yang-Mills (sYM) theory, is exact for four and five point amplitudes, while it receives non-trivial corrections (the so-called remainder function) for higher point amplitudes. } Indeed, it has been shown through explicit computations through four loops~\cite{Caron-Huot:2017fxr,Caron-Huot:2017zfo,Caron-Huot:2020grv,Caron-Huot:2020vlo,Falcioni:2020lvv,Falcioni:2021buo,Falcioni:2021dgr} that while multi-Reggeon interactions ($t$-channel exchange of multiple $W$ fields) contain contributions that are leading in the large-$N_c$ limit, the latter are universal: they do not depend on the partonic process considered, and hence do not lead to Regge-pole factorization violations.  This stands in sharp contrast to contributions that are subleading in the large-$N_c$ limit, which do differ between partonic processes (starting at two loops) and necessarily generate factorization violation. It is thus only the latter, namely non-planar multi-Reggeon contributions, which generate a Regge cut. 

These theoretical developments allowed~\cite{Falcioni:2021buo,Falcioni:2021dgr,Caola:2021izf,Caola:2021rqz} to disentangle between Regge pole and Regge cut contributions in $2\to 2$ scattering, and hence uniquely determine the Regge pole parameters, the gluon Regge trajectory and the impact factors, from three-loop results of $2\to 2$ QCD amplitudes for $qq$, $qg$ and $gg$ scattering~\cite{Caola:2022dfa,Caola:2021rqz,Caola:2021izf}. With these parameters fixed, NNLL corrections associated with the Regge pole have been fixed to all orders in perturbation theory, while the corresponding Regge cut corrections have been shown to be non-planar to all orders, and have been explicitly determined to four loops. 

The aim of the present paper is to take a step towards extending these results to $n$-point scattering amplitudes in the so-called Multi-Regge Kinematic (MRK) limit~\cite{Lipatov:1976zz,Bartels:1978fc,Bartels:1980pe,Fadin:1993wh,DelDuca:1995hf}, where the final-state particles are all strongly ordered in rapidity, while their transverse momenta are unconstrained. 
It has been established~\cite{Fadin:2006bj} through NLL accuracy that the dispersive part of any $n$-point $2\to n-2$ amplitude admits a similar factorization structure in MRK to that of $2\to 2$ amplitudes in the Regge limit, with the same universal impact factors and gluon Regge trajectory,  with one additional ingredient, namely the vertex by which a real gluon is emitted from the Reggeon. This Reggeon-gluon-Reggeon vertex~\cite{Lipatov:1976zz}, referred to as the Lipatov vertex, is a key element in BFKL theory, which is needed in particular to determine the BFKL kernel. 
The Lipatov vertex appears for the first time in five-point amplitudes, providing the natural setting to extract it.
This quantity was determined in the tree approximation in~\cite{Lipatov:1976zz}, and at one loop in~\cite{Fadin:1993wh,Fadin:1994fj,Fadin:1996yv,DelDuca:1998cx,DelDuca:2009ac,DelDuca:2009ae,Fadin:2023roz}. In this paper we determine it at two loops.\footnote{An independent computation by another group has recently appeared, see ref.~\cite{Buccioni:2024gzo}.}      

Given the recent completion of the calculation of 
all $2\to 3$ amplitudes in QCD~\cite{Abreu:2018zmy,Abreu:2019odu,Abreu:2021oya,Agarwal:2023suw,DeLaurentis:2023nss,DeLaurentis:2023izi,DeLaurentis:2024arp}, and given that the impact factor and gluon Regge trajectory have already been fixed based on $2\to 2$ amplitudes~\cite{Falcioni:2021buo,Falcioni:2021dgr,Caola:2022dfa,Caola:2021rqz,Caola:2021izf}, one might naively expect that the determination of the two-loop vertex would be straightforward. 
However, this is not so. Similarly to the case of $2\to 2$ scattering, at two loops the relevant (odd-odd signature, octet) component of the amplitude, starts to receive contributions from multiple Reggeon exchange, giving rise to Regge cuts in the non-planar theory. Thus, in order to isolate the single Reggeon contribution, which factorizes in the MRK limit at NNLL accuracy, one needs once again to separately determine the non-factorizing non-planar (and process-dependent) multi-Reggeon contributions and set them aside before using the usual Regge-pole factorization formula (which is fully governed by a Regge pole) to extract the Lipatov vertex. 
To this end we first need to set up the shockwave formalism in the context of $2\to 3$ scattering (for previous work in this direction, see~\cite{Caron-Huot:2013fea,Caron-Huot:2020vlo}), and then compute all multi-Reggeon contributions to the relevant component of the amplitude. The final results of this computation have been reported in~\cite{Abreu:2024mpk} (see also \cite{Buccioni:2024gzo}); their derivation will be presented here, before turning to extract the two-loop vertex from the $gg\to ggg$, $gq\to ggq$ and $qq\to qgq$ amplitudes in MRK. 

The structure of the paper is as follows. In section~\ref{sec:preliminaries} we define the MRK limit and set up our notations for $2\to 3$ massless scattering momenta, helicities, colour and signature. In section~\ref{MRK-QCD} we briefly explain how we compute the two-loop QCD $gg\to ggg$, $gq\to ggq$ and $qq\to qgq$ amplitudes in MRK. 
In section~\ref{sec:Reggeization} we discuss the factorization structure in $2\to 3$ scattering, define the Lipatov vertex, and analyze its symmetries and analytic properties based on general considerations. Next, in section~\ref{sec:MatchingNLL} we determine the QCD Lipatov vertex at one loop through ${\cal O}(\epsilon^4)$ and discuss its properties. At this point, in section~\ref{MRK-shockwave}, we turn to discuss the theory of multi-Reggeon interactions within the shock-wave formalism. This theory is then used in section~\ref{sec:MRE} to compute all multi-Reggeon contributions to $2\to 3$ amplitudes at one loop, and the odd-odd component at two loops, in order to facilitate the extraction of the two-loop Lipatov vertex in section~\ref{sec:scheme}. The result is presented in section~\ref{VertTwoLoopStruct}. We also provide the expressions for the MRK amplitude and the vertex components in ancillary files~\cite{Ancillary}.


\section{\texorpdfstring{$2\to 3$}{2->3} amplitudes in multi-Regge kinematics}
\label{sec:preliminaries}


\subsection{Kinematics} \label{sec:kinematics}

\begin{figure}[t]
\begin{center}
 \includegraphics[width=0.52\textwidth]{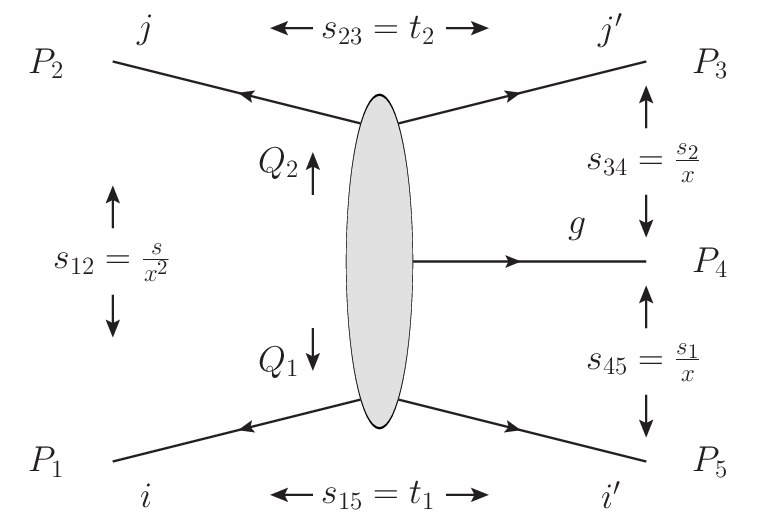}
\end{center}
\caption{Momentum label for $2 \to 3$ scattering 
processes. Capital letters are used to indicate 
4-vector momenta.}
\label{2to3MomentumLabels}
\end{figure}
We consider $2\to 3$ scattering amplitudes 
in any massless gauge theory in multi-Regge 
kinematics (MRK), as depicted in 
figure~\ref{2to3MomentumLabels}. 
The amplitude $\mathcal{M}_{ij\to i'gj'}$ 
corresponds to a scattering process 
$ij\to i'gj'$ where $g$ is an emitted gluon 
while $i$ ($i'$) and $j$  ($j'$) may each be 
a quark ($q$) or a gluon ($g$). We will often 
drop the subscript when we do not need to 
specify the process. We label four-vector 
momenta in terms of capital letters 
$P_i$, and take all momenta to be 
outgoing: $\sum_i P_i = 0$. With this 
notation the scattering process in 
figure~\ref{2to3MomentumLabels} reads 
\begin{align}
\label{kin_def}
i\big[(-P_1)^{-\lambda_1},a_1\big] + j\big[(-P_2)^{-\lambda_2},a_2\big] 
\to j'\big[P_3^{\lambda_3},a_3\big] + g\big[P_4^{\lambda_4},a_4\big] 
+ i'\big[P_5^{\lambda_5},a_5\big],
\end{align}
where $P_i^2=0$, $a_i$ is the colour index 
of parton $i$ and the superscript $\lambda_i 
= \oplus$ or $\ominus$ labels the helicity of 
particle $i$. Using momentum 
conservation, the amplitudes in generic 
kinematics depend on five independent 
Mandelstam invariants. 
Defining $s_{kl} = (P_k + P_l)^2$, the 
physical scattering region is defined by
\begin{align}\label{MandelstamDef}
s_{12} > 0 ,  
\qquad s_{34} > 0 , \qquad s_{45} > 0 , 
\qquad s_{23} < 0 , \qquad s_{15} < 0 , 
\qquad \Delta < 0,
\end{align}
where $\Delta = \text{det}(s_{kl})_{k,l=1\dots 4}$
is the Gram determinant of the external 
momenta and is also related to the 
pseudoscalar invariant
\begin{equation}\label{tr5}
{\rm tr}_5\equiv -4\,i\,\epsilon_{\mu\nu\sigma\rho} 
P_1^{\mu}P_2^{\nu}P_3^{\sigma}P_5^{\rho}
\end{equation}
through $\Delta = ({\rm tr}_5)^2$, and where the convention for the Levi-Civita tensor $\epsilon_{\mu\nu\sigma\rho}$ is $\epsilon_{0123} = +1$. It implies 
that ${\rm tr}_5$ is purely imaginary in the 
physical region, and while the absolute value 
of the imaginary part is dependent on $s_{ij}$, 
the sign of ${\rm tr}_5$ is required to fully 
specify a phase-space point.

In the centre of mass frame $P_1$ and $P_2$ are 
back to back, and we choose coordinates such that
\begin{equation} \label{p1p2def1}
P_1 = \frac{\sqrt{s_{12}}}{2}{(1,0,0,-1)},  
\qquad \qquad 
P_2 = \frac{\sqrt{s_{12}}}{2}{(1,0,0,1)}.
\end{equation}
In this way we define two light-like directions 
along which a generic momentum $Q$ can be 
decomposed as follows: 
\begin{equation}\label{Sudakov1} 
Q = (q^+ , q^- ; \mathbf{q}), 
\qquad\qquad \mbox{where} \qquad 
q^{\pm} = q^0\pm q^3.
\end{equation}
Therefore $P_1$ ($P_2$) is solely in the 
$-$ ($+$) direction. The transverse momentum 
$\mathbf{q}$ can be expressed in terms of the 
complex-valued coordinates $\mathbf{q} = q^1 + i\,q^2$ 
with ${\mathbf{\bar{q}}} = q^1 - i\,q^2$ and then 
$Q^2= q^{+}q^{-}-\bar{\mathbf{q}}\mathbf{q}$.

The multi-Regge kinematic (MRK) limit is defined 
by strong ordering in rapidity, which is alternatively 
described by the conditions
\begin{equation}\label{hel_def0} 
|p_{3}^{+}| \gg |p_{4}^{+}| \gg |p_{5}^{+}|, 
\qquad 
|p_{3}^{-}| \ll |p_{4}^{-}| \ll |p_{5}^{-}|, 
\qquad 
|\mathbf{p_3}| \simeq |\mathbf{p_4}| 
\simeq |\mathbf{p_5}|. 
\end{equation}
The limit can also be described by defining 
an auxiliary parameter $x$ which relates a point
in generic kinematics to a point in the MRK limit.
This can be done through
\begin{align} \label{kinhel1}
s_{12} = \frac{s}{x^2}\,, \quad
s_{15} = t_1\,, \quad
s_{45} = \frac{s_1}{x}\,, \quad
s_{23} = t_2,   \quad
s_{34} = \frac{s_2}{x}\,, 
\end{align}
and then considering the limit $x\to 0$.
We will also find it convenient to introduce the
variables $z$ and $\bar z$, defined through
\begin{equation}\label{t12}
    t_1 = -(1-z)(1-\bar z)\frac{s_1s_2}{s}\,,\qquad
    t_2 = -z \bar z\frac{s_1s_2}{s}\,.
\end{equation}
The variables $\{s, s_1, s_2, t_1, t_2\}$,
or equivalently $\{s, s_1, s_2, z, \bar z\}$,
describe a point in generic kinematics, which
can be mapped to a point in the MRK limit
by taking $x\to0$ in eq.~\eqref{kinhel1}.
In this limit, the invariant ${\rm tr}_5$ 
in eq.~(\ref{tr5}) becomes
\begin{equation}
\label{tr5MRK}
{\rm tr}_5 \simeq \frac{s_1s_2}{x^2}(z-\bar z)\,,
\end{equation} 
where we use $\simeq$ to denote equalities that are 
valid at leading order in $x\to 0$.
As mentioned after eq.~(\ref{tr5}),
${\rm tr}_5$ is purely imaginary in the physical 
scattering region, which is consistent with
taking $\bar z = z^{*}$.
Furthermore, because the definition of 
eq.~\eqref{t12} is symmetric under 
$z \leftrightarrow \bar{z}$, there remains a residual 
ambiguity in distinguishing which solution to the quadratic 
is $z$ and which is $\bar{z}$.
Eq.~\eqref{tr5MRK} effectively removes this 
ambiguity.

In the MRK limit, we can relate
the transverse components of the various final
state momenta using the $z$ and $\bar z$ variables.
Explicitly, at leading order in $x\to0$, we have:
\begin{equation}\label{def:zzbar}
\mathbf{p}_{3} \simeq -z\,\mathbf{p}_{4}\,, 
\quad 
\mathbf{\bar p}_{3} \simeq -\bar z\,
\mathbf{\bar{p}}_{4}\,,
\quad 
\mathbf{p}_{5} \simeq -(1-z) \,\mathbf{p}_{4}\,, 
\quad 
\mathbf{\bar p}_{5} \simeq -(1-\bar z) \,\mathbf{\bar{ p}}_{4}\,.
\end{equation}
Throughout this paper, we will by default choose a frame
where $\mathbf{p}_{4}$ only has a real component, $\mathbf{\bar{p}}_{4} =\mathbf{p}_{4}$.
Momentum conservation in the transverse plane then
implies that
\be\label{momentumconservation1} 
\mathbf{p}_{4} = 
- \mathbf{p}_{3} - \mathbf{p}_{5}
\simeq - \sqrt{\frac{s_1 s_2}{s}}\,, \qquad
t_1 \simeq -|\mathbf{p}_5|^2\,,\qquad
t_2 \simeq - |\mathbf{p}_3|^2\,.
\ee
Note that eq.~(\ref{def:zzbar}) is consistent with
$\bar z = z^{*}$, as also argued below eq.~\eqref{tr5MRK}.

In what follows it will be convenient 
to introduce a notation also for the 
momenta exchanged in the $t$-channel, 
that we define as follows:
\be\label{q1q2defA} 
Q_i \equiv Q_1 = P_1 + P_5, 
\qquad 
Q_j \equiv Q_2 = P_2 + P_3,
\ee
such that
\beq\label{q1q2defB} 
{\bf q}_1 = {\bf p}_5,
\qquad 
{\bf q}_2 = {\bf p}_3, 
\qquad \qquad
Q_1^2 = t_1, \qquad Q_2^2 = t_2.
\eeq

It can be shown that the tree-level amplitude 
${\cal M}_{ij \to i'gj'}^{\rm tree}$ in MRK 
assumes a factorized structure 
\cite{Fadin:1993wh,DelDuca:1995zy}
\begin{align}
\label{Mtree} 
\begin{split}
{\cal M}_{ij \to i'gj'}^{\rm tree} &= 2s 
 \Big[g_s \,\delta_{-\lambda_5, \lambda_1} 
{\bf T}_{a_5a_1}^x \, 
C_{i}^{(0)}(P_1^{\lambda_1},P_5^{\lambda_5}) \Big]
\, \frac{1}{t_1} \\
&\hspace{0.0cm}\times\,
\Big[g_s {\bf T}_{yx}^{a_4}\, 
V^{(0)}_\mu(Q_1,P_4,Q_2)\,\varepsilon^\mu_{\lambda_4}(P_4) \Big] 
\, \frac{1}{t_2} \,
\Big[g_s \, \delta_{-\lambda_3, \lambda_2} 
{\bf T}_{a_3a_2}^y \,
C_{j}^{(0)}(P_2^{\lambda_2},P_3^{\lambda_3}) \Big].
\end{split}
\end{align}

In this equation, $C_{i/j}^{(0)}$ represent
tree-level impact factors, describing the 
coupling between the external states $i/j$ 
and a gluon exchanged in the 
$t$-channel. In appendix~\ref{Tree5g}, we provide their expression for the gluon amplitude. 
The factors 
$\delta_{-\lambda_3, \lambda_2}$, 
$\delta_{-\lambda_5, \lambda_1}$ 
represent helicity conservation along the lines 
$i$ and $j$, and ${\bf T}_{a_ia_{i'}}^x$ are the 
color generator associated to incoming/outgoing
particles $a_i$, $a_{i'}$: 
${\bf T}_{a_ia_{i'}}^x = (T_F^x)_{a_ia_{i'}}$
for quarks, 
${\bf T}_{a_ia_{i'}}^x = -(T_F^x)_{a_{i'}a_i}$
for anti-quarks, 
${\bf T}_{a_ia_{i'}}^x = i\,f^{a_i x a_{i'}}$
for gluons. 

The vector $V^{(0)}_\mu(Q_1,P_4,Q_2)$ is the Lipatov vertex 
at tree level, which accounts for the emission 
of a gluon at central rapidity \cite{Lipatov:1976zz}:
\begin{align}
\label{eq:treeLipatov}
V^{(0)}_\mu(Q_1,P_4,Q_2)&= {\cal P}_\mu+{\cal Q}_\mu,
\end{align}
where ${\cal P}^\mu$ and ${\cal Q}^\mu$ are defined by
\begin{align}
\label{eq:treeQPtens}
\begin{split} %
    {\cal Q}^\mu&=(Q_1-Q_2)_{\perp}^{\mu} + 2P_1^\mu\frac{|\mathbf{q_1}|^2}{s_{45}} - 2P_2^\mu\frac{|\mathbf{q_2}|^2}{s_{34}},\\
    {\cal P}^\mu&=-|\mathbf{p_4}|^2\left(\frac{P_1^\mu}{s_{45}}-\frac{P_2^\mu}{s_{34}}\right),
\end{split}
\end{align}
with $(Q_i)_{\perp}=(0,0;\mathbf{q_i})$.
Note that ${\cal P}^\mu$ and ${\cal Q}^\mu$ are separately gauge invariant, as they vanish upon contraction with ${P_4}_\mu$. These two also furnish the complete set of tensors that may appear in the vertex $V_\mu(Q_1,P_4,Q_2)$ at higher orders.
For definiteness, we give the expression of the gluon polarisation $\varepsilon^\mu_\lambda(P_4)$ 
in the light-cone 
gauge $\varepsilon^-(P_4) = 0$, the left gauge (see~(\ref{PolVecLeftOurPL})), which will be our default choice of gauge in this paper,
\beqa \label{PolVector4DefA}
\varepsilon^{\mu}_{\oplus}(P_4) &=& \bigg( 
\frac{\mathbf{p}_4}{\sqrt{2} \,p_4^-}, 
\frac{1}{\sqrt{2}}, \frac{i}{\sqrt{2}}, 
\frac{\mathbf{p}_4}{\sqrt{2} \,p_4^-} \bigg)\,,\\
\varepsilon^{\mu}_{\ominus}(P_4) &=& \bigg( 
\frac{\mathbf{p}_4}{\sqrt{2} \,p_4^-}, 
\frac{1}{\sqrt{2}}, -\frac{i}{\sqrt{2}}, 
\frac{\mathbf{p}_4}{\sqrt{2} \,p_4^-} \bigg)\,.
\eeqa
It is convenient to express the polarisation 
vector in lightcone components, using the 
notation of \eqn{Sudakov1},
\beq \label{PolVector4DefB}
\varepsilon^{\mu}_{\oplus}(P_4) = 
\bigg( \frac{\sqrt{2}\, \mathbf{p}_4}{p_4^-},0; 
\boldsymbol{\varepsilon}_{\oplus}(P_4)\bigg), 
\qquad {\rm with } \qquad 
\boldsymbol{\varepsilon}_{\oplus}(P_4) = 
\bigg(\frac{1}{\sqrt{2}}, \frac{i}{\sqrt{2}}\bigg).
\eeq
With this notation one has 
\begin{align}\label{polvectprod}
{\boldsymbol{\varepsilon}}_{\oplus}(P_4)
\cdot {\mathbf p}_k = \, 
\frac{1}{\sqrt{2}}\begin{cases}
\mathbf{p}_k\, ,\,\, k\in\{3,4,5\}\,, \\
0 \, , \,\, k\in\{1,2\}\,.
\end{cases}
\end{align}

In what follows we consider the helicity 
configuration $(\ominus,\ominus,\oplus,
\oplus,\oplus)$. By contracting the Lipatov 
vertex, eq. (\ref{eq:treeLipatov}) with the 
polarisation vector above, we get the scalar 
vertex
\be\label{LipatovTree}
V^{(0)}(Q_1,P_4^{\oplus},Q_2) \equiv 
V^{(0)}_\mu(Q_1,P_4,Q_2)\varepsilon^\mu_{\oplus}(P_4) =
\sqrt{2} \frac{\mathbf{\bar q}_2 \mathbf{q}_1}{\mathbf{\bar p}_4}
= \sqrt{2} \frac{\mathbf{\bar p}_3\mathbf{p}_5}{\mathbf{\bar p}_4}\,,
\ee
such that the tree level amplitude takes the form
\begin{align} \label{Mtree2}
{\cal M}_{ij \to i'gj'}^{\rm tree}
=  C_{i}^{(0)}(P_1^{\ominus},P_5^{\oplus})\,
\mathcal{K}^{(0)}\,\mathcal{C}_{ij}^{(0)}
\, C_{j}^{(0)}(P_2^{\ominus},P_3^{\oplus}),
\end{align}
where for future use we have introduced 
the kinematic and colour factors
\begin{align}
\label{def:Kij}
\mathcal{K}^{(0)} &= 2\sqrt{2} \, g_s^3 \,
\frac{s_{12}}{\mathbf{p_3}\mathbf{\bar p_4}\mathbf{\bar p_5}}, \\
\label{def:Cij}
\mathcal{C}_{ij}^{(0)} &= \,
i f^{ya_4x}\, (\T_i^x)_{a_5a_1}(\T_j^y)_{a_3a_2}.
\end{align}


\subsection{Colour}\label{sec:colour}

The tree amplitude in \eqn{Mtree} is 
the leading term in the coupling expansion  
\begin{align}
\mathcal{M}_{ij \to i'gj'} = \sum_{n=0}^\infty 
\left(\frac{\al_s}{\pi}\right)^n\,\mathcal{M}^{(n)},
\end{align}
i.e. $\mathcal{M}^{(0)} = {\cal M}_{ij \to i'gj'}^{\rm tree}$. 
In general, the coefficients $\mathcal{M}^{(n)}$
develop a non-trivial color structure for $n\geq 1$. 
This is conveniently described in the colour space 
formalism \cite{Bassetto:1983mvz,Catani:1996jh,Catani:1996vz}. 
We begin by expanding the amplitude coefficients 
on a basis of colour tensors
\begin{equation}\label{coloramplitudes}
{\cal M}^{(n)} = \sum_i c^{[i]}_{a_1\dots a_5} {\cal M}^{(n),[i]},
\end{equation}
where the sum includes all linearly independent 
tensors $c^{[i]}_{a_1\dots a_5}$ with indices 
$a_1\dots a_5$ in the representation of the five 
external particles. Denoting the colour operator 
in the representation of the $k$-th parton as 
$\T_k$, we have
\begin{equation}
\label{eq:color_generator}
\T_k^b\,c^{[i]}_{a_1\dots a_5} = 
(\T^b_k)_{a_k\,a_k'}\,c^{[i]}_{a_1\dots a_k'\dots a_5},
\end{equation}
for $k=1\dots 5$, with
\begin{equation}\label{eq:generatorDef}
(\T^b_k)_{a\,a'} = \left\{
\begin{array}{lcl}
    (T_F^b)_{a\,a'} &  & \text{for }k=\text{quark,} \\
    -(T_F^b)_{a'\,a}&  & \text{for }k=\text{antiquark,} \\
    i\,f^{aba'}     &  & \text{for }k=\text{gluon,}
\end{array}
\right.
\end{equation}
where $T_F$ are the SU($N_c$) generators 
in the fundamental representation and $f$ 
are the structure constants. In this language, 
colour conservation is guaranteed by the condition
\begin{equation}
\sum_{k=1}^5 (\T^b_k)\,c^{[i]}_{a_1\dots a_5} = 0, 
\qquad \forall c^{[i]}.
\end{equation}
Following from eq.~(\ref{eq:color_generator}) it 
is clear that generators associated with different 
particles trivially commute, so one has
\begin{equation}
\label{eq:commTaTb}
[\T^a_k,\T^b_l] = i\,f^{abc}\,\T^c_k\,\delta_{kl}    
\end{equation}
and
\begin{equation}
(\T^b_k\,\T^b_k) \equiv \T_k\cdot\T_k = C_k \, {\bf 1},
\end{equation}
where $C_k$ is the quadratic Casimir operator 
of the corresponding colour representation, 
i.e.~$C_q = C_F=(N_c^2-1)/(2N_c)$ for quarks 
and~$C_g = C_A = N_c$ for gluons.

In the Multi-Regge limit, it is 
natural to use a $t$-channel colour 
basis~\cite{DelDuca:2011ae,DelDuca:2011wkl}, 
which diagonalises the operators
\bea
\label{def:Tt1}\T_{t_1}^2 &\equiv& (\T_1 + \T_5)^2 
= 2\, \T_1 \cdot \T_5 + C_1\, {\bf 1} + C_5\, {\bf 1}\,,\\
\label{def:Tt2}\T_{t_2}^2 &\equiv& (\T_2 + \T_3)^2 
= 2\, \T_2 \cdot \T_3 + C_2\, {\bf 1} + C_3\, {\bf 1}\,.
\eea
Since $\T_{t_1}^2$ and $\T_{t_2}^2$ are Hermitian and, 
by eq.~(\ref{eq:commTaTb}), $[\T_{t_1}^2,\T_{t_2}^2] = 0$, 
we choose the tensors $c^{[i]}_{a_1\dots a_5}$ such that 
{\textit{both}} $t$-channel operators are simultaneously 
diagonal \cite{DelDuca:2011ae}. Therefore, the basis 
elements are identified by the eigenvalues of $\T_{t_1}^2$ 
and $\T_{t_2}^2$, which measure the colour charge exchanged 
in the $t$-channel. We modify the notation of 
eq.~(\ref{coloramplitudes}) accordingly
\begin{equation}
\label{t-channel_basis-colour_decomposition}
{\cal M}^{(n)} = 
\sum_{(r_1,r_2),r} c^{[r_1,r_2]_r}_{a_1\dots a_5} 
{\cal M}^{(n),[r_1,r_2]_r},
\end{equation}
where $r_1$ and $r_2$ label the pairs of irreducible 
representations of SU($N_c$) that can flow in the 
$t$-channel and an additional subscripts $r$ is 
introduced to distinguish degenerate cases. 
The bases for $qq\to qqg$, $qg\to qgg$ and 
$gg\to ggg$ scattering are given in 
Appendix~\ref{sec:colourBasis}, where we 
also provide the eigenvalues of $\T_{t_1}^2$ 
and $\T_{t_2}^2$.

In what follows it will be useful to introduce 
also a colour operator associated to the $s$-channel, 
as done in Ref. \cite{DelDuca:2011ae}
\begin{equation}
\label{def:Ts}
\T_s^2=(\T_1+\T_2)^2 = 2\,\T_1\cdot\T_2 + C_1\, {\bf 1}  + C_2\, {\bf 1}.
\end{equation}
The matrix representations of the operator 
$\T_s$ are constructed using the color basis 
in appendix~\ref{sec:colourBasis}. These matrices are non-diagonal and provided in an ancillary file~\cite{Ancillary}. 


\subsection{Signature}\label{sec:signature}

Another important feature that characterises 
the structure of amplitudes in the high energy 
limit is the so-called signature symmetry. 
It is well known that in case of $2\to 2$ 
scattering, the signature symmetry, i.e. 
the way the amplitude transforms under 
the exchange of the kinematic variables 
$s \leftrightarrow u$, plays a relevant 
role in determining the analytic structure 
of the amplitude itself. In order to derive 
the generalization of the signature 
symmetry to $2\to 3$ scattering, 
we start by invoking momentum 
conservation, which allows us to 
determine all kinematic invariants 
in terms of the parameterization 
in \eqn{kinhel1}: considering the 
leading terms in the small $x$ limit, 
one has
\begin{align} \label{kinhel2}
 & s_{12} = \frac{s}{x^2}, \quad
   s_{45} = \frac{s_1}{x}, \quad
   s_{34} = \frac{s_2}{x}, \quad
   s_{15} = t_1, \quad
   s_{23} = t_2, \quad
   s_{35} = \frac{s}{x^2},
\nonumber \\[0.1cm]
&  s_{14} = - \frac{s_1}{x}, \quad
   s_{24} = - \frac{s_2}{x}, \quad
   s_{13} = -\frac{s}{x^2},  \quad
   s_{25} = - \frac{s}{x^2},
\end{align}
which immediately reveals that the natural 
generalization of signature symmetry to 
the $2\to 3$ case is realised via two 
independent permutations:
\begin{subequations}
\label{SigPerm}
\begin{align}
(1\leftrightarrow 5) &  \qquad \qquad 
\rightarrow   \qquad \qquad 
\{s\rightarrow -s,\quad s_1\rightarrow -s_1\},  \\
(2\leftrightarrow 3) & \qquad \qquad 
\rightarrow  \qquad \qquad 
\{s\rightarrow -s,\quad s_2 \rightarrow -s_2\}.
\end{align}
\end{subequations}
As for the $2\to 2$ scattering, it 
proves useful to decompose the $2\to 3$
amplitude into even and odd components 
under the exchange in \eqn{SigPerm}: 
\bea\label{sum-signature-amplitudes}
{\cal M}_{ij \to i'gj'} =
{\cal M}^{(+,+)}_{ij \to i'gj'} 
+{\cal M}^{(+,-)}_{ij \to i'gj'} 
+{\cal M}^{(-,+)}_{ij \to i'gj'} 
+{\cal M}^{(-,-)}_{ij \to i'gj'},
\eea
where
\bea\label{signature-amplitudes} \nn
{\cal M}^{(\sigma_1,\sigma_2)}_{ij \to i'gj'} 
(P_1^{\lambda_1},
P_2^{\lambda_2},P_3^{\lambda_3},
P_4^{\lambda_4},P_5^{\lambda_5}) && \\[0.1cm]
&&\hspace{-5.8cm}=\, \frac{1}{4}\Big[
{\cal M}_{ij \to i'gj'}(P_1^{\lambda_1},
P_2^{\lambda_2},P_3^{\lambda_3},
P_4^{\lambda_4},P_5^{\lambda_5}) 
+ \sigma_{1}\,
{\cal M}_{ij \to i'gj'}(P_1^{\lambda_1},
P_3^{\lambda_3},P_2^{\lambda_2},
P_4^{\lambda_4},P_5^{\lambda_5}) \\ \nn
&&\hspace{-5.4cm}+\, \sigma_{2}\,
{\cal M}_{ij \to i'gj'}(P_5^{\lambda_5},
P_2^{\lambda_2},P_3^{\lambda_3},
P_4^{\lambda_4},P_1^{\lambda_1}) 
+ \sigma_{1}\sigma_{2}\,
{\cal M}_{ij \to i'gj'}(P_5^{\lambda_5},
P_3^{\lambda_3},P_2^{\lambda_2},
P_4^{\lambda_4},P_1^{\lambda_1}) \Big]. 
\eea
The two signature indices $\sigma_{1}$ and $\sigma_{2}$ 
can be separately $+1$ or $-1$, and determine the 
transformation properties of the partial amplitudes 
${\cal M}^{(\sigma_1,\sigma_2)}_{ij \to i'gj'}$
under the permutations in \eqn{SigPerm}. 
Comparing with \eqn{Mtree} one finds that the overall 
factor of $s$ implies ${\cal M}_{ij \to i'gj'}^{\rm tree}
= {\cal M}^{(-,-)}_{ij \to i'gj'}|_{ \rm tree}$. 
In case of gluon scattering, because of Bose 
symmetry, the upper and lower vertices acquire 
a corresponding colour symmetry: at tree 
level the upper vertex is antisymmetric 
upon interchanging the colour indices $a_2$ 
and $a_3$, while the lower vertex is 
antisymmetric upon interchanging $a_1$ 
and $a_5$. At higher orders, all amplitudes 
$(\sigma_1,\sigma_2)$ are nonzero: in case 
of the $gg \to ggg$ amplitude the upper 
vertex is thus symmetric/antisymmetric 
upon interchanging the colour indices 
$a_2$ and $a_3$, for $\sigma_1 = +1$ and 
$\sigma_1 = -1$, respectively. Similarly, 
the lower vertex is symmetric/antisymmetric 
upon interchanging the colour indices $a_1$ 
and $a_5$, for $\sigma_2 = +1$ and 
$\sigma_2 = -1$, respectively. Let us 
stress, however, that the correspondence 
between color symmetry and signature 
holds only for the all gluon amplitude,
because of Bose Symmetry; it does not 
hold for scattering amplitudes 
involving quarks.

The operators $\T^2_{t_1}$ and $\T^2_{t_2}$ introduced 
in Sec~\ref{sec:colour} commute with the operation 
of exchanging particles $2\leftrightarrow 3$ and 
$1\leftrightarrow 5$, thus they are {\textit{even}} 
under both exchanges and preserve the symmetry of 
the colour tensor on which they act.
The operator $\T_s^2$ instead is clearly not invariant 
under $1\leftrightarrow 5$ and $2\leftrightarrow 3$. 
Therefore, the operator $\T_s^2$ mixes colour structures 
with different signatures. As we will see, it proves 
useful to describe the amplitude in terms of colour 
operators with a definite symmetry under the exchange 
$1\leftrightarrow 5$ and $2\leftrightarrow 3$, matching 
the symmetry properties of the amplitudes
${\cal M}^{(\sigma_1,\sigma_2)}_{ij \to i'gj'}$. 
To this end let us define the set of operators 
\begin{align}\label{TpmDef}
\begin{split}
 &  \T_{(++)} =(\T_1^a+\T_5^a)\cdot(\T_2^a+\T_3^a):\,\,
 \text{signature-preserving operator on line $i,j$ ;}   \\   
 &  \T_{(+-)} =(\T_1^a+\T_5^a)\cdot(\T_2^a-\T_3^a):\,\, 
 \text{signature-preserving on line $i$, inverting on $j$ ;}\\ 
 &  \T_{(-+)} =(\T_1^a-\T_5^a)\cdot(\T_2^a+\T_3^a):\,\,
 \text{signature-preserving on line $j$, inverting on $i$ ;}\\ 
 &  \T_{(--)} =(\T_1^a-\T_5^a)\cdot(\T_2^a-\T_3^a):\,\,
 \text{signature-inverting operator on lines $i,j$.}
 \end{split}
\end{align}
Using the definition above we get
\begin{equation}\label{Ts2ToTpm}
\T_s^2 = \frac{1}{2}\,\Big[\T_{(++)}+\T_{(+-)}+\T_{(-+)}+\T_{(--)}\Big] 
+ C_1\, {\bf 1} + C_2\, {\bf 1}.
\end{equation}
In the forthcoming sections will see how colour 
operators with distinct signature symmetry, and 
corresponding colour structure $\T_{(\sigma_1\sigma_2)}$ 
arise. We begin in Sec.~\ref{sec:Reggeization} by 
reviewing the Reggeization property of the signature 
odd-odd amplitude ${\cal M}^{(-,-)}_{ij \to i'gj'}$.


\section{Taking the MRK limit of QCD amplitudes}
\label{MRK-QCD}

In order to compute the required amplitudes
in the MRK limit, we will start from
the amplitudes for 3-jet production at hadron colliders
in general kinematics, which are known up to two loops~\cite{Abreu:2018zmy,Abreu:2019odu,
Abreu:2021oya,DeLaurentis:2023nss,DeLaurentis:2023izi,Agarwal:2023suw}.
Depending on the physical channel and the number of loops,
the various QCD amplitudes are known in slightly different
formats. In all cases, however, they are written as a 
decomposition of the form 
\begin{equation}\label{eq:genDecomp}
    \mathcal{F}=\sum_k r_k\, f_k\,,
\end{equation}
where $\mathcal{F}$ might be either an amplitude
or a finite remainder (see eq.~\eqref{IRfacteq}),
the $r_k$ are rational coefficients,
and the $f_k$ are transcendental functions, 
such as Feynman integrals.
If $\mathcal{F}$ is an amplitude, then it also
depends on the dimensional regulator $\epsilon$.
In order to compute the MRK limit of the QCD amplitudes,
we must then compute the limit of both the
rational coefficients and the transcendental functions.

Our QCD amplitudes in the MRK limit will be constructed
out of the one- and two-loop remainders given in 
refs.~\cite{DeLaurentis:2023nss, DeLaurentis:2023izi},
and as such we focus our discussion on these
expressions.
However, the procedure can be easily adapted to other cases
(such as, e.g., the expressions for one-loop amplitudes given
in ref.~\cite{Abreu:2019odu}, which are written as a 
decomposition in terms of one-loop master integrals, 
and valid to all orders in $\epsilon$).
In refs.~\cite{DeLaurentis:2023nss, DeLaurentis:2023izi}, 
the one- and two-loop remainders 
$\mathcal{H}$ were expressed in the form
\begin{equation}\label{eq:remaindersGen}
    \mathcal{H} = \sum_{j,k} r_{k} \, M_{kj} \, h_j \, ,
\end{equation}
where the $r_{k}$ are rational coefficients of 
(contractions of) 
the spinors $\lambda_\alpha$ and $\lambda_{\dot\alpha}$ 
($\alpha, \dot\alpha \in \{1, \dots, 5\}$), $M_{kj}$ is a rectangular matrix of rational numbers, and the $h_j$ are 
a set of transcendental functions appearing in five-point
massless scattering up to two loops, 
known as pentagon functions \cite{Chicherin:2020oor}.


\subsection{Expansion of the transcendental functions}
\label{sec:expPFs}

The pentagon functions are functions of 
five Mandelstam variables $\{s_{12},s_{23},s_{34},
s_{45},s_{15}\}$ and of the (sign of the) 
pseudoscalar invariant ${\rm tr}_5$ defined
in eq.~\eqref{tr5}. 
This set of functions is sufficient to express
any master integral appearing in five-point massless
scattering at two loops~\cite{Chicherin:2020oor}.
A convenient way to expand them in the MRK limit
is to start from the differential equation they satisfy.
By definition, this set of functions is closed
under differentiation, and 
if one assembles all the pentagon functions
into a vector $\vec h$, one can write a system of 
first-order differential equations for the pentagon 
functions of the form
\begin{equation}\label{eq:DE_PF}
    \mathrm{d}\, \vec h=\epsilon\, M(x,y)\,\vec h\,,
    \qquad\quad
    M(x,y)=\sum_{k}m_k\,\mathrm{d}\log W_k(x,y)\,,
\end{equation}
where the $m_k$ are matrices of rational numbers,
and the $W_k$ are the 31 letters of the alphabet
for five-point massless scattering at two 
loops.\footnote{We thank Vasily Sotnikov
for discussions on this topic, and for providing us
with his own calculation of the differential equations 
satisfied by the pentagon functions.}
In our case, the vector $\vec h$ contains 1083 entries,
each one being a pure function of weight smaller or
equal to 4. In eq.~\eqref{eq:DE_PF}, the parameter
$\epsilon$ is strictly speaking not the dimensional
regulator (since, by definition, the pentagon functions
do not depend on the dimensional regulator), 
but rather a bookkeeping parameter that keeps track of the 
transcendental weight of each object.

The steps to expand the functions $\vec h$ in
the MRK limit starting from eq.~\eqref{eq:DE_PF}
are well known \cite{Asymptotic} (see also 
ref.~\cite{Caron-Huot:2020vlo} for an application 
in the context of the MRK limit of five-point 
amplitudes). We summarise them here for completeness. 
The first step is to rewrite the letters $W_k$ as functions
of $x$, the variable that controls the approach to 
the MRK limit, and $\{s,s_1,s_2,t_1,t_2\}$ or
$\{s,s_1,s_2,z,\bar{z}\}$ 
(see eq.~\eqref{kinhel1} and \eqref{t12}), which
for simplicity we collectively denote as $y$ in what follows.
A point in generic kinematics corresponds to $(x=1,y)$
and a point in the MRK limit corresponds to $(x=0,y)$.
From eq.~\eqref{eq:DE_PF}, we can construct two differential
equations that allow us to move in the $x$ direction and
in the $y$ space, namely
\begin{align}\label{eq:DE_PF_dec}
    \frac{d \vec h}{d x}=\epsilon\, M_x(x,y)\, \vec h\,,\qquad
    \frac{d \vec h}{d y}=\epsilon\, M_y(x,y)\, \vec h\,,
\end{align}
with 
\begin{equation}
    M_\kappa(x,y)=\sum_j m_j \frac{d\log W_j(x,y)}{d\kappa}\,.
\end{equation}
Given that the $\vec h$ are pure functions, they
have at most logarithmic divergences as $x\to0$,
which are generated by single poles at $x=0$ in 
$M_x$. That is,
\begin{equation}
    M_x(x,y) =\frac{M^{(-1)}_x}{x}+
    \sum_{k=0}M^{(k)}_x(y)\,x^k\,,
\end{equation}
where $M^{(-1)}_x$ is a matrix of rational numbers.
In contrast, the matrix $M_y(x,y)$ is regular
at $x=0$.

The solution to eq.~\eqref{eq:DE_PF} can be
written 
as \cite{Caron-Huot:2020vlo,Asymptotic}
\begin{equation}\label{eq:solAllX}
    \vec h(x,y,\epsilon)=
    T(x,y,\epsilon)\,
    \vec{\mathfrak{h}}(x,y,\epsilon)
\end{equation}
where 
\begin{equation}\label{eq:leadingOrderPF}
    \vec{\mathfrak{h}}(x,y,\epsilon)
    =x^{\epsilon M^{(-1)}_x}
    \mathbb{P}\exp\left(
    \epsilon\int_{y_0}^y M_y(0,v)dv
    \right)
    \vec{\mathfrak{h}}_0(\epsilon)
\end{equation}
is a solution to the simpler system of equations
\begin{align}
    \frac{d \vec{\mathfrak{h}}}{d x}=\epsilon\, \frac{M^{(-1)}_x}{x}\, \vec{\mathfrak{h}}\,,\qquad
    \frac{d \vec{\mathfrak{h}}}{d y}=\epsilon\, M_y(0,y)\, \vec{\mathfrak{h}}\,\,,
\end{align}
and it captures the non-analytic behaviour for $x\to0$\,.
The matrix $T(x,y,\epsilon)$ can be expanded in powers
of $x$ and $\epsilon$, and be chosen to be the
identity matrix at $x=0$. We then write
\begin{equation}\label{eq:TExp}
    T(x,y,\epsilon)=\mathbb{I}
    +\sum_{k,j\ge1}x^k\,\epsilon^j\,T_{k,j}(y)
\end{equation}
and the coefficients $T_{k,j}(y)$ can be
recursively computed with
\begin{align}\begin{split}\label{eq:TCoeffs}
    T_{k,1}(y)&=\frac{M_x^{(k)}(y)}{k}\,,\\
    T_{k,j}(y)&=\frac{1}{k}\left(
    \left[M^{(-1)}_x,T_{k,j-1}(y)\right]
    +\sum_{l=1}^{k-1}M_x^{(k-l)}(y)T_{l,j-1}(y)
    \right)\,,\quad \textrm{for } j>1\,.
\end{split}\end{align}
In other words, we can write
\begin{equation}
    \vec h(x,y,\epsilon)=\vec{\mathfrak{h}}(x,y,\epsilon)
    +\sum_{k\geq 1}x^k \, \vec h^{(k)}(x,y,\epsilon)\,.
\end{equation}
$\vec{\mathfrak{h}}(x,y,\epsilon)$ gives the leading
behaviour in the $x\to0$ limit, which contains (powers of) logarithms
of $x$, and the higher-order terms $h^{(k)}(x,y,\epsilon)$ 
can be systematically generated through multiplication by 
(the expansion of) the matrix $T(x,y,\epsilon)$,
which is explicitly given above.

In order to determine $\vec{\mathfrak{h}}(x,y,\epsilon)$, we must
compute the boundary condition $\vec{\mathfrak{h}}_0(\epsilon)$.
This corresponds to the evaluation of the pentagon functions at a point
$(0,y_0)$ in the MRK limit. We start from the point
\begin{equation}
    X_0=\{x=1;s=3,s_1=1,s_2=1,t_1=-1,t_2=-1\}\,,
\end{equation}
which is the point chosen as a boundary value for the pentagon functions
$\vec h(1,y,\epsilon)$ in ref.~\cite{Chicherin:2020oor},
and use the differential equation \eqref{eq:DE_PF} within
\texttt{DiffExp} \cite{Hidding:2020ytt} to compute an expansion
around
\begin{equation}
    X_1=\{x=0;s=1,s_1=1,s_2=1,t_1=-1,t_2=-1\}\,.
\end{equation}
This point is in the MRK limit, and as such the expansion
near $X_1$ contains powers of $\log(x)$. 
These logarithms correspond to the expansion of the
$x^{\epsilon M_{x}^{(-1)}}$ factor in eq.~\eqref{eq:leadingOrderPF}, and
we can indeed verify that the expansion we obtain from \texttt{DiffExp}
contains the expected logarithms. The point $X_1$ was the boundary point
chosen in ref.~\cite{Caron-Huot:2020vlo}, and we could validate our setup
by comparing our results with the ones given there. We note, however, that $X_1$ corresponds to a singular limit within MRK, namely the limit where the centrally emitted gluon is soft. To proceed with the computation of $\vec{\mathfrak{h}}(x,y,\epsilon)$ we therefore use a regular point within MRK 
at which we evaluate the boundary condition:
\begin{equation}
    X_2=\{x=0;s=1,s_1=1,s_2=1,z=i,\bar z=-i\}\,.
\end{equation}
Given that both $X_1$ and $X_2$ lie in the MRK limit (with $x=0$), 
we can simply use the second equation in \eqref{eq:DE_PF_dec}
within \texttt{DiffExp} to obtain a numerical evaluation at
$X_2$ starting from the one at $X_1$. 
In summary, we can use
\texttt{DiffExp} to obtain $\vec{\mathfrak{h}}_0(\epsilon)$
in eq.~\eqref{eq:leadingOrderPF}. At our boundary point $X_2$
these numbers do not correspond to simple constants, so we keep
$\vec{\mathfrak{h}}_0(\epsilon)$ as a vector of numbers with
60 digit precision.

To complete the calculation of $\vec{\mathfrak{h}}(x,y,\epsilon)$,
we must compute the integrals in the path-ordered exponential
in eq.~\eqref{eq:leadingOrderPF}, starting at the point $X_2$. 
As already noted in ref.~\cite{Caron-Huot:2020vlo}, it turns out 
that the alphabet
greatly simplifies in the MRK limit, and in particular the letters
either depend on $\{s,s_1,s_2\}$ or on $\{z,\bar z\}$. 
This makes the integration easy to perform in terms of 
multiple polylogarithms using e.g.~\texttt{PolyLogTools}~\cite{Duhr:2019tlz}.

The most complicated functions appearing in the remainders we 
are interested in are of transcendental weight 4. We thus
require the calculation of $\vec{\mathfrak{h}}_0(\epsilon)$ and 
of the integrals in the path-ordered exponentials
in eq.~\eqref{eq:leadingOrderPF} up to weight 4. 
Given the size of the differential equation \eqref{eq:DE_PF} this
is not a trivial calculation, but we found that 
\texttt{DiffExp}~\cite{Hidding:2020ytt} and
a combination of \texttt{PolyLogTools}~\cite{Duhr:2019tlz} and 
\texttt{HyperInt}~\cite{Panzer:2014caa} was sufficient to 
handle it.

Having determined $\vec{\mathfrak{h}}(x,y,\epsilon)$ with
the steps described above, we can then compute the
subleading terms in the $x\to0$ limit using 
eqs.~\eqref{eq:solAllX}, \eqref{eq:TExp} and \eqref{eq:TCoeffs}.
As described in the next section, for the two-loop remainders
of the QCD amplitudes we require expansions of the pentagon
functions up to terms of order $x^2$ in the MRK limit.

We close this discussion with two comments. First,
while our discussion was in the context of the MRK
limit of pentagon functions, exactly the same procedure is
applicable to Feynman integrals in a pure basis, since
they satisfy a differential equation of the same form
as in eq.~\eqref{eq:DE_PF} 
(and this is indeed what was done in \cite{Caron-Huot:2020vlo}).  
In fact, this procedure is simpler
when applied to a basis of master integrals, because the system 
of equations are typically smaller, and the solutions typically
have more constrained analytic structure than individual pentagon
functions. Second, given the current knowledge of the two-loop
QCD amplitudes we cannot compute them beyond the finite
term in the expansion in the dimensional regulator. However, at
one-loop we do have expressions valid to all orders in the dimensional
regulator \cite{Abreu:2019odu}. We
have thus implemented the same steps described above 
to compute the one-loop integrals in the MRK limit up
to weight 6, allowing us to compute one-loop amplitudes
in the MRK limit up to order $\epsilon^4$.


\subsection{Reconstruction of the rational coefficients in the MRK limit}\label{sec:padicreconstruction}

One might expect that the expansion in the MRK limit of the (rational) 
coefficients $r_k$ in eq.~\eqref{eq:remaindersGen} 
should be much simpler than that
of the transcendental functions $h_j$. This would indeed be the case
if the $r_k$ were simply rational functions of the
Mandelstam variables
$\{s_{12},s_{23},s_{34}, s_{45},s_{15}\}$, and if there were no spurious-pole cancellations. In such a scenario, one could directly substitute the relations in 
eq.~\eqref{kinhel1} and analytically expand around $x = 0$. 
However, as already noted below eq.~\eqref{eq:remaindersGen}, the $r_k$ 
given in refs.~\cite{DeLaurentis:2023nss, DeLaurentis:2023izi}
are rational functions of contractions of spinors,
making this procedure more involved. Furthermore, if an $r_k$ were to exhibit any spurious poles in $x$, retaining additional orders 
in the $x$ expansion to manifest these cancellations analytically could become computationally prohibitive.

In this section, we describe how the expansion of the rational coefficients
around $x=0$ can be obtained through analytic reconstruction of 
$p\kern0.2mm$-adic numerical evaluations \cite{DeLaurentis:2022otd, Chawdhry:2023yyx}. This has two benefits. First, spinors no longer pose a complication since we only need to numerically evaluate the expressions. Second, any cancellation, if present, will happen numerically, which is computationally efficient. Fortunately, due to the simplicity of the rational functions given in refs.~\cite{DeLaurentis:2023nss, DeLaurentis:2023izi}, 
spurious-pole cancellations do not pose an obstacle to the present calculation.  Nevertheless, this technique lays the 
groundwork for more complex scenarios. A preliminary investigation \cite{ByrneWIP} of a similar 
calculation with higher-multiplicity amplitudes \cite{Laurentis:2019bjh} suggests that such cancellations 
become very relevant, and the $p\kern0.2mm$-adic approach offers a scalable solution.

In the computation of the Laurent expansion of the rational functions around $x=0$, we employ a somewhat different minimal set of kinematic variables, which we denote by $w, \bar{w}, X_{34}$ and $X_{45}$ (see also refs.~\cite{Byrne:2022wzk, Byrne:2023nqx}). We define them as
\begin{equation}
    w = - \frac{\mathbf{p}_3}{\mathbf{p}_4} 
    \, , \quad 
    \bar{w} = - \frac{\bar{\mathbf{p}}_3}{\bar{\mathbf{p}}_4} 
    \, , \quad
    X_{34} = \, \frac{p_3^+}{p_4^+} 
    \, , \quad 
    X_{45} = \, \frac{p_4^+}{p_5^+} 
    \, .
\end{equation}
The definition for $w$ and $\bar{w}$ is none other than eq.~\eqref{def:zzbar}, now taken as exact. Given that $z$ and $\bar{z}$ are defined by eq.~(\ref{t12}) instead, we have $w = z + \mathcal{O}(x)$ and $\bar{w} = \bar{z} + \mathcal{O}(x)$ as we move away from the MRK limit. Similarly, $X_{34}$ and $X_{45}$ are closely related to $s_{34}=s_2/x$ and $s_{45}=s_1/x$. In appendix~\ref{app:vars_conversion}, we provide explicit formulae to relate $(w, \bar{w}, X_{34}, X_{45})$ to the set of variables $(z, \bar z, s_{34}, s_{45})$, and therefore $(z, \bar z, s_{1}, s_{2})$, through the third order in the power expansion in $x$. In order to work with $\mathcal{O}(x^0)$ variables, we perform one more change of variables to
\begin{equation}
    \tilde X_{34} = x \, X_{34} 
    \, , \quad 
    \tilde X_{45} = x \, X_{45} \, .
\end{equation}
We shall work with the minimal set of independent variables $(w, \bar{w}, \tilde X_{34}, \tilde X_{45})$.

We consider the MRK limit of the $r_k$ 
in eq.~\eqref{eq:remaindersGen} normalised 
by a factor that removes the spinor weight.
We choose such a factor to be the leading-colour
tree amplitude in a reference permutation, which we
denote ${\mathcal{M}^{\text{tree}}}$.
We write 
\begin{equation}
\tilde{r}_k=\frac{r_k}{\mathcal{M}^{\text{tree}}}\,,
\end{equation}
and the expansion of the $\tilde r_k$ takes the form
\begin{equation}\label{eq:coeffs_Laurent_expansion}
    \tilde r_k = \frac{\tilde{r}_k^{(-2)}(w, \bar{w}, \tilde X_{34}, \tilde X_{45})}{x^2} + \frac{\tilde{r}_k^{(-1)}(w, \bar{w}, \tilde X_{34}, \tilde X_{45})}{x} + \, \tilde{r}_k^{(0)}(w, \bar{w}, \tilde X_{34}, \tilde X_{45}) + \mathcal{O}(x) \, .
\end{equation}
Our objective is to determine the expansion coefficients $\tilde r_k^{(-2)}$, $\tilde r_k^{(-1)}$ and $\tilde r_k^{(0)}$.

The structure of $p\kern0.2mm$-adic numbers
lends itself well to the present task as, unlike
finite-field numbers, they allow us to approach
limits. 
Indeed, they can be thought of as a Laurent expansion in powers of the prime $p$. 
We compute the expansion of eq.~\eqref{eq:coeffs_Laurent_expansion} numerically over $p\kern0.2mm$-adic numbers, by setting $x=p$ (we recall that while $p$ is generally chosen as a large prime to avoid accidental cancellations, $p\kern0.2mm$-adically it is actually a small quantity, $|p|_p \ll 1$), while the four minimal variables are chosen as random $\mathcal{O}(1)$ $p\kern0.2mm$-adic integers.
We interpret the digits of the $p\kern0.2mm$-adic number as finite field evaluations of the corresponding terms in the $x$ expansion of $\tilde r_k$, namely $\tilde r_k^{(-2)}, \tilde r_k^{(-1)}$ and $\tilde r_k^{(0)}$. Note that a direct computation over finite fields of these expansion coefficients would be impossible, since $x$ would need to be either exactly zero, causing divisions by zero, or $\mathcal{O}(1)$, meaning it would be away from the limit. One may rely on a univariate interpolation in $x$, but this would require obtaining the full $x$ dependence before truncating at $\mathcal{O}(x)$.

With $p\kern0.2mm$-adic evaluations at hand, we sequentially reconstruct each digit with finite field interpolation methods \cite{vonManteuffel:2014ixa, Peraro:2016wsq}. Specifically, we obtain the denominators from a univariate slice by matching factors obtained from a univariate Thiele rational interpolation to a list of expected singularities related to the letters of the symbol alphabet~\cite{Abreu:2018zmy}. The set of possible denominator factors that we consider is
\begin{equation}
\big\{ w, \, \bar w, \, w+1, \, w-1, \, \bar w+1, \, \bar w-1, \, 1-w-\bar w, \, w- \bar w, \, \tilde{X}_{34}, \, \tilde{X}_{45} \big\} \, .
\end{equation}
Subsequently, we perform a multivariate Newton interpolation of the numerators. Given the simplicity of the $r_i$, this reconstruction is trivially performed in least common denominator form.

Several components of this calculation are available in public software. The general kinematic results can be loaded using \texttt{antares} \cite{antares} from \texttt{antares-results} \cite{antares-results} and evaluated over $p\kern0.2mm$-adic numbers using the phase-space implementation of \texttt{lips} \cite{lips} and the $p\kern0.2mm$-adic number type implementation in \texttt{pyadic} \cite{pyadic}. Newton and Thiele interpolation algorithms are implemented in \texttt{pyadic.interpolation}.

We close this section with a comment on one-loop amplitudes.
As already noted previously, in refs.~\cite{Abreu:2018zmy,Abreu:2019odu}
expressions are given for the one-loop amplitudes (and not remainders)
that are valid to all orders in the dimensional regulator
$\epsilon$. After normalisation by a factor that removes
the spinor weight, the coefficients are written in terms of
rational functions in the Mandelstam variables and ${\rm tr}_5$,
and as such computing their MRK limit does not require the
more advanced technology described here.
We note, however, that for the amplitudes involving two and four
quarks, the results of refs.~\cite{Abreu:2018zmy,Abreu:2019odu}
are only sufficient for calculations in the leading-colour limit, so we computed the remaining amplitudes to order
$\epsilon^2$ with \texttt{Caravel} \cite{Abreu:2020xvt}.
This allowed us to explicitly construct all the 
required amplitudes from the remainders.


\subsection{Assembly of the remainders}

After discussing the expansion of the pentagon functions
in the MRK limit in section \ref{sec:expPFs} and the expansion
of the coefficients in \ref{sec:padicreconstruction},
in this section we briefly comment on the assembly of the 
the one- and two-loop remainders.

For each remainder $\mathcal H$ as in eq.~\eqref{eq:remaindersGen}, 
we combine the coefficients $r_k$ with the corresponding transcendental
functions $h_k$.
In the most complicated cases the coefficients include terms 
starting at $\mathcal{O}(x^{-2})$, and we must
include the expansion of the functions
up to order $\mathcal{O}(x^{2})$.
As discussed in section \ref{sec:expPFs}, the
boundary conditions $\vec{\mathfrak{h}}_0(\epsilon)$ 
were kept as numbers with 60 digit precision.
The reason why we did not try to identify these numbers is
that we do not expect the pentagon functions to only involve
simple transcendental numbers. However, by the time
we have assembled the remainders, we expect 
this to be the case.
After constructing each remainder, we thus combine all
numeric terms at each weight
 (which are linear combinations of the
$\vec{\mathfrak{h}}_0(\epsilon)$ and the evaluation of the 
integrals in the path-ordered exponentials at $y_0$, 
see eq.~\eqref{eq:leadingOrderPF}) and 
use the PSLQ algorithm to fit them to a basis of
zeta values. Up to weight four, this basis is rather small:
$\{i \pi\}$ for weight 1, $\{\zeta_2\}$ for weight 2,
$\{\zeta_3,i\pi^3\}$ for weight 3 and $\{\zeta_4,i\pi \zeta_3\}$
for weight 4. As such, the 60-digit evaluations 
of $\vec{\mathfrak{h}}_0(\epsilon)$ are sufficient to successfully
fit the remaining numeric terms and obtain fully analytic expressions.

As a cross-check of our results, we compared the analytic
expressions we obtained for the MRK limit of the one- and
two-loop remainders with a numerical evaluation of the
expressions in refs.~\cite{DeLaurentis:2023nss, DeLaurentis:2023izi}
at a point very close to the MRK limit (we take
$x=10^{-10}$ and we run the pentagon functions to octuple
precision). 
The agreement of these two calculations for at least 6 
significant digits 
(which is consistent with the value of $x$ chosen in the 
numerical calculations),
provides a non-trivial check of the expansion procedures 
outlined in sections \ref{sec:expPFs} and 
\ref{sec:padicreconstruction}.

The analytic expressions for all components of the amplitudes of the three process (in the colour basis described in appendix~\ref{sec:colourBasis}) at both one and two loops are provided in the ancillary files~\cite{Ancillary}. 


\section{Reggeization and the Lipatov vertex}\label{sec:Reggeization}

In the high-energy limit the scattering 
amplitude develops large logarithms
proportional to the rapidity differences 
between the final-state particles. 
These contributions are dominated 
by the exchange of a single Reggeized
gluon~\cite{Grisaru:1973ku,Grisaru:1973vw,Grisaru:1973wbb,Lipatov:1976zz}. 
In this section we discuss the amplitude 
in the Regge-pole approximation, where is 
its assumed that a single Reggeized gluon 
(or Reggeon) is exchanged across each of 
the rapidity spans, that is in each of 
the $t_i$ channels. 
While this is not the complete description 
of high-energy amplitudes in MRK, it 
fully captures the planar limit of 
four and five-point amplitudes, and 
in the full non-planar theory it 
holds through next-to-leading 
logarithms for the dispersive 
part of the amplitude~\cite{Fadin:2006bj}.  

In terms of colour structure, the Reggeon 
has the same quantum number of the gluon, 
i.e. an antisymmetric octet. Thus, the 
amplitude in the Regge-pole approximation 
has the same colour structure of the tree-level 
one\footnote{Other components in the basis, which 
appear due to multi-Reggeon exchange across either 
or both of these rapidity spans will be discussed 
in Sections \ref{MRK-shockwave} and~\ref{sec:MRE}.}, 
$\mathcal{C}_{ij}^{(0)}$ defined in eq.~(\ref{def:Kij}).
In terms of signature, each Reggeon has negative parity. 
Hence in this whole section we shall refer to the 
Regge-pole contribution as 
${\cal M}^{(-,-)}_{ij \to i'gj'}\big|^{\mbox{\scriptsize 1-Reggeon}}$.


\subsection{Regge-pole factorization in $2\to 3$ scattering}

At leading logarithmic (LL) accuracy,
Reggeization of the $t$-channel gluons 
across each rapidity interval amounts 
to dressing the corresponding propagators 
in \eqn{Mtree} as follows: 
\be
\frac{1}{t_1} \to \frac{1}{t_1} 
\bigg(\frac{s_{45}}{\tau}\bigg)^{C_A\alpha_{g}(t_1)}, 
\qquad \qquad 
\frac{1}{t_2} \to \frac{1}{t_2} 
\bigg(\frac{s_{34}}{\tau}\bigg)^{C_A\alpha_{g}(t_2)}, 
\ee
where the scale $\tau \sim - t_1 \sim - t_2 
\sim |\mathbf{p}_4|^2 > 0$ is a reference 
scale, and $\alpha_{g}(t)$ is the so-called 
\emph{gluon Regge trajectory} \cite{Lipatov:1976zz}
\be
\label{Regge_traj}
\alpha_g(t_i) = \frac{\as(\mu^2)}{\pi} \alpha_g^{(1)}(t_i,\mu^2,\eps)
+ \bigg(\frac{\as(\mu)}{\pi}\bigg)^2 \alpha_g^{(2)}(t_i,\mu^2,\eps) 
+ \ldots.
\ee
At LL accuracy, only the one-loop term 
in the Regge trajectory is relevant 
\be
\label{alphagOneLoop}
\alpha_g^{(1)}(t_i,\mu^2,\eps) = \frac{r_{\Gamma}}{2\eps} 
\bigg(\frac{\mu^2}{-t_i}\bigg)^{\eps}, \qquad
r_{\Gamma}(\epsilon) = \frac{e^{\eps\gamma_E}\Gamma(1-\eps)^2
\Gamma(1+\eps)}{\Gamma(1-2\eps)}= 
1-\frac{\zeta_2}{2}\eps^2
-\frac{7\zeta_3}{3}\eps^3+\ldots,
\ee
and the LL amplitude reads 
\be\label{ggLL}
{\cal M}_{ij \to i'gj'}^{\rm LL} =
\bigg(\frac{s_{45}}{\tau}\bigg)^{C_A\alpha_{g}(t_1)} 
\bigg(\frac{s_{34}}{\tau}\bigg)^{C_A\alpha_{g}(t_2)} 
{\cal M}_{ij \to i'gj'}^{\rm tree}.
\ee
Extending the Reggeization hypothesis 
beyond leading logarithms requires 
corrections to the gluon Regge trajectory 
(\ref{Regge_traj}), the impact factors
and the Lipatov vertex. 
The impact factors are the effective 
vertices describing the emission of a 
Reggeon from the scattered partons
$i$ and $j$, while the Lipatov vertex 
is the emission vertex of a real gluon 
$g$ from the Reggeon. Below we briefly 
define our notation for these quantities, 
while we delegate a more detailed exposition 
of their perturbative expansion to 
appendix~\ref{sec:ReggeCoefficients}. 

Let us parametrize the 
loop correction to the impact 
factors and Lipatov vertex as 
follows: for the impact 
factors we write 
\be\label{ImpactPertDef1}
C_{i}(P_i,P_{i'},\tau) =
g_s \, C^{(0)}_{i}(P_i,P_{i'}) \, c_{i}(t_i,\tau), 
\ee
where $t_i = Q_i^2 = (P_i+P_{i'})^2$
(cf. \eqns{q1q2defA}{q1q2defB}), with
\be\label{ImpactPertDef2}
c_{i}(t_i,\tau) = 1 + \sum_{n= 1}^{\infty} 
\bigg(\frac{\as}{\pi}\bigg)^n c^{(n)}_{i}(t_i,\tau,\mu^2).
\ee
In the rest of this section we will keep 
the dependence on the renormalization scale 
$\mu$ implicit, i.e. $c_{i/j}^{(n)}(t_{k},\tau) 
\equiv c_{i/j}^{(n)}(t_{k},\tau,\mu^2)$. We discuss 
the complete renormalization and factorization 
scale dependence of the impact factors in 
appendix \ref{sec:ReggeCoefficients}.

Concerning the Lipatov vertex, 
in a similar fashion we define
\be\label{LipatovPertDef1B}
V_\mu(Q_1,P_4,Q_2,\tau) = 
g_s \, V^{(0)}(Q_1,P_4,Q_2) \, 
v_\mu(z,\bar{z},|\mathbf{p}_4|^2,\tau), 
\ee
with
\begin{equation}
v_\mu(z,\bar{z},|\mathbf{p_4}|^2,\tau) 
= \frac{V^{(0)}_\mu(Q_1,P_4,Q_2) }{V^{(0)}(Q_1,P_4,Q_2)}
+\sum_{n=1}^{\infty}\left(\frac{\alpha_s}{\pi}\right)^n 
v^{(n)}_\mu(z,\bar{z},|\mathbf{p_4}|^2,\tau),
\end{equation}
where $V^{(0)}_\mu(Q_1,P_4,Q_2)$ and 
$V^{(0)}(Q_1,P_4,Q_2$ introduced in eq.~(\ref{eq:treeLipatov}) 
and (\ref{LipatovTree}), respectively. Note that this 
definition implies that $v_\mu$ is normalized similarly 
to (\ref{ImpactPertDef1}) such that
\begin{equation}
v_{\oplus}(z,\bar{z},|\mathbf{p_4}|^2,\tau)\equiv 
v_\mu(z,\bar{z},|\mathbf{p_4}|^2,\tau) \varepsilon^\mu_{\oplus}(P_4) 
= 1+ \sum_{n=1}^{\infty}\left(\frac{\alpha_s}{\pi}\right)^n \,v^{(n)}_\oplus\,.
\end{equation}
where we omitted the kinematic dependence 
for the coefficients, $v^{(n)}_\oplus = 
\varepsilon^\mu_{\oplus}(P_4)v^{(n)}_\mu$\,.

Under the Reggeization hypothesis, 
the amplitude takes the form
\begin{align}\label{REggNLL}
\begin{split}
{\cal M}^{(-,-)}_{ij \to i'gj'}\Big|^{\mbox{\scriptsize 1-Reggeon}} 
& = 
c_{i}(t_{1},\tau)\, 
e^{\omega_1\eta_1}\,
v_{\mu}(t_{1},t_{2},|\mathbf{p}_4|^2,\tau)\,
e^{\omega_2\eta_2}\, c_{j}(t_{2},\tau)\,
\,\varepsilon^\mu_{\lambda_4}  \,{\cal M}_{ij \to i'gj'}^{\rm tree} \\
& = 
c_{i}(t_{1},\tau)\, 
e^{\omega_1\eta_1}\,
v_{\lambda_4}(t_{1},t_{2},|\mathbf{p}_4|^2,\tau)\,
e^{\omega_2\eta_2}\, c_{j}(t_{2},\tau)\,
 \,{\cal M}_{ij \to i'gj'}^{\rm tree}\,,
\end{split}
\end{align}
with
\begin{equation}
\label{complexLogsB}
\eta_1\equiv \log\left(\frac{s_{45}}{\tau}\right)-i\frac{\pi}{4}\,,
\quad \qquad 
\eta_2 \equiv \log\left(\frac{s_{34}}{\tau}\right)-i\frac{\pi}{4}\,,
\end{equation}
where the exponents are $\omega_k\equiv C_A \alpha_{g}(t_k)$ 
for $k=1,2$ and where the phases associated with the Reggeized 
gluons have been fixed based on the infrared singularity 
structure (see appendix~\ref{sec:DipoleFormula}): 
these are the terms proportional to $\T_{t_1}^2$ 
and $\T_{t_2}^2$ in eqs.~(\ref{TppToTt12}), which accompany 
the energy logarithms in (\ref{ZetaHatDef}).


\subsection{The analytic structure and symmetries of the Lipatov Vertex}
\label{sec:VertexSymmetry}

Because the impact factors $c_i$ and $c_j$ 
in (\ref{REggNLL}) are real, the remaining 
phase of the single Reggeon amplitude~(\ref{REggNLL}), 
must be attributed to the Lipatov vertex $v_\mu$. 
This phase is tightly constrained by \emph{signature} 
and \emph{analyticity}. The former requires the 
{\textit{antisymmetry}} of the single Reggeon 
amplitude upon either $\{s_{12}\to-s_{12},s_{45}\to-s_{45}\}$, 
or $\{s_{12}\to-s_{12},s_{34}\to-s_{34}\}$, eq.~(\ref{SigPerm}). 
The latter forbids sequential discontinuities in 
partially overlapping channels, such as 
$s_{34}^{\text{power}}\,s_{45}^{\text{power}}$. 
Accounting for this, the Regge-pole ansatz takes 
the form 
\begin{align}\label{ggNLLsig2} 
\begin{split}
\left.{\cal M}_{ij \to igj}^{(-,-)}
\right|^{\mbox{\scriptsize 1-Reggeon}} 
&=  c_{i}(t_{1},\tau) \, c_{j}(t_{2},\tau) \, 
{\cal M}_{ij \to igj}^{\rm tree}\,\varepsilon^\mu_{\lambda_4}(P_4) 
{\cal A}_{\mu}\big(z,\bar{z},s_{45},s_{34},|\mathbf{p}_4|^2,\tau\big),
\end{split}
\end{align}
where the production amplitude ${\cal A}_\mu$ is~\cite{Drummond:1969jv,Bartels:1974tj,Fadin:1993wh,Bartels:1980pe}
\begin{align}\label{AmuB} 
\begin{split}
 {\cal A}_{\mu}&\big(z,\bar{z},s_{45},s_{34},|\mathbf{p}_4|^2,\tau\big)=\\
&\frac{1}{4}\, \Bigg[\bigg(\frac{s_{34}}{\tau}\bigg)^{\omega_2-\omega_1} 
\!\!+ \bigg(\frac{-s_{34}-i\delta}{\tau}\bigg)^{\omega_2-\omega_1}\Bigg] 
\,\Bigg[\bigg(\frac{s_{12}}{\tau}\bigg)^{\omega_1} 
\!\!+ \bigg(\frac{-s_{12}-i\delta}{\tau}\bigg)^{\omega_1} \Bigg] 
R_\mu\big(z,\bar{z},|\mathbf{p}_4|^2,\tau\big)\,  \\
+&
\frac{1}{4}\,\Bigg[\bigg(\frac{s_{45}}{\tau}\bigg)^{\omega_1-\omega_2}
\!\!+ \bigg(\frac{-s_{45}-i\delta}{\tau}\bigg)^{\omega_1-\omega_2} \Bigg] 
\, 
\Bigg[\bigg(\frac{s_{12}}{\tau}\bigg)^{\omega_2} 
\!\!+ \bigg(\frac{-s_{12}-i\delta}{\tau}\bigg)^{\omega_2} \Bigg]
L_\mu\big(z,\bar{z},|\mathbf{p}_4|^2,\tau\big)\,,
\end{split}
\end{align}
where the relation~(\ref{momentumconservation1})   
${s_{12}}=s_{34}s_{45}/|\mathbf{p_4}|^2$ can be 
used to restore the overall dependence on the 
powers of $s_{45}$ and $s_{34}$, consistently 
with eq.~(\ref{complexLogsB}).
The {\emph{right}} and {\emph{left}} vertices, 
$R_\mu$ and $L_\mu$, are \emph{real} functions, 
obeying
\begin{subequations}
\begin{align}
\label{eq:RealityVRVLB}
&R_\mu(z,\bar{z},|\mathbf{p_4}|^2,\tau) 
= R_\mu^*(z,\bar{z},|\mathbf{p_4}|^2,\tau)
= R_\mu(\bar{z},z,|\mathbf{p_4}|^2,\tau), \\
&L_\mu(z,\bar{z},|\mathbf{p_4}|^2,\tau) 
= L_\mu^*(z,\bar{z},|\mathbf{p_4}|^2,\tau)
=L_\mu(\bar{z},z,|\mathbf{p_4}|^2,\tau).
\end{align} 
\end{subequations}
Note that this reality property is no more 
manifest once the vertex is contracted into 
the gluon complex polarization vector in 
(\ref{ggNLLsig2}), defining the conventional 
left and right  vertices 
\beq
\label{vRL}
v_R^{(\lambda_4)}\equiv  \varepsilon^\mu_{\lambda_4}(P_4)R_\mu\big(z,\bar{z},|\mathbf{p}_4|^2,\tau\big),\qquad 
v_L^{(\lambda_4)}\equiv \varepsilon^\mu_{\lambda_4}(P_4)L_\mu\big(z,\bar{z},|\mathbf{p}_4|^2,\tau\big)\,.
\eeq 
Hence, for now we proceed to consider 
the vectors quantities $R_\mu$ and $L_\mu$.

In practice, $R_\mu$ and $L_\mu$ are 
written in terms of a basis of transcendental 
functions\footnote{Example of such a basis, 
in which we express the result for the vertex 
through two loop is provided in 
section~\ref{VertTwoLoopStruct}.} 
and their associated rational 
coefficients, so for example for $R_{\mu}$
\begin{align}\label{RTransSCoeff}
\left.
R_\mu(z,\bar{z},|\mathbf{p_4}|^2,\tau)\right\vert_{\tau=|\mathbf{p_4}|^2} 
=  {\cal Q}_\mu\sum_iS_{\cal Q}^{(i)}(z,\bar{z})\,\Phi_i(z,\bar{z}) 
+{\cal P}_\mu\sum_i S_{\cal P}^{(i)}(z,\bar{z})\,\Phi_i(z,\bar{z}),
\end{align}
where ${\cal P}_\mu$ and ${\cal Q}_\mu$ are 
the gauge invariant tensors in eq.~(\ref{eq:treeQPtens}), 
$\Phi_i$ are the elements of the transcendental basis and 
$S_{\cal P}^{(i)}$, $S_{\cal Q}^{(i)}$ are rational 
coefficients. A similar expansion (with different 
coefficients) holds for $L_{\mu}$.
Given the reality property, it is natural to choose 
a basis in which neither the transcendental functions 
nor theirvrational coefficients would have any explicit 
imaginary $i$ factors. Then, upon complex conjugation, 
equation (\ref{RTransSCoeff}) becomes
\begin{align}
\left.
R^*_\mu(z,\bar{z},|\mathbf{p_4}|^2,\tau)\right\vert_{\tau=|\mathbf{p_4}|^2} 
= {\cal Q}_\mu\sum_iS_{\cal Q}^{(i)}(\bar{z},z)\,\Phi_i(\bar{z},z) 
+{\cal P}_\mu\sum_i S_{\cal P}^{(i)}(\bar{z},z)\,\Phi_i(\bar{z},z).
\end{align}
The reality condition (\ref{eq:RealityVRVLB}) 
can then be satisfied if the functions and 
their respective rational coefficients are 
all either real, namely,
\beq
\label{realfunctionsandcoef}
S_{\cal Q/P}^{(i)}(\bar{z},z) 
= S_{\cal Q/P}^{(i)}(z, \bar{z}),\qquad \Phi_i(\bar{z},z) 
= \Phi_i(z,\bar{z})\,,
\eeq
or purely imaginary 
\beq
\label{imaginaryfunctionsandcoef}
S_{\cal Q/P}^{(i)}(\bar{z},z) 
= - S_{\cal Q/P}^{(i)}(z, \bar{z}),\qquad \Phi_i(\bar{z},z) 
= - \Phi_i(z,\bar{z})\,.
\eeq
We recall that the amplitude will eventually be 
contracted with the complex polarization vector 
$\varepsilon^\mu_{\lambda_4}(P_4)$ in (\ref{ggNLLsig2}). 
Thus, the rational coefficients in the amplitude 
would no longer obey the symmetry conditions above. 
However, the transcendental functions $\Phi_i(z,\bar{z})$ 
cannot be affected by the contraction with $\varepsilon^\mu$, 
and therefore can always chosen such that each is either 
purely real (i.e. symmetric under $z\leftrightarrow \bar{z}$) 
or purely imaginary (antisymmetric under 
$z\leftrightarrow \bar{z}$). Equivalently, 
one may chose a basis of complex functions, 
but such that they are all hermitian:
\begin{equation}
\label{herm}
    \Phi_i^*(z,\bar{z}) = \Phi_i(\bar{z},z).
\end{equation}
In section~\ref{sec:scheme} we will choose 
such a basis to express the vertex through 
two loops.

By comparing eq.~(\ref{REggNLL}) with 
eqs.~(\ref{ggNLLsig2}) and (\ref{AmuB}), 
it follows that
\begin{align}
\label{LipatovAnalytics1muB}
\begin{split}
v_{\mu}\left(t_{1},t_{2},|\mathbf{p}_4|^2,\tau\right)
=&\,\,
\bigg(\frac{|\mathbf{p}_4|^2}{\tau}\bigg)^{-\omega_1}\,
\exp{\left\{i\frac{\pi}{4}(\omega_1-\omega_2)\right\}}
\,\,\widetilde{R}_{\mu}\left(z,\bar{z},{|\mathbf{p}_4|^2},{\tau}\right)
\\ 
&+\bigg(\frac{|\mathbf{p}_4|^2}{\tau}\bigg)^{-\omega_2}
\,\exp{\left\{i\frac{\pi}{4}(\omega_2-\omega_1)\right\}}
\,\,\widetilde{L}_{\mu}\left(z,\bar{z},{|\mathbf{p}_4|^2},{\tau}\right) ,
\end{split}
\end{align}
where we introduce the real functions
\begin{subequations}
\label{LipatovAnalytics2muB}
\begin{align}
\widetilde{R}_{\mu}\left(z,\bar{z},{|\mathbf{p}_4|^2},{\tau}\right)
& = R_{\mu}\left(z,\bar{z},{|\mathbf{p}_4|^2},{\tau}\right) \, 
\cos\Big[\frac{\pi}{2}\omega_1\Big] \, 
\cos\Big[\frac{\pi}{2}\left(\omega_1-\omega_2\right)\Big], \\ 
\widetilde{L}_{\mu}\left(z,\bar{z},{|\mathbf{p}_4|^2},{\tau}\right)
& = L_{\mu}\left(z,\bar{z},|\mathbf{p}_4|^2,\tau\right) \, 
\cos\Big[\frac{\pi}{2}\omega_2\Big] \, 
\cos\Big[\frac{\pi}{2}\left(\omega_1-\omega_2\right)\Big].
\end{align}
\end{subequations}
By expanding the complex exponentials, we may
now read off eq.~(\ref{LipatovAnalytics1muB}) 
the dispersive and the absorptive 
parts~\cite{Fadin:1993wh} of the complex 
Lipatov vertex of eq.~(\ref{REggNLL}): 
\begin{subequations}
\begin{align}
\label{eq:DisVmu}
\text{Disp}\left[v_\mu(z,\bar{z},|\mathbf{p_4}|^2,\tau)\right]
&=\left[\left(\frac{\tau}{|\mathbf{p_4}|^2}\right)^{\omega_1}\widetilde{R}_\mu
+\left(\frac{\tau}{|\mathbf{p_4}|^2}\right)^{\omega_2}\widetilde{L}_\mu\right]
\cos\left[\frac{\pi}{4}(\omega_1-\omega_2)\right],\\
\label{eq:AbsVmu}
\text{Absorp}\left[v_\mu(z,\bar{z},|\mathbf{p_4}|^2,\tau)\right]
&=\left[\left(\frac{\tau}{|\mathbf{p_4}|^2}\right)^{\omega_1}\widetilde{R}_\mu
-\left(\frac{\tau}{|\mathbf{p_4}|^2}\right)^{\omega_2}\widetilde{L}_\mu\right]
\sin\left[\frac{\pi}{4}(\omega_1-\omega_2)\right],
\end{align}
\end{subequations}
where we drop the arguments of $\widetilde{R}_\mu$ 
and $\widetilde{L}_\mu$ to keep the expressions 
compact. We now contract $\text{Disp}[v_\mu]$ and 
$\text{Absorp}[v_\mu]$ with the gluon polarisations, 
given in eq.~(\ref{PolVector4DefA}), to compute the 
Lipatov vertex for positive and negative helicity 
of the emitted gluon. 

By introducing
\begin{align}
\label{def:vTildePlus}
&\widetilde{v}_R(z,\bar{z},|\mathbf{p_4}|^2,\tau)
\equiv
\widetilde{R}_\mu(z,\bar{z},|\mathbf{p_4}|^2,\tau)
\varepsilon^\mu_\oplus, \qquad  
\widetilde{v}_L(z,\bar{z},|\mathbf{p_4}|^2,\tau)
\equiv\widetilde{L}_\mu(z,\bar{z},|\mathbf{p_4}|^2,\tau)
\varepsilon^\mu_\oplus,
\end{align}
and noting that the negative helicity polarization 
vector is the complex conjugate of the positive 
helicity one, $\varepsilon^{\mu}_{\ominus}(P_4) 
=(\varepsilon^{\mu}_{\oplus}(P_4))^*$, using the 
reality of the functions~(\ref{LipatovAnalytics2muB}) 
we get:
\begin{align}
\label{def:vTildeMin}
&\widetilde{R}_\mu(z,\bar{z},|\mathbf{p_4}|^2,\tau)
\varepsilon^\mu_\ominus 
=\widetilde{v}_R^*(z,\bar{z},|\mathbf{p_4}|^2,\tau), \qquad 
\widetilde{L}_\mu(z,\bar{z},|\mathbf{p_4}|^2,\tau)
\varepsilon^\mu_\ominus =
\widetilde{v}_L^*(z,\bar{z},|\mathbf{p_4}|^2,\tau)\,,
\end{align}
The complex vertex of eq.~(\ref{REggNLL}), after 
contraction with a positive helicity polarization 
vector, may be expressed as
\begin{align}
\begin{split}
    v_\oplus(z,\bar{z},|\mathbf{p_4}|^2,\tau) &\equiv 
v_\mu(z,\bar{z},|\mathbf{p_4}|^2,\tau)\varepsilon^\mu_\oplus \\
&= \text{Disp}\left[v_\oplus(z,\bar{z},|\mathbf{p_4}|^2,\tau)\right] + i \text{Absorp}\left[v_\oplus(z,\bar{z},|\mathbf{p_4}|^2,\tau)\right],
\end{split}
\end{align}
where the definition of $\text{Disp}\left[v_\oplus\right]$ 
and $\text{Absorp}\left[v_\oplus\right]$ follows from 
eqs.~(\ref{eq:DisVmu}) and (\ref{eq:AbsVmu})
\begin{subequations}
\begin{align}
\label{def:voplusDisp}
\text{Disp}\left[v_\oplus(z,\bar{z},|\mathbf{p_4}|^2,\tau)\right]
&=\left[\left(\frac{\tau}{|\mathbf{p_4}|^2}\right)^{\omega_1}\widetilde{v}_{R}
+\left(\frac{\tau}{|\mathbf{p_4}|^2}\right)^{\omega_2}\widetilde{v}_{L}\right]
\cos\left[\frac{\pi}{4}(\omega_1-\omega_2)\right], \\
\label{def:voplusAbs}
\text{Absorp}\left[v_\oplus(z,\bar{z},|\mathbf{p_4}|^2,\tau)\right]
&=\left[\left(\frac{\tau}{|\mathbf{p_4}|^2}\right)^{\omega_1}\widetilde{v}_R
-\left(\frac{\tau}{|\mathbf{p_4}|^2}\right)^{\omega_2}\widetilde{v}_L\right]
\sin\left[\frac{\pi}{4}(\omega_1-\omega_2)\right].
\end{align}
\end{subequations}
Since the polarisation vector is complex, both 
$\text{Disp}[v_\oplus]$ and $\text{Absorp}[v_\oplus]$ 
are complex-valued functions of $z$ and $\bar{z}$. 
However, they do not include explicit factors of $i$, 
namely, as a consequence of eq.~(\ref{eq:RealityVRVLB}), 
one may take their complex conjugate using
\begin{equation}
\widetilde{v}_R^*(z,\bar{z}) 
= \widetilde{v}_R(\bar{z},z),\qquad \widetilde{v}_L^*(z,\bar{z}) 
= \widetilde{v}_L(\bar{z},z)\,.
\end{equation}
Using eq.~(\ref{def:vTildeMin}), we find the dispersive 
and absorptive part of the Lipatov vertex of negative 
helicity in terms of the quantities defined for positive 
helicity in eqs.~(\ref{def:voplusDisp}) 
and~(\ref{def:voplusAbs}),
\begin{align}
&\text{Re}\left\{\text{Disp}\left[v_\ominus(z,\bar{z},|\mathbf{p_4}|^2)\right]\right\} 
= \text{Re}\left\{\text{Disp}\left[v_\oplus(z,\bar{z},|\mathbf{p_4}|^2)\right]\right\}, 
\nonumber \\
&\text{Im}\left\{\text{Disp}\left[v_\ominus(z,\bar{z},|\mathbf{p_4}|^2)\right]\right\} 
= -\text{Im}\left\{\text{Disp}\left[v_\oplus(z,\bar{z},|\mathbf{p_4}|^2)\right]\right\},
\\
&\text{Re}\left\{\text{Absorp}\left[v_\ominus(z,\bar{z},|\mathbf{p_4}|^2)\right]\right\} 
= \text{Re}\left\{\text{Absorp}\left[v_\oplus(z,\bar{z},|\mathbf{p_4}|^2)\right]\right\}, 
\nonumber \\
&\text{Im}\left\{\text{Absorp}\left[v_\ominus(z,\bar{z},|\mathbf{p_4}|^2)\right]\right\} 
= -\text{Im}\left\{\text{Absorp}\left[v_\oplus(z,\bar{z},|\mathbf{p_4}|^2)\right]\right\}.
\end{align}
In the rest of the paper, we will focus on 
the determination of the dispersive and of 
the absorptive parts of $v_\oplus$.

Another important property of the amplitude 
and the vertex is target-projectile symmetry. 
It is defined by the simultaneous swap of 
partons $1\leftrightarrow 2$ along with 
$3\leftrightarrow 5$ so $s_{34}\leftrightarrow s_{45}$ 
and $t_1\leftrightarrow t_2$. 
This amounts to following transformation of 
the emitted gluon polarization vector:
${\varepsilon}_{L\perp}^{\oplus}(P_4) 
\leftrightarrow {\varepsilon}_{R\perp}^{\ominus}(P_4)$, 
i.e. changing from the $L$ gauge to the $R$ 
gauge and picking the negative helicity 
instead of the positive one.  
According to eqs.~(\ref{PolVecRightPerpOur}) 
and (\ref{PolVecLeftPerpOur}) this produces 
an overall minus sign that compensates in 
(\ref{LipatovTree}) for the change of sign 
of the two vectors $P_{\mu}$ and $Q_{\mu}$ 
of (\ref{eq:treeQPtens}). In terms of the 
variables $z$ and $\bar{z}$ appearing in 
the vertex, target-projectile symmetry 
is a symmetry under 
\beq
\label{target_projectile}
z\to 1-\bar{z},\qquad\quad \bar{z}\to 1-z\,.
\eeq
Note that this operation swaps $t_1\leftrightarrow t_2$ 
according to eq.~(\ref{t12}), while preserving 
the sign of ${\text{tr}}_5$ in (\ref{tr5MRK}). 
One may readily verify that (\ref{target_projectile}) 
preserves the contracted expression for the 
tree-level vertex in (\ref{LipatovTree}). 
The same must happen at higher orders.

Considering now eqs.~(\ref{eq:DisVmu}) and 
(\ref{eq:AbsVmu}) under target-projectile 
symmetry, we observe that $t_1\leftrightarrow t_2$ 
implies to  $\omega_1 \leftrightarrow \omega_2$, 
and since the left hand side flips sign, on the 
right hand side we must have 
$\widetilde{R}_{\mu}\leftrightarrow  -\widetilde{L}_{\mu}$. 
Equivalently, after contraction with the polarization 
vector we observe that under target-projectile symmetry 
both the dispersive (\ref{def:voplusDisp}) and absorptive 
parts (\ref{def:voplusAbs}) of the vertex remain invariant, 
while  $\widetilde{v}_R(z,\bar{z}) \leftrightarrow 
\widetilde{v}_L(z,\bar{z})$, which means
\begin{equation}
\label{vRLtargetProjectile}
\widetilde{v}_R(1-\bar{z},1-z) = \widetilde{v}_L(z,\bar{z})\,.
\end{equation}
This has immediate implications for the basis of 
transcendental functions and their coefficients. 
Rather than considering the separate expressions 
for $\widetilde{v}_R$ and $\widetilde{v}_L$, it 
is now convenient to consider directly the quantities $\text{Disp}\left[v_\oplus(z,\bar{z},|\mathbf{p_4}|^2,\tau)\right]$ 
and $\text{Absorp}\left[v_\oplus(z,\bar{z},|\mathbf{p_4}|^2,\tau)\right]$ 
in terms of transcendental functions: both are symmetric 
under (\ref{target_projectile}), hence, it is natural to 
express them both in terms of functions that admit 
\begin{equation}
\label{TargetProjectileSymmetryPhi}
\Phi_i(1-\bar{z},1-z) = \pm \Phi_i(z,\bar{z})\,,
\end{equation}
with rational coefficients that admit the same relation, 
respectively, so that the  sign would cancel in the product, 
restoring the exact symmetry of the vertex. 

We conclude that in expressing the dispersive 
and absorptive parts of the complex vertex 
$v_\oplus(z,\bar{z},|\mathbf{p_4}|^2,\tau)$ 
it should be possible, and indeed natural, 
to choose a basis of transcendental functions 
satisfying the two properties of hermiticity 
(\ref{herm}) and target-projectile symmetry 
(\ref{TargetProjectileSymmetryPhi}). In the 
next section we shall see that these properties 
are realized straightforwardly for the one-loop 
vertex. In  section~\ref{sec:scheme} these 
properties will guide our choice of basis 
of transcendental functions for the 
two-loop vertex. 


\section{The QCD Lipatov vertex at one loop}
\label{sec:MatchingNLL}

In this section we derive the Lipatov vertex $v_{\oplus}(t_1,t_2,|\mathbf{p}_4|^2,\tau,\mu^2)$ at one loop in QCD through  ${\cal{O}}(\epsilon^4)$. The gluon-loop contribution to the one-loop Lipatov vertex was originally computed in Ref.~\cite{Fadin:1993wh} through finite terms as $\epsilon\to 0$. The terms proportional to $n_f$, originating from quark loop diagrams, were calculated to all orders in $\epsilon$ soon afterwards~\cite{Fadin:1994fj}. The complete vertex to ${\cal{O}}(\epsilon^0)$ was also extracted from the one-loop five-gluon amplitude in~\cite{DelDuca:1998cx}, finding agreement with the previous results. In addition, the soft limit of the one-loop vertex has been computed in Ref.~\cite{Fadin:1996yv} to all orders in $\epsilon$. A representation of the vertex in general kinematics, which is valid to all orders in $\epsilon$, is given in Ref.~\cite{Fadin:2000yp}. The latter has been evaluated recently up to ${\cal{O}}(\epsilon^2)$~\cite{Fadin:2023roz}, in terms of classical and Nielsen polylogarithms and of double sums, denoted as $\mathcal{M}$-functions, originating from the $\epsilon$-expansion of Appell $F_4$ and Kamp\'{e} de F\'{e}riet functions~\cite{DelDuca:2009ac}. The same functions appear in the Lipatov vertex in planar $\mathcal{N}=4$ super Yang Mills (sYM) theory, which has been determined to ${\cal{O}}(\epsilon^2)$ \cite{DelDuca:2009ae}. Here we follow the general strategy of ref.~\cite{DelDuca:1998cx}, that is, we determine $v^{(1)}$ by computing the $2\to 3$ amplitude in the multi-Regge kinematic limit and then matching it with relevant factorization formula.
However, while Ref.~\cite{DelDuca:1998cx} applied Regge factorisation only to the dispersive part of the amplitude, here we factorise the full, signaturised one, $\mathcal{M}_{gg\to ggg}^{(-,-)}$, which at one loop,  features both dispersive and absorptive parts.\footnote{Similarly to the tree amplitude, $\mathcal{M}_{gg\to ggg}^{(-,-)}$ at one loop consists exclusively of the $[8_a,8_a]$ colour component, and corresponds to a single Reggeon exchange. However, in contrast to the tree amplitude, it features both dispersive and absorptive parts, as is evident from eq.~(\ref{AmuB}).}  

To extract the vertex we match $\mathcal{M}_{gg\to ggg}^{(-,-)}$ with the factorization formula in eq.~(\ref{REggNLL}). Upon expansion we obtain
\begin{align}
\label{eq:REggNLLexpanded}
    {\cal M}^{(-,-)}_{ij \to i'gj'}\Big|^{\mbox{\scriptsize NLL}}&={\mathcal{M}}_{\text{tree}}\left(\frac{s_{45}}{\tau}\right)^{\omega_1}\left(\frac{s_{23}}{\tau}\right)^{\omega_2}\Big[1+\left(\frac{\alpha_s}{\pi}\right)\bigg(c_i^{(1)}(-t_1,\tau,\mu^2)+c_j^{(1)}(-t_2,\tau,\mu^2)\nonumber\\
   &\hspace{-25pt}+v_{\oplus}^{(1)}(t_1,t_2,|\mathbf{p}_4|^2,\tau,\mu^2)-i\frac{C_A\pi}{4}(\alpha_g^{(1)}(\mu^2,-t_1)+\alpha_g^{(1)}(\mu^2,-t_2))\bigg)\nonumber\\
    &\hspace{-25pt}+C_A\left(\frac{\alpha_s}{\pi}\right)^2\left(\alpha_g^{(2)}(\mu^2,-t_1)\log\frac{s_{45}}{\tau}+\alpha_g^{(2)}(\mu^2,-t_2)\log\frac{s_{23}}{\tau}\right)\Big].
\end{align}   
In the equation above, the one-loop Regge trajectory $\alpha^{(1)}(-t,\mu^2)$, eq.~(\ref{alphagOneLoop}), and the one-loop gluon impact factor are known to all orders in $\epsilon$. The latter reads 
\begin{align}
    c_i^{(1)}(-t,\tau,\mu^2)&=\left(\frac{\mu^2}{-t}\right)^\epsilon \left[c_i^{(1)}-C_A\frac{\alpha_g^{(1)}}{2}\log\left(\frac{-t}{\tau}\right)\right] - \frac{b_0}{8\epsilon}\left(1-\left(\frac{\mu^2}{-t}\right)^\epsilon\right),
\end{align}
where $b_0$ is given in (\ref{eq:betacoeffs}), $\alpha_g^{(1)}=\alpha_g^{(1)}(-t,-t)=\frac{r_{\Gamma}}{2\epsilon}$~\cite{Lipatov:1976zz}  where $r_{\Gamma}$ is given in eq.~(\ref{alphagOneLoop}) and $c_i^{(1)}=c_i^{(1)}(-t,-t,-t)$. The latter has been computed in Ref.~\cite{Bern:1998sc} and is given by
\begin{align}
c_g^{(1)}&=r_\Gamma(\epsilon) \Bigg[N_c\bigg(-\frac{1}{2\epsilon^2}+\frac{\psi(1+\epsilon)-2\psi(1-\epsilon)+\psi(1)}{4\epsilon}-\frac{11-7\epsilon}{8\epsilon(3-2\epsilon)(1-2\epsilon)}\bigg)\nonumber\\&+n_f\frac{(1-\epsilon)}{4\epsilon(1-2\epsilon)(3-2\epsilon)}\Bigg]-\frac{b_0}{8\epsilon}.
\end{align}

From eq.~(\ref{eq:REggNLLexpanded}) it follows that the one-loop vertex is
\begin{align}
v_{\oplus}^{(1)}(t_1,t_2,&|\mathbf{p}_4|^2,\tau,\mu^2)=\frac{{\cal M}^{(-,-),(1)}_{gg \to ggg}}{{\cal M}_{gg \to ggg}^{\rm tree}}-\bigg[C_A\left(\alpha^{(1)}(-t_1,\mu^2)\log\frac{s_{45}}{\tau}+\alpha^{(1)}(-t_2,\mu^2)\log\frac{s_{34}}{\tau}\right)\nonumber\\
    &+c_g^{(1)}(-t_1,\tau,\mu^2)+c_g^{(1)}(-t_2,\tau,\mu^2)-iC_A\frac{\pi}{4}\left(\alpha^{(1)}(-t_1,\mu^2)+\alpha^{(1)}(-t_2,\mu^2)\right)\bigg].
\end{align}
The signaturised amplitude $\mathcal{M}^{(-,-),(1)}_{gg\to ggg}$ is the colour projection of the five gluon scattering amplitude $\mathcal{M}^{(1)}_{gg\to ggg}$ onto the component $c^{[8_a,8_a]}$ of the orthonormal basis in appendix~\ref{sec:colourBasis}. To compute the amplitude we used the results of ref.~\cite{Abreu:2019odu}
for the five-gluon amplitudes, which are given as 
a decomposition in terms of one-loop master integrals
valid to all orders in $\epsilon$. They are written as
\begin{align}
\label{eq:RationalMaster}
    \mathcal{M}^{(1)}_{gg\to ggg} = \sum_{\sigma\in\mathcal{S}_5} \sum_{i=1}^{11} R_{\sigma,i}\, I_{\sigma,i},
\end{align}
where $\sigma$ is a permutation of the 5 external gluons and the $I_{\sigma,i}$ are a basis of one-loop master integrals.
This is a particular case of the type of decompositions
we discussed in section~\ref{MRK-QCD}, 
and these expressions can be expanded in the MRK
limit with the techniques described there (see in 
particular the comments at the end of sections~\ref{sec:expPFs}
and \ref{sec:padicreconstruction}).
All the master integrals have been written in terms of powers of logarithms of $s_{34}>0$, $s_{45}>0$, $|\mathbf{p}_4|^2>0$, and Goncharov multiple polylogarithms (GPLs) \cite{Goncharov:1998kja,Goncharov:2001iea} with the symbol alphabet
\begin{align}
\label{eq:alphabet}
&\{z,\bar{z},1-z,1-\bar{z},z-{\bar{z}},1-z-{\bar{z}}\},
\end{align}
where we recall the relations in eq.~(\ref{t12}) in the physical region, which to leading power in the MRK limit read
\[-t_1 \simeq (1-z)(1-\bar z)|\mathbf{p}_4|^2>0 \qquad \text{and}\qquad  -t_2 \simeq  z\bar z|\mathbf{p}_4|^2>0\,,\] so $z$ and $\bar{z}$ are complex conjugate.
Notably, the latter transcendental functions appear only in single-valued combinations, and we used the Maple code \texttt{HyperlogProcedures}~\cite{Schnetz:2021ebf} to write all the master integrals, up to transcendental weight 6, in terms of Single Valued Multiple Polylogarithms (SVMPLs)~\cite{Brown:2004ugm,Pennington:2012zj,Dixon:2012yy,Brown:2013gia,Schnetz:2013hqa,DelDuca:2013lma,Duhr:2019tlz,Caron-Huot:2020grv,Caron-Huot:2020vlo,Schnetz:2021ebf}. 
Note that the notation used for SVMPLs (${\cal I}$) in the \texttt{HyperlogProcedures} code, relates to the one we use here~(${\cal{G}}$) via
\begin{equation}
\label{HyperlogNotation}
{\cal{G}}(a_1,\ldots, a_n,z)
\equiv {\cal I}(z,a_1,\ldots, a_n,0)\,
\end{equation}
where the heading $z$ and the trailing $0$ represent the limits of the final (single-valued) integration. 
By plugging these results for the master integrals into eq.~(\ref{eq:RationalMaster}), we obtained the amplitude $\mathcal{M}^{(-,-),(1)}$ up to ${\cal{O}}(\epsilon^4)$ and, via eq.~(\ref{eq:REggNLLexpanded}), the one-loop Lipatov vertex $v^{(1)}(t_1,t_2,|\mathbf{p}_4|^2,\tau,\mu^2)$ to the same order in $\epsilon$.  
We provide a replacement list for the single-valued functions ${\cal G}$ into ordinary GPLs, $G$, which may evaluated numerically using the GiNaC library \cite{Bauer:2000cp} via PolyLogTools~\cite{Duhr:2019tlz}.

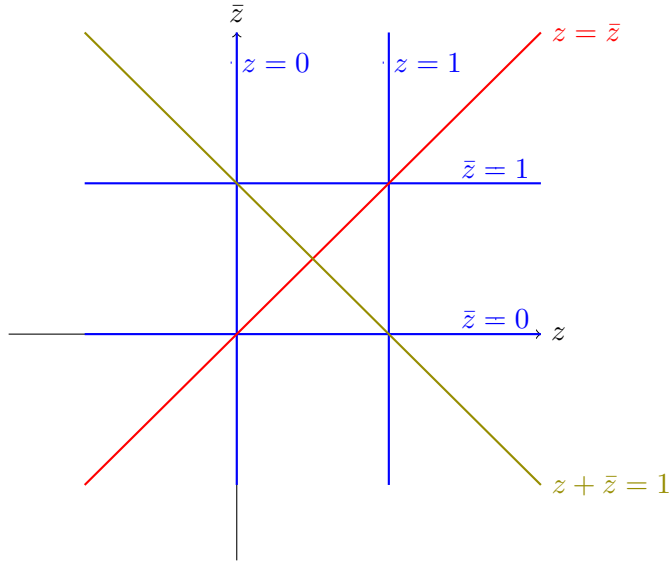
\begin{figure}
    \centering
    \begin{tikzpicture}[scale=2.]
        \draw[->] (-1.5,0) -- (2,0) node[right] {$z$};
        \draw[->] (0,-1.5) -- (0,2) node[above] {$\bar{z}$};
        
        \draw[blue, thick] (-.04,1.8)   -- (-0.04,1.8) node[right]{$z=0$};
        \draw[blue, thick] (.96,1.8)   -- (.96,1.8) node[right]{$z=1$};
        \draw[blue, thick] (1.7,1.1)   -- (1.7,1.1) node[behind path]{$\bar{z}=1$};
        \draw[blue, thick] (1.7,0.1)   -- (1.7,0.1) node[behind path]{$\bar{z}=0$};
        \draw[blue, thick] (-1,0) -- (2,0);
        \draw[blue, thick] (0,-1) -- (0,2);
         \draw[blue, thick] (-1,1) -- (2,1);
        \draw[blue, thick] (1,-1) -- (1,2);
        \draw[red, thick] (-1,-1) -- (2,2) node[right] {$z=\bar{z}$};
        \draw[olive, thick] (-1,2) -- (2,-1) node[right] {$z+\bar{z}=1$};
    \end{tikzpicture}
    \caption{Lines in the $z$ and $\bar{z}$ plane identifying the position of physical ($z=0,1$ and $\bar{z}=0,1$) and spurious ($z=\bar{z}$ and $z+\bar{z}=1$) singularities of the Lipatov vertex (or the amplitude at MRK). The physical singularities (drawn in blue) represent the zeros of the squared transverse momentum $t_{1}$ or $t_2$ of (\ref{t12}) carried by the Reggeized gluons (additional singularities appear when these invariant diverge, at $z\to \infty$ and $\bar{z}\to \infty$). The spurious singularity at $z=\bar{z}$ (in red) represents, at the level of the amplitude, the limit where ${\rm tr}_5\equiv -4\,i\,\epsilon_{\mu\nu\sigma\rho} 
P_1^{\mu}P_2^{\nu}P_3^{\sigma}P_5^{\rho}=0$, as can be seen from eq.~(\ref{tr5MRK}). Finally, spurious singularity at $z+\bar{z}=1$ (in olive green) corresponds to the soft limit where $|\mathbf{p}_4|^2\to 0$ and $t_1=t_2\equiv t$. }
    \label{fig:sing_lines}
\end{figure}

The dependence of $v^{(1)}(t_1,t_2,|\mathbf{p}_4|^2,\tau,\mu^2)$ on the factorisation and the renormalisation scales can be described to all-orders in $\epsilon$ by 
\begin{align}
\label{eq:v1scales}
    v^{(1)}(t_1,t_2,|\mathbf{p}_4|^2,\tau,\mu^2) &= \left(\frac{\mu^2}{|\mathbf{p}_4|^2}\right)^\epsilon\,v^{(1)}(t_1,t_2,|\mathbf{p}_4|^2)+\frac{b_0}{8\epsilon}\left[\left(\frac{\mu^2}{|\mathbf{p_4}|^2}\right)^\epsilon-1\right]\nonumber\\
    &-\frac{\alpha_g^{(1)}C_A}{2}\log\left(\frac{|\mathbf{p_4}|^2}{\tau}\right)\left[\left(\frac{\mu^2}{-t_1}\right)^\epsilon+\left(\frac{\mu^2}{-t_2}\right)^\epsilon\right]\,.
\end{align}
The procedure of obtaining eq.~(\ref{eq:v1scales}) is described in appendix~\ref{sec:ReggeCoefficients}, and it can by straightforwardly extended to higher-loop orders.
Using appendix~\ref{sec:ReggeCoefficients} and eq.~(\ref{eq:v1scales}) we have defined 
\begin{equation}
v^{(1)}(z,\bar{z})\equiv 
v^{(1)}(t_1,t_2,|\mathbf{p}_4|^2)\equiv 
v^{(1)}(t_1,t_2,|\mathbf{p}_4|^2,\tau=|\mathbf{p}_4|^2,\mu^2=|\mathbf{p}_4|^2).
\end{equation}
This single complex function captures the complexity of the one-loop Lipatov vertex, due to its
dependence on the three transverse momenta $\mathbf{q}_1, \mathbf{q}_2$ and $\mathbf{p}_4$. $v^{(1)}(z,\bar{z})$ is an analytic function of the complex conjugate pair of variables $z$ and $\bar{z}$, introduced in eq.~(\ref{def:zzbar}). Figure \ref{fig:sing_lines} describes the location of singularities and potential singularities of the vertex as a function of $z$ and $\bar{z}$.
The lines $z=0,1$ and $\bar{z}=0,1$ in the $(z,\bar{z})$ plane mark the physical singularities of the amplitude. These lines correspond to the poles associated to the Reggeized gluon propagators with virtualities $t_1$ and $t_2$ going on shell. There are two more regions in the $(z,\bar{z})$ plane associated to potential singularities of the amplitude. One is the line $z=\bar{z}$, where the invariant $\text{tr}_5$, defined in eq.~(\ref{tr5}), vanishes, following eq.~(\ref{tr5MRK}). 
This corresponds to the situation where all 5 parton 3-momenta are lying in a plane, or alternatively 
that the three transverse momenta of the vertex are aligned, i.e. they admit \begin{equation}
    (\mathbf{q}_1\cdot \mathbf{q}_2)^2-|\mathbf{q}_1|^2|\mathbf{q}_2|^2=\frac{|\mathbf{p}_4|^4(z-\bar{z})^2}{4}\to 0\,.
\end{equation}
The second potential singularity is the line $1-z-\bar{z}=0$, corresponding to the soft limit, $|\mathbf{p}_4|^2\to 0$, where all components of the emitted gluon momentum vanish and the two $t$ channel invariants become equal, $t_1=t_2\equiv t$.

It is interesting to examine 
the soft limit of the vertex in more detail. 
In Figure \ref{fig:sing_lines} the line $z+\bar{z}=1$ corresponds to this limit. We note that reflection about this line, eq.~(\ref{target_projectile}), correspond to target-projectile symmetry. 
Because the $|\mathbf{p}_4|^2\to 0$ limit is taken keeping $t_1=-|\mathbf{q}_1|^2$ and $t_2=-|\mathbf{q}_2|^2$ finite, the variables 
 $z$ and $\bar{z}$ develop in this limit large imaginary parts (see eq.~(\ref{t12})):
\begin{equation}
\label{eq:SoftLimit2}
    z(\bar{z})  \to  \frac{1+\delta}{2} \pm  i \sqrt{\frac{|\mathbf{q}_2|^2}{|\mathbf{p}_4|^2} - \frac{(1-\delta)^2}{4}} \,\,,\qquad \quad |\mathbf{p}_4|\to 0,\quad \delta=\frac{|\mathbf{q}_1|^2-|\mathbf{q}_2|^2}{|\mathbf{p}_4|^2}\to0\,.
\end{equation}
In ref.~\cite{Fadin:1996yv,DelDuca:1998cx,Fadin:2023roz} the sum and the difference of the right and left vertices\footnote{Equivalently, one may use $\widetilde{v}_R$ and $\widetilde{v}_L$ defined here in (\ref{def:vTildePlus}), since the expansion of the cosine functions in (\ref{LipatovAnalytics2muB}) does not contribute at one loop.} of eq.~(\ref{vRL}), have been computed to all orders in $\epsilon$. 
Using these results in eqs.~(\ref{def:voplusDisp}) and (\ref{def:voplusAbs}) respectively, we derive the soft limit of the dispersive and absorptive parts of the complex vertex $v_{\oplus}$ at one loop (defined by eq.~(\ref{REggNLL}) for positive helicity) to all orders in $\epsilon$:
\begin{align}
\label{softOneLoopVertex}
\begin{split}
\lim_{|\mathbf{p}_4|^2\to 0} \!v_{\oplus}^{(1)}(t_{1},t_{2},|\mathbf{p}_4|^2,\tau)=1+\frac{\alpha_s(\mu^2)}{\pi}
\left\{
-\frac{b_0}{8\epsilon}
+\frac{N_c}{4}\frac{r_\Gamma(\epsilon)}{\epsilon}
\left[
2\left(\frac{\mu^2}{-t}\right)^\epsilon 
\ln\left(\frac{\tau}{|\mathbf{p}_4|^2}\right)
\right.
\right.
\\
\left.
\left.
-\left(\frac{\mu^2}{|\mathbf{p}_4|^2}\right)^\epsilon
\bigg(
\frac{1}{\epsilon}
+\Psi(1-\epsilon)-\Psi(1+\epsilon)+\frac{i\pi}{2}
\bigg)
\right]
\right\}
\end{split}
\end{align}
where
\[
\lim_{|\mathbf{p}_4|^2\to 0} t_1 = \lim_{|\mathbf{p}_4|^2\to 0} t_2\equiv t\,,
\]
and $r_\Gamma(\epsilon)$ is defined in eq.~(\ref{alphagOneLoop}). 
This result in the soft limit (\ref{softOneLoopVertex}) provides a useful check of our general kinematics calculation below.

We stress that while there are alphabet letters (\ref{eq:alphabet}) which vanish on $z=\bar{z}$ or on $z+\bar{z}=1$, there are
no physical singularities in these kinematic limits, only spurious poles. In fact, spurious poles only appear as $z+\bar{z}\to 1$, and the terms containing these poles are known to all orders in $\epsilon$~\cite{Fadin:1994fj,Fadin:2000yp,Fadin:2023roz} and it is convenient to single them out as follows, and separate the one-loop vertex as follows
\begin{equation}
\label{eq:vREGvsSPU}
v_{\oplus}^{(1)}(t_1,t_2,|p_4|^2) = v^{(1)}_{\text{sYM}}(t_1,t_2,|p_4|^2) + v^{(1)}_{\text{spurious}}(t_1,t_2,|p_4|^2)+v_{\beta}^{(1)}(\epsilon),
\end{equation}
where we isolate the renormalisation term
\begin{equation}
    v_\beta^{(1)}(\epsilon)=-\frac{11N_c-2n_f}{24\epsilon},
\end{equation}
and where $v^{(1)}_{\text{spurious}}$ reads
\begin{equation}
\label{v1spur}
 v^{(1)}_{\text{spurious}}(t_1,t_2,|p_4|^2) = c_a\,v^{(1)}_a(t_1,t_2,|p_4|^2) + c_b\,v^{(1)}_b(t_1,t_2,|p_4|^2),  
\end{equation}
with 
\begin{equation}
\label{def:SpuriousVColour}
    c_a = N_c\,(1-\epsilon)-n_f,\qquad c_b=-(11-7\epsilon)\,N_c+2(1-\epsilon)\,n_f
\end{equation}
and
\begin{subequations}
\label{vab}
\begin{align}
   v^{(1)}_a &= e^{\epsilon\,\gamma_\text{E}} 
\frac{\Gamma(\epsilon)\Gamma(1\epsilon)^2
(|\mathbf{p}_4|^2)^\epsilon}{\Gamma(4-2\epsilon)
\bar{z}(1-z)}
\Bigg[
   -\frac{\epsilon}{4}\frac{(-t_1)^{1-\epsilon}-(-t_2)^{1-\epsilon}}{(t_1-t_2)}\nn\\
   &+\frac{|\mathbf{p}_4|^2(1+z-\bar{z})}{2}\left(\frac{t_1t_2\,((-t_1)^{-\epsilon}-(-t_2)^{-\epsilon})}{(t_1-t_2)^3}
   - \epsilon\frac{(-t_1)^{1-\epsilon}+(-t_2)^{1-\epsilon}}{2(t_1-t_2)^2}\right)
   \Bigg]
   \\
   v^{(1)}_b&= -e^{\epsilon\,\gamma_{\text{E}}} |\mathbf{p}_4|^2(1+z-\bar{z})\,\frac{\Gamma(\epsilon)\Gamma(1-\epsilon)\Gamma(2-\epsilon)(|\mathbf{p}_4|^2)^\epsilon}{\Gamma(4-2\epsilon)}\frac{(-t_1)^{-\epsilon}-(-t_2)^{-\epsilon}}{4\,(t_1-t_2)}\,.
\end{align}
\end{subequations}
The expansion in $\epsilon$ of eq.~(\ref{vab}) is straightforwards: $v^{(1)}_{\text{spurious}}$ is finite as $\epsilon\to0$ and the only transcendental functions appearing are (single-valued) logarithms of $|z|^2$ or of $|1-z|^2$. 

Importantly, both $v^{(1)}_a$ and $v^{(1)}_b$ vanish in the soft limit, eq.~(\ref{eq:SoftLimit2}), despite the denominators $t_1-t_2$ in eq.~(\ref{vab}).
This can be checked to all orders in $\epsilon$, by writing $t_1$ and $t_2$ in terms of $z$ and $\bar{z}$ and taking the (commuting) expansions around $|\mathbf{p}_4|\to0$ and $\delta\to0$ to find that both $v^{(1)}_a$ and $v^{(1)}_b$ are power suppressed. 

We note that $v^{(1)}_{\text{spurious}}$ does not contribute at maximal transcendental weight, hence it should not~\cite{Kotikov:2001sc,Kotikov:2002ab} contribute to the $\mathcal{N}=4$ super Yang Mills (sYM) vertex.
Indeed, by inspecting  the limit $\epsilon\to0$ of the colour factors entering $v^{(1)}_{\text{spurious}}$, eq.~(\ref{def:SpuriousVColour}), we recover the supersymmetric decomposition presented in Ref.~\cite{DelDuca:1998cx} 
\begin{equation}
\label{eq:cacbEPS0}
    c_a\to N_c-n_f,\qquad c_b\to -(11N_c-2n_f).
\end{equation}
This shows that neither $c_a$ or $c_b$ can contribute to the sYM vertex: the former vanishes in the supersymmetric limit, while the latter is proportional to the running of the coupling, much like $v^{(1)}_{\beta}$, which vanishes for ${\cal N}=4$ sYM. 

The remaining contribution to the one-loop vertex in eq.~(\ref{eq:vREGvsSPU}) 
is $v^{(1)}_{\text{sYM}}(t_1,t_2,|p_4|^2)$, which can be identified as the Lipatov vertex in $\mathcal{N}=4$ sYM. This contribution is free of spurious poles, and is expressed exclusively in terms of SVMPLs of maximal weight, in line with general expectations~\cite{Kotikov:2001sc,Kotikov:2002ab}. We extracted this function through ${\cal{O}}(\epsilon^4)$ (weight 6).
We also present below the explicit expression for $v^{(1)}_{\text{sYM}}$ through ${\cal{O}}(\epsilon^2)$, separated into its dispersive and absorptive components:
\begin{subequations}
    \begin{align}
    \label{eq:Dispvreg}
    \text{Disp.}&\{v^{(1)}_{\text{sYM}}\}=-\frac{N_c}{4\epsilon^2}+\frac{N_c}{8}\left[\frac{5\pi^2}{6}-\log^2\left(\frac{|z|^2}{|1-z|^2}\right)\right]\nonumber\\
    &+\epsilon\,N_c\,\Big[\frac{{\cal{G}}^3(0,z)+{\cal{G}}^3(1,z)}{12}-\frac{1}{4}\Big({\cal{G}}(0,0,1,z)-{\cal{G}}(0,1,0,z)+{\cal{G}}(0,1,1,z)\nonumber\\
    &+2{\cal{G}}(0,1-\bar{z},0,z)-2{\cal{G}}(0,1-\bar{z},1,z)+{\cal{G}}(1,0,0,z)+{\cal{G}}(1,0,1,z)+{\cal{G}}(1,1,0,z)\Big)\nonumber\\
    &+\frac{\pi^2}{24}\,\log\left(|z|^2|1-z|^2\right)+\frac{13}{12}\,\zeta_3\Big]\nonumber\\
    &+\epsilon^2\,N_c\,\Big[\frac{1}{4}\Big({\cal{G}}(0,0,0,1,z)-{\cal{G}}(0,0,1,0,z)-{\cal{G}}(0,1,0,0,z)+{\cal{G}}(0,0,1,1,z)\nonumber\\
    &+2{\cal{G}}(0,0,1-\bar{z},0,z)-2{\cal{G}}(0,0,1-\bar{z},1,z)+{\cal{G}}(0,1,0,1,z)-{\cal{G}}(0,1,1,0,z)\nonumber\\
    &+{\cal{G}}(0,1,1,1,z)+2{\cal{G}}(0,1,1-\bar{z},0,z)-2{\cal{G}}(0,1,1-\bar{z},1,z)+2{\cal{G}}(0,1-\bar{z},0,0,z)\nonumber\\
    &-2{\cal{G}}(0,1-\bar{z},1,1,z)+{\cal{G}}(1,0,0,0,z)+{\cal{G}}(1,0,0,1,z)-{\cal{G}}(1,0,1,0,z)+{\cal{G}}(1,0,1,1,z)\nonumber\\
    &+2{\cal{G}}(1,0,1-\bar{z},0,z)-2{\cal{G}}(1,0,1-\bar{z},1,z)+{\cal{G}}(1,1,0,0,z)+{\cal{G}}(1,1,0,1,z)\nonumber\\
    &+{\cal{G}}(1,1,1,0,z),+2{\cal{G}}(\bar{z},0,1,0,z)-2{\cal{G}}(\bar{z},0,1-\bar{z},0,z)+2{\cal{G}}(\bar{z},0,1-\bar{z},1,z)\nonumber\\
    &-2{\cal{G}}(\bar{z},1,0,1,z)-2{\cal{G}}(\bar{z},1,1-\bar{z},0,z)+2{\cal{G}}(\bar{z},1,1-\bar{z},1,z)\Big)-\frac{{\cal{G}}^4(0,z)+{\cal{G}}^4(1,z)}{32}\nonumber\\
    &-\frac{\pi^2}{96}\Big({\cal{G}}^4(0,z)+{\cal{G}}^2(1,z)+2(9{\cal{G}}(0,1,z)+{\cal{G}}(1,0,z)-8{\cal{G}}(\bar{z},1,z))\Big)\nonumber\\
    &-\frac{\zeta_3}{4}({\cal{G}}(0,z)+{\cal{G}}(1,z))+\frac{13\pi^4}{1920}\Big],
    \end{align}
    \begin{align}
    \label{eq:Absovreg}
    \text{Absorp.}&\{v^{(1)}_{\text{sYM}}\}=\frac{\pi N_c}{8}\Big[\frac{1}{\epsilon}+\epsilon\left({\cal{G}}(0,1,z)-{\cal{G}}(1,0,z)+\frac{\pi^2}{12}\right)-\epsilon^2\Big({\cal{G}}(0,0,1,z)-{\cal{G}}(0,1,0,z)\nonumber\\
    &+{\cal{G}}(0,1,1,z)-{\cal{G}}(1,0,0,z)+{\cal{G}}(1,0,1,z)-{\cal{G}}(1,1,0,z)-2{\cal{G}}(\bar{z},0,1,z)\nonumber\\
    &+2{\cal{G}}(\bar{z},1,0,z)-\frac{7\zeta_3}{3}\Big)\Big],
\end{align}
\end{subequations}
where the functions ${\cal{G}}$ are SVMPLs, which can be converted to multi-valued Goncharov multiple polylogarithms using the package~\texttt{HyperlogProcedures} \cite{Schnetz:2021ebf}
(recall the relation between notations in (\ref{HyperlogNotation})).
In the ancillary files~\cite{Ancillary}, we provide both expressions of $v^{(1)}_{\text{sYM}}$ up to ${\cal{O}}(\epsilon^4)$. The expression written in terms of SVMPL is very compact. Writing it in terms of these functions manifests the absence of branch points in the physical region, where $z$ and $\bar{z}$ are complex conjugates. 

In order to compare with the literature~\cite{Fadin:2000yp,Fadin:2023roz} it is useful to relate the dispersive (absorptive) part of the vertex to the sum (difference) of the left and right vertices $\widetilde{v}_L$ and $\widetilde{v}_R$ using eq. (\ref{def:voplusDisp}) and (\ref{def:voplusAbs}), respectively.
These, we recall, are related through contraction with the polarization vector 
${v}_R\pm v_L= \varepsilon^{\mu}_{\oplus}(P_4)\,(R_\mu\pm L_\mu)$ to the quantities $R^\mu\pm L^\mu$ that were determined in Ref.~\cite{Fadin:2023roz} through ${\cal{O}}(\epsilon^2)$, in terms of polylogarithmic functions and of $\mathcal{M}$-functions~\cite{DelDuca:2009ac}. The latter arise in one of the master integrals computed in Refs.~\cite{Fadin:2000yp,Fadin:2023roz}, named ${\cal{I}}_5-{\cal{L}}_3$, which involves the 4-dimensional scalar pentagon integral. To better compare with our results in eqs.~(\ref{eq:Dispvreg}) and (\ref{eq:Absovreg}), we derived a different representation of the 4-dimensional pentagon integral in terms of the master integrals $I_{\sigma,i}$ of eq.~(\ref{eq:RationalMaster}).
We find that $R^\mu+ L^\mu$ contains spurious poles at $t_1=t_2$, which match $v^{(1)}_{\text{spurious}}$ of eq.~(\ref{def:SpuriousVColour}). The remaining terms in $R^\mu + L^\mu$ and $R^\mu- L^\mu$ agree with eqs.~(\ref{eq:Dispvreg}) and (\ref{eq:Absovreg}), respectively.

It is interesting to see how the two properties we discussed in section~\ref{sec:VertexSymmetry}, reality and target-projectile symmetry, are realised in the one loop vertex. 
We have seen that the reality property is only expected to hold prior to contraction with the polarization vector, so while the transcendental functions can be chosen hermitian (\ref{TargetProjectileSymmetryPhi}), the rational coefficients would not be so. 
In turn, target-projectile symmetry holds for the complex vertex, and the expectation  is that if the transcendental functions are chosen to be invariant under
\begin{equation}
\label{TPzzbar}
\{z\to 1-\bar{z},\,\,\,\bar{z}\to 1-{z}\}.
\end{equation}
so will be their rational coefficients. 

Let's now examine the results for the 
one-loop vertex. Consider first the contribution $v^{(1)}_{\text{spurious}}$. We see immediately that this term is real: it trivially obeys $z\leftrightarrow \bar{z}$ symmetry, since it depends only on $t_i$ of (\ref{t12}). We also observe that this contribution is manifestly symmetric under the target-projectile symmetry (\ref{TPzzbar}): in eq.~(\ref{vab}) both the rational prefactors and each of the terms depending on $t_1$ and $t_2$ separately have this symmetry. 

Consider next $v^{(1)}_{\text{sYM}}(t_1,t_2,|p_4|^2)$. Also here our expectations hold, but the way this is realised is more interesting, as the symbol alphabet~(\ref{eq:alphabet}) transforms non-trivially under both $z\leftrightarrow \bar{z}$ and under $z\leftrightarrow 1-\bar{z}$. 
While $\text{Disp.}\{v^{(1)}_{\text{sYM}}\}$ and $\text{Absorp.}\{v^{(1)}_{\text{sYM}}\}$ 
separately satisfy the target-projectile symmetry (\ref{TPzzbar}), they transform nontrivally under $z\leftrightarrow \bar{z}$, as indeed expected, and they both have a an even (symmetric) as well as an odd (antisymmetric) components under this transformation. 

We note in passing that while both the even and odd components of the dispersive part 
of $v^{(1)}_{\text{sYM}}(z,\bar{z})$, as well as 
the odd component of the absorptive part of $v^{(1)}_{\text{sYM}}(z,\bar{z})$, are all non-trivial (uniform weight) single-valued GPLs functions of the alphabet in eq.~(\ref{eq:alphabet}),  $\,\text{Absorp.}
\left\{v^{(1)}_{\text{sYM.Even}}(z,\bar{z})\right\}$ is merely a uniform weight transcendental constant at any given order in $\epsilon$, consisting of both even and odd $\zeta$ values. We obtain:
\begin{equation}
\text{Absorp.}
\left\{v^{(1)}_{\text{sYM.Even}}(z,\bar{z})\right\}
=\frac{\pi^3}{96}\epsilon+\frac{7\pi}{24}\zeta_3 
\epsilon^2 + \frac{47\pi^5}{11520} \epsilon^3
+\left(
-\frac{7\pi^3}{288} \zeta_3 + \frac{31\pi}{40}  \zeta_5\right)\epsilon^4\,+\cdots\,.
\end{equation}


\section{The theory of multi-Reggeon interactions}
\label{MRK-shockwave}

As discussed in section \ref{sec:Reggeization}
and \ref{sec:MatchingNLL}, up to NLL accuracy 
the $(-,-)$ component of the amplitude is 
entirely captured by the exchange of a single 
Reggeon. Beyond this logarithmic accuracy for 
${\cal M}^{(-,-)}$, and in general for the 
other signature components of the amplitude, 
a correct description requires to take into 
account the exchange of multiple Reggeons. 
A framework to calculate their contribution 
efficiently has been developed in 
Refs.~\cite{Caron-Huot:2013fea,Caron-Huot:2017fxr,Gardi:2019pmk,Falcioni:2020lvv,Falcioni:2021buo,Falcioni:2021dgr}.
The method is based on an effective-theory 
approach, in which rapidity evolution equations 
in the shockwave formalism are used to determine 
all multi-Reggeon contributions to the amplitude.
Subsequently, by matching to the high-energy 
expansion of the full amplitude, it is possible 
to extract matching coefficients which are not
determined by the rapidity evolution, namely  
the impact factors, the Regge trajectory and 
(starting with $2\to 3$ amplitudes) the 
Lipatov vertex. In this section we review the 
formalism and extend its application to the 
calculation of $2\to 3$ amplitudes, building 
on earlier work in Ref.~\cite{Caron-Huot:2013fea,Caron-Huot:2020vlo}.


\subsection{From the shock-wave formalism to Reggeon fields}

In the shockwave formalism~\cite{Balitsky:1995ub,
Jalilian-Marian:1996mkd,JalilianMarian:1996xn,JalilianMarian:1997gr,Caron-Huot:2013fea,Caron-Huot:2017fxr,Falcioni:2021buo} 
one describes the projectile as a product 
of (an indefinite number of) Wilson-line 
operators, $U({z_1}_{\perp})U({z_2}_{\perp})\ldots$, 
each of which extends over the infinite $+$ 
lightcone direction and located as a distinct 
transverse position $z_{\perp}$:
\begin{equation}
\label{Udef}
U(z_{\perp})={\cal P}\,\exp\left[ \int_{-\infty}
^{\infty}T^a A_{+}^a(x^{+} ,x^{-}=0, z_\perp) 
dx^+ \right]\,.
\end{equation}
The Wilson lines (\ref{Udef}) have 
rapidity divergences, so the operators 
$U({z_1}_{\perp})U({z_2}_{\perp})\ldots$ 
admit rapidity evolution equations 
\begin{equation}
\label{H_acting_on_UUU}
-\frac{d}{d\eta} \left[U({z_1}_{\perp})U({z_2}_{\perp})\ldots\right] 
= H \left[U({z_1}_{\perp})U({z_2}_{\perp})\ldots\right]\,,
\end{equation}
known as the Balitsky-JIMWLK 
equations~\cite{Balitsky:1995ub,Jalilian-Marian:1996mkd,JalilianMarian:1996xn,JalilianMarian:1997gr}, 
where the leading-order Hamiltonian 
is given by
\begin{equation}
\label{B-JIMWLK-H}
H=\frac{\alpha_s}{2\pi^2} \mu^{2\epsilon}\int 
[d{ z}_0]
\frac{{ z}_{0i} \cdot { z}_{0j}}{
\big({ z}_{0i}^2 { z}_{0j}^2\big)^{1-\epsilon}} 
\left(T_{i,L}^a T_{j,L}^a +T_{i,R}^a T_{j,R}^a 
- U_{\rm adj}^{ab}({ z}_0) 
(T_{i,L}^a T_{j,R}^b +T_{j,L}^a T_{i,R}^b )\right)\,,
\end{equation}
where $z_{0i}\equiv z_0-z_i$ and $[dz]\equiv d^{2{-}2\eps}z$.
Here we work exclusively in the $2-2\eps$ dimensional
transverse plane, and therefore we drop the boldface
notation we used earlier for the transverse components.
In eq.~(\ref{B-JIMWLK-H}) we used functional derivative 
operators, $T_{i,L}^a$ and $T_{i,R}^a$, which generate 
colour rotation. They are defined by
\begin{equation}
\label{Tder}
T_{i,L}^a \equiv T^aU({ z}_i)\frac{\delta}{\delta U({ z}_i)},
\qquad 
T_{i,R}^a \equiv U({ z}_i)T^a\frac{\delta}{\delta U({ z}_i)}\,.
\end{equation}
The action of the Hamiltonian $H$ on the product 
of Wilson lines involves an additional Wilson line 
in the adjoint representation $U_{\rm adj}^{ab}({ z}_0)$ 
generated through the interaction with the target 
shockwave. Thus this system of equations is 
non-linear and each iteration of the evolution 
involves an increasing number of Wilson lines. 
Clearly it is very hard to solve this system 
in general. However, in the perturbative regime 
where fields are weak, each Wilson $U(z_{\perp})$ 
is close to unity. The dynamics is then best 
described in terms of its logarithm $W$, defined 
via~\cite{Caron-Huot:2013fea}
\begin{equation}
U(z_{\perp}) = e^{ig_s T^a W^a(z_{\perp})}\,.
\end{equation}
The field $W$ can be identified as sourcing a 
single Reggeon exchanged in the $t$ channel. 
Wilson lines $U$ (and products thereof) can 
be thus described perturbatively, order by 
order in $g_s$ as an expansion in $W$ fields, 
all at the same transverse position,  
\bea\label{UexpansionW} \nn
U= e^{ig_s \,W^a T^a} &=& 1+ ig_s \, W^a \, T^a 
- \frac{g_s^2}{2} W^a W^b \, T^a T^b
-i\frac{g^3_s}{6}W^a W^b W^c\, T^a T^b T^c \\
&& + \, \frac{g^4_s}{24}W^a W^bW^cW^d\, T^aT^b T^c T^d 
+ {\cal O}(g_s^5 \,W^5),
\eea
where $n$ fields $W$ source $n$ 
Reggeons~\cite{Caron-Huot:2013fea}. 
We can also express the action of 
the colour rotation operators 
$iT_{j,L/R}^{a}$ defined in 
(\ref{Tder}) in terms of 
functional derivatives of $W$
\bea\label{CBH_formula}
iT_{j,L/R}^{a} &=& \frac{1}{g_s}\frac{\delta}{\delta W_j^a}
\pm\frac{1}{2} f^{abx} W_j^x \frac{\delta}{\delta W_j^b} 
-\frac{g_s}{12} W_j^x W_j^y (F^xF^y)^{a}{}_{b}
\frac{\delta}{\delta W_j^b} \nn \\ 
&&\hspace{0.0cm}-\,\frac{g_s^3}{720} W_j^x W_j^y W_j^z W_j^t 
(F^xF^yF^zF^t)^{a}{}_b\frac{\delta}{\delta W_j^b} + \ldots,
\eea
where we call $W_j\equiv W(z_j)$. 
In~\eqn{CBH_formula} we have introduced 
the Hermitian colour matrix 
$(F^x)^{a}{}_b\equiv if^{axb}$ 
for simplicity. 

Through this expansion, 
Refs.~\cite{Caron-Huot:2017fxr,Falcioni:2021buo} 
have translated the action of the Balitsky-JIMWLK 
Hamiltonian in eq.~(\ref{H_acting_on_UUU}) to an 
action on a vector of increasing order in the 
number of $W$ fields, which schematically 
takes the form
\begin{eqnarray}
\label{H_mat_form}
\begin{split}
\def\bra#1{\langle#1|}
\def\ket#1{|#1\rangle}
H  \left( 
\begin{array}{c}
  W      \\  WW     \\  WWW  \\  \cdots
\end{array}
\right) &\equiv&
\left(
\begin{array}{cccc}
 H_{1{\to}1} & 0  & H_{3{\to}1} & \ldots \\
 0 & H_{2{\to}2}  & 0  &  \ldots \\
 H_{1{\to}3} & 0  & H_{3{\to}3} & \ldots\\
 \cdots & \cdots  & \cdots & \cdots\\
\end{array}
\right)\left(
\begin{array}{c}
  W      \\  WW     \\  WWW  \\  \cdots\end{array}
\right)
\\&\sim&
\left(
\begin{array}{cccc}
 g_s^2 & 0  & g_s^4 & \ldots \\
 0 & g_s^2  & 0  &  \ldots \\
 g_s^4 & 0  & g_s^2 & \ldots\\
 \cdots & \cdots  & \cdots & \cdots\\
\end{array}
\right) \left(
\begin{array}{c}
  W      \\  WW     \\  WWW  \\  \cdots\end{array}
\right),    
\end{split}
\end{eqnarray}
where the first line defines the components 
of the Hamiltonian $H_{k\to m}$ mediating 
between a state with $k$ Reggeons and one 
with $m$ Reggeons (note that transitions 
between odd and even number of Reggeons 
are forbidden) while the second line 
indicates the order in $g_s$ at which 
each such component of the Hamiltonian 
begins. In this way the perturbative 
expansion of the Wilson lines in effect 
linearises the non-linear evolution, and 
upon working to a given order in the 
coupling, the system involves a restricted 
number of Reggeons.

In $2\to 2$ scattering $ij\to i'j'$ one 
defines (see~\cite{Falcioni:2021buo} for 
more details\footnote{Notice that we 
divide by the unobservable phase 
associated with the tree-level impact 
factors $C_{i/j}^{(0)}$, see e.g. 
\eqn{impactfactorscalcH}. The removal 
of this phase is taken into account 
also in~\cite{Falcioni:2021buo}, 
although not explicitly stated.}):
\bea \label{bracketDef}
\big| \psi_i \big\rangle \equiv 
\frac{\Big(Z_i C_i^{(0)}\Big)^{-1}}{2p_{i}^+} 
a_{i'}(p_{i'}) a^{\dag}_i(p_{i}) | 0\rangle,  
\qquad
\big\langle \psi_j \big| \equiv
\frac{\Big(Z_j C_j^{(0)}\Big)^{-1}}{2p_{j}^-} 
\langle 0 | a_{j'}(p_{j'}) a^{\dag}_j(p_j) ,
\eea
where $a^{\dag}_i$ and $a_{i'}$ are 
respectively creation and annihilation 
operators associated with the projectile, 
while $a^{\dag}_j$ and $a_{j'}$ are 
associated with the target. Next, 
modelling these states in terms of 
products of (an indefinite number of) 
Wilson lines $U$, one may represent them 
as an expansion in terms of states with a 
definite number $n$ of Reggeons $W$,
\bea \label{bracketDef2}
\big| \psi_i \big\rangle 
= \sum_{n=1}^{\infty}\ket{\psi_{i,n}},  
\qquad\quad
\big\langle \psi_j \big|
= \sum_{n=1}^{\infty}\bra{\psi_{j,n}}. 
\eea
An $n$ Reggeon state $\big| \psi_i \big\rangle$ can in turn be expanded  
\be \label{eq:ketexp}
\ket{\psi_{i,n}} =\ket{\psi_{i,n}}^{\text{LO}}
+\sum_{k=1}^{\infty}
\ket{\psi_{i,n}}^{\text{N}^k\text{LO}}\,.
\ee
Here leading order (LO) contributions 
arise from the expansion of a single 
Wilson line~$U(z_{\perp})$, in which 
case
\begin{subequations}
\label{123Reggeons} 
\beqa 
\ket{\psi_{i,1}}^{\text{LO}} &=& 
ig_s\,\T_i^a \, W^a(p)|0\rangle, \label{eq:psi1} \\
\ket{\psi_{i,2}}^{\text{LO}} &=& 
- \frac{g_s^2}{2} \T_i^a\T_i^b 
\int [\dbar q] \, W^a(q)W^b(p{-}q)|0\rangle, \label{eq:psi2} \\ 
\ket{\psi_{i,3}}^{\text{LO}} &=& 
-\, \frac{ig_s^3}{6} \T_i^a\T_i^b\T_i^c 
\int [\dbar q_1] [\dbar q_2] \, 
W^a(q_1)W^b(q_2)W^c(p{-}q_1{-}q_2)|0\rangle, \label{eq:psi3}
\eeqa
\end{subequations}
where we use the following convention 
for the Fourier transform of the $W$ fields:
\be\label{Wtransform}
W^a(p) = \int [dz] \,e^{-ipz}\,W^a(z), \qquad
W^a(z) = \int [\dbar p] \,e^{ipz}\, W^a(p).
\ee
and the measure is
\beq
[dz]\equiv d^{2{-}2\eps}z, \qquad \quad 
[\dbar p] \equiv \frac{d^{2-2\eps} p}{(2\pi)^{2-2\eps}}.
\eeq 
At next-to-leading order (NLO) two Wilson 
lines at distinct transverse positions play 
a role, and then non-trivial impact factors 
for the emission of a given number of Reggeons 
arise: 
\begin{subequations}
\beqa \label{NLO_wavefunction} 
\ket{\psi_{i,1}}^{\text{NLO}} &=& ig_s\,\T_i^a \,
 \frac{\as}{\pi} d_i^{(1)}(-p^2)\, W^a(p)|0\rangle, \\
\ket{\psi_{i,2}}^{\text{NLO}} &=& 
- \frac{g_s^2}{2} \T_i^a\T_i^b \int [\dbar q] 
\, \frac{\as}{\pi} \psi^{(1)}_i(p,q) \, W^a(q)W^b(p{-}q)|0\rangle\,.
\eeqa
\end{subequations}
Similarly, at next-to-next-to-leading order (NNLO) one has
\beq \label{NNLO_wavefunction}
\ket{\psi_{i,1}}^{\text{NNLO}} = ig_s\, \T_i^a \, 
\left(\frac{\as}{\pi}\right)^2 
d_i^{(2)}(-p^2,\tau)\, W^a(p)|0\rangle,
\eeq
and so on. In particular, the all orders 
generalization of the 1-Reggeon wavefunction 
reads
\beq \label{1Reggeon_wavefunction}
\ket{\psi_{i,1}} = ig_s\, \T_i^a \, 
d_i(- p^2,\tau)\, W^a(p)|0\rangle,
\eeq
where the impact factor $d_i(-p^2,\tau)$ 
has perturbative expansion
\beq \label{DImpactPertDef2}
d_{i}(-p^2,\tau) = 1 + 
\sum_{n} \left(\frac{\as}{\pi}\right)^n d_i^{(n)}(-p^2,\tau)\, .
\eeq
Comparing with eqs.~(\ref{ImpactPertDef1}), 
(\ref{ImpactPertDef2}) and~(\ref{bracketDef}), 
it is easy to see that the functions $d_i$ 
represent the perturbative correction to the
impact factors with collinear singularities 
removed: 
\beq \label{ImpFactIRsubtract}
C^{\rm SR}_i(p_i,p_{i'},\tau) = C^{(0)}_i(p_i,p_{i'}) \,
Z_{i}(t_i,\tau)\, d_{i}(t_i,\tau)\, ,
\eeq
or,  equivalently, 
\beq \label{ImpFactIRsubtract2}
c^{\rm SR}_{i}(t_i,\tau) = Z_{i}(t_i,\tau)\, d_{i}(t_i,\tau)\, .
\eeq
Notice that in \eqns{ImpFactIRsubtract}{ImpFactIRsubtract2}
and in what follows we add a superscript ``SR'', for ``single 
Reggeon'', in order to stress that the perturbative correction 
to the impact factors are defined as matching coefficients that arise naturally
within the multi-Reggeon effective theory, identifying a Reggeon with a $W$ field. To be precise, they are defined such that any contribution of the amplitude associated exclusively with a single $W$ field, throughout the entire rapidity span between the target and the projectile, is considered a SR transition, while all other contributions to the amplitude in which more than one $W$ field is involved in the transition (at any stage, even over part of the rapidity span or through mixing under evolution), is considered a MR transition. Matching in the SR/MR scheme means factorizing only the SR component of the amplitude when defining the impact factors, the Regge trajectory and the Reggeon-gluon-Reggeon vertex, while excluding all MR transitions.
The alternative, pole/cut scheme originally defined 
in~\cite{Falcioni:2021dgr,Falcioni:2021buo} will be discussed 
in section~\ref{sec:scheme}.

Having expressed the projectile and the 
target in terms of states consisting of 
a definite number of Reggeons, and 
additionally knowing how the Hamitonian acts 
on these states (see~(\ref{H_mat_form})) to 
generate rapidity evolution, we can determine 
how the projectile (and likewise the target) 
evolves, namely
\begin{equation}
\label{eq:PsiEvo}
-\frac{d}{d\eta}\big| \psi_i \big\rangle 
= H \big| \psi_i \big\rangle 
\end{equation}
With this we can compute $2 \to 2$ amplitudes 
in the  high-energy limit via
\begin{equation}
\label{Regge2to2FactBasic}
\frac{i}{2s} {\cal \bar M}_{ij\to ij}  \equiv 
\frac{i \big(Z_i Z_j C_i^{(0)} C_j^{(0)}\big)^{-1}}{2s}
{\cal M}_{ij\to ij} \equiv \braket{\psi_j|e^{-H \eta}|\psi_i}\, 
\end{equation}
where the factor $e^{-H \eta}$ corresponds to 
the evolution of the target to the rapidity of 
the projectile (or vice versa) where $\eta$ is 
logarithm of the energy ratio. This technique was 
developed and used in Ref.~\cite{Caron-Huot:2017fxr} 
to compute the NNLL tower of logarithms of 
$2 \to 2$ amplitudes to three loops and in
Refs.~\cite{Falcioni:2020lvv,Falcioni:2021buo,Falcioni:2021dgr} 
to four loops.


\subsection{Real emission from Reggeons}

In eq.~(\ref{Regge2to2FactBasic}) we expressed $2 \to 2$ high-energy amplitude through a rapidity evolution of target and projectile states that can be expanded in terms of a fixed number of Reggeons.
In a similar way, one writes the $2 \to 3$ 
amplitude $i(p_1)j(p_2)\to i'(p_5)g(p_4) j'(p_3)$ 
in MRK as follows~\cite{Caron-Huot:2013fea,Caron-Huot:2017fxr,Caron-Huot:2020vlo}:
\be\label{ReggeFactBasic}
\frac{i}{2s_{12}} {\cal \bar M}_{ij \to i'gj'} \equiv 
\frac{i \big(Z_i Z_j C_i^{(0)} C_j^{(0)}\big)^{-1}}{2s_{12}}
{\cal M}_{ij \to i'gj'} = 
\big\langle \psi_j \big| 
e^{-H \eta_{2}} \, a_4(p_4) \,
e^{-H \eta_{1}} \big| \psi_i \big\rangle, 
\ee
where $\eta_{1,2}\gg 1$ represents the large 
energy logarithms. These can be chosen to 
coincide with the logarithms in \eqn{complexLogsB},
namely
\begin{equation} \label{RapidityLogs}
\eta_1 = \log\left(\frac{s_{45}}{\tau}\right)-i\frac{\pi}{4}\,,\quad \qquad 
\eta_2 = \log\left(\frac{s_{34}}{\tau}\right)-i\frac{\pi}{4}\,,
\end{equation}
The new element of~\eqn{ReggeFactBasic} 
compared to the case of $2\to 2$ scattering 
is the presence of the annihilation operator 
for the gluon $a_4(p_4)$. The first evolution 
operator $e^{-H \eta_{1}}$ evolves the Reggeons 
appearing in the expansion of $|\psi_i\rangle$ 
to the rapidity of the produced gluon with 
momentum~$p_4$ (similarly, $e^{-H \eta_{2}}$
evolves the Reggeons appearing in $|\psi_j\rangle$ 
to the rapidity of the gluon $p_4$). Therefore,
we need to determine the action of the operator 
$a_4(p_4)$ on a set of Wilson lines $U(z_i)$, 
and the corresponding linearised action of 
$a_4(p_4)$ on Reggeon fields~$W$. This 
has been determined in \cite{Caron-Huot:2013fea},
and we review the derivation in what follows 
before proceeding to use it in  evaluating
(\ref{ReggeFactBasic}).

First of all, we need the action of the 
operator $a_4$ on a Wilson line $U(z_1)$, 
which reads~\cite{Caron-Huot:2013fea}
\be\label{Uona4}
U(z_1) \, a^a(p_2)  = 2 g_s 
\int [\dbar q] [dz_2] \frac{q\cdot\varepsilon(p_2)}{q^2}
e^{i q\cdot (z_2-z_1)}e^{-ip_2\cdot z_2} \,
\Big( U^{ab}_{\rm ad}(z_2) T_{1,R}^b - T_{1,L}^a \Big)
U(z_1) ,
\ee
where the polarization vector 
$\varepsilon(p_2) \equiv \boldsymbol{\varepsilon}(p_2)$
refers to the transverse component of the 
full gluon polarization vector\footnote{Taking 
into account our convention for Fourier 
transformation in \eqn{Wtransform}, 
\eqn{Uona4} is equivalent to the definition 
given in eq.~(5.7) of \cite{Caron-Huot:2013fea}. 
We check in section \ref{SRamplitude} that starting 
from \eqn{Uona4} we reproduce the correct tree level 
amplitude.}. Recall that 
$U^{ab}_{\rm ad}(z_2)$ is a Wilson line in 
the adjoint representation at position $z_2$, 
where the gluon crosses the shockwave, and 
the colour rotations $T^a_{R/L,i}$ are defined 
in eq.~(\ref{Tder}). Using eqs. (\ref{UexpansionW}) 
and (\ref{CBH_formula}) to expand 
$U^{ab}_{\rm ad}(z_2)$ and $T^a_{R/L,i}$ 
we obtain the master equation
\begin{align}\label{Uona4b}\nn
&U(z_1) \, a^a(p_2) \sim 2 i \, g_s 
\int [\dbar q] [dz_2] 
\frac{q\cdot\varepsilon(p_2)}{q^2}
e^{i q\cdot (z_2-z_1)}e^{-i p_2 \cdot z_2}
\bigg\{ i (F^{c})^a{}_b \Big[W_1^c - W_2^c\Big] \\\nn
&-\,\frac{g_s}{2} (F^{cd})^a{}_b
\Big[W_2^c \Big(W_1^d - W_2^d\Big)\Big] 
+\,\frac{g_s^2}{12} i (F^{cde})^a{}_b
\Big[W_2^c W_1^d W_1^e 
-3 W_2^c W_2^d W_1^e 
+ 2 W_2^c W_2^dW_2^e\Big] \\ 
& -\frac{g_s^3}{24} (F^{cdef})^a{}_b
\Big[ W_2^cW_2^dW_1^eW_1^f 
- 2 W_2^cW_2^dW_2^e W_1^f 
+ W_2^cW_2^dW_2^eW_2^f \Big]
\bigg\}\frac{\delta}{\delta W_1^{b}} U(z_1) \,.
\end{align}
where we included in the expansion up to 
terms of $\ord({W^4})$, using the shorthand 
notation for $W_i^c=W^a(z_i)$ and $F^{a_1a_2\dots a_n}
\equiv F^{a_1}F^{a_2}\dots F^{a_n}$. Eq.~(\ref{Uona4}) 
determines the action of the operator $a^a(p_2)$ on 
an arbitrary number of Reggeon fields. The 
corresponding vertices are obtained by 
expanding the Wilson line $U(z_1)$ up 
to the required number of Reggeons. 

\begin{figure}[t]
\begin{center}
\includegraphics[width=0.88\textwidth]{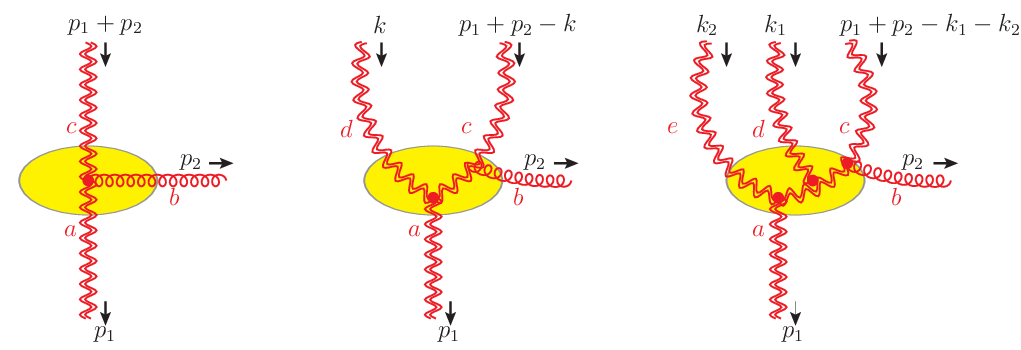}
\end{center}
\caption{Diagrams describing the three vertices 
as one Reggeon into one, two or three Reggeons 
plus an emitted real gluon. }
\label{2to3EffectiveVertices}
\end{figure}

\begin{figure}[t]
\begin{center}
	\includegraphics[width=0.78\textwidth]{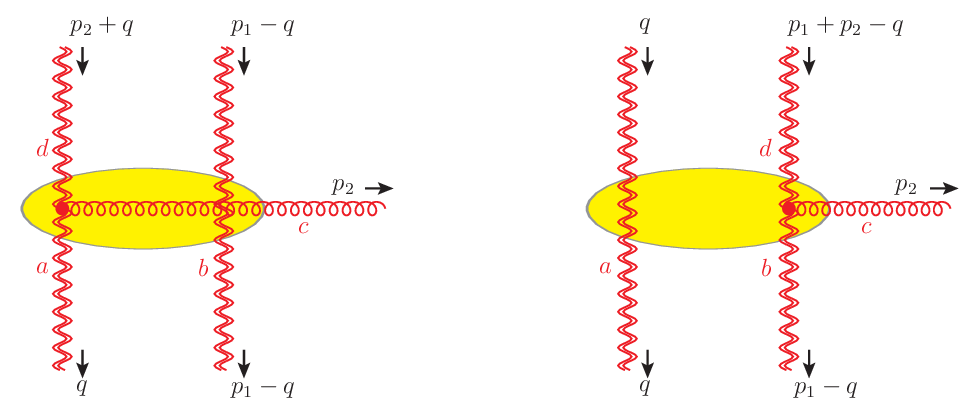}
\end{center}
\caption{A Diagram describing the two-to-two Reggeon 
vertex associated with the emission of a real gluon. }
\label{2to3EffectiveVertices2to2}
\end{figure}

\begin{figure}[t]
\begin{center}
	\includegraphics[width=0.98\textwidth]{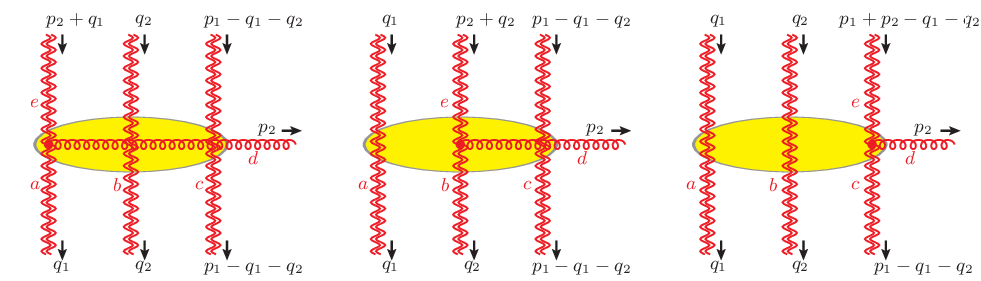}
\end{center}
\caption{A Diagram describing the three-to-three Reggeon 
vertex associated with the emission of a real gluon. }
\label{2to3EffectiveVertices3to3}
\end{figure}

First, we consider the action of $a^a(p_2)$ 
on a single Reggeon, by expanding $U(z_1)$ to 
the linear order in $W^a(z_1)$ and then taking 
the Fourier transform according to 
eq.~(\ref{Wtransform}). We get
\bea\label{Wona4}\nn
W^a(p_1) \, a^b(p_2) &\sim& \bigg\{ 
-2 i \, g_s \, f^{abc} W^c(p_1 + p_2) 
\bigg[\frac{p^{\mu}_1}{p_1^2}
+\frac{p_2^{\mu}}{p_2^2}\bigg] \\ \nn
&&\hspace{-1.0cm}-\, i\, g^2_s \, f^{bce}f^{ead}
\int [\dbar k] \,  W^c(p_1 + p_2-k) W^d(k) 
\bigg[\frac{p_1^{\mu}}{p_1^2}
-\frac{(p_1-k)^{\mu}}{(p_1-k)^2}\bigg] \\ \nn
&&\hspace{-1.0cm}+\,i\,\frac{g^3_s}{6}\,f^{bcx}f^{xdy}f^{yae}
\int [\dbar k_1][\dbar k_2] \, 
W^c(p_1 + p_2-k_1-k_2) W^d(k_1) W^e(k_2) \\ 
&&\hspace{1.0cm} \cdot \, \bigg[
\frac{(p_1-k_1-k_2)^{\mu}}{(p_1-k_1-k_2)^2}
-3\frac{(p_1-k_2)^{\mu}}{(p_1-k_2)^2}
+2\frac{p_1^{\mu}}{p_1^2}\bigg] 
\bigg\}\, \varepsilon_{\mu}(p_2)
+\ldots,
\eea
These contributions are described by 
figure~\ref{2to3EffectiveVertices}. 
The terms of higher order, which are 
omitted here, involve the coupling 
between a gluon and five or more 
Reggeons. These vertices start to 
contribute to N\textsuperscript{3}LL 
and therefore are beyond the scope 
of the present paper.
Next, we consider the term containing
two $W$ fields from the expansion of $U(z_1)$. 
At leading order we obtain
\bea\label{WWona4A} \nn
\int [\dbar q] W^{a}(q) W^b(p_1 - q) 
\, a^c(p_2) &\sim & \\ 
&&\hspace{-5.1cm}- 2 i \, g_s \int [\dbar q] 
\Big\{f^{acd} \, W^d(p_2+q)W^b(p_1  - q)
\bigg[\frac{p_2^{\mu}}{p_2^2}
+\frac{q^{\mu}}{q^2} \bigg] \\ \nn
&&\hspace{-3.3cm}+\,f^{bcd} \, 
W^a(q) W^d(p_1 + p_2 - q) 
\bigg[\frac{p_2^{\mu}}{p_2^2}
+\frac{(p_1-q)^{\mu}}{(p_1-q)^2}
 \bigg] \bigg\} \varepsilon_{\mu}(p_2)
 +\ldots,
\eea
where the ellipses stand for $WW \to WWW$ 
and $WW \to WWWW$ transitions not yet calculated, 
which will enter only at N\textsuperscript{3}LL.  
The two-to-two Reggeon vertex is shown in 
figure~\ref{2to3EffectiveVertices2to2}.
The last contribution necessary to achieve 
NNLL accuracy is given by the exchange of 
three Reggeons with emission of a gluon, 
which originates from considering the 
term containing three $W$ fields from 
the expansion of $U(z_1)$. At leading 
order this gives
\bea\label{WWWona4A} \nn
\int [\dbar q_1]\int [\dbar q_2] 
W^{a}(q_1) W^{b}(q_2) W^c(p_1 - q_1 - q_2) 
\, a^d(p_2) &\sim & \\ \nn
&&\hspace{-9.1cm}- 2 i \, g_s \int [\dbar q_1]\int [\dbar q_2] 
\Big\{f^{ade} \, W^e(p_2+q_1)W^b(q_2)W^c(p_1 - q_1- q_2)
\bigg[\frac{p_2^{\mu}}{p_2^2}
+\frac{q_1^{\mu}}{q_1^2} \bigg] \\ 
&&\hspace{-8.8cm}+\,f^{bde} \, W^a(q_1) W^e(p_2+q_2) W^c(p_1 - q_1- q_2)
\bigg[\frac{p_2^{\mu}}{p_2^2}
+\frac{q_2^{\mu}}{q_2^2} \bigg] \\ \nn
&&\hspace{-8.8cm}+\,f^{cde} \, W^a(q_1) W^b(q_2) W^e(p_1 + p_2 - q_1 - q_2)
\bigg[\frac{p_2^{\mu}}{p_2^2}
+\frac{(p_1-q_1-q_2)^{\mu}}{(p_1-q_1-q_2)^2} 
\bigg] \bigg\}\varepsilon_{\mu}(p_2)  +\ldots,
\eea
where the ellipses stand for $WWW \to WWWW$ 
and $WWW \to WWWWW$ transitions not yet calculated, 
which will enter starting at N\textsuperscript{3}LL.  
The three-to-three Reggeon vertex is shown in 
figure~\ref{2to3EffectiveVertices3to3}. 
Note that, as discussed in \cite{Caron-Huot:2013fea}, 
at this logaritmic accuracy, \eqns{WWona4A}{WWWona4A} 
corresponds to summing the application of the first 
line of \eqn{Wona4} respectively on $W^{a}$, $W^{b}$ 
and $W^{a}$, $W^{b}$ and $W^{c}$, with the Reggeons 
not involved in the emission acting as spectators. 

So far, all the vertices given in eqs.~({\ref{Wona4}}) 
and (\ref{WWona4A}) are derived in the framework of the 
shockwave formalism to leading order, see eq.~(\ref{Uona4}). 
In general, each vertex will receive new contributions 
beyond the leading order. In particular, the 
Reggeon-gluon-Reggeon vertex, which we denote by 
${\cal R} g {\cal R}$, has the structure
\beq\label{Wona4B}
W^a(p_1) \, a^b(p_2) \sim - 2 i \, g_s \, f^{abc} \, 
v^{\rm SR}\big[(p_1+p_2)^2, p_1^2, p_2^2,\tau\big] 
\, W^c(p_1 + p_2) \bigg[\frac{p_1^{\mu}}{p_1^2}
+\frac{p_2^{\mu}}{p_2^2}\bigg] \varepsilon_{\mu}(p_2),
\eeq
where the function $v^{\rm SR}[p_1^2,(p_1+p_2)^2, p_2^2,\tau]
= 1 + \ord(\as)$ identifies loop corrections to 
the Lipatov vertex: as we will see below, it 
coincides with the function defined in 
eq.~(\ref{REggNLL}), barring 
the scheme in which Regge factorization is applied:
as explained following  
\eqns{ImpFactIRsubtract}{ImpFactIRsubtract2}, here, within the context of the multi-Reggeon effective theory, we naturally use the SR/MR scheme, hence the superscript ``SR'' on $v^{\rm SR}$. In section~\ref{sec:scheme} the vertex will be related, along with the impact factors and Regge trajectory, to the  alternative, pole/cut scheme of~\cite{Falcioni:2021dgr,Falcioni:2021buo}.
Note also that in the present section we use simplified notation, 
omitting the polarization index ($\lambda_4$) corresponding to 
the helicity of the emitted gluon. The one-loop correction 
contributes to the $2\to3$ amplitude at NLL in the high-energy 
limit; the two-loop corrections will become relevant at NNLL.


\subsection{Identifying single Reggeon  contributions to $2\to 3$ amplitudes}
\label{SRamplitude}

We are now in a position to consider again 
the amplitude in~\eqn{ReggeFactBasic}, and  
separate single-Reggeon from multiple-Reggeon 
exchanges, with the aim to reproduce the Regge 
factorisation formula in \eqn{REggNLL}.
To this end we split the amplitude in 
\eqn{ReggeFactBasic} into single Reggeon (SR)
and \emph{Multi-Reggeon} (MR) components:
\beq\label{ReggeFactBasicbis}
{\cal \bar M}_{ij \to i'gj'} =
{\cal \bar M}^{\rm SR}_{ij \to i'gj'} + 
{\cal \bar M}^{\rm MR}_{ij \to i'gj'},
\eeq
and we decompose the Hamiltonian~$H$ 
in \eqn{eq:PsiEvo} accordingly, i.e. we 
separate the evolution of a single Reggeon 
from the remaining Multi-Reggeon components:
\begin{align}
H = H_{1\to 1} + H_{\rm MR}, 
\qquad\text{where} \qquad  
H_{\rm MR}\equiv\sum_{n,m} H_{n\to m}(1-\delta_{n1}\delta_{m1}),
\end{align}
where we identify the single-Reggeon term 
$H_{1\to 1}$ with the gluon Regge trajectory 
$\al^{\rm SR}_g(t)$ times the $t$-channel 
colour charge $-\T_{t_i}^2$:
\beq \label{eq:H11}
H_{1\to 1}\,\eta_i = -\T_{t_i}^2\al^{\rm SR}_g(t_i)\,\eta_i.
\eeq 
Let us stress once again that the 
superscript ``SR'' indicates that 
$\al^{\rm SR}_g(t)$ identifies the Regge 
trajectory in the SR/MR scheme, as explained following  
\eqns{ImpFactIRsubtract}{ImpFactIRsubtract2}. 
The relation will the pole/cut scheme 
\cite{Falcioni:2021dgr,Falcioni:2021buo},  
will be presented in section \ref{sec:scheme}.

The single Reggeon amplitude involves exclusively 
the exchange of a single Reggeon in both the $\eta_1$ 
and $\eta_2$ rapidity spans, thus we have
\beq\label{ReggeFactBasicb}
\frac{i}{2s_{12}}
{\cal \bar M}^{\rm SR}_{ij \to i'gj'} = 
\big\langle \psi_{j,1} \big| 
e^{-H_{1\to 1} \eta_{2}} \, a_4(p_4) \,
e^{-H_{1\to 1} \eta_{1}} \big| \psi_{i,1} \big\rangle.
\eeq
Instead, the Multi-Reggeon amplitude 
${\cal \bar M}^{\rm MR}_{ij \to i'gj'}$
involves $H_{\rm MR}$ in either or both 
of the rapidity spans, and we postpone 
its analysis to section \ref{sec:MRE}.
Here we proceed by elaborating the SR component. 
Using \eqn{eq:H11}, as well as 
\eqns{1Reggeon_wavefunction}{ImpFactIRsubtract} 
in eq.~(\ref{ReggeFactBasic}) we obtain
\beqa\label{ReggeFactBasicc} \nonumber
\frac{i}{2s_{12}}
{\cal M}^{\rm SR}_{ij \to i'gj'} &=& 
- \Big[g_s\,\T_i^x \, C^{\rm SR}_i(p_i,p_i',\tau) \Big]
e^{C_A \, \al^{\rm SR}_g(t_1)\,\eta_1}
\Big[ g_s \, \T_j^y \, C^{\rm SR}_j(p_j,p_j',\tau) \Big]
e^{C_A \, \al^{\rm SR}_g(t_2)\,\eta_2} \\[0.1cm]
&&\hspace{4.0cm}\times\,
\big\langle 0 \big| W^y(-q_2) 
 \, a_4(p_4) \, W^x(q_1) \big| 0 \big\rangle.
\eeqa
We can now take into account the emission 
of a central gluon by means of 
\eqn{Wona4B}. This gives
\beqa\label{RgRtree1} \nonumber
\big\langle 0 \big| W^y(-q_2) 
 \, a^{a_4}_4(p_4) \, W^x(q_1) \big| 0 \big\rangle &=&  
- 2 i \, g_s \, f^{xa_4y'} \, v^{\rm SR}(t_1,t_2, p_4^2,\tau) 
\bigg[\frac{q_1^{\mu}}{q_1^2}
+\frac{p_4^{\mu}}{p_4^2}\bigg]\varepsilon_{\mu}(p_4) \\ 
&&\hspace{1.5cm}\times \, 
\big\langle 0 \big| W^{y}(-q_2) \, W^{y'}(q_1+p_4) \big| 0 \big\rangle\,.
\eeqa
At this point we are left with the expectation 
value in the last line, which simply gives the free 
Reggeon propagator\footnote{Note that we follow 
the convention that Reggeons emitted from the 
projectile $i$ have incoming momentum, while
Reggeons Absorbed into the target $j$ have 
outgoing momentum, such that momentum in
the transverse plane flows from $j$ to $i$.} 
\cite{Caron-Huot:2013fea,Caron-Huot:2017fxr}: 
\beq
\label{Reggeon_prop}
\big\langle 0 \big| W^{y}(-q_2) \, W^{y'}(q_1+p_4) \big| 0 \big\rangle
= \frac{i\delta^{yy'}}{q_2^2}(2\pi)^{2-2\eps}\delta^{(2-2\eps)}(q_1 + q_2 + p_4)\,.
\eeq
Inserting this result into \eqn{RgRtree1}
we obtain 
\beqa\label{RgRtree2} \nonumber
\big\langle 0 \big| W^y(-q_2) \, 
a^{a_4}_4(p_4) \, W^x(q_1) \big| 0 \big\rangle &=&  
2 \, g_s \, f^{xa_4y} \, v(t_1,t_2, p_4^2,\tau)
\frac{(2\pi)^{2-2\eps} \delta^{(2-2\eps)}(q_1 + q_2 + p_4)}{q_1^2\, q_2^2} \\
&&\hspace{4.0cm}\times \,
\bigg[\frac{q_1^{\mu}}{q_1^2}
+\frac{p_4^{\mu}}{p_4^2}\bigg]\varepsilon_{\mu}(p_4) \,.
\eeqa
We can easily check that this is indeed 
proportional to the tree-level Lipatov 
vertex. Taking into account that 
$v(t_1,t_2, p_4^2,\tau) = 1 + \ord(\as)$, 
at tree level we can evaluate 
\eqn{RgRtree2} in $d = 2$, such that the 
relation in \eqn{polvectprod} applies; 
one has $(q_1^{\mu}/q_1^2
+p_4^{\mu}/p_4^2)\varepsilon_{\mu}(p_4)
= -1/\sqrt{2} ({\bf \bar q}_2 {\bf q_1})/{\bf \bar p_4}$;
enforcing momentum conservation according 
to \eqns{momentumconservation1}{q1q2defB}
we get 
\beqa\label{RgRtree3} \nonumber
\big\langle 0 \big| W^y(-q_2) \, 
a^{a_4}_4(p_4) \, W^x(q_1) 
\big| 0 \big\rangle\big|_{\rm tree} && \\[0.2cm] \nn
&&\hspace{-4.0cm}=\,
\sqrt{2} \, g_s \, f^{ya_4x} \, 
\frac{{\bf \bar p}_3 {\bf p_5}}{\bf \bar p_4} \, 
\frac{(2\pi)^{2-2\eps} \delta^{(2-2\eps)}({\bf p}_3
+{\bf p}_4+{\bf p}_5)}{t_1\, t_2} \\
&&\hspace{-4.0cm}=\, g_s \, f^{ya_4x} 
V^{(0)}(Q_1,P_4^{\oplus},Q_2) 
\, \frac{(2\pi)^{2-2\eps} \delta^{(2-2\eps)}({\bf p}_3
+{\bf p}_4+{\bf p}_5)}{t_1\, t_2} \,,
\eeqa
where in the last line we have identified the
tree-level Lipatov vertex $V^{(0)} = \sqrt{2}\,   
{\bf \bar p}_3 {\bf p_5} / {\bf \bar p_4}$ 
as given in \eqn{LipatovTree}.
Beyond tree level, comparing \eqn{RgRtree2}
with \eqn{RgRtree3} we readily obtain
\beqa\label{RgRtree4} \nn
\big\langle 0 \big| W^y(-q_2) \, 
a^{a_4}_4(p_4) \, W^x(q_1) \big| 0 \big\rangle && \\[0.2cm] 
&&\hspace{-5.0cm}=\, g_s \, f^{ya_4x} \, 
V^{(0)}(Q_1,P_4^{\oplus},Q_2) \, v^{\rm SR}(t_1,t_2, {\bf p}_4^2,\tau)\,
\frac{(2\pi)^{2-2\eps} \delta^{(2-2\eps)}({\bf p}_3 
+{\bf p}_4+{\bf p}_5)}{t_1\, t_2} \,.
\eeqa
Inserting \eqn{RgRtree4} into \eqn{ReggeFactBasicc}
we finally obtain
\beqa\label{eq:ampDecomp} \nonumber
{\cal M}^{\rm SR}_{ij \to i'gj'} &=& 
2 s_{12} \, \Big[g_s\,\T_i^x \, C^{\rm SR}_i(p_i,p_i',\tau) \Big]
e^{C_A \, \al^{\rm SR}_g(t_1)\,\eta_1}
\Big[ g_s \, \T_j^y \, C^{\rm SR}_j(p_j,p_j',\tau) \Big]
e^{C_A \, \al^{\rm SR}_g(t_2)\,\eta_2} \\[0.1cm]  \nn
&&\hspace{-1.0cm}\times\,
g_s ( i \, f^{ya_4x} ) V^{(0)}(Q_1,P_4^{\oplus},Q_2) 
\, v^{\rm SR}(t_1,t_2, {\bf p}_4^2,\tau)\, 
\frac{(2\pi)^{2-2\eps} \delta^{(2-2\eps)}({\bf p}_3 
+{\bf p}_4+{\bf p}_5)}{t_1\, t_2} \\[0.2cm] \nn
&=& c^{\rm SR}_i(p_i,p_{i'},\tau)\,
e^{C_A \, \al^{\rm SR}_g(t_1)\,\eta_1} \,
v^{\rm SR}(t_1,t_2, {\bf p}_4^2,\tau) \\
&&\hspace{3.0cm}\times\, 
e^{C_A \, \al^{\rm SR}_g(t_2)\,\eta_2} \,
c^{\rm SR}_j(p_j,p_{j'},\tau) \,
{\cal M}^{\rm tree}_{ij \to i'gj'}\, ,
\eeqa
where in the second line we have identified
the tree-level amplitude according to 
\eqns{Mtree}{Mtree2}. We see that the 
1-Reggeon transition as obtained within 
the shockwave formalism  reproduces 
the classical Regge factorization formula 
of \eqn{REggNLL}, when the rapidity factors 
are identified as in \eqn{RapidityLogs}. 
Notice, however, that within the shockwave 
formalism (at least in the present formulation)
one only determines rapidity evolution, therefore 
the impact factors and the Lipatov vertex 
appears as effective matching coefficients.
In particular, this means that the 
shockwave formalism cannot predict the 
analytic structure in \eqn{ggNLLsig2}.
The impact factors and Lipatov vertex
are not predicted within the theory, 
and need to be extracted by matching the 
factorization formula on the r.h.s. to 
the amplitude on the l.h.s of 
\eqn{eq:ampDecomp}. In this respect, 
we see that the contribution due to 
the multi-Reggeon transitions in the 
last line of \eqn{eq:ampDecomp} plays
a relevant role. In particular, they 
become essential, starting at NNLL
accuracy, for a consistent extraction 
of the Lipatov vertex. The shockwave 
formalism provides a method to 
systematically compute multi-Reggeon 
transitions, expressing them in terms 
of integrals which solve iteratively 
the Balitsky-JIMWLK evolution equation,
as discussed above. We will compute 
explicitly their contribution at one
and two loops in section \ref{sec:MRE}.
The consistent extraction of the Lipatov 
vertex at two loops will be then 
discussed in Section~\ref{sec:scheme}.


\section{Multi-Reggeon computations}
\label{sec:MRE}

In section \ref{sec:Reggeization} we have discussed the factorized structure of the amplitude 
in presence of a single Reggeon exchange. In section 
\ref{MRK-shockwave} we presented the formalism for computing multi-Reggeon 
exchanges. We now  compute their contribution to the $2\to 3$ scattering 
amplitude. As an illustration, in section~\ref{sec:MREOneLoop} we calculate the complete
multi-Reggeon contribution at one loop. Subsequently, in section~\ref{sec:MRETwoLoop} we describe the structure of all multi-Reggeon contributions to the two-loop amplitude, and finally in section~\ref{sec:oddOdd2} compute explicitly  
the respective $(-,-)$ component.


\subsection{Multi-Reggeon contributions to the one-loop amplitude}\label{sec:MREOneLoop}

\begin{figure}[t]
\centering
    \includegraphics[width=0.64\textwidth]{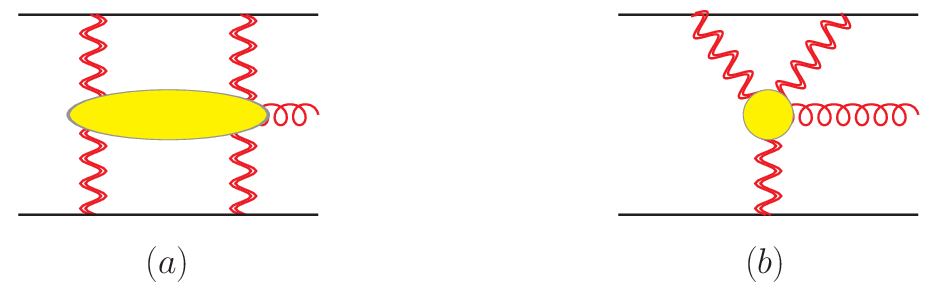}
\caption{Multi-Reggeon exchange contribution to the one-loop 
$2\to 3$ amplitude. Diagram (a) involves the exchange of two 
Reggeons between the target and the projectile, labelled as 
${\cal R}^2 g {\cal R}^2$ in the main text. Diagram (b) 
represents the one-to-two Reggeon transition amplitude, 
denoted by ${\cal R} g {\cal R}^2$ in the main text.
At one loop there is an additional diagram, involving the 
two-to-one Reggeon transition, or ${\cal R}^2 g {\cal R}$
amplitude. The latter is not shown here, because it is identical to the 
${\cal R} g {\cal R}^2$ amplitude by 
target-projectile symmetry.}
\label{fig:oneloop}
\end{figure}

The expansion of \eqn{ReggeFactBasic} to one 
loop shows that there are two possible transitions 
involving multiple Reggeons. One entails the 
exchange of two Reggeons between the target and 
the projectile, as shown in Fig.~\ref{fig:oneloop} 
(a). In this case the emission of the central gluon 
is governed by \eqn{WWona4A}, whose diagrammatic 
representation is given in 
Fig.~\ref{2to3EffectiveVertices2to2}. 
The other transition involves the emission of 
a single Reggeon from the projectile $i$ which 
subsequently, upon emission of the central gluon, 
is turned into two Reggeons on the target side, 
as shown in Fig.~\ref{fig:oneloop} (b), or its 
symmetric counterpart under target-projectile 
interchange.
This transition is governed by the second 
line of \eqn{Wona4}, which in turn is represented 
diagrammatically by the central diagram in 
Fig.~\ref{2to3EffectiveVertices}. In what follows 
we evaluate both transitions in some detail, to be able to discuss the techniques 
involved in the calculation. 


\paragraph{The ${\cal R}^2 g {\cal R}^2$ transition.}

Using \eqn{eq:psi2} in \eqn{ReggeFactBasic} one has
\bea \label{R2_calculationA} \nn
\frac{i}{2s_{12}}
{\cal \bar M}_{ij \to i'gj'} 
\Big|^{\mbox{\scriptsize 1-loop}}_{{\cal R}^2 g {\cal R}^2}
&=& \left(\frac{g_s^2}{2}\right)^2
\T_i^a\T_i^b  \, \T_j^{a'}\T_j^{b'}
\int [\dbar k_1] \int [\dbar k_2]  \\[0.2cm]
&&\hspace{-1.0cm} \times \,
\langle 0 | W^{a'}(k_2)W^{b'}(-q_2-k_2) \, a_4^{c}(p_4) 
\, W^a(k_1)W^b(q_1-k_1)|0\rangle, 
\eea
Next, using \eqn{WWona4A} we get 
\bea \label{R2_calculationB} \nn
\frac{i}{2s_{12}}
{\cal \bar M}_{ij \to i'gj'} 
\Big|^{\mbox{\scriptsize 1-loop}}_{{\cal R}^2 g {\cal R}^2}
&=& - 2 i \, g_s \left(\frac{g_s^2}{2}\right)^2
\T_i^a\T_i^b \, \T_j^{a'}\T_j^{b'}
\int [\dbar k_1] \int [\dbar k_2] \\ 
&&\hspace{-4.0cm} \times \Bigg\{
f^{acd} \langle 0 | W^{a'}(k_2)W^{b'}(-q_2-k_2)
\, W^d(k_1+p_4)W^b(q_1-k_1) |0\rangle 
\bigg[\frac{p_4^{\mu}}{p_4^2}
+\frac{k_1^{\mu}}{k_1^2} \bigg] \\ \nn
&&\hspace{-4.0cm} +\, f^{bcd} \langle 0 | 
W^{a'}(k_2) W^{b'}(-q_2-k_2) \,W^a(k_1)W^d(-q_1-k_1+p_4)|0\rangle 
\bigg[\frac{p_4^{\mu}}{p_4^2}
+\frac{(q_1-k_1)^{\mu}}{(q_1-k_1)^2} 
\bigg] \Bigg\} \varepsilon_{\mu}(p_4)\,.
\eea
We contract a Reggeon in the target 
with one in the projectile according to 
\beq \label{Reggeon_prop1}
\big\langle 0 \big| W^{x}(p_1) \, W^y(p_2) \big| 0 \big\rangle
= \frac{i\delta^{xy}}{p_1^2}\,(2\pi)^{2-2\epsilon}\,
\delta^{(2-2\eps)}(p_1-p_2)\,,
\eeq
which identifies the momenta defined in the 
same orientation. Wick contractions between 
two $W$s both originating from $i$ or both 
originating with $j$ lead to vanishing 
(scaleless) integrals. Integrating 
over $k_2$ using the $\delta$ functions
from the Reggeon propagators gives
\bea \label{R2_calculationC} \nn
\frac{i}{2s_{12}}
{\cal \bar M}_{ij \to i'gj'} 
\Big|^{\mbox{\scriptsize 1-loop}}_{{\cal R}^2 g {\cal R}^2}
&=& 2 i \, g_s \left(\frac{g_s^2}{2}\right)^2
\T_i^a\T_i^b \, \T_j^{a'}\T_j^{b'} \, 
(2\pi)^{2-2\eps} \delta^{(2-2\eps)}(p_4+q_1+q_2) \\ \nn
&&\hspace{-3.6cm}
\times\, \int [\dbar k_1] \Bigg\{
\left(f^{ac b'} \delta^{a'b} + f^{aca'}\delta^{bb'}\right)
\frac{1}{(q_1-k_1)^2 (q_1+q_2-k_1)^2} 
\bigg[\frac{p_4^{\mu}}{p_4^2}
+\frac{k_1^{\mu}}{k_1^2} \bigg] \\ 
&&\hspace{-2.0cm}
+\, \left(f^{bca'} \delta^{ab'} + f^{bcb'}\delta^{aa'}\right)
\frac{1}{k_1^2 (k_1+q_2)^2} \bigg[\frac{p_4^{\mu}}{p_4^2}
+\frac{(q_1-k_1)^{\mu}}{(q_1-k_1)^2} 
\bigg] \Bigg\} \varepsilon_{\mu}(p_4).
\eea
The kinematic factor in first term in the 
curly brackets is equal to the second one, 
upon performing the shift of the loop 
momentum $q_1 - k_1 \to k_1$. After this 
manipulation we combine terms and get 
\bea \label{R2_calculationD} 
\frac{i}{2s_{12}}
{\cal \bar M}_{ij \to i'gj'} 
\Big|^{\mbox{\scriptsize 1-loop}}_{{\cal R}^2 g {\cal R}^2}
&=& 2 i \, g_s \left(\frac{g_s^2}{2}\right)^2
\T_i^a\T_i^b \, \T_j^{a'}\T_j^{b'} \, 
(2\pi)^{2-2\eps} \delta^{(2-2\eps)}(p_4+q_1+q_2)\\ \nn
&&\hspace{-3.8cm}
\times\,  \Big(f^{ac b'} \delta^{a'b} + f^{aca'}\delta^{bb'} 
+ f^{bca'} \delta^{ab'} + f^{bcb'}\delta^{aa'} \Big)
\int \frac{[\dbar k_1]}{k_1^2 (k_1+q_2)^2} 
\bigg[\frac{p_4^{\mu}}{p_4^2}
+\frac{(q_1-k_1)^{\mu}}{(q_1-k_1)^2} 
\bigg] \Bigg\} \varepsilon_{\mu}(p_4).
\eea
Thus, we showed that the result  factorises as a single colour
factor multiplying the entire integral. Defining 
\be\label{TabDef1} 
\T_i^{\{a,b\}} = \frac12\left(\T_i^a\T_i^b+\T_i^b\T_i^a\right),
\ee
we may write this colour factor as follows:
\beq
i\,\T_i^a\T_i^b \, \T_j^{a'}\T_j^{b'} 
\left(f^{ac b'} \delta^{a'b} + f^{aca'}\delta^{bb'} 
+ f^{bca'} \delta^{ab'} + f^{bcb'}\delta^{aa'} \right) 
= - 4i\,f^{a'ca} \T_i^{\{a,b\}} \T_j^{\{a',b\}} \,,
\eeq
hence we obtain
\bea \label{R2_calculationE} \nn 
\frac{i}{2s_{12}}
{\cal \bar M}_{ij \to i'gj'} 
\Big|^{\mbox{\scriptsize 1-loop}}_{{\cal R}^2 g {\cal R}^2}
&=& - 8 i \pi^2 \, g_s^3 \, \frac{\as}{\pi} \,
f^{a'ca} \T_i^{\{a,b\}} \T_j^{\{a',b\}} \, 
(2\pi)^{2-2\eps} \delta^{(2-2\eps)}(p_4+q_1+q_2) \\ 
&&\hspace{1.0cm}
\times\, \int \frac{[\dbar k_1]}{k_1^2 (k_1+q_2)^2} 
\bigg[\frac{p_4^{\mu}}{p_4^2}+\frac{(q_1-k_1)^{\mu}}{(q_1-k_1)^2} 
\bigg] \varepsilon_{\mu}(p_4).
\eea
As for the case of $2\to 2$ amplitude
(see \cite{Caron-Huot:2017fxr,Falcioni:2021buo}), 
we want to express this result in terms of loop 
integrals and color operators acting on the 
tree-level amplitude. Indeed, we observe that 
the color structure in \eqn{R2_calculationE} 
can be written as follows: 
\bea\label{eq:TmmColourFactor} \nn
\T_i^{\{a,b\}} \, i \, f^{a'ca} \, \T_j^{\{a',b\}}
&=& \frac{1}{4}\big(\T_1\cdot \T_2 -\T_1\cdot \T_3 
- \T_2\cdot \T_5 + \T_3\cdot \T_5\big)
\big[\T_i^{a} \, i \, f^{a'ca} \, \T_j^{a'} \big] \\
&=& \frac14\T_{(--)}\,\mathcal{C}_{ij}^{(0)},
\eea
where in the first line we 
have used techniques discussed 
in Refs.~\cite{Caron-Huot:2017fxr,Falcioni:2021buo}
to express the color factor in terms of operators 
acting on the tree-level color structure. In the 
second line we have then identified the operators
with $\T_{(--)}$, defined in \eqn{TpmDef}, and 
the tree-level colour structure with the factor 
$\mathcal{C}_{ij}^{(0)}$, defined in \eqn{def:Cij}.

We can now consider the full amplitude, 
switching from ${\cal \bar M}_{ij \to i'gj'}$
to ${\cal M}_{ij \to i'gj'}$ according to 
\eqn{ReggeFactBasic}. We take into account 
\eqn{eq:TmmColourFactor} and also 
the fact that $Z_{i/j} = 1+\ord(\as)$.
Furthermore, in order to evaluate the loop 
integral, it is useful to notice that 
$q_1 = p_5$, $q_2 = p_3$, and momentum 
conservation fixes $p_4 = -p_3 - p_5$. 
Also, we work in dimensional regularization 
in the $\overline{\rm MS}$ scheme, thus 
we make now the renormalization scale explicit 
by replacing $\as \to \as \big[\mu^{2} 
e^{\gamma_E}/(4\pi)\big]^{\eps}$. In the
end we get
\bea \label{R2_calculationG} \nn 
{\cal M}_{ij \to i'gj'} 
\Big|^{\mbox{\scriptsize 1-loop}}_{{\cal R}^2 g {\cal R}^2}
&=& 4 i \pi^2 \, s_{12} \, g_s^3 \, \frac{\as}{\pi} 
\, C_i^{(0)} C_j^{(0)} \, \T_{(--)}\,\mathcal{C}_{ij}^{(0)} 
\, (2\pi)^{2-2\eps}\delta^{(2-2\eps)}(p_3+p_4+p_5) \\ 
&&\hspace{0.0cm}
\times\, \bigg(\frac{\mu^{2} e^{\gamma_E}}{4\pi}\bigg)^{\eps}
\int \frac{[\dbar k_1]}{k_1^2(k_1+p_3)^2} \bigg[\frac{p_4^{\mu}}{p_4^2} 
- \frac{(k_1-p_5)^{\mu}}{(k_1-p_5)^2} 
\bigg] \varepsilon_{\mu}(p_4).
\eea
The loop integration can be reduced to a 
set of master integrals by means of IBPs. 
One has
\bea \label{IntToMIs}\nn
\pi \bigg(\frac{\mu^{2} e^{\gamma_E}}{4\pi}\bigg)^{\eps}
\int \frac{[\dbar k_1]}{k_1^2(k_1+p_3)^2} 
\bigg[\frac{p_4^{\mu}}{p_4^2} 
-\frac{(k_1-p_5)^{\mu}}{(k_1-p_5)^2} 
\bigg] \varepsilon_{\mu}(p_4) && \\[0.3cm] \nn 
&&\hspace{-8.0cm}=\, \Bigg\{ 
\frac{{\cal I}_{(1,1,0)}}{p_3^2} \bigg[ 
\frac{p_4^{\mu}}{p_4^2}
+\frac{1}{2}\frac{p_3\cdot p_5 \,p_3^{\mu}
-p_3^2 \, p_5^{\mu}}{(p_3\cdot p_5)^2 
- p_3^2 \,p_5^2} \bigg] + 
\frac{{\cal I}_{(1,0,1)}}{2p_5^2}\frac{p_5^2 \,p_3^{\mu}
-p_3\cdot p_5 \, p_5^{\mu}}{(p_3\cdot p_5)^2 
- p_3^2 \,p_5^2} \\ 
&&\hspace{-7.5cm}-\, \frac{{\cal I}_{(0,1,1)}}{2 p_3^2 \, p_5^2} 
\frac{(p_3\cdot p_5+p_5^2) p_3^{\mu}
- (p_3^2 + p_3\cdot p_5)p_5^{\mu}}{(p_3\cdot p_5)^2 
- p_3^2 \,p_5^2} \\ \nn
&&\hspace{-7.5cm}+\, \frac{{\cal I}_{(1,1,1)}}{2 p_3^2 \, p_5^2} 
\frac{\big[ 2 (p_3\cdot p_5)^2 
+ p_3^2 (p_3\cdot p_5 - p_5^2)\big] p_5^{\mu}
- (p_3^2 + p_3\cdot p_5)p_5^2 \, p_3^{\mu}}{(p_3\cdot p_5)^2 
- p_3^2 \,p_5^2} \Bigg\} \varepsilon_{\mu}(p_4),
\eea
where we have introduced the family of master integrals
\be\label{MIdef}
{\cal I}_{(a,b,c)} = \pi
\bigg(\frac{\mu^{2} e^{\gamma_E}}{4\pi}\bigg)^{\eps}
\int [\dbar k] \, \frac{\left(p_5^2\right)^b 
\left(p_3^2\right)^c}{\big[ k^2 \big]^a 
\big[(k+p_3)^2\big]^{b+ b'\eps} \big[(k-p_5)^2\big]^{c+c'\eps}} .
\ee
All the scalar products on the r.h.s.~of 
\eqn{IntToMIs} can be evaluated in two 
transverse dimension by means of eqs.~(\ref{def:zzbar}), 
(\ref{momentumconservation1}) and (\ref{polvectprod}),
except for the integrals ${\cal I}_{(a,b,c)}$
which needs to be evaluated in dimensional 
regularization in $d = 2-2\eps$. Inserting 
\eqn{IntToMIs} into \eqn{R2_calculationG}, 
after some elaboration we get
\bea \label{R2_calculationI} \nn 
{\cal M}_{ij \to i'gj'} 
\Big|^{\mbox{\scriptsize 1-loop}}_{{\cal R}^2 g {\cal R}^2}
&=& - i \pi \, \frac{\as}{\pi} \frac{1}{z-\bar z}
\bigg[(1 - \bar z)\,{\cal I}_{(1, 1, 0)} 
+ z  \, {\cal I}_{(1, 0, 1)}  
-\frac{{\cal I}_{(0, 1, 1)}}{|{\bf p}_4|^2 \bar z (1 - z)} 
- {\cal I}_{(1, 1, 1)} \Big] \\[0.2cm]
&&\hspace{6.0cm} \times\, 
\T_{(--)}\,{\cal M}_{ij \to igj}^{\rm tree} \,,
\eea
where we have identified 
${\cal M}_{ij \to igj}^{\rm tree}$ 
according to \eqn{Mtree2}.

\Eqn{R2_calculationI} contains three bubble integrals 
and a triangle integral. The bubbles are easy 
to calculate by means of the formula 
\be\label{Bubble}
B_{a,b}^{(2-2\eps)}(p^2,\eps) 
= \pi \bigg(\frac{\mu^{2} e^{\gamma_E}}{4\pi}\bigg)^{\eps}
\int [\dbar k]\, \frac{1}{[k^2]^a [(k+p)^2]^b} 
= \frac{B_{a,b}(\eps)}{2\eps}  
\bigg(\frac{\mu^2}{p^2}\bigg)^{\eps} (p^2)^{1-a-b},
\ee
where 
\be\label{Bubble2}
B_{a,b}(\eps) = 2\eps \frac{e^{\eps \gamma_E}}{4}
\frac{\Gamma(1-a-\eps)\Gamma(1-b-\eps)
\Gamma(a+b-1+\eps)}{\Gamma(a)\Gamma(b)\Gamma(2-2\eps-a-b)}.
\ee
It is easy to identify 
\bea \nn
{\cal I}_{(1, 1, 0)}
&=& p_3^2 \,B_{(1,1)}^{(2-2\eps)}(p_3^2,\eps), \\[0.2cm]  
{\cal I}_{(1, 0, 1)} &=& 
p_5^2 \,B_{(1,1)}^{(2-2\eps)}(p_5^2,\eps),  \\[0.2cm]  \nn
{\cal I}_{(0, 1, 1)} 
&=& p_3^2 \,p_5^2 \,B_{(1,1)}^{(2-2\eps)}[(p_3+p_5)^2,\eps] 
= p_3^2 \, p_5^2  \, B_{(1,1)}^{(2-2\eps)}(p_4^2,\eps).
\eea
The triangle integral for generic powers $a$, $b$, 
$c$ has been provided to all order in $\eps$ in 
\cite{Anastasiou:1999cx,Anastasiou:1999ui}. Here 
we need the case $a = b = c = 1$, which allows 
for some simplification. In the end one has 
\bea \label{TriangleMI} \nn
{\cal I}_{(1,1,1)} &=& \pi
\bigg(\frac{\mu^{2} e^{\gamma_E}}{4\pi}\bigg)^{\eps}
\int [\dbar k]\, \frac{p_5^2\, p_3^2 }{k^2 \, 
(k+p_3)^2\, (k-p_5)^2 } \\ \nn
&=& -\frac{e^{\gamma_E \eps} \Gamma(1 - \eps)  \Gamma(1 + \eps)^2}{4 \eps}
(z-\bar z) 
\bigg(\frac{\mu^2 (z-\bar z)^2}{z \bar z (1-z)(1-\bar z) p_4^2} 
\bigg)^{\eps} \\ \nn
&&+\frac{e^{\gamma_E \eps} (1+2\eps) \Gamma(1 - \eps)^2
\Gamma(1 + \eps)}{2 \eps(1+\eps)\Gamma(1-2\eps)}
 \, z \bar z (1-z)(1-\bar z) \,\bigg(\frac{\mu^2}{p_4^2}\bigg)^{\eps} \\ 
&&\hspace{0.5cm}\times \bigg\{\frac{1}{z(1-\bar z)}\, 
_2F_1\bigg(1,-\eps,2+\eps,
\frac{\bar z(1-z)}{z(1-\bar z)}\bigg) \\ \nn
&&\hspace{1.5cm}-\, \frac{(z\bar z)^{-1-\eps}}{z-\bar z}\, 
_2F_1\bigg(1,2(1+\eps),2+\eps,
-\frac{1-z}{z-\bar z}\bigg) \\ \nn
&&\hspace{2.0cm}-\, \frac{[(1-z)(1-\bar z)]^{-1-\eps}}{z-\bar z}\, 
_2F_1\bigg(1,2(1+\eps),2+\eps,
-\frac{\bar z}{z-\bar z}\bigg) \bigg\}.
\eea
Inserting \eqn{TriangleMI} as well as 
the expression for the bubble integrals 
from \eqn{Bubble} into \eqn{R2_calculationI}, 
expanding in powers of $\eps$, after 
some elaboration one has 
\be \label{R2_calculationJ} 
{\cal M}_{ij \to i'gj'} 
\Big|^{\mbox{\scriptsize 1-loop}}_{{\cal R}^2 g {\cal R}^2}
= i \pi \, \frac{\as}{4\pi} \, 
G^{(1)}_{{\cal R}^2 g {\cal R}^2}(z,\bar z, |\mathbf{p}_4|^2,\mu^2) 
\, \T_{(--)}\, {\cal M}_{ij \to igj}^{\rm tree},
\ee
where the function $G^{(1)}_{{\cal R}^2 g {\cal R}^2}$ reads
\bea \label{Gmm1}  \nn
G^{(1)}_{{\cal R}^2 g {\cal R}^2}(z,\bar z, |\mathbf{p}_4|^2,\mu^2)
&=& \frac{1}{\eps}+\ln \bigg(\frac{\mu^2 
|\mathbf{p}_4|^2}{|\mathbf{p}_3|^2 |\mathbf{p}_5|^2} \bigg)
+\eps \bigg[-\frac{\pi^2}{12} - 2D_2(z,\bar z) \\[0.2cm] 
&&\hspace{-2.0cm}+\,
\frac{1}{2} \ln\bigg( \frac{\mu^2}{|\mathbf{p}_3|^2} \bigg)^2 
- \frac{1}{2} \ln\bigg( \frac{\mu^2}{|\mathbf{p}_4|^2} \bigg)^2 
+ \frac{1}{2} \ln\bigg( \frac{\mu^2}{|\mathbf{p}_5|^2} \bigg)^2
\bigg] + \ord(\eps^2),
\eea
with $|\mathbf{p}_3|^2 = z \bar z |\mathbf{p}_4|^2$,
$|\mathbf{p}_5|^2 = (1-z) (1-\bar z) |\mathbf{p}_4|^2$ 
and the function $D_2(z,\bar z)$ represents a 
single-valued combination of dilogarithms
\be \label{eq:D2def}
D_2(z,\bar z) = {\rm Li}_2(z) - {\rm Li}_2(\bar z)
+ \frac{1}{2}\ln \bigg( \frac{1 - z}{1 - \bar z}\bigg) \ln (z \bar z),
\ee
which is referred to as the 
Bloch-Wigner dilogarithm~\cite{BlochWigner}.

Before concluding, it is instructive to 
have an explicit look at the action of the 
color operator $\T_{(--)}$ on the tree
level amplitude ${\cal M}_{ij \to igj}^{\rm tree}$. 
Let us recall that in color space the 
amplitude is written as a vector, according 
to \eqn{coloramplitudes}. An operator $\T_X$ 
then acts as a matrix, according to (see 
appendix \ref{Colorflow} for more details)
\begin{equation}\label{colorOpMatrix}
\sum_{ji}  c^{[j]} \, \T_X^{[j][i]} \,{\cal M}^{[i]}
= \sum_{j} c^{[j]} \, {\cal M}^{[j]}.
\end{equation}
Decomposing the amplitude on an 
orthonormal basis in the $t$-channel, 
as in \eqn{t-channel_basis-colour_decomposition},
the tree level amplitudes ${\cal M}_{ij \to igj}^{\rm tree}$
are given by a single element: explicitly one has
\bea \label{TreeAmpColor} \nn
{\cal M}_{qq \to qgq}^{\rm tree} &=& c^{[8,8]_a}\, 
\big({\cal M}_{qq \to qgq}^{\rm tree}\big)^{[8,8]_a}, \\ 
{\cal M}_{qg \to qgg}^{\rm tree} &=& c^{[8,8_a]_a}\, 
\big({\cal M}_{qg \to qgg}^{\rm tree}\big)^{[8,8_a]_a}, \\ \nn 
{\cal M}_{gg \to ggg}^{\rm tree} &=& c^{[8_a,8_a]}\, 
\big({\cal M}_{gg \to ggg}^{\rm tree}\big)^{[8_a,8_a]}, 
\eea
where the basis elements $c^{[r_1,r_2]_r}$
for the respective scattering processes 
are defined in eqs.~(\ref{basisqgq}),~(\ref{basisqgg}),
(\ref{basisggq}) and~(\ref{basisggg}), and the explicit 
coefficients $\big({\cal M}_{ij \to igj}^{\rm tree}\big)^{[J]}$
are provided in \eqn{TreeAmpColorExplicit}. Taking 
this into account, the action of the operator 
$\T_{(--)}$ on these tree level amplitudes 
gives
\bea\label{eq:TmmColourFactorExplicit} 
\sum_{j}  c^{[j]} \, \T_{(--)}^{[j][8,8]_a} \,
\big({\cal M}_{qq \to qgq}^{\rm tree}\big)^{[8,8]_a}
&=&  \frac{N_c^2-4}{2N_c} \, 
c^{[8, 8]_a}\, \big({\cal M}_{qq \to qgq}^{\rm tree}\big)^{[8,8]_a}, \\ 
\sum_{j}  c^{[j]} \, \T_{(--)}^{[j][8,8_a]_a} \,
\big({\cal M}_{qg \to qgg}^{\rm tree}\big)^{[8,8_a]_a}
&=& \frac{\sqrt{N_c^2-4}}{2}\,
c^{[8, 8_s]_a} \, 
\big({\cal M}_{qg \to qgg}^{\rm tree}\big)^{[8,8_a]_a}, \\ 
\sum_{j}  c^{[j]} \, \T_{(--)}^{[j][8_a,8]_a} \,
\big({\cal M}_{gq \to ggq}^{\rm tree}\big)^{[8_a,8]_a}
&=& \frac{\sqrt{N_c^2-4}}{2}\,
c^{[8_s,8]_a} \, 
\big({\cal M}_{qg \to qgg}^{\rm tree}\big)^{[8_a,8]_a}, \\ \nn
\sum_{j}  c^{[j]} \, \T_{(--)}^{[j][8_a,8_a]} \,
\big({\cal M}_{gg \to ggg}^{\rm tree}\big)^{[8_a,8_a]}
&=& 2 \bigg(\sqrt{\frac{N_c-3}{2N_c}} \,c^{[0, 0]} 
+\sqrt{\frac{N_c+3}{2N_c}} \,c^{[27, 27]} \\  
&& \hspace{1.0cm} 
+\, \frac{N_c}{4} \, c^{[8_s, 8_s]}\bigg)
\big({\cal M}_{gg \to ggg}^{\rm tree}\big)^{[8_a,8_a]}.
\eea
Notice that for completeness we provide both the case 
in which $i = q$, $j = g$, namely ${\cal M}_{qg \to qgg}$, 
and the case in which $i = g$, $j = q$, i.e. 
${\cal M}_{gq \to ggq}$. As it is evident, the action 
of $\T_{(--)}$ on these amplitudes is obviously symmetric.
Comparing with the color basis in \eqn{basisggg}
it is interesting to note that, for the $gg \to ggg$
amplitude, the color operator $\T_{(--)}$ projects 
onto the color basis elements which are even both 
under $1\leftrightarrow 5$ and $2\leftrightarrow 3$, 
as expected from Bose symmetry. 


\paragraph{The ${\cal R} g {\cal R}^2$ 
and ${\cal R}^2 g {\cal R}$ transitions.}

At one loop there are two additional contributions, 
namely, the one-to-two (${\cal R} g {\cal R}^2$) and 
the two-to-one (${\cal R}^2 g {\cal R}$) Reggeon 
transition amplitudes. These transitions are related
by the target-projectile (or top-bottom) symmetry, 
therefore we focus on the calculation of the 
${\cal R} g {\cal R}^2$ amplitude, represented
in Fig.~\ref{fig:oneloop}~(b). In the end 
we will obtain the ${\cal R}^2 g {\cal R}$ 
amplitude by symmetry. 

The ${\cal R} g {\cal R}^2$ amplitude
at one loop reads
\bea \label{RgR2_calculationA} \nn
\frac{i}{2s_{12}}
{\cal \bar M}_{ij \to i'gj'} 
\Big|^{\mbox{\scriptsize 1-loop}}_{{\cal R} g {\cal R}^2}
&=& - i\, g_s \, \frac{g_s^2}{2}
\T_i^{a} \, \T_j^{a'}\T_j^{b'} 
\int [\dbar k_1]  \\[0.2cm]
&&\hspace{0.0cm} \times \,
\langle 0 | W^{a'}(k_2)W^{b'}(-q_2-k_2)
\, a_4^{c}(p_4) \, W^a(q_1) |0\rangle. 
\eea
In this case we need the second line of 
\eqn{Wona4}, represented by the central 
diagram in Fig.~\ref{2to3EffectiveVertices}. 
We get 
\bea \label{RgR2_calculationB} 
\frac{i}{2s_{12}}
{\cal \bar M}_{ij \to i'gj'} 
\Big|^{\mbox{\scriptsize 1-loop}}_{{\cal R} g {\cal R}^2}
&=& - \frac{g_s^5}{2}\, f^{cc'e} f^{ead}
\T_i^{a} \, \T_j^{a'}\T_j^{b'}
\int [\dbar k_1] \int [\dbar k_2] \\ \nn
&&\hspace{-3.5cm} \times 
\langle 0 | W^{a'}(k_2)W^{b'}(-q_2-k_2) \, W^{c'}(q_1-k_1+p_4)W^d(k_1)|0\rangle 
\bigg[\frac{q_1^{\mu}}{q_1^2}
-\frac{(q_1-k_1)^{\mu}}{(q_1-k_1)^2} 
\bigg] \varepsilon_{\mu}(p_4)\,.
\eea
Contracting the Reggeon fields according to 
\eqn{Reggeon_prop1} and integrating over $k_2$ 
using the $\delta$ functions from the Reggeon 
propagators, after some elaboration gives
\bea \label{RgR2_calculationD} \nn
\frac{i}{2s_{12}}
{\cal \bar M}_{ij \to i'gj'} 
\Big|^{\mbox{\scriptsize 1-loop}}_{{\cal R} g {\cal R}^2}
&=& - 4 \pi^2 \,g^3_s \, \frac{\as}{\pi} \, f^{a'ce} f^{eab'}
\,\T_i^{a} \, \T_j^{\{a',b'\}}\, (2\pi)^{2-2\eps}
\delta^{(2-2\eps)}(p_4+q_1+q_2) \\ 
&&\hspace{0.5cm}
\times\, \int \frac{[\dbar k_1]}{k_1^2(k_1+q_2)^2}
\Bigg[\frac{q_1^{\mu}}{q_1^2} + \frac{(k_1-q_1)^{\mu}}{(k_1-q_1)^2} 
\Bigg] \varepsilon_{\mu}(p_4),
\eea
where we have expressed the color structure 
on the line $i$ in terms of the operator 
defined in \eqn{TabDef1}. As for the 
${\cal R}^2 g {\cal R}^2$ transition 
amplitude, we want to write the result in 
\eqn{RgR2_calculationD} in terms of 
loop integrals and color operators 
acting on the tree level amplitude. 
To this end we first use the Jacobi 
identity as follows:
\be
f^{eab'} \, \T_i^{a}
= i \big(\T_i^{e} \T_i^{b'} - \T_i^{b'} \T_i^{e}\big).
\ee
Inserting this result into the full color
structure of \eqn{RgR2_calculationD}, 
and proceeding as before with methods 
described in \cite{Caron-Huot:2017fxr,Falcioni:2021buo}
we obtain 
\bea \label{eq:TpmColourFactor} \nn
i f^{a'ce} \, \big(\T_i^{e} \T_i^{b'} 
- \T_i^{b'} \T_i^{e}\big) \, \T_j^{\{a',b'\}}
&=& \frac{1}{2}\big(
\T_3\cdot \T_5 - \T_2\cdot \T_5
+\T_1\cdot \T_3 - \T_1\cdot \T_2 \big)
\big[\T_i^{a} \, i \, f^{a'ca} \, \T_j^{a'} \big] \\
&=& \frac{1}{2}\T_{(+-)}\,\mathcal{C}_{ij}^{(0)},
\eea
where in the second line we have identified 
the color operators in the first line with 
$\T_{(+-)}$, defined in \eqn{TpmDef}, and 
the tree-level colour structure with the factor 
$\mathcal{C}_{ij}^{(0)}$, defined in \eqn{def:Cij}.

We can now consider the full amplitude, 
switching from ${\cal \bar M}_{ij \to i'gj'}$
to ${\cal M}_{ij \to i'gj'}$ according to 
\eqn{ReggeFactBasic}. Taking into account
\eqn{eq:TmmColourFactor}, and also that 
$Z_{i/j} = 1+\ord(\as)$, at one loop we 
have
\bea \label{RgR2_calculationF} \nn 
{\cal M}_{ij \to i'gj'} 
\Big|^{\mbox{\scriptsize 1-loop}}_{{\cal R} g {\cal R}^2}
&=& 4 i \pi^2 \, s_{12} \, g_s^3 \, \frac{\as}{\pi} 
\, C_i^{(0)} C_j^{(0)} \, \T_{(+-)}\,\mathcal{C}_{ij}^{(0)} 
\, (2\pi)^{2-2\eps}
\delta^{(2-2\eps)}(p_4+q_1+q_2) \\ 
&&\hspace{1.0cm}
\times\, \int \frac{[\dbar k_1]}{k_1^2(k_1+q_2)^2}
\Bigg[\frac{q_1^{\mu}}{q_1^2} + \frac{(k_1-q_1)^{\mu}}{(k_1-q_1)^2} 
\Bigg] \varepsilon_{\mu}(p_4).
\eea
As before, we set $q_1 = p_5$,
$q_2 = p_3$, and make the renormalization 
scale explicit by replacing $\as \to \as 
\big[\mu^{2} e^{\gamma_E}/(4\pi)\big]^{\eps}$, 
obtaining
\bea \label{RgR2_calculationG} \nn 
{\cal M}_{ij \to i'gj'} 
\Big|^{\mbox{\scriptsize 1-loop}}_{{\cal R} g {\cal R}^2}
&=& 4 i \pi^2 \, s_{12} \, g_s^3 \, \frac{\as}{\pi} 
\, C_i^{(0)} C_j^{(0)} \, \T_{(+-)}\,\mathcal{C}_{ij}^{(0)} 
\, (2\pi)^{2-2\eps} \delta^{(2-2\eps)}(p_3+p_4+p_5) \\ 
&&\hspace{0.5cm}
\times\, \bigg(\frac{\mu^{2} e^{\gamma_E}}{4\pi}\bigg)^{\eps}
\int \frac{[\dbar k_1]}{k_1^2(k_1+p_3)^2}
\bigg[\frac{p_5^{\mu}}{p_5^2} + \frac{(k_1-p_5)^{\mu}}{(k_1-p_5)^2} 
\bigg] \varepsilon_{\mu}(p_4).
\eea
The loop integration in 
\eqn{RgR2_calculationG} can be reduced 
to a set of master integrals by means 
of IBPs. In this case we obtain
\bea \label{IntToMIsb}\nn
\pi \bigg(\frac{\mu^{2} e^{\gamma_E}}{4\pi}\bigg)^{\eps}
\int \frac{[\dbar k_1]}{k_1^2(k_1+p_3)^2} 
\bigg[\frac{p_5^{\mu}}{p_5^2} 
+\frac{(k_1-p_5)^{\mu}}{(k_1-p_5)^2} 
\bigg] \varepsilon_{\mu}(p_4) && \\[0.3cm] 
&&\hspace{-8.0cm}=\, \Bigg\{ 
\frac{{\cal I}_{(1,1,0)}}{2p_3^2} 
\bigg[ \frac{2 (p_3\cdot p_5)^2 
- p_3^2 \, p_5^2}{(p_3\cdot p_5)^2 
- p_3^2 \,p_5^2} \frac{p_5^{\mu}}{p_5^2}
- \frac{p_3\cdot p_5\, p_3^{\mu}}{(p_3\cdot p_5)^2 
- p_3^2 \,p_5^2} \bigg]  \\ \nn
&&\hspace{-7.5cm}+\, 
\frac{{\cal I}_{(1,0,1)}}{2p_5^2}
\frac{p_3\cdot p_5\,p_5^{\mu}
- p_5^2 \,p_3^{\mu}}{(p_3\cdot p_5)^2 
- p_3^2 \,p_5^2} 
+ \frac{{\cal I}_{(0,1,1)}}{2p_3^2\, p_5^2} 
\frac{(p_3\cdot p_5+p_5^2)p_3^{\mu}
- (p_3^2 + p_3\cdot p_5) p_5^{\mu}}{(p_3\cdot p_5)^2 
- p_3^2 \,p_5^2}  \\ \nn
&&\hspace{-7.5cm}+\, 
\frac{{\cal I}_{(1,1,1)}}{2p_3^2\, p_5^2}
\frac{(p_3^2 + p_3\cdot p_5) p_5^2 p_3^{\mu}
- \big[ 2 (p_3\cdot p_5)^2 + p_3^2 
(p_3\cdot p_5- p_5^2) \big] p_5^{\mu}}{(p_3\cdot p_5)^2 
- p_3^2 \,p_5^2} \Bigg\} \varepsilon_{\mu}(p_4),
\eea
where the family of integrals is the 
same introduced in \eqn{MIdef}. All 
the scalar products on the r.h.s.~of 
\eqn{IntToMIsb} can be evaluated in two 
transverse dimension by means of eqs.~(\ref{def:zzbar}), 
(\ref{momentumconservation1}) and (\ref{polvectprod}), 
except for the integrals ${\cal I}_{(a,b,c)}$
which needs to be evaluated in dimensional 
regularization in $d = 2-2\eps$. Inserting 
\eqn{IntToMIsb} into \eqn{RgR2_calculationG}, 
after some elaboration we obtain
\bea \label{RgR2_calculationH} \nn 
{\cal M}_{ij \to i'gj'} 
\Big|^{\mbox{\scriptsize 1-loop}}_{{\cal R} g {\cal R}^2}
&=& i \pi \, \frac{\as}{\pi} \,
\frac{1}{z-\bar z}
\Big[(1 - z)\,{\cal I}_{(1, 1, 0)} 
+ z  \, {\cal I}_{(1, 0, 1)}  \\ 
&&\hspace{2.0cm}-\, 
\frac{{\cal I}_{(0, 1, 1)}}{|{\bf p}_4|^2 \bar z (1 - z)} 
- {\cal I}_{(1, 1, 1)} \Big]\, \T_{(+-)} \, 
{\cal M}_{ij \to igj}^{\rm tree}.
\eea
Inserting \eqn{TriangleMI} as well as 
the expression for the bubble integrals 
from \eqn{Bubble} into \eqn{RgR2_calculationH}, 
expanding in powers of $\eps$ we arrive at
\be \label{RgR2_calculationI} 
{\cal M}_{ij \to i'gj'} 
\Big|^{\mbox{\scriptsize 1-loop}}_{{\cal R} g {\cal R}^2}
= i \pi \, \frac{\as}{4\pi} \,  
G^{(1)}_{{\cal R} g {\cal R}^2}(z,\bar z, |\mathbf{p}_4|^2,\mu^2) 
\, \T_{(+-)}\, {\cal M}_{ij \to igj}^{\rm tree},
\ee
where
\bea \label{Gpm1} \nn
G^{(1)}_{{\cal R} g {\cal R}^2}(z,\bar z, |\mathbf{p}_4|^2,\mu^2) 
&=& \frac{1}{\eps} +\ln \bigg(\frac{\mu^2 
|\mathbf{p}_5|^2}{|\mathbf{p}_3|^2 |\mathbf{p}_4|^2} \bigg)
+\eps \bigg[- \frac{\pi^2}{12} + 2D_2(z,\bar z) \\[0.2cm] 
&&\hspace{-2.0cm}+\,
\frac{1}{2} \ln\bigg( \frac{\mu^2}{|\mathbf{p}_3|^2} \bigg)^2 
+ \frac{1}{2} \ln\bigg( \frac{\mu^2}{|\mathbf{p}_4|^2} \bigg)^2 
- \frac{1}{2} \ln\bigg( \frac{\mu^2}{|\mathbf{p}_5|^2} \bigg)^2
\bigg] + \ord(\eps^2).
\eea

As for the ${\cal R}^2 g {\cal R}^2$ 
transition, it is instructive to have an 
explicit look at the action of the 
color operator $\T_{(+-)}$ on the tree
level amplitude ${\cal M}_{ij \to igj}^{\rm tree}$. 
Using the same notation as in \eqn{eq:TmmColourFactor}, 
we have
\bea\label{eq:TpmColourFactorB} \nn
\sum_{j}  c^{[j]} \, \T_{(+-)}^{[j][8,8]_a} \,
\big({\cal M}_{qq \to qgq}^{\rm tree}\big)^{[8,8]_a}
&=& - \bigg(\sqrt{2}\, c^{[8, 1]} \\ 
&&\hspace{0.5cm}
+\,\frac{\sqrt{N_c^2-4}}{2}\,c^{[8, 8]_s}\bigg)
\big({\cal M}_{qq \to qgq}^{\rm tree}\big)^{[8,8]_a}, \\ \nn 
\sum_{j}  c^{[j]} \, \T_{(+-)}^{[j][8,8_a]_a} \,
\big({\cal M}_{qg \to qgg}^{\rm tree}\big)^{[8,8_a]_a}
&=& - \bigg( \frac{2 N_c}{\sqrt{N_c^2-1}}\, c^{[8, 1]} 
+\frac{N_c}{2}\, c^{[8, 8_s]_s} \\ 
&&\hspace{-2.5cm}+\, \sqrt{\frac{N_c + 3}{N_c +1}}\, 
c^{[8, 27]} +\sqrt{\frac{N_c - 3}{N_c -1}}\, 
c^{[8,0]} \bigg)
\big({\cal M}_{qg \to qgg}^{\rm tree}\big)^{[8,8_a]_a}, \\ \nn 
\sum_{j}  c^{[j]} \, \T_{(+-)}^{[j][8_a,8]_a} \,
\big({\cal M}_{gq \to ggq}^{\rm tree}\big)^{[8_a,8]_a}
&=& \bigg(\sqrt{2}\, c^{[8_a,1]}  
+ \frac{\sqrt{N_c^2-4}}{2}\,c^{[8_a,8]_s}\bigg)
\big({\cal M}_{gq \to ggq}^{\rm tree}\big)^{[8_a,8]_a}, 
\\ && \\ \nn 
\sum_{j}  c^{[j]} \, \T_{(+-)}^{[j][8_a,8_a]} \,
\big({\cal M}_{gg \to ggg}^{\rm tree}\big)^{[8_a,8_a]}
&=& - \bigg( \frac{2 N_c}{\sqrt{N_c^2-1}}\, c^{[8_a,1]} 
+\frac{N_c}{2}\, c^{[8_a, 8_s]} \\ 
&&\hspace{-2.5cm}+\, \sqrt{\frac{N_c + 3}{N_c +1}}\, 
c^{[8_a, 27]} +\sqrt{\frac{N_c - 3}{N_c -1}}\, 
c^{[8_a, 0]} \bigg)
\big({\cal M}_{gg \to ggg}^{\rm tree}\big)^{[8_a,8_a]}.
\eea
Comparing with the color basis in \eqn{basisggg}
it is interesting to note that, for the $gg \to ggg$
amplitude, the color operator $\T_{(+-)}$ projects 
onto the color basis elements which are odd under 
$1\leftrightarrow 5$ and even under $2\leftrightarrow 3$, 
as expected from Bose symmetry. 

With the result for the ${\cal R} g {\cal R}^2$
in \eqn{RgR2_calculationI}, we can immediately 
obtain the ${\cal R}^2 g {\cal R}$ transition 
amplitude by applying projectile-target symmetry.
For the color operator, the symmetry implies 
$\T_{(+-)} \to \T_{(-+)}$, while the kinematic 
dependence is obtained by applying the 
target-projectile symmetry $z \to 1-\bar z$, 
$\bar z \to 1-z$ to \eqn{RgR2_calculationI}. 
Taking into account the property 
\be\label{D2symmetry}
D_2(1-\bar z,1- z) = D_2(z,\bar z),
\ee
we have
\be \label{R2gR_calculationI} 
{\cal M}_{ij \to i'gj'} 
\Big|^{\mbox{\scriptsize 1-loop}}_{{\cal R}^2 g {\cal R}}
= i \pi \, \frac{\as}{4\pi} \, 
G^{(1)}_{{\cal R}^2 g {\cal R}}(z,\bar z, |\mathbf{p}_4|^2,\mu^2) 
\, \T_{(-+)}\, {\cal M}_{ij \to igj}^{\rm tree},
\ee
where
\bea \label{Gmp1} \nn
G^{(1)}_{{\cal R}^2 g {\cal R}}(z,\bar z, |\mathbf{p}_4|^2,\mu^2) 
&=& \frac{1}{\eps} +\ln \bigg(\frac{\mu^2 
|\mathbf{p}_3|^2}{|\mathbf{p}_4|^2 |\mathbf{p}_5|^2} \bigg)
+\eps \bigg[-\frac{\pi^2}{12} + 2D_2(z,\bar z) \\[0.2cm] 
&&\hspace{-2.0cm}-\, 
\frac{1}{2} \ln\bigg( \frac{\mu^2}{|\mathbf{p}_3|^2} \bigg)^2
+ \frac{1}{2} \ln\bigg( \frac{\mu^2}{|\mathbf{p}_4|^2} \bigg)^2 
+\frac{1}{2} \ln\bigg( \frac{\mu^2}{|\mathbf{p}_5|^2} \bigg)^2 
\bigg] + \ord(\eps^2).
\eea

The action of the $\T_{(-+)}$ color
operator on the tree level reads
\bea\label{eq:TmpColourFactor} 
\sum_{j}  c^{[j]} \, \T_{(-+)}^{[j][8,8]_a} \,
\big({\cal M}_{qq \to qgq}^{\rm tree}\big)^{[8,8]_a}
&=& \bigg(\sqrt{2}\, c^{[1, 8]} 
+\frac{\sqrt{N_c^2-4}}{2}\,c^{[8, 8]_s}\bigg)
\big({\cal M}_{qq \to qgq}^{\rm tree}\big)^{[8,8]_a}, \\ \nn
\sum_{j}  c^{[j]} \, \T_{(-+)}^{[j][8,8_a]_a} \,
\big({\cal M}_{qg \to qgg}^{\rm tree}\big)^{[8,8_a]_a}
&=& \bigg(\sqrt{2}\, c^{[1, 8_a]}  
+ \frac{\sqrt{N_c^2-4}}{2}\,c^{[8, 8_a]_s}\bigg)
\big({\cal M}_{qg \to qgg}^{\rm tree}\big)^{[8,8_a]_a}, 
\\ && \\ \nn 
\sum_{j}  c^{[j]} \, \T_{(-+)}^{[j][8_a,8]_a} \,
\big({\cal M}_{gq \to ggq}^{\rm tree}\big)^{[8_a,8]_a}
&=& - \bigg( \frac{2 N_c}{\sqrt{N_c^2-1}}\, c^{[1,8]} 
+\frac{N_c}{2}\, c^{[8_s,8]_s} \\ 
&&\hspace{-2.5cm}+\, \sqrt{\frac{N_c + 3}{N_c +1}}\, 
c^{[27,8]} +\sqrt{\frac{N_c - 3}{N_c -1}}\, 
c^{[0,8]} \bigg)
\big({\cal M}_{gq \to ggq}^{\rm tree}\big)^{[8_a,8]_a}, \\ \nn 
\sum_{j}  c^{[j]} \, \T_{(-+)}^{[j][8_a,8_a]} \,
\big({\cal M}_{gg \to ggg}^{\rm tree}\big)^{[8_a,8_a]}
&=& \bigg( \frac{2 N_c}{\sqrt{N_c^2-1}}\, c^{[1, 8_a]} 
+\frac{N_c}{2}\, c^{[8_s, 8_a]} \\ 
&&\hspace{-2.5cm}+\, \sqrt{\frac{N_c + 3}{N_c +1}}\, 
c^{[27, 8_a]} +\sqrt{\frac{N_c - 3}{N_c -1}}\, 
c^{[0, 8_a]} \bigg)
\big({\cal M}_{gg \to ggg}^{\rm tree}\big)^{[8_a,8_a]}.
\eea
As expected, for the $gg \to ggg$
amplitude, the color operator $\T_{(-+)}$ projects 
onto the color basis elements which are even under 
$1\leftrightarrow 5$ and odd under $2\leftrightarrow 3$, 
as expected from Bose symmetry. 


\paragraph{The complete MR amplitude at one loop.}

The full one-loop multiple-Reggeon contribution is given 
by the sum of eqs.~(\ref{R2_calculationJ}),~(\ref{RgR2_calculationI}) and~(\ref{R2gR_calculationI}), and reads 

\bea \label{MR1loop} \nn
{\cal M}_{ij \to i'gj'}^{\rm MR}\Big|^{\mbox{\scriptsize 1-loop}}
&=& {\cal M}_{ij \to i'gj'} 
\Big|^{\mbox{\scriptsize 1-loop}}_{{\cal R}^2 g {\cal R}^2}
+{\cal M}_{ij \to i'gj'} 
\Big|^{\mbox{\scriptsize 1-loop}}_{{\cal R} g {\cal R}^2}
+{\cal M}_{ij \to i'gj'} 
\Big|^{\mbox{\scriptsize 1-loop}}_{{\cal R}^2 g {\cal R}} \\ \nn
&&\hspace{-2.5cm}=\, i \pi \, \frac{\as}{4\pi} \, \Bigg\{
\frac{1}{\eps} \Big[\T_{(--)}+\T_{(-+)}+\T_{(+-)}\Big] \\ \nn
&&\hspace{-2.5cm}
+ \ln \bigg(\frac{\mu^2 |\mathbf{p}_4|^2}{|\mathbf{p}_3|^2 
|\mathbf{p}_5|^2} \bigg) \T_{(--)}
+ \ln \bigg(\frac{\mu^2 |\mathbf{p}_3|^2}{|\mathbf{p}_4|^2 
|\mathbf{p}_5|^2} \bigg) \T_{(-+)} 
+\ln \bigg(\frac{\mu^2 |\mathbf{p}_5|^2}{|\mathbf{p}_4|^2 
|\mathbf{p}_3|^2} \bigg) \T_{(+-)} \\ \nn
&&\hspace{-2.5cm}+\,\eps \Bigg[
\Bigg(-\frac{\pi^2}{12} -2D_2(z,\bar z) 
+ \frac{1}{2} \ln^2\bigg( \frac{\mu^2}{|\mathbf{p}_3|^2} \bigg) 
- \frac{1}{2} \ln^2\bigg( \frac{\mu^2}{|\mathbf{p}_4|^2} \bigg) 
+ \frac{1}{2} \ln^2\bigg( \frac{\mu^2}{|\mathbf{p}_5|^2} \bigg)
\Bigg) \T_{(--)} \\ \nn
&&\hspace{-2.1cm}+\,\Bigg(-\frac{\pi^2}{12} +2D_2(z,\bar z)
- \frac{1}{2} \ln^2\bigg( \frac{\mu^2}{|\mathbf{p}_3|^2} \bigg) 
+ \frac{1}{2} \ln^2\bigg( \frac{\mu^2}{|\mathbf{p}_4|^2} \bigg) 
+ \frac{1}{2} \ln^2\bigg( \frac{\mu^2}{|\mathbf{p}_5|^2} \bigg)
\Bigg) \T_{(-+)} \\ \nn
&&\hspace{-2.1cm} +\,\Bigg(-\frac{\pi^2}{12} +2D_2(z,\bar z) 
+ \frac{1}{2} \ln^2\bigg( \frac{\mu^2}{|\mathbf{p}_3|^2} \bigg)
+ \frac{1}{2} \ln^2\bigg( \frac{\mu^2}{|\mathbf{p}_4|^2} \bigg)
- \frac{1}{2} \ln^2\bigg( \frac{\mu^2}{|\mathbf{p}_5|^2} \bigg) 
\Bigg) \T_{(+-)}  \Bigg] \\ 
&&\hspace{6.0cm}+\, \ord(\eps^2)\Bigg\} 
\, {\cal M}_{ij \to i'gj'}^{\rm tree}.
\eea


\subsection{Multi-Reggeon contributions to the two-loop amplitude\label{sec:MRETwoLoop}}

\begin{figure}[t]
    \centering
    \includegraphics[width=0.88\textwidth]{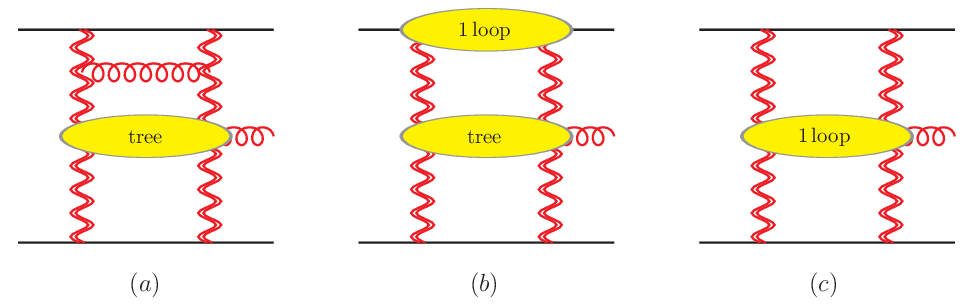}
    \caption{Multi-Reggeon exchange contribution to 
    two-loop even-even amplitude ${\cal M}^{(+,+)}$.
    Adopting the notation from \cite{Caron-Huot:2017fxr,Caron-Huot:2020vlo,Falcioni:2021buo},
    the horizontal gluon in the ladder represents one 
    application of the BFKL kernel, while the blob
    on the upper line $j$ represents the two-Reggeon 
    impact factor. There are two additional diagrams, 
    in which the BFKL kernel is applied below the 
    emission vertex, and one where the Two-Reggeon 
    impact factor is taken on the leg $i$.}
    \label{fig:TwoLoops-Even}
\end{figure}
We are now ready to consider the amplitude at two 
loops. At this order we need to take into account 
several multi-Reggeon transitions, which contribute 
to all the amplitude components ${\cal M}^{(\pm,\pm)}$.
As discussed in the introduction, purpose of this 
paper is to disentangle the Regge pole and cut 
contribution, such as to achieve an unambiguous 
definition of the Litpatov vertex at two loops. 
To this end we need to focus on the multi-Reggeon 
contribution to ${\cal M}^{(-,-)}$ only, that we
will calculate in detail in what follows. For
illustration, however, let us start by identifying 
all the contributions that appear at two loops.

As usual, we obtain such terms by expanding 
\eqn{ReggeFactBasic} to second order in the strong 
coupling constant. We recall at this point that 
the signature symmetry, i.e. the symmetry of the 
amplitude under the exchange of $1\leftrightarrow 5$
for the projectile~$i$, and $2\leftrightarrow 3$ for 
the target~$j$, is in direct correspondence with the 
parity of the number of Reggeons emitted: one Reggeon 
belong to the odd ($-$) sector, two Reggeons to the 
even ($+$) sector, three Reggeons to the odd sector 
again, and so on. Following this criterion, we can 
unambiguously assign all contributions to
a specific amplitude component ${\cal M}^{(\pm,\pm)}$.
Starting from the even-even amplitude, or 
${\cal M}^{(+,+)}$, we have 
\bea\label{TwoLoops-Even} \nn
\frac{i}{2s_{12}} {\cal \bar M}^{(+,+)}_{ij \to i'gj'}
|^{\mbox{\scriptsize 2-loops}} &=& 
\frac{i}{2s_{12}} {\cal \bar M}_{ij \to i'gj'} 
\Big|^{\mbox{\scriptsize 2-loops}}_{{\cal R}^2 g {\cal R}^2} \\[0.1cm] \nn
&=& - \eta_{2} \,\big\langle \psi_{j,2} \big| 
H_{2\to 2} \, a_4(p_4) \big| \psi_{i,2} \big\rangle^{\text{LO}}
- \eta_{1} \, \big\langle \psi_{j,2} \big| a_4(p_4) \,
H_{2\to 2}  \big| \psi_{i,2} \big\rangle^{\text{LO}} \\[0.1cm] 
&&+\, \big\langle \psi_{j,2} \big| a_4(p_4) 
\big| \psi_{i,2} \big\rangle^{\text{NLO}}.
\eea
In this equation, the term proportional to $\eta_2$
involves an application of the leading order BFKL kernel 
$H_{2\to 2}$ between the target $j$ and the emission of 
the central gluon, as represented in Fig.~\ref{fig:TwoLoops-Even} 
(a); in the second term, proportional to $\eta_1$, the BFKL 
kernel is applied between the central emission gluon and 
the projectile $i$. The last term, instead, involves the 
sum of three NLO contributions: the first is given by the 
insertion of the one-loop two-Reggeon impact factor 
on the target, as in Figure Fig.~\ref{fig:TwoLoops-Even} 
(b); the second contribution is obtained by inserting 
the two-Reggeon impact factor at one loop on the 
projectile; the third term is given by the one-loop 
version of the two-Reggeon-gluon emission vertex in 
\eqn{WWona4A}, which is yet to be calculated. Notice that 
the two terms involving $H_{2\to 2}$ in \eqn{TwoLoops-Even}
contributes at NLL to the amplitude, as evident from the 
rapidity factors $\eta_1$, $\eta_2$ multiplying them, while
the last term contributes at NNLL. 

\begin{figure}[t]
    \centering
    \includegraphics[width=0.88\textwidth]{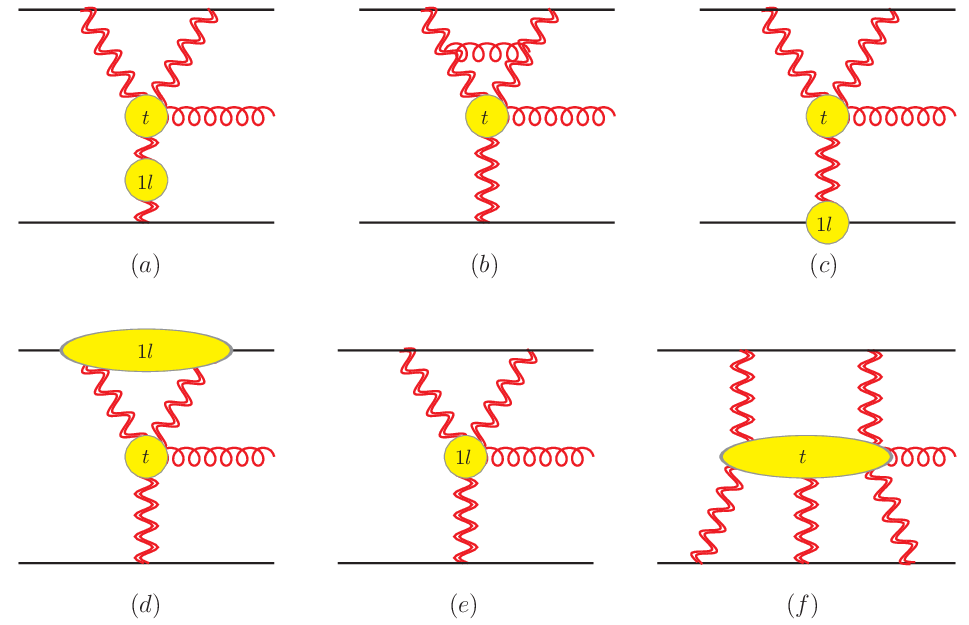}
    \caption{Multi-Reggeon exchange contribution to 
    the two-loop odd-even amplitude ${\cal M}^{(-,+)}$.
    A small $t$ indicates that the corresponding vertex
    is taken at tree level, while the label $1l$ indicates
    that one needs to consider the corresponding insertion 
    at one loop. The two-loop even-odd amplitude 
    ${\cal M}^{(+,-)}$ can be obtained from the 
    diagrams above by exploiting the target-projectile 
    (or top-bottom) symmetry.}
    \label{fig:TwoLoops-Odd-Even}
\end{figure}
Next, we consider the two-loop odd-even amplitude ${\cal M}^{(-,+)}$. 
In contrast to the one-loop case, 
where this amplitude only receives a
contribution from the 
${\cal R} g {\cal R}^2$ transition, at two loops we also need to take 
into account the transition ${\cal R}^3 g {\cal R}^2$, 
which involves three Reggeons emission from the target. This is in
correspondence to what happens in case of the $2\to 2$
scattering amplitudes: in that case, in fact, in the 
odd sector ${\cal M}_{ij \to ij}^{(-)}$, starting
at two loops one needs to take into account the mixing
between one and three Reggeons \cite{Caron-Huot:2017fxr}.
One has thus 
\be\label{TwoLoops-Odd-Even} 
\frac{i}{2s_{12}} {\cal \bar M}^{(-,+)}_{ij \to i'gj'}
|^{\mbox{\scriptsize 2-loops}} =
\frac{i}{2s_{12}} \bigg( {\cal \bar M}_{ij \to i'gj'} 
\Big|^{\mbox{\scriptsize 2-loops}}_{{\cal R} g {\cal R}^2}
+ {\cal \bar M}_{ij \to i'gj'} 
\Big|^{\mbox{\scriptsize 2-loops}}_{{\cal R}^3 g {\cal R}^2}
\bigg),
\ee
where in turn the two transitions reads  
\bea\label{TwoLoops-RgR2} \nn
\frac{i}{2s_{12}} {\cal \bar M}_{ij \to i'gj'} 
\Big|^{\mbox{\scriptsize 2-loops}}_{{\cal R} g {\cal R}^2}
&=& - \eta_{2} \,\big\langle \psi_{j,2} \big| 
\, a_4(p_4) H_{1\to 1} \big| \psi_{i,1} \big\rangle^{\text{LO}}
- \eta_{1} \, \big\langle \psi_{j,2} \big| H_{2\to 2} \, a_4(p_4) 
\big| \psi_{i,1} \big\rangle^{\text{LO}} \\[0.1cm] 
&&+\, \big\langle \psi_{j,2} \big| a_4(p_4) 
\big| \psi_{i,1} \big\rangle^{\text{NLO}},
\eea
and 
\bea\label{TwoLoops-R3gR2} \nn
\frac{i}{2s_{12}} {\cal \bar M}_{ij \to i'gj'} 
\Big|^{\mbox{\scriptsize 2-loops}}_{{\cal R}^3 g {\cal R}^2}
&=& \big\langle \psi_{j,2} \big| 
a_4(p_4) \big| \psi_{i,3} \big\rangle^{\text{LO}}.
\eea
The first term in \eqn{TwoLoops-RgR2} is represented 
in fig.~\ref{fig:TwoLoops-Odd-Even} (a), and involves 
the insertion of the $H_{1\to 1}$ evolution kernel 
(i.e., the Regge trajectory at one loop) between 
the projectile and the gluon emission. The second term
is represented by fig.~\ref{fig:TwoLoops-Odd-Even} 
(b), and involves the insertion of the leading order 
BFKL kernel $H_{2\to 2}$ between the central 
emission gluon and the target~$j$. 
The third term denoted by the superscript NLO
involves three contributions: the first term 
is represented in fig.~\ref{fig:TwoLoops-Odd-Even} 
(c), and involves the insertion of the one loop 
single Reggeon impact factor on the projectile; the 
second term, represented in fig.~\ref{fig:TwoLoops-Odd-Even} 
(d), is obtained by inserting the two-Reggeon 
impact factor on the target. Last, 
the third term is represented in
fig.~\ref{fig:TwoLoops-Odd-Even} (e), 
and involves the one- to two-Reggeon-gluon 
emission vertex at one loop, which, as in 
case of the two-Reggeon-gluon emission vertex
in Fig.~\ref{fig:TwoLoops-Even}, is yet to be 
calculated, and it would involve the BFLK 
kernel at NLO. The term in \eqn{TwoLoops-R3gR2}
is given instead by the three- to 
two-Reggeon-gluon emission vertex. 
This term can be obtained either by 
expanding \eqn{WWona4A} one additional 
order in $g_s$, thus taking into account 
the three-gluon terms on the r.h.s. of the 
equation; alternatively, this term is simply
given by all possible combinations of the 
one- to -two-Reggeon-gluon vertex given
in the second line of \eqn{Wona4} (central 
diagram in Fig.~\ref{2to3EffectiveVertices}),
with an additional Reggeon propagating between 
the target and the projectile. 

\begin{figure}[t]
    \centering
    \includegraphics[width=0.64\textwidth]{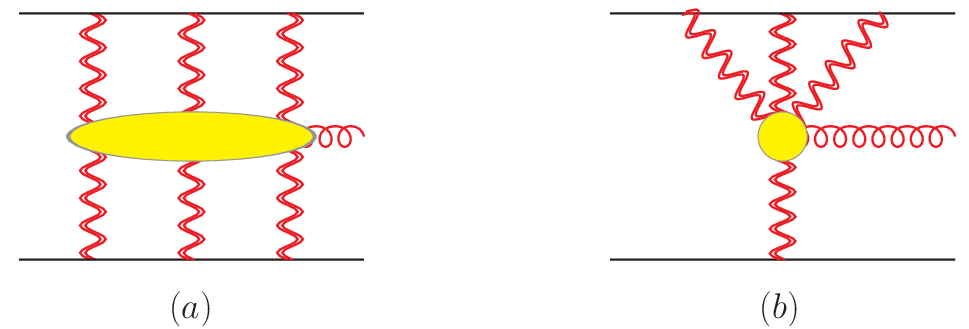}
    \caption{Multi-Reggeon exchange contribution to 
    the two-loop odd-even amplitude ${\cal M}^{(-,-)}$.}
    \label{fig:TwoLoop-Odd}
\end{figure}
Last, we have the odd-odd amplitude, namely ${\cal M}^{(-,-)}$. 
In this case we need to take into account 
one-to-three, three-to-one and three-to three Reggeon 
transitions, on top of the single-Reggeon transition 
we described in section \ref{SRamplitude}. We have
\bea\label{TwoLoops-Odd-odd}  \nn
\frac{i}{2s_{12}} {\cal \bar M}^{(-,-)}_{ij \to i'gj'}
|^{\mbox{\scriptsize 2-loops}} &=&
\frac{i}{2s_{12}} \bigg( {\cal \bar M}_{ij \to i'gj'} 
\Big|^{\mbox{\scriptsize 2-loops}}_{{\cal R} g {\cal R}}
+ {\cal \bar M}_{ij \to i'gj'} 
\Big|^{\mbox{\scriptsize 2-loops}}_{{\cal R} g {\cal R}^3} \\
&&\hspace{1.0cm}+\, {\cal \bar M}_{ij \to i'gj'} 
\Big|^{\mbox{\scriptsize 2-loops}}_{{\cal R}^3 g {\cal R}}
+ {\cal \bar M}_{ij \to i'gj'} 
\Big|^{\mbox{\scriptsize 2-loops}}_{{\cal R}^3 g {\cal R}^3}\bigg).
\eea
The first term in this equation represents the single
Reggeon transition, namely, considering the complete 
amplitude we have
\be
{\cal M}_{ij \to i'gj'} 
\Big|^{\mbox{\scriptsize 2-loops}}_{{\cal R} g {\cal R}}
= {\cal M}^{\rm SR}_{ij \to i'gj'} \Big|^{\mbox{\scriptsize 2-loops}},
\ee
which can be obtained by expanding 
\eqn{eq:ampDecomp} consistently to second
order in the strong coupling constant. 
The second term in \eqn{TwoLoops-Odd-odd}
reads 
\be\label{TwoLoops-RgR3} 
\frac{i}{2s_{12}} {\cal \bar M}_{ij \to i'gj'} 
\Big|^{\mbox{\scriptsize 2-loops}}_{{\cal R} g {\cal R}^3}
= \big\langle \psi_{j,3} \big| 
a_4(p_4) \big| \psi_{i,1} \big\rangle^{\text{LO}},
\ee
and is represented in Fig.~\ref{fig:TwoLoop-Odd}
(b), where the single-Reggeon to three-Reggeon-gluon
vertex is given by the last term in \eqn{Wona4},
also represented by the diagram on the right in 
Fig.~\ref{2to3EffectiveVertices}. Next, the 
third term reads
\be\label{TwoLoops-R3gR} 
\frac{i}{2s_{12}} {\cal \bar M}_{ij \to i'gj'} 
\Big|^{\mbox{\scriptsize 2-loops}}_{{\cal R}^3 g {\cal R}}
= \big\langle \psi_{j,1} \big| 
a_4(p_4) \big| \psi_{i,3} \big\rangle^{\text{LO}},
\ee
and can be obtained by target-projectile symmetry 
from the ${\cal R} g {\cal R}^3$ transition
that we just described. Last, the term involving 
three Reggeons reads 
\be\label{TwoLoops-R3gR3} 
\frac{i}{2s_{12}} {\cal \bar M}_{ij \to i'gj'} 
\Big|^{\mbox{\scriptsize 2-loops}}_{{\cal R}^3 g {\cal R}^3}
= \big\langle \psi_{j,3} \big| a_4(p_4) 
\big| \psi_{i,3} \big\rangle^{\text{LO}}.
\ee
It is represented in Fig.~\ref{fig:TwoLoop-Odd} 
(a), and is obtained by using \eqn{WWWona4A} 
onto $a_4(p_4) \big| \psi_{i,3} \big\rangle$. 
In the following section we proceed to evaluate 
explicitly the terms giving rise \eqn{TwoLoops-Odd-odd}. 


\subsection{Odd-odd Amplitude}\label{sec:oddOdd2}

\paragraph{The ${\cal R}^3 g {\cal R}^3$ transition.}

We start with the calculation of the three-Reggeon 
transition amplitude. Inserting \eqn{eq:psi3} into 
\eqn{TwoLoops-RgR2} one has
\bea \label{R3_calculationA} 
\frac{i}{2s_{12}} \
{\cal \bar M}_{ij \to i'gj'} 
\Big|^{\mbox{\scriptsize 2-loops}}_{{\cal R}^3 g {\cal R}^3}
&=& - \left(\frac{g_s^3}{6}\right)^2
(\T^a\T^b\T^c)_i  \, (\T^{a'}\T^{b'}\T^{c'})_j
\int [\dbar k_1] [\dbar k_2] 
\int [\dbar k_3] [\dbar k_4] \\[0.2cm] \nn
&&\hspace{-3.5cm} \times \,
\langle 0 | W^{a'}(k_3)W^{b'}(k_4)W^{c'}(-q_2-k_3-k_4) 
\, a_4^{d}(p_4) \, W^a(k_1)W^b(k_2)W^c(q_1-k_1-k_2)|0\rangle, 
\eea
Using \eqn{WWWona4A} with $q_1=k_1$, 
$q_2=k_2$, $p_1 = q_1$, $p_2=p_4$, 
we get 
\bea \label{R3_calculationB} \nn
\frac{i}{2s_{12}} \
{\cal \bar M}_{ij \to i'gj'} 
\Big|^{\mbox{\scriptsize 2-loops}}_{{\cal R}^3 g {\cal R}^3}
&=& 2 i \, g_s\left(\frac{g_s^3}{6}\right)^2
(\T^a\T^b\T^c)_i  \, (\T^{a'}\T^{b'}\T^{c'})_j \\ \nn
&&\hspace{-3.5cm} \times \int [\dbar k_1] [\dbar k_2] 
\int [\dbar k_3] [\dbar k_4]\, 
\langle 0 | W^{a'}(k_3)W^{b'}(k_4)W^{c'}(-q_2-k_3-k_4) \\ \
&&\hspace{-3.0cm} \times \Bigg\{
f^{ade} \, \bigg[\frac{p_4^{\mu}}{p_4^2}
+\frac{k_1^{\mu}}{k_1^2} \bigg] 
W^{e}(k_1+p_4)W^b(k_2)W^c(q_1-k_1-k_2)|0\rangle \\ \nn
&&\hspace{-2.8cm} +\, f^{bde} \, 
\bigg[\frac{p_4^{\mu}}{p_4^2}
+\frac{k_2^{\mu}}{k_2^2} \bigg] 
W^{a}(k_1)W^e(k_2+p_4)W^c(q_1-k_1-k_2)|0\rangle  \\ \nn
&&\hspace{-2.8cm} +\, f^{cde} \,
\bigg[\frac{p_4^{\mu}}{p_4^2}
+\frac{(q_1-k_1-k_2)^{\mu}}{(q_1-k_2-k_2)^2} \bigg] 
W^{a}(k_1)W^b(k_2)W^c(q_1-k_1-k_2+p_4)|0\rangle
\Bigg\} \, \varepsilon_{\mu}(p_4)\,.
\eea
From here one proceeds as illustrated 
in the previous sections, namely, we 
contract the Reggeon fields according 
to \eqn{Reggeon_prop1}, which gives rise 
to three Dirac deltas for each terms; 
subsequently we integrate over $k_3$ 
and $k_4$, thus enforcing momentum
conservation. In the end one can exploit 
the freedom to shift the remaining two 
integration variables, i.e. $k_1$ and 
$k_2$, to express the momentum 
dependent part in a compact form. 
After some work one gets
\bea \label{R3_calculationD} 
\frac{i}{2s_{12}} \nn
{\cal \bar M}_{ij \to i'gj'} 
\Big|^{\mbox{\scriptsize 2-loops}}_{{\cal R}^3 g {\cal R}^3}
&=& i \,\frac{16\pi^4}{3} \, g^3_s \left(\frac{\as}{\pi}\right)^2 
(2\pi)^{2-2\eps}\delta^{(2-2\eps)}(p_4+q_1+q_2) \\[0.1cm] 
&&\hspace{-2.0cm} \times\, 
\Big\{(\T^a\T^b\T^c)_i + 
(\T^b\T^a\T^c)_i + (\T^c\T^b\T^a)_i \Big\}  
\,(i f^{a'da})\,\T_{j}^{\{a',b,c\}}  \\[0.1cm] \nn
&&\hspace{-2.0cm} \times\, 
\int [\dbar k_1] [\dbar k_2] 
\frac{1}{k_1^2k_2^2(k_1+k_2+q_2)^2}
\bigg[\frac{p_4^{\mu}}{p_4^2} - 
\frac{(k_1 + k_2 - q_1)^{\mu}}{(k_1 + k_2 - q_1)^2} 
\bigg] \varepsilon_{\mu}(p_4), 
\eea
where, in analogy to \eqn{TabDef1}, 
we have used the notation
\be\label{TabDef2}
\T_i^{\{a,b,c\}} = \frac1{3!}
\sum_{\sigma\in\mathcal{S}_3}
\T_i^{\sigma(a)}\T_i^{\sigma(b)}\T_i^{\sigma(c)}.
\ee
In order to express this result in terms of 
loop integrals and color operators acting on 
the tree level amplitude, we first notice 
that the color structure in \eqn{R3_calculationD} 
can be written as follows:
\bea\label{eq:TmmColourFactor2} \nn
\Big\{(\T^a\T^b\T^c)_i + 
(\T^b\T^a\T^c)_i + (\T^c\T^b\T^a)_i \Big\}
\,(i f^{a'da})\,\T_{j}^{\{a',b,c\}} && \\ \nn
&& \hspace{-8.0cm}=\, \frac{1}{6}
\bigg\{\T_1\cdot \T_2 \big(+2 \T_1\cdot \T_2 
+ \T_1\cdot \T_3 - \T_2\cdot \T_5 - \T_3\cdot \T_5 \big) \\ \nn
&& \hspace{-7.0cm}+\,
\T_1\cdot \T_3 \big(- \T_1\cdot \T_2 
+2\T_1\cdot \T_3 + \T_2\cdot \T_5 - \T_3\cdot \T_5 \big) \\[0.2cm] \nn
&& \hspace{-7.0cm}+\,
\T_2\cdot \T_5 \big(- \T_1\cdot \T_2 
+2 \T_2\cdot \T_5 - \T_3\cdot \T_5 \big) \\[0.2cm] \nn
&& \hspace{-7.0cm}+\,
\T_3\cdot \T_5 \big(- \T_1\cdot \T_3 
- \T_2\cdot \T_5 +2\T_3\cdot \T_5 \big) \\[0.1cm] \nn
&& \hspace{-7.0cm}+\,\frac{C_A}{2}
\big(\T_1+\T_5\big)\big(\T_2+\T_3\big) \bigg\}
\big[\T_i^{a} \, i f^{a'da} \, \T_j^{a'} \big] \\ 
&& \hspace{-8.0cm}=\, 
\frac{1}{48} \bigg[9\T_{(--)}^2 +\T_{(++)}^2
+3 \Big(\T_{(+-)}^2 + \T_{(-+)}^2 \Big)
+ 4 C_A \T_{(++)}\bigg]\,\mathcal{C}_{ij}^{(0)}.
\eea 
Inserting this result into \eqn{R3_calculationD},
switching from ${\cal \bar M}_{ij \to i'gj'}$
to ${\cal M}_{ij \to i'gj'}$ according to 
\eqn{ReggeFactBasic}, taking into account 
momentum conservation and the fact that
$Z_{i/j} = 1+\ord(\as)$, at two loops 
we obtain
\bea \label{R3_calculationE} \nn 
{\cal M}_{ij \to i'gj'} 
\Big|^{\mbox{\scriptsize 2-loops}}_{{\cal R}^3 g {\cal R}^3}
&=& \frac{2\pi^4}{9} \, s_{12} \, g^3_s 
\left(\frac{\as}{\pi}\right)^2\,
(2\pi)^{2-2\eps}\delta^{(2-2\eps)}(p_3+p_4+p_5) 
\, C_i^{(0)} C_j^{(0)}  \\[0.1cm] 
&&\hspace{-1.0cm}  \times\,
\bigg[9\T_{(--)}^2 +\T_{(++)}^2
+3 \Big(\T_{(+-)}^2 + \T_{(-+)}^2 \Big)
+ 4 C_A \T_{(++)}\bigg] \, 
\mathcal{C}_{ij}^{(0)} \\[0.1cm] \nn
&&\hspace{-1.0cm} 
\times\, \int [\dbar k_1] [\dbar k_2] 
\frac{1}{k_1^2k_2^2(k_1+k_2+p_3)^2}
\bigg[\frac{p_4^{\mu}}{p_4^2} - 
\frac{(k_1 + k_2 - p_5)^{\mu}}{(k_1 + k_2 - p_5)^2} 
\bigg] \varepsilon_{\mu}(p_4), 
\eea
This expression can be elaborates such that 
one of the two loop integrals reduces to 
a simple bubble integral, of the form given 
in \eqn{Bubble}, and the second is given in 
terms of the same family of integrals defined 
in \eqn{MIdef}. To this end, we shift $k_1+k_2 \to k_2$, 
then perform the $k_1$ bubble integral, 
obtaining 
\bea \nn
\pi^2 \int [\dbar k_1] [\dbar k_2] 
\frac{1}{k_1^2k_2^2(k_1+k_2+p_3)^2}
\bigg[\frac{p_4^{\mu}}{p_4^2} - 
\frac{(k_1 + k_2 - p_5)^{\mu}}{(k_1 + k_2 - p_5)^2} 
\bigg] \varepsilon_{\mu}(p_4) && \\ 
&&\hspace{-10.0cm}=\, 
\pi \frac{\mu^{2\eps} B_{1,1}(\eps)}{2\eps} 
\int [\dbar k_2] 
\frac{1}{(k_2^2)^{1+\eps}(k_2+p_3)^2}
\bigg[\frac{p_4^{\mu}}{p_4^2} - 
\frac{(k_2 - p_5)^{\mu}}{(k_2 - p_5)^2} 
\bigg] \varepsilon_{\mu}(p_4).
\eea
The integral over $k_2$ can be reduced 
by means of IBPs to a set of master 
integrals belonging to the topology
in \eqn{MIdef}.  
After some elaboration we obtain
\bea \label{R3_calculationF} \nn
{\cal M}_{ij \to i'gj'} 
\Big|^{\mbox{\scriptsize 2-loops}}_{{\cal R}^3 g {\cal R}^3}
&=& (i \pi)^2 \, \left(\frac{\as}{\pi}\right)^2\, 
\frac{\mu^{2\eps} B_{1,1}(\eps)}{2\eps} 
\frac{1}{18(z-\bar z)}  \\[0.1cm] 
&&\hspace{-1.0cm}
\times\,\bigg\{ (1 - \bar z)\,{\cal I}_{(1+\eps, 1, 0)} 
+ z  \, {\cal I}_{(1+\eps, 0, 1)} 
-\frac{{\cal I}_{(\eps, 1, 1)}}{|{\bf p}_4|^2 \bar z (1 - z)} 
- {\cal I}_{(1+\eps, 1, 1)} \bigg\} \\[0.1cm] \nn
&&\hspace{0.0cm}\times\,  
\bigg[9\T_{(--)}^2 +\T_{(++)}^2   
+3 \Big(\T_{(+-)}^2 + \T_{(-+)}^2 \Big)  
+ 4 C_A \T_{(++)}\bigg] \,
{\cal M}_{ij \to igj}^{\rm tree}.
\eea
The first two master integrals are
easily evaluated by means of \eqn{Bubble},
while the last two integrals require a 
generalization of \eqn{TriangleMI}.
For conciseness, let us define the 
color operator involved in the 
${\cal R}^3 g {\cal R}^3$ transition 
as follows:
\be \label{Cmm2La}
{\bf {\cal C}}_{{\cal R}^3 g {\cal R}^3}
\equiv 9\T_{(--)}^2 +\T_{(++)}^2   
+3 \Big(\T_{(+-)}^2 + \T_{(-+)}^2 \Big)  
+ 4 C_A \T_{(++)}.
\ee
After some elaboration we obtain
\be \label{R3_calculationG} 
{\cal M}_{ij \to i'gj'} 
\Big|^{\mbox{\scriptsize 2-loops}}_{{\cal R}^3 g {\cal R}^3}
= (i \pi)^2  \left(\frac{\as}{\pi}\right)^2 \, 
G^{(2)}_{{\cal R}^3 g {\cal R}^3}(z,\bar z, |\mathbf{p}_4|^2,\mu^2)
\, {\bf {\cal C}}_{{\cal R}^3 g {\cal R}^3} \,
{\cal M}_{ij \to igj}^{\rm tree}.
\ee
where, setting for simplicity $\mu^2=|\mathbf{p}_4|^2$, 
$G^{(2)}_{{\cal R}^3 g {\cal R}^3}$ reads
\bea \label{Gmm2La} 
G^{(2)}_{{\cal R}^3 g {\cal R}^3}(z,\bar z, 
|\mathbf{p}_4|^2,|\mathbf{p}_4|^2) &=&  
\frac{1}{288} \bigg\{
\frac{1}{\eps^2} - \frac{2}{\eps} 
\log \big[z \bar z (1-z)(1-\bar z) \big]
-6\,D_2(z,\bar z) - \zeta_2 \\[0.1cm] \nn 
&&\hspace{-3.0cm}
+\, 2\log^2\big[z\bar z\big]
+ 2\log^2\big[(1-z)(1-\bar z) \big] 
+ \log \big[z\bar z\big]
\log \big[(1-z)(1-\bar z) \big]
+\ord(\eps) \bigg\}.
\eea

To conclude the evaluation of the
${\cal R}^3 g {\cal R}^3$ transition, 
we report the explicit action of the 
color operator on the tree level 
amplitude, using the same notation 
used in \eqns{eq:TmmColourFactor}{eq:TpmColourFactor}:
we have
\bea\label{eq:ColourFactor2LoopsR3gR3} 
\sum_{j}  c^{[j]} \, 
{\bf {\cal C}}^{[j][8,8]_a}_{{\cal R}^3 g {\cal R}^3} \,
\big({\cal M}_{qq \to qgq}^{\rm tree}\big)^{[8,8]_a}
&=& \bigg( 2 N_c^2 -12 + \frac{36}{N_c^2} \bigg)c^{[8, 8]_a}
\big({\cal M}_{qq \to qgq}^{\rm tree}\big)^{[8,8]_a}, \\ 
\sum_{j}  c^{[j]} \, 
{\bf {\cal C}}^{[j][8,8_a]_a}_{{\cal R}^3 g {\cal R}^3} \,
\big({\cal M}_{qg \to qgg}^{\rm tree}\big)^{[8,8_a]_a}
&=& \big(2 N_c^2 + 12 \big) c^{[8, 8_a]_a}
\big({\cal M}_{qg \to qgg}^{\rm tree}\big)^{[8,8_a]_a}, \\ 
\sum_{j}  c^{[j]} \, 
{\bf {\cal C}}^{[j][8_a,8]_a}_{{\cal R}^3 g {\cal R}^3} \,
\big({\cal M}_{gq \to ggq}^{\rm tree}\big)^{[8_a,8]_a}
&=& \big(2 N_c^2 + 12 \big) c^{[8_a,8]_a}
\big({\cal M}_{gq \to ggq}^{\rm tree}\big)^{[8_a,8]_a}, \\ \nn 
\sum_{j}  c^{[j]} \, 
{\bf {\cal C}}^{[j][8_a,8_a]}_{{\cal R}^3 g {\cal R}^3}\,
\big({\cal M}_{gg \to ggg}^{\rm tree}\big)^{[8_a,8_a]}
&=& \bigg[\big(2N_c^2+72\big) c^{[8_a, 8_a]} \\ 
&&\hspace{0.0cm}-\, 36 \sqrt{N_c -4}\,
c^{[10, \overline{10}]_1} \bigg]
\big({\cal M}_{gg \to ggg}^{\rm tree}\big)^{[8_a,8_a]}.
\eea


\paragraph{The ${\cal R} g {\cal R}^3$ transition.}

We continue with the Reggeon to three Reggeon
transition, represented by the diagram (b) in 
Fig.~\ref{fig:TwoLoop-Odd}, which, to the best 
of our knowledge, has not been considered in 
previous works. This transition involve the 
state \eqn{eq:psi3} for the target $j$ and the 
state \eqn{eq:psi1} for the projectile $i$. 
Inserting these states into \eqn{TwoLoops-RgR3} 
we have 
\bea \label{RgR3_calculationA} 
\frac{i}{2s_{12}} \
{\cal \bar M}_{ij \to i'gj'} 
\Big|^{\mbox{\scriptsize 2-loops}}_{{\cal R} g {\cal R}^3}
&=& g_s \left(\frac{g_s^3}{6}\right) \T^{a}_i \, 
(\T^{a'}\T^{b'}\T^{c'})_j
\int [\dbar k_1] [\dbar k_2] \\[0.2cm] \nn
&&\hspace{-3.5cm} \times \,
\langle 0 | W^{a'}(k_3) W^{b'}(k_4) W^{c'}(-q_2 - k_3- k_4) 
\, a_4^{d}(p_4) \, W^a(q_1)|0\rangle. 
\eea
Using the third line of \eqn{Wona4} with $p_1 = q_1$, 
$p_2=p_4$, we get 
\bea \label{RgR3_calculationB} \nn
\frac{i}{2s_{12}} \
{\cal \bar M}_{ij \to i'gj'} 
\Big|^{\mbox{\scriptsize 2-loops}}_{{\cal R} g {\cal R}^3}
&=& i \, \frac{g_s^7}{36} \, \, 
f^{dgx}f^{xfy}f^{yae} \T^{a}_i \,
(\T^{a'}\T^{b'}\T^{c'})_j \, 
\int [\dbar k_1] [\dbar k_2] 
\int [\dbar k_3] [\dbar k_4] \\[0.2cm] \nn
&&\hspace{-3.0cm} \times \, 
\langle 0 | W^{a'}(k_3) W^{b'}(k_4) W^{c'}(-q_2 - k_3- k_4)
\, W^{g}(q_1+p_4-k_1-k_2)W^f(k_1)W^e(k_2)|0\rangle \\[0.2cm] 
&&\hspace{0.0cm} \times \, 
\bigg[\frac{(q_1-k_1-k_2)^{\mu}}{(q_1-k_1-k_2)^2}
- 3 \frac{(q_1-k_2)^{\mu}}{(q_1-k_2)^2}
+ 2 \frac{q_1^{\mu}}{q_1^2} \bigg] 
\varepsilon_{\mu}(p_4)\,.
\eea
From here we contract the Reggeon fields 
according to \eqn{Reggeon_prop1}; this 
gives rise to three Dirac deltas for 
each terms, and subsequently we integrate 
over $k_3$ and $k_4$, thus enforcing momentum
conservation. In the end we exploit 
the freedom to shift the remaining two 
integration variables, i.e. $k_1$ and 
$k_2$, to express the momentum 
dependent part in a compact form. 
After some work we get
\bea \label{RgR3_calculationD} \nn 
\frac{i}{2s_{12}} 
{\cal \bar M}_{ij \to i'gj'} 
\Big|^{\mbox{\scriptsize 2-loops}}_{{\cal R} g {\cal R}^3}
&=& \frac{8\pi^4}{3} \, g^3_s \left(\frac{\as}{\pi}\right)^2 
(2\pi)^{2-2\eps}\delta^{(2-2\eps)}(p_4+q_1+q_2)  \\[0.2cm] \nn
&&\hspace{-1.0cm} \times\, f^{dgx}f^{xfy}f^{yae} \, 
\T^{a}_i \,\T_{j}^{\{e,f,g\}} \int [\dbar k_1] [\dbar k_2] 
\,\frac{1}{k_1^2k_2^2(k_1+k_2+q_2)^2} \\[0.2cm] 
&&\hspace{0.0cm} \times\,
\bigg[\frac{(q_1-k_1-k_2)^{\mu}}{(q_1-k_1-k_2)^2}
- 3 \frac{(q_1-k_2)^{\mu}}{(q_1-k_2)^2}
+ 2 \frac{q_1^{\mu}}{q_1^2} \bigg] 
\varepsilon_{\mu}(p_4).
\eea
The color structure can be written as follows:
\bea\label{eq:TmmColourFactor2C} \nn
i f^{dgx}f^{xfy}f^{yae} \, \T^{a}_i 
\,\T_{j}^{\{e,f,g\}} && \\ \nn
&& \hspace{-3.0cm}=\, - \frac{1}{6}
\bigg\{\T_1\cdot \T_2 \big(+2 \T_1\cdot \T_2 
- \T_1\cdot \T_3 +2 \T_2\cdot \T_5 - 2 \T_3\cdot \T_5 \big) \\ \nn
&& \hspace{-2.0cm}+\,
\T_1\cdot \T_3 \big( - \T_1\cdot \T_2 
+2\T_1\cdot \T_3 -2 \T_2\cdot \T_5 +2 \T_3\cdot \T_5 \big) \\[0.2cm] \nn
&& \hspace{-2.0cm}+\,
\T_2\cdot \T_5 \big( + 2\T_1\cdot \T_2 
+2 \T_2\cdot \T_5 - \T_3\cdot \T_5 \big) \\[0.2cm] \nn
&& \hspace{-2.0cm}+\,
\T_3\cdot \T_5 \big( + 2 \T_1\cdot \T_3 
- \T_2\cdot \T_5 +2 \T_3\cdot \T_5 \big) \\[0.1cm] \nn
&& \hspace{-2.0cm}+\,\frac{C_A}{2}
\big(\T_1+\T_5\big)\big(\T_2+\T_3\big) \bigg\} 
\big[\T_i^{a} \, i \, f^{a'c a} \, \T_j^{a'} \big] \\ 
&& \hspace{-3.0cm}=\, 
- \frac{1}{12} \bigg[\T_{(++)}^2
+ 3\T_{(+-)}^2 + C_A \T_{(++)}\bigg]\,\mathcal{C}_{ij}^{(0)}.
\eea 
Inserting this result into \eqn{R3_calculationD},
switching from ${\cal \bar M}_{ij \to i'gj'}$
to ${\cal M}_{ij \to i'gj'}$ according to 
\eqn{ReggeFactBasic}, taking into account 
momentum conservation and the fact that 
$Z_{i/j} = 1+\ord(\as)$, at two loops 
we obtain
\bea \label{RgR3_calculationE} \nn 
{\cal M}_{ij \to i'gj'} 
\Big|^{\mbox{\scriptsize 2-loops}}_{{\cal R} g {\cal R}^3}
&=& \frac{4\pi^4}{9} \, s_{12} \, g^3_s 
\left(\frac{\as}{\pi}\right)^2\,
(2\pi)^{2-2\eps}\delta^{(2-2\eps)}(p_3+p_4+p_5) 
\, C_i^{(0)} C_j^{(0)}  \\[0.1cm]  \nn
&&\hspace{-1.0cm}  \times\,
\bigg[\T_{(++)}^2 + 3\T_{(+-)}^2 
+ C_A \T_{(++)}\bigg]\,\mathcal{C}_{ij}^{(0)}
\int [\dbar k_1] [\dbar k_2] \,
\frac{1}{k_1^2k_2^2(k_1+k_2+p_3)^2} \\[0.1cm] 
&&\hspace{2.0cm} \times\,  
\bigg[2 \frac{p_5^{\mu}}{p_5^2}
+ 3 \frac{(k_2-p_5)^{\mu}}{(k_2-p_5)^2}
- \frac{(k_1+k_2-p_5)^{\mu}}{(k_1+k_2-p_5)^2}
\bigg] \varepsilon_{\mu}(p_4). 
\eea
We write the loop integral as follows:
\bea \nn
\pi^2 \int \frac{[\dbar k_1] 
[\dbar k_2]}{k_1^2k_2^2(k_1+k_2+p_3)^2} 
\bigg[3 \bigg(\frac{p_5^{\mu}}{p_5^2}
+ \frac{(k_2-p_5)^{\mu}}{(k_2-p_5)^2} \bigg)
- \bigg(\frac{p_5^{\mu}}{p_5^2}
+ \frac{(k_1+k_2-p_5)^{\mu}}{(k_1+k_2-p_5)^2}
\bigg)\bigg] \varepsilon_{\mu}(p_4) && \\[0.2cm] \nn
&&\hspace{-12.0cm}=\, 
\pi \frac{\mu^{2\eps} B_{1,1}(\eps)}{2\eps} 
\int [\dbar k_2] \Bigg\{ 
\frac{1}{(k_2^2)^{1+\eps}(k_2+p_3)^2}
\bigg[2 \frac{p_5^{\mu}}{p_5^2}
- \frac{(k_2-p_5)^{\mu}}{(k_2-p_5)^2} \bigg] \\
&&\hspace{-7.0cm}+ \frac{3}{k_2^2 [(k_2+p_3)^2]^{1+\eps}}
\frac{(k_2-p_5)^{\mu}}{(k_2-p_5)^2} \Bigg\} 
\varepsilon_{\mu}(p_4).
\eea
The integral over $k_2$ can be reduced 
by means of IBPs to a set of master 
integrals belonging to the topology
in \eqn{MIdef}, obtaining
\bea \label{RgR3_calculationF} \nn
{\cal M}_{ij \to i'gj'} 
\Big|^{\mbox{\scriptsize 2-loops}}_{{\cal R} g {\cal R}^3}
&=& - (i \pi)^2 \, \left(\frac{\as}{\pi}\right)^2\, 
\frac{\mu^{2\eps} B_{1,1}(\eps)}{2\eps} 
\frac{1}{3(z-\bar z)}  \\[0.1cm] \nn
&&\hspace{-3.0cm}
\times\,\bigg\{ (1-z)
\bigg( {\cal I}_{(1, 1+\eps, 0)}
- \frac{1}{3} {\cal I}_{(1, 1+\eps, 0)} \bigg)
+z\bigg( {\cal I}_{(1, \eps, 1)}
- \frac{1}{3} {\cal I}_{(1+\eps, 0, 1)} \bigg) 
\\[0.1cm] \nn
&&\hspace{-2.5cm}+\, 
\frac{1}{|{\bf p}_4|^2 \bar z (1 - z)}
\bigg( \frac{1}{3}{\cal I}_{(\eps, 1, 1)}
- {\cal I}_{(0, 1+\eps, 1)} \bigg) 
-{\cal I}_{(1, 1+\eps, 1)} + \frac{1}{3}
{\cal I}_{(1+\eps, 1, 1)} \bigg\} \\
&&\hspace{2.0cm} \times \, 
\bigg[\T_{(++)}^2 + 3\T_{(+-)}^2 
+ C_A \T_{(++)}\bigg] \,
{\cal M}_{ij \to igj}^{\rm tree}.
\eea
The bubble integrals can be evaluated 
in terms of the general formula 
given in \eqn{Bubble}, while the 
triangle integrals are given in terms 
of a generalization of \eqn{TriangleMI}.
For conciseness, let us define the 
color operator involved in the 
${\cal R} g {\cal R}^3$ transition 
as follows: 
\be \label{Cmm2Lb}
{\bf {\cal C}}_{{\cal R} g {\cal R}^3}
\equiv \T_{(++)}^2 + 3\T_{(+-)}^2 
+ C_A \T_{(++)},
\ee
Setting for simplicity $\mu^2=|\mathbf{p}_4|^2$, 
after some elaboration we obtain
\be \label{RgR3_calculationG} 
{\cal M}_{ij \to i'gj'} 
\Big|^{\mbox{\scriptsize 2-loops}}_{{\cal R} g {\cal R}^3}
= (i \pi)^2  \left(\frac{\as}{\pi}\right)^2 \,
G^{(2)}_{{\cal R} g {\cal R}^3}(z,\bar z, |\mathbf{p}_4|^2,\mu^2)
\, {\bf {\cal C}}_{{\cal R} g {\cal R}^3} \,
{\cal M}_{ij \to igj}^{\rm tree},
\ee
where, setting $\mu^2=|\mathbf{p}_4|^2$,
we have
\bea \label{Gmm2Lb} 
G^{(2)}_{{\cal R} g {\cal R}^3}(z,\bar z, 
|\mathbf{p}_4|^2,|\mathbf{p}_4|^2)
&=& \frac{1}{144} \bigg\{
\frac{1}{\eps^2} - \frac{2}{\eps} 
\log \bigg[\frac{z \bar z}{(1-z)^2(1-\bar z)^2} \bigg]
+12\,D_2(z,\bar z) - \zeta_2 \\[0.1cm] \nn 
&&\hspace{-3.0cm}
+\, 2\log^2\big[z\bar z\big]
- \log^2\big[(1-z)(1-\bar z) \big] 
- 2 \log \big[z\bar z\big]
\log \big[(1-z)(1-\bar z) \big]
+\ord(\eps) \bigg\}.
\eea

The color operator acting on the 
tree level can be evaluated explicitly, 
and following the notation of 
\eqn{eq:ColourFactor2LoopsR3gR3}
we have 
\bea\label{eq:ColourFactor2LoopsRgR3} 
\sum_{j}  c^{[j]} \, 
{\bf {\cal C}}^{[j][8,8]_a}_{{\cal R} g {\cal R}^3} \,
\big({\cal M}_{qq \to qgq}^{\rm tree}\big)^{[8,8]_a}
&=& \bigg( \frac{N_c^2}{2} +3 \bigg)c^{[8, 8]_a}
\big({\cal M}_{qq \to qgq}^{\rm tree}\big)^{[8,8]_a}, \\ \nn
\sum_{j}  c^{[j]} \, 
{\bf {\cal C}}^{[j][8,8_a]_a}_{{\cal R} g {\cal R}^3} \,
\big({\cal M}_{qg \to qgg}^{\rm tree}\big)^{[8,8_a]_a}
&=& \bigg[ \bigg(\frac{N_c^2}{2} + 18 \bigg) c^{[8, 8_a]_a} \\ 
&&\hspace{-1.0cm}
-\, \frac{9\sqrt{N_c^2-4}}{\sqrt{2}} 
c^{[8, 10+\overline{10}]} \bigg]
\big({\cal M}_{qg \to qgg}^{\rm tree}\big)^{[8,8_a]_a}, \\ \nn 
\sum_{j}  c^{[j]} \, 
{\bf {\cal C}}^{[j][8_a,8]_a}_{{\cal R} g {\cal R}^3} \,
\big({\cal M}_{gq \to ggq}^{\rm tree}\big)^{[8_a,8]_a}
&=& \bigg(\frac{N_c^2}{2} + 3 \bigg) c^{[8_a,8]_a} 
\big({\cal M}_{gq \to ggq}^{\rm tree}\big)^{[8_a,8]_a}, \\ \nn 
\sum_{j}  c^{[j]} \, 
{\bf {\cal C}}^{[j][8_a,8_a]}_{{\cal R} g {\cal R}^3}\,
\big({\cal M}_{gg \to ggg}^{\rm tree}\big)^{[8_a,8_a]}
&=& \bigg[ \bigg(\frac{N_c^2}{2} + 18 \bigg) c^{[8_a, 8_a]} \\ 
&&\hspace{-1.0cm}
-\, \frac{9\sqrt{N_c^2-4}}{\sqrt{2}} 
c^{[8_a, 10+\overline{10}]} \bigg]
\big({\cal M}_{gg \to ggg}^{\rm tree}\big)^{[8_a,8_a]}.
\eea

With the result in \eqn{RgR3_calculationG}
we immediately get the amplitude corresponding 
to the transition ${\cal R}^3 g {\cal R}$
by exploiting the target-projectile symmetry. Namely,
starting from \eqn{RgR3_calculationG} we replace 
$\T_{(+-)} \to \T_{(-+)}$, as well as
$z \to 1-\bar z$, $\bar z \to 1-z$ 
for the kinematic factors. Taking into 
account \eqn{D2symmetry} we get 
\be \label{R3gR_calculationG}
{\cal M}_{ij \to i'gj'} 
\Big|^{\mbox{\scriptsize 2-loops}}_{{\cal R}^3 g {\cal R}}
= (i \pi)^2  \left(\frac{\as}{\pi}\right)^2 \,
G^{(2)}_{{\cal R}^3 g {\cal R}}(z,\bar z, |\mathbf{p}_4|^2,\mu^2)
\, {\bf {\cal C}}_{{\cal R}^3 g {\cal R}} \,
{\cal M}_{ij \to igj}^{\rm tree},
\ee
where
\bea \label{Gmm2Lc} 
G^{(2)}_{{\cal R}^3 g {\cal R}}(z,\bar z, 
|\mathbf{p}_4|^2,|\mathbf{p}_4|^2)
&=&  \frac{1}{144} \bigg\{
\frac{1}{\eps^2} - \frac{2}{\eps} 
\log \bigg[\frac{(1-z) (1-\bar z)}{z^2 \bar z^2} \bigg]
+12\,D_2(z,\bar z) - \zeta_2 \\[0.1cm] \nn 
&&\hspace{-3.0cm}
-\, \log^2\big[z\bar z\big]
+ 2 \log^2\big[(1-z)(1-\bar z) \big] 
- 2 \log \big[z\bar z\big]
\log \big[(1-z)(1-\bar z) \big]
+\ord(\eps) \bigg\},
\eea
and the color operator reads 
\be\label{Cmm2Lc}
{\bf {\cal C}}_{{\cal R}^3 g {\cal R}}
\equiv \T_{(++)}^2 + 3\T_{(-+)}^2 
+ C_A \T_{(++)}.
\ee
Its action on the tree level 
amplitude explicitly gives 
\bea\label{eq:ColourFactor2LoopsR3gR} 
\sum_{j}  c^{[j]} \, 
{\bf {\cal C}}^{[j][8,8]_a}_{{\cal R}^3 g {\cal R}} \,
\big({\cal M}_{qq \to qgq}^{\rm tree}\big)^{[8,8]_a}
&=& \bigg( \frac{N_c^2}{2} +3 \bigg)c^{[8, 8]_a}
\big({\cal M}_{qq \to qgq}^{\rm tree}\big)^{[8,8]_a}, \\ 
\sum_{j}  c^{[j]} \, 
{\bf {\cal C}}^{[j][8,8_a]_a}_{{\cal R}^3 g {\cal R}} \,
\big({\cal M}_{qg \to qgg}^{\rm tree}\big)^{[8,8_a]_a}
&=& \bigg(\frac{N_c^2}{2} + 3 \bigg) c^{[8, 8_a]_a} 
\big({\cal M}_{qg \to qgg}^{\rm tree}\big)^{[8,8_a]_a}, \\ \nn 
\sum_{j}  c^{[j]} \, 
{\bf {\cal C}}^{[j][8_a,8]_a}_{{\cal R}^3 g {\cal R}} \,
\big({\cal M}_{gq \to ggq}^{\rm tree}\big)^{[8_a,8]_a}
&=& \bigg[ \bigg(\frac{N_c^2}{2} + 18 \bigg) c^{[8_a,8]_a} \\ 
&&\hspace{-1.0cm}
-\, \frac{9\sqrt{N_c^2-4}}{\sqrt{2}} 
c^{[10+\overline{10},8]} \bigg]
\big({\cal M}_{gq \to ggq}^{\rm tree}\big)^{[8_a,8]_a}, \\ \nn 
\sum_{j}  c^{[j]} \, 
{\bf {\cal C}}^{[j][8_a,8_a]}_{{\cal R}^3 g {\cal R}}\,
\big({\cal M}_{gg \to ggg}^{\rm tree}\big)^{[8_a,8_a]}
&=& \bigg[ \bigg(\frac{N_c^2}{2} + 18 \bigg) c^{[8_a, 8_a]} \\ 
&&\hspace{-1.0cm}
+\, \frac{9\sqrt{N_c^2-4}}{\sqrt{2}} 
c^{[8_a, 10+\overline{10}]} \bigg]
\big({\cal M}_{gg \to ggg}^{\rm tree}\big)^{[8_a,8_a]}.
\eea


\paragraph{The complete MR odd-odd amplitude at two loops.}

The complete multi-Reggeon contribution to 
the odd-odd amplitude is given by the sum 
of eqs.~(\ref{R3_calculationG}), 
(\ref{RgR3_calculationG}) and 
(\ref{R3gR_calculationG}):
\bea \label{MRKmm2L} 
{\cal M}^{(-,-)}_{ij \to i'gj'} 
\Big|^{\mbox{\scriptsize 2-loops}}_{\rm MR}
&=& (i \pi)^2  \left(\frac{\as}{\pi}\right)^2
\bigg\{ 
G^{(2)}_{{\cal R}^3 g {\cal R}^3}(z,\bar z, |\mathbf{p}_4|^2,\mu^2)
\, {\bf {\cal C}}_{{\cal R}^3 g {\cal R}^3} \\[0.1cm] \nn 
&&\hspace{-2.0cm}+\, 
G^{(2)}_{{\cal R} g {\cal R}^3}(z,\bar z, |\mathbf{p}_4|^2,\mu^2)
\, {\bf {\cal C}}_{{\cal R} g {\cal R}^3} +
G^{(2)}_{{\cal R}^3 g {\cal R}}(z,\bar z, |\mathbf{p}_4|^2,\mu^2)
\, {\bf {\cal C}}_{{\cal R}^3 g {\cal R}} \bigg\}
{\cal M}_{ij \to igj}^{\rm tree},
\eea
where the functions $G^{(2)}_{i}$ are given 
in eqs.~(\ref{Gmm2La}), (\ref{Gmm2Lb}) and 
(\ref{Gmm2Lc}), and the color opeartors are 
listed in eqs.~(\ref{Cmm2La}), (\ref{Cmm2Lb}) 
and (\ref{Cmm2Lc}).

For later convenience, let us provide the 
octet-octet component of \eqn{MRKmm2L},
i.e. the coefficient of $c^{[8,8]_a}$, 
$c^{[8,8_a]_a}$ and $c^{[8_a,8_a]}$, 
respectively for $qq \to qgq$, $qg \to qgg$
and $gg \to ggg$ scattering, which can be 
obtained using eqs.~(\ref{eq:ColourFactor2LoopsR3gR3}), 
(\ref{eq:ColourFactor2LoopsRgR3}) and 
(\ref{eq:ColourFactor2LoopsR3gR}). 
We obtain
\bea\label{eq:M2MRoctet} \nn
{\cal M}^{[8,8]_a}_{qq \to qgq} 
\Big|^{\mbox{\scriptsize 2-loops}}_{\rm MR} 
&=& \frac{(i \pi)^2}{72} \left(\frac{\as}{\pi}\right)^2 
\bigg[N_c^2 \, 
F_{\rm fact}^{(2)}(z,\bar z, |\mathbf{p}_4|^2,\mu^2) \\ 
&&\hspace{3.0cm}+\, 
F_{\rm non\,\,fact}^{qq (2)}(z,\bar z, |\mathbf{p}_4|^2,\mu^2)
\bigg]  {\cal M}_{qq \to qgq}^{\rm tree}, \\[0.2cm] \nn 
{\cal M}^{[8,8_a]_a}_{qg \to qgg} 
\Big|^{\mbox{\scriptsize 2-loops}}_{\rm MR} 
&=& \frac{(i \pi)^2}{72} \left(\frac{\as}{\pi}\right)^2 
\bigg[ N_c^2 \, 
F^{(2)}_{\rm fact}(z,\bar z, |\mathbf{p}_4|^2,\mu^2) \\ 
&&\hspace{3.0cm}+\, 
F^{qg(2)}_{\rm non\,\, fact}(z,\bar z, |\mathbf{p}_4|^2,\mu^2)
\bigg]  {\cal M}_{qg \to qgg}^{\rm tree}, \\[0.2cm] 
{\cal M}^{[8_a,8_a]}_{gg \to ggg} 
\Big|^{\mbox{\scriptsize 2-loops}}_{\rm MR} 
&=& \frac{(i \pi)^2}{72}  \left(\frac{\as}{\pi}\right)^2 
(N_c^2+36) 
F^{(2)}_{\rm fact}(z,\bar z, |\mathbf{p}_4|^2,\mu^2)
\,  {\cal M}_{gg \to ggg}^{\rm tree},
\eea
where 
\bea\label{FFact2L} 
F^{(2)}_{\rm fact}(z,\bar z, 
|\mathbf{p}_4|^2, |\mathbf{p}_4|^2)
&=& \frac{1}{\eps^2} 
- \frac{1}{2\eps}  \log \big[z \bar z (1-z) (1-\bar z)\big]
+3\,D_2(z,\bar z) - \zeta_2 \\[0.1cm] \nn 
&&\hspace{-3.0cm}
+\, \frac{5}{4}\log^2\big[z\bar z\big]
+ \frac{5}{4} \log^2\big[(1-z)(1-\bar z) \big] 
- \frac{1}{2} \log \big[z\bar z\big]
\log \big[(1-z)(1-\bar z) \big]
+\ord(\eps),
\eea
and 
\bea\label{FNonFact2Lqq} \nn 
F^{qq(2)}_{\rm non\,\, fact}(z,\bar z, 
|\mathbf{p}_4|^2, |\mathbf{p}_4|^2)
&=& \frac{9}{\eps} \log \big[z \bar z (1-z) (1-\bar z)\big]
+54\,D_2(z,\bar z)  \\[0.1cm] 
&&\hspace{-4.0cm}
-\, \frac{9}{2}\log^2 \big[z \bar z (1-z) (1-\bar z)\big] 
+ \frac{9}{N_c^2} \bigg\{
\frac{1}{\eps^2} 
- \frac{2}{\eps}  \log \big[z \bar z (1-z) (1-\bar z)\big]
-6\,D_2(z,\bar z)  \\[0.1cm]  \nn
&&\hspace{-4.0cm}
+ 2\log^2\big[z\bar z\big]
+ 2\log^2\big[(1-z)(1-\bar z) \big] 
+ \log \big[z\bar z\big]
\log \big[(1-z)(1-\bar z) \big] 
- \zeta_2 \bigg\} +\ord(\eps), \\[0.2cm] \nn
\label{FNonFact2Lqg}
F^{qg(2)}_{\rm non\,\, fact}(z,\bar z, 
|\mathbf{p}_4|^2, |\mathbf{p}_4|^2)
&=& \frac{27}{2\eps^2} 
- \frac{9}{\eps} \Big( 2 \log \big[z \bar z \big]
-3 \log \big[(1-z) (1-\bar z)\big] \Big)
+108\,D_2(z,\bar z)  \\[0.1cm]  
&&\hspace{-3.0cm}
+\, \frac{45}{2}\log^2\big[z\bar z\big]
- 18 \log \big[z\bar z\big]
\log \big[(1-z)(1-\bar z) \big]
- \frac{27}{2}\zeta_2
+\ord(\eps).
\eea
Given that no other color component 
is leading in color, as can be seen 
by inspecting eqs.~(\ref{eq:ColourFactor2LoopsR3gR3}), 
(\ref{eq:ColourFactor2LoopsRgR3}) and 
(\ref{eq:ColourFactor2LoopsR3gR}),
we immediately deduce that the 
leading color planar limit is 
universal, i.e. it does not depend on
the representation of the scattered 
particles, and it is proportional to 
the function $F^{(2)}_{\rm fact}$:
\be\label{eq:M2MRoctetPlanar}
{\cal M}^{[8_a,8_a]}_{ij \to i'gj'} 
\Big|^{\mbox{\scriptsize 2-loops}}_{\rm MR,planar} =
(i \pi)^2  \left(\frac{\as}{\pi}\right)^2 
\frac{N_c^2}{72} \, 
F^{(2)}_{\rm fact}(z,\bar z, |\mathbf{p}_4|^2,\mu^2)
\,  {\cal M}_{ij \to igj}^{\rm tree}.
\ee
This property will be important for extracting 
the two-loop vertex in the next section.


\section{The QCD Lipatov vertex at two loops}
\label{sec:scheme}


\subsection{Extracting the Reggeon-gluon-Reggeon vertex}
\label{sec:extracting2LoopVertex}

As described in section \ref{MRK-shockwave}, 
within the theory of multi-Reggeon interactions 
we express the amplitude as a sum over 
multi-Reggeon exchanges in the $t_i$ channels 
\beq\label{ReggeFactBasicC}
{\cal M}_{ij \to i'gj'} =
{\cal M}^{\rm SR}_{ij \to i'gj'} + 
{\cal M}^{\rm MR}_{ij \to i'gj'},
\eeq
where we split the pure single-Reggeon (SR) exchange (involving a single $W$ field throughout) from all other contributions involving multi-Reggeon (MR) exchange (i.e. multiple $W$ fields). 
In particular, the SR component 
factorizes according to \eqn{eq:ampDecomp}, 
namely, one has 
\beq\label{SingleReggeon}
{\cal M}_{ij \to i'gj'}^{(-,-)\,{\rm SR}} 
= c^{\rm SR}_{i}(t_{1},\tau)\, 
e^{C_A\alpha^{\rm SR}_{g}(t_1)\eta_1}\,
v^{\rm SR}(t_{1},t_{2},|\mathbf{p}_4|^2,\tau)\,
e^{C_A \alpha^{\rm SR}_{g}(t_2)\eta_2}\, c^{\rm SR}_{j}(t_{2},\tau)\,
\,{\cal M}_{ij \to i'gj'}^{\rm tree},
\eeq
where $v^{\rm SR}(t_{1},t_{2},|\mathbf{p}_4|^2,\tau)$ 
defines the perturbative corrections to 
the Reggeon-gluon-Reggeon (${\cal R}g {\cal R}$)
vertex within the so-called SR/MR scheme. 
With the perturbative corrections
to the impact factors $c^{\rm SR}_{i}(t_{1},\tau)$, 
$c^{\rm SR}_{j}(t_{2},\tau)$ and Regge Trajectory 
$\alpha^{\rm SR}_{g}(t_i)$ extracted from $2\to 2$ 
parton scattering
\cite{Falcioni:2021dgr,Falcioni:2021buo}, 
the calculation of ${\cal M}_{ij \to i'gj'}$
on the l.h.s. of \eqn{ReggeFactBasicC} to two 
loops, as discussed in section \ref{MRK-QCD}, 
allows one to extract 
$v^{\rm SR}(t_{1},t_{2},|\mathbf{p}_4|^2,\tau)$
to the same loop accuracy. 

While from a calculation perspective the separation 
of the amplitude according to (\ref{ReggeFactBasicC}) 
is convenient, physically, the amplitude 
${\cal M}_{ij \to i'gj'}$  is expected to feature 
both a Regge pole and a Regge cut, with different 
factorization and exponentiation properties. Hence 
a separation into the pole and cut components is
desirable~\cite{DelDuca:2001gu,DelDuca:2013ara,DelDuca:2013dsa,DelDuca:2014cya,Caron-Huot:2017fxr,Fadin:2016wso,Fadin:2017nka,Fadin:2020lam,Fadin:2021csi,Falcioni:2020lvv,Falcioni:2021buo}.

The pole and cut components of the amplitude do not 
immediately correspond to the single- and multi-Reggeon 
transitions. Furthermore, given that the high-energy 
analytic properties are only manifest upon resumming 
the perturbative series, it is a priori not obvious 
how to disentangle the Regge pole from the Regge cut 
in a fixed-order computation. Addressing this issue, 
a criterion has been proposed 
in~\cite{Falcioni:2021dgr,Falcioni:2021buo},
building upon the fact that four and five point 
amplitudes are expected to have only a Regge pole 
in the large-$N_c$ (planar) 
limit~\cite{Eden:1966dnq,Collins:1977jy}. 
The fact that MR exchanges are present in the 
planar limit is well 
known~\cite{Bartels:1980pe,Kwiecinski:1980wb,Lipatov:1993yb,Faddeev:1994zg,Derkachov:2001yn,Derkachov:2002wz,Caron-Huot:2017fxr}.
From this it follows~\cite{Falcioni:2021dgr,Falcioni:2021buo} 
that, while the Regge cut arises in $2\to2$ and $2\to 3$ 
amplitudes exclusively from MR exchanges, planar MR 
exchanges contribute instead (only) to the Regge pole, 
along with the SR exchange.
In turn, the Regge cut is associated exclusively to 
the non-planar\footnote{An interesting precursor of 
this is the observation, due to 
Mandelstam~\cite{Mandelstam:1963cw,Collins:1977jy}, 
analysing the high-energy asymptotic behaviour of 
individual scalar Feynman integrals, is that Regge 
cuts are present only in (particular) 
non-planar ones.} MR exchanges.

In $2\to 2$ scattering amplitudes, this criterion 
has been shown~\cite{Falcioni:2021dgr,Falcioni:2021buo} 
to be consistent order by order in pertubation theory, 
to four loops. This consistency is tested by two
properties:
\begin{enumerate}
    \item The planar MR contributions are universal, i.e. 
    they are the same in the three partonic processes, 
    $gg$, $qg$ and $qq$ scattering, hence they cannot 
    violate Regge-pole factorization. This stands in 
    sharp contrast to the non-planar MR contributions, 
    which do differ between partonic channels and 
    violate factorization.
    \item The planar MR contributions in $2\to 2$ 
    scattering vanishes identically at four loops, 
    and conjecturally beyond (any non-vanishing 
    contribution would invalidate the factorization 
    structure at NNLL, since all parameters at this 
    logarithmic accuracy are fixed at 3 loops).  
\end{enumerate}
We can now further test and then immediately use 
the same criterion to disentangle the Regge pole 
and cut in the $2\to 3$ scattering amplitudes.
The test has already been presented in the previous 
section, namely, the universality of the planar MR 
contributions in the three $2\to 3$ scattering 
processes -- see eq.~(\ref{eq:M2MRoctetPlanar}) 
above. This universality is essential for the 
planar MR contributions to factorize together 
with the single Reggeon, as will be clear 
from what follows.

Applying the criterion 
of~\cite{Falcioni:2021dgr,Falcioni:2021buo} 
to $2\to 3$ scattering, the pole part should 
be separated from the cut as follows:
\begin{align}
\label{pole-cut}
{\cal M}_{ij \to i'gj'}^{(-,-)} &\!= \underbrace{{\cal M}_{ij \to i'gj'}^{(-,-)\,{\rm SR}} 
+ \left.{\cal M}_{ij \to i'gj'}^{(-,-)\,{\rm MR}}\right\vert_{\text{planar}}} 
+
\left.{\cal M}_{ij \to i'gj'}^{(-,-)\,{\rm MR}}\right\vert_{\text{nonplanar}} \nn \\
&\!= \hspace*{56pt} {\cal M}_{ij \to i'gj'}^{(-,-)\,{\rm pole}} \hspace*{30pt} 
+ \hspace*{1pt} {\cal M}_{ij \to i'gj'}^{(-,-)\,{\rm cut}},
\end{align}
where, crucially, 
${\cal M}_{ij \to i'gj'}^{(-,-)\,{\rm pole}}$
has formally the same factorization
structure of ${\cal M}_{ij \to i'gj'}^{(-,-)\,{\rm SR}}$
in \eqn{SingleReggeon}, i.e.: 
\beq\label{ReggePole}
{\cal M}_{ij \to i'gj'}^{(-,-)\,{\rm pole}}
= c_{i}(t_{1},\tau)\, 
e^{C_A \alpha_{g}(t_1)\eta_1}\,
v(t_{1},t_{2},|\mathbf{p}_4|^2,\tau)\,
e^{C_A\alpha_{g}(t_2)\eta_2}\, c_{j}(t_{2},\tau)\,
\,{\cal M}_{ij \to i'gj'}^{\rm tree},
\eeq
where $c_{i}(t_{1},\tau)$, 
$c_{j}(t_{2},\tau)$, $\alpha_{g}(t_k)$
and $v(t_{1},t_{2},|\mathbf{p}_4|^2,\tau)$ 
represent respectively the impact factors, 
the Regge trajectory and the 
Reggeon-gluon-Reggeon Lipatov 
vertex in the~\emph{pole/cut scheme}. 

The relation between the two schemes for 
the Regge trajectory and the impact factors
has been determined from $2\to 2$ scattering 
up to NNLL accuracy in \cite{Falcioni:2021dgr,Falcioni:2021buo}:
one has 
\bea\label{eq:Ctilde} \nn
{c}_{i/j}^{(1)} &=& c^{\rm SR\,(1)}_{i/j}, \\ 
{c}_{i/j}^{(2)} &=& c^{\rm SR\,(2)}_{i/j} 
+ \frac{\pi^2}{12} N_c^2 \, \left(r_\Gamma(\epsilon)\right)^2 \, S^{(2)}(\epsilon),
\eea
where $r_\Gamma(\epsilon)$ is given in (\ref{alphagOneLoop}) and 
\be\label{rgammaandS2}
S^{(2)}(\epsilon) = -\frac{1}{8\epsilon^2}+\frac{3}{4}\epsilon\zeta_3
+\frac{9}{8}\epsilon^2\zeta_4+{\cal{O}}(\epsilon^3).
\ee
For the Regge trajectory we have 
\bea \label{eq:reggeTilde} \nn
\al_g^{(1)} &=& \al_g^{\rm SR \,(1)}, \\ \nn
\al_g^{(2)} &=& \al_g^{\rm SR \,(2)}, \\ 
\al_g^{(3)} &=& \al_g^{\rm SR \,(3)} 
- \left(r_\Gamma(\epsilon)\right)^3N_c^2\frac{\pi^2}{18}\left(S^{(3)}_A(\epsilon)-S^{(3)}_B(\epsilon)\right),
\eea 
where 
\bea \label{SAB} \nn
S^{(3)}_A(\epsilon) &=& \frac{1}{48\epsilon^3} + \frac{37}{24}\zeta_3 
+ \frac{37}{16}\epsilon\,\zeta_4 + {\cal{O}}(\epsilon^2), \\
S^{(3)}_B(\epsilon) &=& \frac{1}{24\epsilon^3} + \frac{1}{12}\zeta_3 
+ \frac{1}{8}\epsilon\,\zeta_4 + {\cal{O}}(\epsilon^2).
\eea 

Taking into account these results, first of all
we are able to determine the relation between 
the Lipatov vertex in the SR/MR and in the 
pole/cut scheme; then, using the explicit 
calculation of the two-loop $2\to 3$ amplitudes 
together with the results for the MR contribution 
to the odd-odd $[8,8]$ amplitude in 
section~\ref{sec:oddOdd2}, we will determine 
the Lipatov vertex explicitly. We will present 
the result for the two-loop vertex and discuss 
the structure in section \ref{VertTwoLoopStruct}. 

Let us start here by determining the relation 
between $ v^{\rm SR}(t_{1},t_{2},|\mathbf{p}_4|^2,\tau)$ 
in the SR/MR scheme and $v(t_{1},t_{2},|\mathbf{p}_4|^2,\tau)$ 
in the pole/cut scheme.
Expanding eqs.~(\ref{pole-cut}),~(\ref{SingleReggeon})
and~(\ref{ReggePole}) to second order in
the coupling constant, and restricting to 
the NNLL terms, the matching equations read
\bea\label{matching_SR_Lipatov} \nn
{\cal M}^{[8_a,8_a]}_{ij \to i'gj'} 
\Big|^{\rm NNLL}_{\mbox{\scriptsize 2-loops}} -
{\cal M}^{[8_a,8_a]\,{\rm MR}}_{ij \to i'gj'} 
\Big|^{\rm NNLL}_{\mbox{\scriptsize 2-loops}} && \\ \nn
&&\hspace{-5.0cm}=\, \left(\frac{\as}{\pi}\right)^2 
\bigg[ c_{i}^{\rm SR \,(2)} + c_{j}^{\rm SR \,(2)} 
+ c_{i}^{\rm SR \,(1)} c_{j}^{\rm SR \,(1)} \\ 
&&\hspace{-3.0cm}+\,
\Big(c_{i}^{\rm SR \,(1)} + c_{j}^{\rm SR \,(1)}\Big) 
v^{\rm SR \,(1)} + v^{\rm SR \,(2)}\bigg]
{\cal M}^{[8_a,8_a]}_{ij \to i'gj'}\big|_{\rm tree},
\eea
to extract the Lipatov vertex in the SR/MR scheme, and 
\bea\label{matching_Cut_Lipatov} \nn
{\cal M}^{[8_a,8_a]}_{ij \to i'gj'} 
\Big|^{\rm NNLL}_{\mbox{\scriptsize 2-loops}} -
{\cal M}^{[8_a,8_a]\,{\rm cut}}_{ij \to i'gj'} 
\Big|^{\rm NNLL}_{\mbox{\scriptsize 2-loops}} && \\ 
&&\hspace{-5.0cm}=\, \left(\frac{\as}{\pi}\right)^2 
\bigg[ c_{i}^{(2)} + c_{j}^{(2)} 
+ c_{i}^{(1)} c_{j}^{(1)}
+\Big(c_{i}^{(1)} + c_{j}^{(1)}\Big) 
v^{(1)} + v^{(2)} \bigg]
{\cal M}^{[8_a,8_a]}_{ij \to i'gj'}\big|_{\rm tree},
\eea
in the pole/cut scheme. All functions on the r.h.s.~of 
\eqns{matching_SR_Lipatov}{matching_Cut_Lipatov}
are known, except for $v^{\rm SR \,(2)}$ and $v^{(2)}$. 
More in detail, ${\cal M}^{[8_a,8_a]\,{\rm MR}}_{ij \to i'gj'} 
\big|^{\rm NNLL}_{\mbox{\scriptsize 2-loops}}$ 
in \eqn{matching_SR_Lipatov} has been given 
for all cases $qq \to qgq$, $qg \to qgg$, and 
$gg \to ggg$ in \eqn{eq:M2MRoctet}, while 
${\cal M}^{[8_a,8_a]\,{\rm cut}}_{ij \to i'gj'} 
\big|^{\rm NNLL}_{\mbox{\scriptsize 2-loops}}$
in \eqn{matching_Cut_Lipatov} can be easily 
determined according to \eqn{pole-cut},
namely one has 
\be
{\cal M}^{[8_a,8_a]\,{\rm cut}}_{ij \to i'gj'} 
\Big|_{\mbox{\scriptsize 2-loops}} 
= {\cal M}^{[8_a,8_a]\,{\rm MR}}_{ij \to i'gj'} 
\Big|_{\mbox{\scriptsize 2-loops}} - 
{\cal M}^{[8_a,8_a]\,{\rm MR}}_{ij \to i'gj'} 
\Big|^{\rm planar}_{\mbox{\scriptsize 2-loops}}, 
\ee
where ${\cal M}^{[8_a,8_a]\,{\rm MR}}_{ij \to i'gj'} 
\big|^{\rm planar}_{\mbox{\scriptsize 2-loops}}$
has been given in \eqn{eq:M2MRoctetPlanar}. 
Exploiting this information, it is easy to 
realize that the relation between the Lipatov 
vertex in the two schemes is independent 
on the exact form of the two loop amplitude:
it relies solely on 
${\cal M}^{[8_a,8_a]\,{\rm MR}}_{ij \to i'gj'} 
\big|^{\rm planar}_{\mbox{\scriptsize 2-loops}}$
and the relation between the impact factors in 
the two schemes: by taking the difference of 
\eqn{matching_SR_Lipatov} with 
\eqn{matching_Cut_Lipatov}, taking 
into account the fact that $v^{\rm SR \,(1)} 
= v^{(1)}$, as well as \eqn{eq:Ctilde}, we have 
\be
{\cal M}^{[8_a,8_a]\,{\rm MR}}_{ij \to i'gj'} 
\Big|^{\rm planar}_{\mbox{\scriptsize 2-loops}} = 
\left(\frac{\as}{\pi}\right)^2 
\bigg[  v^{\rm SR \,(2)}  - v^{(2)}
-\frac{\pi^2}{6} N_c^2 \,r_\Gamma^2 \,S^{(2)}(\epsilon) \bigg]
{\cal M}^{[8_a,8_a]}_{ij \to i'gj'}\big|_{\rm tree},
\ee 
and upon using \eqn{eq:M2MRoctetPlanar}
we get 
\bea\label{v2v2tilde} \nn
v^{(2)} &=& v^{\rm SR \,(2)} + \pi^2 N_c^2 \bigg( 
\frac{F^{(2)}_{\rm fact}(z,\bar z, |\mathbf{p}_4|^2,\mu^2)}{72}
- \frac{r_\Gamma^2 \,S^{(2)}(\epsilon)}{6}\bigg) \\  \nn
&=& v^{\rm SR \,(2)} + \frac{\pi^2 N_c^2}{144} \bigg\{ \frac{5}{\eps^2}
- \frac{1}{\eps}  \log \big[z \bar z (1-z) (1-\bar z)\big]
+ 6 D_2(z,\bar z) - 2\zeta_2 
+ \frac{5}{2}\log^2\big[z\bar z\big]\\ 
&&\hspace{1.2cm}
+\, \frac{5}{2} \log^2\big[(1-z)(1-\bar z) \big] 
- \log \big[z\bar z\big]
\log \big[(1-z)(1-\bar z) \big]
+\ord(\eps)\bigg\}.
\eea

Let us stress that the relations in 
eqs.~(\ref{eq:Ctilde}),~(\ref{eq:reggeTilde})
and~(\ref{v2v2tilde}) completely fix 
the factorization structure up to NNLL. That is, 
given the three-loop information from $2\to2$ scattering, 
and the two-loop information determined here for $2\to 3$ 
scattering (i.e. the two-loop lipatov vertex) at this 
logarithmic accuracy the theory (i.e. the exponentiation 
of rapidity logarithms associated with the pole) becomes 
predictive for $2\to 3$ scattering at three loops and 
beyond. To see this, let us consider the NNLL three-loops 
contributions to the amplitude. In the SR/MR scheme one has
\bea\label{ThreeLoopsSRMR} \nn
{\cal M}^{[8_a,8_a]}_{ij \to i'gj'} 
\Big|^{\rm NNLL}_{\mbox{\scriptsize 3-loops}} &=&  
\left(\frac{\as}{\pi}\right)^3 
\sum_{k=1}^2
\bigg\{ \eta_k\, C_A \bigg[ \al_g^{\rm SR \,(1)}(t_k) 
\bigg( v^{\rm SR \,(2)} + v^{\rm SR \,(1)} 
\Big(c_{i}^{\rm SR \,(1)} + c_{j}^{\rm SR \,(1)}\Big) \\ \nn  
&&\hspace{-0.7cm}+\, c_{i}^{\rm SR \,(1)} c_{j}^{\rm SR \,(1)} 
+ c_{i}^{\rm SR \,(2)} + c_{j}^{\rm SR \,(2)} \bigg) +\, \al_g^{\rm SR \,(2)}(t_k) 
\bigg( v^{\rm SR \,(1)} + c_{i}^{\rm SR \,(1)} 
+ c_{j}^{\rm SR \,(1)}\bigg)
\\ && \hspace{-0.7cm} + \al_g^{\rm SR \,(3)}(t_k) \bigg] \bigg\} {\cal M}^{[8_a,8_a]}_{ij \to i'gj'}\big|_{\rm tree}
+ {\cal M}^{[8_a,8_a]\,{\rm MR}}_{ij \to i'gj'} 
\Big|^{\rm NNLL}_{\mbox{\scriptsize 3-loops}},
\eea
while in the pole/cut scheme one has 
\bea\label{ThreeLoopsPoleCut} \nn
{\cal M}^{[8_a,8_a]}_{ij \to i'gj'} 
\Big|^{\rm NNLL}_{\mbox{\scriptsize 3-loops}} &=&  
\left(\frac{\as}{\pi}\right)^3 
\sum_{k=1}^2\bigg\{ \eta_k\, C_A \bigg[ \al_g^{(1)}(t_k) \Big( 
 v^{(2)} +  v^{(1)} \big( c_{i}^{(1)} 
+  c_{j}^{(1)}\big) 
\\ \nn  
&&\hspace{-0.5cm}
+\,  c_{i}^{(1)}  c_{j}^{(1)} 
+  c_{i}^{(2)} +  c_{j}^{(2)}\Big) +\, \al_g^{(2)}(t_k) 
\Big(  v^{(1)} +  c_{i}^{(1)} 
+  c_{j}^{(1)}\Big) 
\\ 
&&\hspace{-0.5cm}
+  \al_g^{(3)}(t_k) \bigg] \bigg\} 
{\cal M}^{[8_a,8_a]}_{ij \to i'gj'}\big|_{\rm tree}+\, {\cal M}^{[8_a,8_a]\,{\rm cut}}_{ij \to i'gj'} 
\Big|^{\rm NNLL}_{\mbox{\scriptsize 3-loops}}.
\eea
Comparing \eqn{ThreeLoopsSRMR} with 
\eqn{ThreeLoopsPoleCut}, we deduce the 
following: first of all, all terms in the 
pole component of \eqn{ThreeLoopsPoleCut}, 
namely $ \al_g^{(1)}$, $ \al_g^{(2)}$, 
$ c_{i/j}^{(1)}$, $ c_{i/j}^{(2)}$, 
$ v^{(1)}$ and $ v^{(2)}$ have been 
determined from previous loop orders,
therefore, there is no room for a further 
redefinitions of these coefficients. 
Given that the amplitude in 
\eqns{ThreeLoopsSRMR}{ThreeLoopsPoleCut}
are the same, equating them gives
\bea\label{ThreeLoopsPrediction1} \nn
{\cal M}^{[8_a,8_a]\,{\rm MR}}_{ij \to i'gj'} 
\Big|^{\rm NNLL}_{\mbox{\scriptsize 3-loops}} &=&  
\left(\frac{\as}{\pi}\right)^3 N_c \bigg\{ 
\bigg[ \eta_1 \,\al_g^{(1)}(t_1) 
+ \eta_2 \, \al_g^{(1)}(t_2)\bigg] \\ \nn
&&\hspace{1.0cm}\times\, 
\bigg[ \Big(v^{(2)}- v^{\rm SR\, (2)} \Big)
+ \Big(c_{i}^{(2)} + c_{j}^{(2)} 
- c_{i}^{\rm SR\, (2)} - c_{j}^{\rm SR\, (2)}\Big) \bigg] \\ \nn
&&\hspace{-1.0cm}+\, 
\eta_1 \Big( \al_g^{(3)}(t_1) - \al_g^{\rm SR\, (3)}(t_1) \Big)  
+\eta_2 \Big( \al_g^{(3)}(t_2) - \al_g^{\rm SR\, (3)}(t_2) \Big)
\bigg\}{\cal M}^{[8_a,8_a]}_{ij \to i'gj'}\big|_{\rm tree} \\
&&\hspace{4.0cm}+\,
{\cal M}^{[8_a,8_a]\,{\rm cut}}_{ij \to i'gj'} 
\Big|^{\rm NNLL}_{\mbox{\scriptsize 3-loops}}.
\eea
Expressing $c_{i/j}^{(2)}$ and  
$v^{(2)}$ in terms of $c_{i/j}^{\rm SR\, (2)}$
and $v^{\rm SR\, (2)}$ as given in 
\eqns{eq:Ctilde}{v2v2tilde}, we have 
\bea\label{ThreeLoopsPrediction2} \nn
{\cal M}^{[8_a,8_a]\,{\rm MR}}_{ij \to i'gj'} 
\Big|^{\rm NNLL}_{\mbox{\scriptsize 3-loops}} &=&  
\left(\frac{\as}{\pi}\right)^3 
\frac{\pi^2 N_c^3}{18}
\Bigg\{
\bigg[ \eta_1\,\al_g^{(1)}(t_1) 
+ \eta_2 \,\al_g^{(1)}(t_2)\bigg] 
\frac{1}{4} 
F^{(2)}_{\rm fact}(z,\bar z, |\mathbf{p}_4|^2,\mu^2) \\ 
&&\hspace{-3.0cm} -\, 
\big(\eta_1 + \eta_2 \big) (r_\Gamma)^3 
\left(S^{(3)}_A(\epsilon)-S^{(3)}_B(\epsilon)\right)\bigg\}
{\cal M}^{[8_a,8_a]}_{ij \to i'gj'}\big|_{\rm tree} 
+{\cal M}^{[8_a,8_a]\,{\rm cut}}_{ij \to i'gj'} 
\Big|^{\rm NNLL}_{\mbox{\scriptsize 3-loops}},
\eea
and recalling that, by definition,
\be
{\cal M}^{[8_a,8_a]\,{\rm cut}}_{ij \to i'gj'} 
\Big|_{\mbox{\scriptsize 3-loops}} 
= {\cal M}^{[8_a,8_a]\,{\rm MR}}_{ij \to i'gj'} 
\Big|_{\mbox{\scriptsize 3-loops}} - 
{\cal M}^{[8_a,8_a]\,{\rm MR}}_{ij \to i'gj'} 
\Big|^{\rm planar}_{\mbox{\scriptsize 3-loops}}, 
\ee
we finally obtain 
\bea\label{ThreeLoopsPrediction3} \nn
{\cal M}^{[8_a,8_a]\,{\rm MR}}_{ij \to i'gj'} 
\Big|^{\rm planar}_{\mbox{\scriptsize 3-loops}}
&=&  
\left(\frac{\as}{\pi}\right)^3 
\frac{\pi^2 N_c^3}{18}
\Bigg\{ \bigg[ \eta_1 \,\al_g^{(1)}(t_1) 
+ \eta_2 \,\al_g^{(1)}(t_2)\bigg] 
\frac{1}{4}
F^{(2)}_{\rm fact}(z,\bar z, |\mathbf{p}_4|^2,\mu^2) \\ 
&&\hspace{1.0cm} -\, \big(\eta_1 + \eta_2 \big) (r_\Gamma)^3
\left(S^{(3)}_A(\epsilon)-S^{(3)}_B(\epsilon)\right)\bigg\}
{\cal M}^{[8_a,8_a]}_{ij \to i'gj'}\big|_{\rm tree},
\eea
i.e., we obtain a prediction for the three
loop planar MR amplitude at NNLL, which, if  
verified, would in turn provide further evidence
for the conjecture in \eqn{pole-cut}, besides
the equivalent check already satisfied 
in case of the four loop $2\to 2$
scattering amplitude, see 
\cite{Falcioni:2021dgr,Falcioni:2021buo}.

This observation completes our discussion
concerning the extraction of the Lipatov 
vertex at two loops. One uses \eqn{matching_SR_Lipatov}, 
with ${\cal M}^{[8_a,8_a]\,{\rm MR}}_{ij \to i'gj'} 
\big|^{\rm NNLL}_{\mbox{\scriptsize 2-loops}}$
calculated as in section \ref{sec:oddOdd2}, 
and ${\cal M}^{[8_a,8_a]}_{ij \to i'gj'} 
\big|^{\rm NNLL}_{\mbox{\scriptsize 2-loops}}$
as discussed in section \ref{MRK-QCD},
to obtain $v^{\rm SR\, (2)}$. Subsequently, 
$v^{(2)}$ is obtained by simply resorting 
to \eqn{v2v2tilde}. In the next section 
we are going to discuss the structure 
and analytic properties of the Lipatov 
vertex in the pole/cut scheme, 
i.e. $v^{(2)}$.


\subsection{Vertex structure through two loop}
\label{VertTwoLoopStruct}

The sYM vertex constitutes the (pure, maximal weight) weight-4 component of the QCD vertex. We have already provided the full expression for 
$v^{(1)}_{\text{sYM}}=  v^{(1)}_{\text{reg}}$
in section ~\ref{sec:MatchingNLL}, 
see eqs.~(\ref{eq:Dispvreg}) and~(\ref{eq:Absovreg})
 there. The result was expressed in terms of the single-valued GPL functions ${\cal G}$ (see
eq.~(\ref{HyperlogNotation}) above). The ancillary file~\cite{Ancillary} 
{\tt{svRL.m}} provide a replacement list of these functions in terms of combinations of ordinary (multi-valued) GPLs of $z$ and $\bar{z}$, which can be evaluated using the GiNaC library \cite{Bauer:2000cp} via PolyLogTools~\cite{Duhr:2019tlz}. Alternatively, one may use the replacement list {\tt{svpdRL.m}} to obtain GPLs that are fibrated in terms of the variables $p$ and $q$ where
\begin{equation}
\label{pq}
    p=1-z-\bar{z},\qquad\quad q=z-\bar{z}\,.
\end{equation}
The latter set of variables is convenient to express the results and understand their symmetry properties under $z\to 1-\bar{z}$, which corresponds to $p\to -p$ (with fixed $q$), and  $z\to \bar{z}$, which corresponds to $q\to -q$ (with fixed $p$).

Here we proceed to provide a similar expressions for the two-loop vertex in sYM. The results in QCD will be presented in the following subsections. 

\begin{align}
\label{eq:v2sYM}
\begin{split}
v_{\text{sYM}}^{(2)}(t_1,t_2,|p_4|^2)  = v_{\text{sYM}}^{(2)}(z,\bar{z}) 
= &\,\, {\text{Disp}}\left\{v_{\text{sYM}}^{(2)}(z,\bar{z}) \right\}
+i  {\text{Absorp}} \left\{ v_{\text{sYM}}^{(2)}(z,\bar{z})\right\}
\\\equiv &\,\,
v_{\text{sYM},\, \text{Disp}}^{(2)}(z,\bar{z}) 
+i \pi \,\,  v_{\text{sYM},\, {\text{Absorp}}}^{(2)}(z,\bar{z}) 
\end{split}
\end{align}   
Note that in the second line we extracted a power of $\pi$ from the absorptive part of the vertex, which always appears together with the explicit factor of $i$. This implies in particular that $v_{\text{sYM},\, {\text{Absorp}}}$ is of weight 3, in contrast to $v_{\text{sYM},\, \text{Disp}}^{(2)}(z,\bar{z})$, which is of weight 4.

\begin{align}
\label{eq:v2sYMAbsorb}
\begin{split}
v_{\text{sYM},\, {\text{Absorp}}}^{(2)}&=
\frac{N_c^2}{32 \epsilon^3}+
\frac{N_c^2}{{\epsilon}} \left(\frac{1}{32}
   \,{\cal G}(0,0,z)-\frac{1}{16}
   \,{\cal G}(0,1,z)+\frac{1}{32}
   \,{\cal G}(1,1,z)-\frac{\pi
   ^2}{96}\right)
   \\& +\frac{N_c^2}{16}
   \left(
   \,{\cal G}(0,1-\bar{z},0,z)-
   \,{\cal G}(0,1-\bar{z},1,z)-
   \,{\cal G}(\bar{z},0,1,z)+\,{\cal G}(\bar{z},1,0,z)
   \right.
   \\&\left.
  -\frac{\pi ^2}{12}
    \,{\cal G}(0,z)-\frac{\pi
   ^2 }{12} \,{\cal G}(1,z)-
   \,{\cal G}(0,0,0,z)+
   \,{\cal G}(0,0,1,z)\right.
   \\&\left.-
   \,{\cal G}(0,1,0,z)+
   \,{\cal G}(0,1,1,z)+
   \,{\cal G}(1,0,1,z)-
   \,{\cal G}(1,1,1,z)-\frac{17 \zeta
   (3)}{6}\right)
\end{split}
\end{align}   

\begin{align}
\label{eq:v2sYMDisp}
\begin{split}
v_{\text{sYM},\, \text{Disp}}^{(2)}&=
\frac{N_c^2}{32\epsilon^4}+
   \frac{N_c^2}{48 \epsilon^2} 
   \left( 3 \,{\cal G}(0,0,z)-3
   \,{\cal G}(0,1,z)-3 \,{\cal G}(1,0,z)+3
   \,{\cal G}(1,1,z)-\pi ^2\right)
    \\& 
   +\frac{N_c^2}{192 \epsilon}
   \left( 24
   \,{\cal G}(0,1-\bar{z},0,z)-24
   \,{\cal G}(0,1-\bar{z},1,z)-5 \pi ^2
   \,{\cal G}(0,z)-5 \pi ^2
   \,{\cal G}(1,z)
\right.\\&\left. 
-24 \,{\cal G}(0,0,0,z)+12
   \,{\cal G}(0,0,1,z)-12
   \,{\cal G}(0,1,0,z)+12
   \,{\cal G}(0,1,1,z)
   \right.\\& \left.+12
   \,{\cal G}(1,0,0,z)+12
   \,{\cal G}(1,0,1,z)+12
   \,{\cal G}(1,1,0,z)-24
   \,{\cal G}(1,1,1,z)-46 \zeta
   (3)\right)
    \\& 
   +\frac{N_c^2}{11520}
   \left( -480 \pi ^2
   \,{\cal G}(\bar{z},1,z)-1440
   \,{\cal G}(0,0,1-\bar{z},0,z)\right.\\& \left.+1440
   \,{\cal G}(0,0,1-\bar{z},1,z)-1440
   \,{\cal G}(0,1,1-\bar{z},0,z)
   +1440
   \,{\cal G}(0,1,1-\bar{z},1,z)
   \right.\\& \left.
   -1440
   \,{\cal G}(0,1-\bar{z},0,0,z)+1440
   \,{\cal G}(0,1-\bar{z},1,1,z)-1440
   \,{\cal G}(1,0,1-\bar{z},0,z)
   \right.\\& \left.
   +1440
   \,{\cal G}(1,0,1-\bar{z},1,z)-1440
   \,{\cal G}(\bar{z},0,1,0,z)+1440
   \,{\cal G}(\bar{z},0,1-\bar{z},0,z)
   \right.\\& \left.
   -1440
   \,{\cal G}(\bar{z},0,1-\bar{z},1,z)+14
   40 \,{\cal G}(\bar{z},1,0,1,z)+1440
   \,{\cal G}(\bar{z},1,1-\bar{z},0,z)
   \right.\\& \left.-1440
   \,{\cal G}(\bar{z},1,1-\bar{z},1,z)+72
   0 \zeta (3) \,{\cal G}(0,z)+720 \zeta (3)
   \,{\cal G}(1,z)+1140 \pi ^2
   \,{\cal G}(0,1,z)
   \right.\\& \left.
   -60 \pi ^2
   \,{\cal G}(1,0,z)+4320
   \,{\cal G}(0,0,0,0,z)-2880
   \,{\cal G}(0,0,0,1,z)-1440
   \,{\cal G}(0,0,1,0,z)
   \right.\\& \left.+1440
   \,{\cal G}(0,0,1,1,z)-1440
   \,{\cal G}(0,1,0,0,z)+1440
   \,{\cal G}(0,1,0,1,z)
   \right.\\& \left.
   +2880
   \,{\cal G}(0,1,1,0,z)
   -2880
   \,{\cal G}(0,1,1,1,z)-2880
   \,{\cal G}(1,0,0,0,z)
   \right.\\& \left.+1440
   \,{\cal G}(1,0,0,1,z)+2880
   \,{\cal G}(1,0,1,0,z)-2880
   \,{\cal G}(1,0,1,1,z)\right.\\& \left.+1440
   \,{\cal G}(1,1,0,0,z)-2880
   \,{\cal G}(1,1,0,1,z)-2880
   \,{\cal G}(1,1,1,0,z)
   \right.\\& \left.+4320
   \,{\cal G}(1,1,1,1,z)-53 
   \pi^4\right)
\end{split}
\end{align}   
We note that the symbol alphabet of the sYM two-loop vertex consists of 6 letters, 
$\{z,\bar{z},1-z,1-\bar{z},z-{\bar{z}}, 1-z-{\bar{z}}\}$
and is identical to the one found at one loop, eq.~(\ref{eq:alphabet}). 
We comment that the letters $z-\bar{z}$ and $1-z-{\bar{z}}$ appear at two loops already at the level of the $1/\epsilon$ singularities. 


\subsubsection{Basis of transcendental functions \label{sec:transcenBases}}

Here we present a basis for the \emph{additional} transcendental functions of weight 1 through 3, which are needed to express the QCD vertex through ${\cal O}(\epsilon^2)$ at one loop and ${\cal O}(\epsilon^0)$ at two loops. 
Similarly to the sYM results discussed above, each of the functions we introduce has the following properties:
\begin{enumerate}
\item It is expressed in terms of \emph{single-valued polylogarithms} $\cal G$ of the 5 letter alphabet 
$\{z,\bar{z},1-z,1-\bar{z},1-z-{\bar{z}}\}$ (the additional letter $z-\bar{z}$ in (\ref{eq:alphabet}), does not appear here);
\item It is finite in the soft limit;
\item It admits either 
\begin{equation}
\label{SunderTP}
\Phi(z,\bar{z}) =+ \Phi(1-\bar{z},1-z) 
\end{equation}
or
\begin{equation}
\label{AunderTP}
\Phi(z,\bar{z}) =- \Phi(1-\bar{z},1-z)
\end{equation}
which, together with a similar property of the corresponding rational coefficients, manifests the exact target-projectile symmetry of the vertex;
\item It admits the hermiticity relation 
\begin{equation}
\label{hermiticity}
\Phi(z,\bar{z}) =\left(\Phi(\bar{z},z)\right)^*\,,
\end{equation}
which manifest the reality properties of the vertex prior to contraction with the polarization vector. Most of the functions are real, $\mathbb{C}\to \mathbb{R}$, so they simply admit $\Phi(z,\bar{z}) =\Phi(\bar{z},z)$, but some are complex valued, $\mathbb{C}\to \mathbb{C}$, and admit (\ref{hermiticity}).  
\end{enumerate}

\begin{subequations}
\label{phiBasis}
\begin{align}
&\phi_1 (z,\bar{z})\,=\, {\cal G}(0,z)+{\cal G}(1,z),\\
&\phi_2 (z,\bar{z})\,=\, 
   {\cal G}(1,z)-{\cal G}(0,z),\\ 
&\phi_3 (z,\bar{z})\,=\, 
   {\cal G}(0,0,z)-{\cal G}(0,1,z)-{\cal
   G}(1,0,z)+{\cal G}(1,1,z),
   \\
&\phi_4 (z,\bar{z})\,=\, 
   {\cal G}(0,0,z)-{\cal G}(1,1,z),\\ 
&\phi_5 (z,\bar{z})\,=\, 
   {\cal G}(0,0,z)+{\cal G}(1,1,z),\\ 
&\phi_6 (z,\bar{z})\,=\,  \frac{1}{4}
   {\cal G}(0,1,z)-\frac{1}{4}
   {\cal G}(1,0,z)+\frac{\pi ^2}{16},\\ 
\begin{split}
&\phi_7 (z,\bar{z})\,=\, 
   {\cal G}(1-{\bar z},0,0,z)-{\cal G}(1
   -{\bar z},0,1,z)-{\cal G}(1-{\bar z},1
   ,0,z)+{\cal G}(1-{\bar z},1,1,z)
   \\
   &
   +{\cal 
   G}(0,0,0,z)-{\cal G}(0,1,1,z)-{\cal
   G}(1,0,0,z)+{\cal G}(1,1,1,z)+\frac{8
   \zeta (3)}{3},
   \end{split}
   \\ 
\begin{split}
    &
\phi_8 (z,\bar{z})\,=\,  \frac{1}{2}
   (-{\cal G}(0,1-{\bar z},0,z)
   +{\cal G}
   (0,1-{\bar z},1,z)-{\cal G}(1,1-
   {\bar z},0,z)+{\cal G}(1,1-{\bar z},1,z)\\
   &+
   {\cal G}(0,1,0,z)-{\cal G}(1,0,1,z))
   +\frac
   {1}{12} \pi ^2
   ({\cal G}(1,z)-{\cal G}(0,z)),
   \end{split}
   \\ 
&\phi_9 (z,\bar{z})\,=\, 
   {\cal G}(1-{\bar z},0,z)-{\cal G}(1-
   {\bar z},1,z)+\frac{1}{2}
   {\cal G}(0,1,z)-\frac{1}{2}
   {\cal G}(1,0,z)-\frac{\pi ^2}{6},\\
&\phi_{10} (z,\bar{z})\,=\, 
   {\cal G}(0,0,0,z)-{\cal G}(1,1,1,z),
   \\ 
\begin{split}
&\phi_{11} (z,\bar{z})\,=\, 3 \bigg[3
   {\cal G}(0,1-{\bar z},0,z)
   -3
   {\cal G}(0,1-{\bar z},1,z)
+3
   {\cal G}(1,1-{\bar z},0,z)
-3
   {\cal G}(1,1-{\bar z},1,z)\\&
+2
   {\cal G}(1-{\bar z},0,0,z)
-4
   {\cal G}(1-{\bar z},0,1,z)
+4
   {\cal G}(1-{\bar z},1,0,z)
-2
   {\cal G}(1-{\bar z},1,1,z)
   \\&
-8
   {\cal G}(1-{\bar z},1-{\bar z},0,z)
+8
   {\cal G}(1-{\bar z},1-{\bar z},1,z)
-4
   {\cal G}(0,0,0,z)
+4
   {\cal G}(0,0,1,z)
+{\cal G}(0,1,0,z)
   \\&
-2
   {\cal G}(0,1,1,z)
+2
   {\cal G}(1,0,0,z)
-{\cal G}(1,0,1,z)
-4
   {\cal G}(1,1,0,z)
+4
   {\cal G}(1,1,1,z)
+4 \zeta (3)\bigg]
   \\&
   +4 \pi^2
   {\cal G}(1-{\bar z},z)-3 \pi^2
   {\cal G}(1,z)
   \end{split}
\end{align}
\end{subequations}

\begin{table}[h!]
\begin{center}
\begin{tabular}{||c || c c c  c c c  c c c  c c||} 
 \hline 
Property & $\phi_1$ & $\phi_2$ & $\phi_3$ & $\phi_4$ & $\phi_5$ & $\phi_6$ & $\phi_7$ & $\phi_8$ & $\phi_9$ & $\phi_{10}$ & $\phi_{11}$  
\\ [0.5ex] 
 \hline\hline
 Weight & 1 & 1 & 2 & 2& 2 &2 & 3 & 3 & 2 & 3& 3 \\
 \hline
 $z\to \bar{z}$ & S & S & S &  S & S &  C & S  & C & S & S & S  \\ 
 \hline
 $z\to 1-\bar{z}$ &  S & A  & S & A & S & S & S & S & A & A & A \\
 \hline
\end{tabular}
\end{center}
\caption{Table summarising the weight and symmetry properties of the transcendental functions entering the non-maximal weight part of the QCD Lipatov vertex through two loops.}
\label{tab:phi_i_properties}
\end{table}

For easy reference we summarize the properties of the~$\phi_i$ functions in table~\ref{tab:phi_i_properties}. The first row in the table represents the transcendental weight. The second row shows how each function realises the hermiticity property of (\ref{hermiticity}), where S indicates real functions, $\phi_i(z,\bar{z}) =\phi_i(\bar{z},z)$, while C stands for complex functions obeying (\ref{hermiticity}). Finally, in the third row S indicates symmetry under target-projectile swap, as in (\ref{SunderTP}), while A indicates antisymmetry under this swap, as in (\ref{AunderTP}).  

Next, we define non-pure, mixed-weight transcendental functions involving rational prefactors. Notably, as we have seen at one loop in eqs.~(\ref{v1spur}) through (\ref{vab}), these contain spurious poles in $p=1-z-\bar{z}$, corresponding to the soft limit of the emitted gluon $p_4$. As already discussed these $1/p^n$ poles cancel in the combinations that enter the vertex. Here we make this manifest by defining a basis of functions where each has a finite $p\to 0$ limit. These are defined in terms of the basis of  $\phi_i$ we defined in (\ref{phiBasis}) above:
\begin{subequations}
\label{muBasis}
\begin{align}
&\mu_1 (z,\bar{z})\,= \frac{\phi_2}{p}\\
&\mu_2 (z,\bar{z})\,= -\frac{\left(q^2-1\right) \phi_2}{p^3}-\frac{4}{p^2}
\\
&\mu_3 (z,\bar{z})\,= \frac{\phi_4}{p}
\\
&\mu_4 (z,\bar{z})\,= \frac{\phi_9}{p}
\\
&\mu_5 (z,\bar{z})\,= \frac{\phi_3}{p^2}
\\
&\mu_6 (z,\bar{z})\,= \frac{2 \phi_1}{p^2}-\frac{\left(q^2-1\right) \phi_4}{p^3}
\\
&\mu_7 (z,\bar{z})\,= \frac{\left(q^2-1\right) (\phi_2-3 \phi_9)}{p^3}+\frac{3 (\phi_1+4)+4}{p^2}
\\
&\mu_8 (z,\bar{z})\,= \frac{\phi_{10}}{p}
\\
&\mu_9 (z,\bar{z})\,=  -\frac{\phi_{11}}{p}
\\
&\mu_{10} (z,\bar{z})\,= \frac{\left(q^2-1\right) \left(\frac{1}{12} \pi ^2 
\phi_{2}+\phi_{10}\right)}{p^3}+\frac{\frac{\pi ^2}{3}-2 \phi_{5}}{p^2}
\\
&\mu_{11} (z,\bar{z})\,=  -\frac{6  \left(2 \phi_1-\phi_3+\phi_5+\pi^2+16\right)
}{p^2 \left(q^2-1\right)}
-\frac{3 \pi ^2 \phi_2+\phi_{11}}{p^3 }
\end{align}
\end{subequations}
Since the functions $\mu_j (z,\bar{z})$ have been constructed as linear combinations of $\phi_i$ with coefficients that are symmetric under $z\to\bar{z}$, it immediately follows that they admit the hermiticity property (\ref{hermiticity}). Furthermore, given that they do not involve the two complex functions ($\phi_6$ and $\phi_8$), these functions are all \emph{real}:
\begin{equation}
\mu_j(z,\bar{z}) =\mu_j(\bar{z},z)\,.
\end{equation}
Finally, these functions are all symmetric under the target-projectile swap $z\to 1-\bar{z}$, (corresponding to $p\to -p$) as in 
(\ref{SunderTP}):
\begin{equation}
\mu_j (z,\bar{z}) =\mu_j(1-\bar{z},1-z) \,.
\end{equation}
This holds since the antisymmetric $\phi_i$ enter eqs.~(\ref{muBasis}) along with odd powers of $p$, while the symmetric $\phi_i$ with even powers.


\subsubsection{Another look 
at the QCD vertex at one loop}

In section~\ref{sec:MatchingNLL} have presented one-loop QCD vertex by separating the terms containing non-maximal weight functions in a closed form, to all order in $\epsilon$ following refs.~\cite{Fadin:1994fj,Fadin:2000yp,Fadin:2023roz} -- see eqs.~(\ref{v1spur}) through (\ref{vab}). 
Here, in preparation to presenting the two-loop case, we expand the result and keep terms through ${\cal O}(\epsilon^2)$ (or weight 4). As explained in the previous subsection, we do so in order to express the one and two loop results using a common basis of transcendental functions and parallel notation for the coefficients. 

At the outset we split the coefficients into dispersive and absorptive parts, 
\begin{align}
\label{eq:vnDispAbsorp}
\begin{split}
v_{\oplus}^{(n)}(t_1,t_2,|p_4|^2)  = 
v_{\oplus}^{(n)}(z,\bar{z}) 
=&\,\, {\text{Disp}}\left\{v_{\oplus}^{(n)}(z,\bar{z}) \right\}
+i  {\text{Absorp}} \left\{v_{\oplus}^{(n)}(z,\bar{z})\right\}
\\\equiv &\,\,
v_{\text{\text{Disp}}}^{(n)}(z,\bar{z}) 
+i \pi \,\,  v_{{\text{Absorp}}}^{(n)}(z,\bar{z} )
\end{split}
\end{align} 

At one loop ($n=1$) we have
\begin{subequations}
\label{one_loopvertex_expanded}
\begin{align}
\begin{split}
\label{one_loopvertex_expanded_Disp}
 v_{\text{\text{Disp}}}^{(1)}(z,\bar{z})
=\,&\,   v_{\text{sYM},\, \text{Disp}}^{(1)}(z,\bar{z})  + {\cal R}^{(1)}_{\text{Disp}}
\\& 
+ \sum_{i\in\{1,2,4,5\}} W_i(p,q) \phi_i(z,\bar{z})
\,\,+ \sum_{j\in\{1,2,3,6,8,10\}}\!\! R_j(p,q) \mu_j(z,\bar{z})
\end{split}
\\
 v_{\text{\text{Absorp}}}^{(1)}(z,\bar{z})
=\,&    v_{\text{sYM},\, \text{Absorp}}^{(1)}(z,\bar{z})  
\end{align}
\end{subequations}
where $p$ and $q$ are given in (\ref{pq}), the uniform weight functions $\phi_i(z,\bar{z})$ are given in eq.~(\ref{phiBasis}) and the non-pure (and mixed-weight) functions $\mu_j(z,\bar{z})$ are given in eq.~(\ref{muBasis}). 
The final three contributions to $v_{\text{\text{Disp}}}^{(1)}(z,\bar{z})$ in ~(\ref{one_loopvertex_expanded_Disp}) correspond to 
$v^{(1)}_{\text{spurious}}(t_1,t_2,|p_4|^2)+v_{\beta}^{(1)}(\epsilon)$ in eq.(\ref{eq:vREGvsSPU}), expanded in $\epsilon$.
The coefficients $W_i$, $R_j$ and ${\cal R}_{\text{Disp}}^{(1)}$ are rational functions of $p$, $q$, $N_c$ and $n_f$ (note that they do contain $\pi^2$ terms, but no other transcendental numbers or functions). 
They read as follows:
\begin{subequations}
\begin{align}
W_1&=\frac{q}{6 (q-1)
   \left((q-1)^2-p^2\right)}
   \left(\epsilon 
   \left(n_f-N_c\right)
+   \frac{1}{3} \epsilon ^2 \left(11
   n_f-8 N_c\right)\right)
   \\
W_2&=\frac{p}{1-q}W_1\\
W_4&=\frac{p q \epsilon ^2 \left(N_c-n_f\right)  }{6 (q-1)^2
   \left((q-1)^2-p^2\right)}\\
W_5&=\frac{q-1}{p}W_4\,.
\end{align}
\end{subequations}
Note that $W_2$ and $W_4$ are antisymmetric under $p\to -p$ (with fixed $q$), precisely compensating for the antisymmetry of the corresponding transcendental functions they multiply, $\phi_2$ and $\phi_4$. In turn, $W_1$ and $W_5$ are symmetric under $p\to -p$, in line with the symmetry of the corresponding $\phi_i$. 

Next, the coefficients of the non-pure, mixed-weight functions $\mu_j$ are given by
\begin{subequations}
\begin{align}
\begin{split}
R_1&=-\frac{1}{16}
   q \left(n_f-7
   N_c\right)+\frac{1}{16}
   \left(7 N_c-n_f\right)
    \\&
+   \epsilon \left(-\frac{1}{16} q
   \left(n_f-14
   N_c\right)-\frac{1}{6}
   \frac{q}{(1-q)^2}
   \left(n_f-N_c\right)+\frac{1}{
   48} \left(44 N_c-5
   n_f\right)\right)
 \\&
  + \epsilon ^2 \left(-\frac{1}{16} q
   \left(n_f-28
   N_c\right)+\frac{1}{192} \pi
   ^2 (q+1) \left(n_f-7
   N_c\right)-\frac{1}{18}
   \frac{q}{(1-q)^2} \left(11 n_f-8
   N_c\right)
   \right.\\&\left.
   +\frac{1}{144}
   \left(268 N_c-31
   n_f\right)\right)
   \end{split}
   \\
R_2&=\frac{(1+q)^2}{1
   -q}\left(\frac{1}{48}
   \left(N_c-n_f\right)
  -\frac{1}{144}
   \epsilon  \left(11 n_f-8
   N_c\right)
  +\frac{1}{432}
   \epsilon ^2 \left(52 N_c-85
   n_f\right)\right)\\
R_3&=
-\frac{\epsilon}{16}
   (q+1)   \left(n_f-7
   N_c\right)
+\epsilon ^2 \left(-\frac{1}{16} q
   \left(n_f-14
   N_c\right)-\frac{q
   \left(n_f-N_c\right)}{6
   (q-1)^2}+\frac{1}{48} \left(44
   N_c-5
   n_f\right)\right)\\
R_6&=\frac{(q+1)^2}{144 (1-q)} \left(
3 \epsilon  \left(N_c-n_f\right)+
\epsilon ^2 \left(8 N_c-11 n_f\right)\right)\\
R_8&=-\frac{1}{16} (q+1) \epsilon ^2 \left(7 N_c-n_f\right)\\
R_{10}&=\frac{(q+1)^2 \epsilon ^2 \left(N_c-n_f\right)}{48 (1-q)}\,,
\end{align}
\end{subequations}
and finally
\begin{align}
\begin{split}
{\cal R}^{(1)}_{\text{Disp}}&=
-\frac{11 N_c - 2 n_f}{24 \epsilon} +
\frac{q}{(q-1)
   \left((q-1)^2-p^2\right)}
\left(
\frac{1}{3}
   \left(N_c-n_f\right)
   -\frac{\epsilon}{9}
     \left(11 n_f-8
   N_c\right)\right)
\\&+\epsilon ^2\left(
   -\frac{1}{27} 
   \left(85 n_f-52 N_c\right)+\frac{\pi ^2}{36}
   \left(n_f-N_c\right)\right)\,.
\end{split}
\end{align}
We emphasise that the transcendental functions entering the one-loop vertex beyond sYM consist exclusively of the subset of functions which correspond to powers of \emph{single-valued logarithms} -- rather than GPLs. These include $\phi_i$ for $i=1,2,4,5$ and  $\mu_j$ for $j=1,2,3,6,8,10$, which are in turn expressible in terms of $\phi_i$ for $i=1,2,4,5$ and $10$. This must indeed be so, as we know that the non-maximal weight component of the one-loop vertex exponentiates to all orders in the $\epsilon$ expansion according to eqs.~(\ref{v1spur}) through (\ref{vab})~\cite{Fadin:1994fj,Fadin:2000yp,Fadin:2023roz}. Next, we consider the two-loop vertex, where no such simplification occurs. In  particular, the non-maximal weight component of the vertex involves the complete basis of GPLs we presented in sec.~\ref{sec:transcenBases} above. 


\subsubsection{QCD vertex at two loop}

In analogy with the one-loop case, eq.~(\ref{one_loopvertex_expanded}) we write the two-loop QCD vertex separated into dispersive and absorptive parts as in eq.~(\ref{eq:vnDispAbsorp}). These two components are expressed using the bases of transcendental functions introduced in section~\ref{sec:transcenBases}:
\begin{subequations}
\label{two_loopvertex_expanded}
\begin{align}
\begin{split}
 v_{\text{\text{Disp}}}^{(2)}(z,\bar{z})
=\,&\,   v_{\text{sYM},\, \text{Disp}}^{(2)}(z,\bar{z})  + {\cal R}^{(2)}_{\text{Disp}}
\\& 
+ \sum_{i\in\{1,2,3,4,5,7,8\}} 
U_i^{\text{\text{Disp}}}(p,q) \phi_i(z,\bar{z})
\,\,+ \sum_{j\in\{1,2,3,4,5,6,7,9,11\}}\!\! S_j^{\text{\text{Disp}}}(p,q) \mu_j(z,\bar{z})
\end{split}
\\
\begin{split}
 v_{\text{\text{Absorp}}}^{(2)}(z,\bar{z})
=\,&    v_{\text{sYM},\, \text{Absorp}}^{(2)}(z,\bar{z})  
+ {\cal R}^{(2)}_{\text{Absorp}}
\\& 
+ \sum_{i\in\{1,2,3,6\}} U_i^{{\text{Absorp}}}(p,q) \phi_i(z,\bar{z})
\,\,+ \sum_{j\in\{1,2,3,4,6,7\}}\!\! S_j^{{\text{Absorp}}}(p,q) \mu_j(z,\bar{z})
\end{split}
\end{align}
\end{subequations}
where, in contrast to the one-loop case, also the absorptive part of the vertex receives non-pure lower-weight functions, including spurious poles, $1/p^n$.

The coefficients ${\cal R}^{(2)}$, $U_i$ and $S_j$ appearing in (\ref{two_loopvertex_expanded}) are rational functions of $p$, $q$, $N_c$ and $n_f$ at each order in $\epsilon$. They contain $\pi^2$ factors, but no other transcendental numbers or functions. 
The coefficients of $\phi_i$ in the dispersive part of the vertex are: 
\begin{subequations}
\begin{align}
\begin{split}
U_1^{\text{Disp}}&=\frac{q}{(q-1)
   \left((q-1)^2-p^2\right)}
\left(-\frac{N_c
   \left(n_f-N_c\right)}{24 \epsilon
   }+\frac{1}{24} q N_c
   \left(N_c-n_f\right)\right.
   \\&\left.
   -\frac{1}{144} \left(2 n_f-N_c\right)
   \left(5 N_c+n_f\right)\right)
   -\frac{1}{576}
   \pi ^2 N_c \left(2 n_f-11 N_c\right)
\end{split}
\\
\begin{split}
U_2^{\text{Disp}}&=
\frac{p q}{(q-1)^2
   \left((q-1)^2-p^2\right)}
\left(\frac{N_c
   \left(n_f-N_c\right)}{24 \epsilon
   }+\frac{1}{24} q N_c
   \left(n_f-N_c\right)\right.\\&\left. +\frac{1}{144} \left(35 N_c n_f-27
   N_c^2-2 n_f^2\right)\right)
   \end{split}
   \\
\begin{split}
U_3^{\text{Disp}}&=
\frac{N_c
   \left(2 n_f-11 N_c\right)}{96 \epsilon
   }
+   \frac{q N_c \left(n_f-N_c\right)}{12 (q-1)
   \left((q-1)^2-p^2\right)}+\frac{1}{576} \left(-7 N_c n_f-144
   N_c^2+4 n_f^2\right)
   \end{split}
   \\
U_4^{\text{Disp}}&=
   \frac{p q}{24
   (q-1)^2 \left((q-1)^2-p^2\right)}
   N_c \left(n_f-N_c\right)
   \\
U_5^{\text{Disp}}&= \frac{q-1}{p} U_4^{\text{Disp}}
\\
U_7^{\text{Disp}}&=U_8^{\text{Disp}}
= \frac{1}{48} N_c \left(11 N_c-2 n_f\right)\,,
\end{align}
\end{subequations}
while those in the absorptive part read:
\begin{subequations}
\begin{align}
\begin{split}
U_1^{\text{Absorp}}&= \frac{q}{(q-1)
   \left((q-1)^2-p^2\right)} \frac{1}{48} N_c
   \left(N_c-n_f\right)
   \\
   U_2^{\text{Absorp}}&=\frac{p}{1-q} U_1^{\text{Absorp}}\\
    U_3^{\text{Absorp}}&= \frac{1}{192} N_c \left(11 N_c-2
   n_f\right)\\
   U_6^{\text{Absorp}}&= \frac{1}{48} N_c \left(11 N_c-2
   n_f\right)\,.
\end{split}
\end{align}
\end{subequations}
Note that the only antisymmetric coefficients under $p\to -p$ are those multiplying the two antisymmetric transcendental functions $\phi_2$ and $\phi_4$ (see Table~\ref{tab:phi_i_properties}), realizing the exact target-projectile symmetry of the vertex. 

Next, the coefficients of the non-pure, mixed weight functions $\mu_j$ entering the dispersive part of the two-loop vertex are:
\begin{subequations}
\begin{align}
\begin{split}
S_1^{\text{Disp}}&=-\frac{(q+1) N_c \left(7
   N_c-n_f\right)}{64 \epsilon
   ^2}
   \\&
   +\frac{1}{\epsilon
   }\left(\frac{q N_c
   \left(n_f-N_c\right)}{24
   (q-1)^2}+\frac{1}{384} (q+1)
   \left(31 N_c n_f-161 N_c^2-2
   n_f^2\right)+\frac{1}{96} N_c
   \left(n_f-N_c\right)\right)
   \\&
   +\frac{(q+1)^2 N_c
   \left(n_f-N_c\right)}{96
   (q-1)}-\frac{1}{96} q N_c
   \left(n_f-N_c\right) -\frac{q
   \left(-41 N_c n_f+33 N_c^2+2
   n_f^2\right)}{144
   (q-1)^2}
   \\&+q
   \left(-\frac{43 N_c
   n_f}{288}+\frac{5 n_f}{128
   N_c}+\frac{133
   N_c^2}{576}+\frac{7
   n_f^2}{576}\right)-\frac{11 N_c
   n_f}{192}-\frac{1}{32} N_c
   \left(n_f-N_c\right)
   \\& +\frac{5
   n_f}{128 N_c}+\frac{11
   N_c^2}{72}+\frac{5 n_f^2}{576}
   \end{split}
   \\
   \begin{split}
S_2^{\text{Disp}}&=\frac{(q+1)^2}{1-q}\left(-\frac{N_c
   \left(N_c-n_f\right)}{192
   \epsilon ^2}+\frac{35 N_c
   n_f-27 N_c^2-2 n_f^2}{1152
   \epsilon }+\frac{1}{192} \pi
   ^2 N_c
   \left(N_c-n_f\right)\right.
   \\ &\left.
   -\frac{2
   N_c n_f^2-68 N_c^3+27
   n_f}{3456 N_c}\right)
   \end{split}
   \\
   \begin{split}
S_3^{\text{Disp}}&=
   -\frac{(q+1) N_c \left(7
   N_c-n_f\right)}{64 \epsilon
   }+\frac{q N_c
   \left(n_f-N_c\right)}{24
   (q-1)^2}+\frac{1}{384} (q+1)
   \left(-13 N_c n_f-13 N_c^2+2
   n_f^2\right)\\&+\frac{1}{96} N_c
   \left(n_f-N_c\right)
   \end{split}
   \\
 S_4^{\text{Disp}}&=  \frac{1}{16} (q+1) N_c
   \left(N_c-n_f\right)
   \\
S_5^{\text{Disp}}&=
   -\frac{(q+1)^2 (q+5) N_c
   \left(N_c-n_f\right)}{192
   (q-1)}
   \\
S_6^{\text{Disp}}&=
   \frac{(q+1)^2}{1-q}\left(\frac{N_c
   \left(n_f-N_c\right)}{192
   \epsilon }+\frac{-9 N_c n_f+13
   N_c^2+2 n_f^2}{1152}\right)
   \\
S_7^{\text{Disp}}&=
   \frac{(q+1)^2}{1-q}\left( -\frac{1}{48} N_c
   \left(N_c-n_f\right)\right)
   \\
S_9^{\text{Disp}}&=
   \frac{1}{192} (q+1) N_c \left(7
   N_c-n_f\right)
    \\
S_{11}^{\text{Disp}}&=
   \frac{1}{576} (q+1)^3 N_c
   \left(N_c-n_f\right)\,,
\end{align}
\end{subequations}
while those entering the absorptive part read:
\begin{subequations}
    \begin{align}
    \begin{split}
        S_{1}^{\text{Absorp}}&=
        -\frac{(q+1) N_c \left(7
   N_c-n_f\right)}{128 \epsilon
   }+\frac{1}{128} (q+1) N_c
   \left(2 n_f-15
   N_c\right)
  \\& +\frac{q N_c
   \left(n_f-N_c\right)}{48
   (q-1)^2}+\frac{1}{192} N_c
   \left(n_f-N_c\right)
   \end{split}
   \\
S_{2}^{\text{Absorp}}&= \frac{(q+1)^2}{1-q} 
\frac{1}{384}\left(\frac{N_c
   \left(n_f-N_c\right)}{\epsilon
   }+N_c^2\right)
\\
S_{3}^{\text{Absorp}}&=-\frac{1}{64} (q+1) N_c \left(7 N_c-n_f\right)\\
S_{4}^{\text{Absorp}}&=-S_{3}^{\text{Absorp}}\\
S_{6}^{\text{Absorp}}&=\frac{(q+1)^2 N_c \left(N_c-n_f\right)}{192 (q-1)}\\
S_{7}^{\text{Absorp}}&=-\frac13 S_{6}^{\text{Absorp}} \,.
    \end{align}
\end{subequations}
Finally,
\begin{align}
\begin{split}
{\cal R}^{(2)}_{\text{Disp}}&=
\frac{5}{192 \epsilon ^3} N_c \left(11 N_c-2
   n_f\right)
   \\&
   +\frac{1}{\epsilon
   ^2}\left(\frac{-q
   N_c \left(N_c-n_f\right)}{12 (q-1)
   \left((q-1)^2-p^2\right)}+\frac{-112 N_c n_f+229
   N_c^2+12 n_f^2}{1152}\right)
\\&
+\frac{1}{\epsilon}\,\, \Bigg(
\frac{q \left(35 N_c n_f-27 N_c^2-2
   n_f^2\right)}{72 (q-1)
   \left((q-1)^2-p^2\right)}
   +\frac{155 N_c n_f}{1728}+\frac{\pi ^2}{288}  N_c
   \left(2 n_f-11 N_c\right)-\frac{n_f}{64
   N_c}\\&
 -\frac{173\, N_c^2}{432}\Bigg) +\frac{11
   \pi ^2 N_c \left(67 N_c-10
   n_f\right)}{3456}-\frac{N_c \left(1142
   N_c-65 n_f\right)}{2592}
 \\&  
      +\frac{q}{(q-
   1)
   \left((q-1)^2-p^2\right)} \left(\left(-\frac{2
   N_c}{3}-\frac{1}{8 N_c}\right)
   n_f+\frac{\pi ^2}{24}  N_c
   \left(N_c-n_f\right)+\frac{53
   N_c^2}{54}-\frac{n_f^2}{108}\right)
   \end{split}
\end{align}
and 
\begin{align}
\begin{split}
{\cal R}^{(2)}_{\text{Absorp}}&=\frac{N_c \left(11 N_c-2 n_f\right)}{96
   \epsilon^2}
   -\frac{N_c
   \left(67 N_c-10 n_f\right)}{576
   \epsilon}-\frac{1}{864} N_c \left(193 N_c-19
   n_f\right)
 \\&
  + \frac{q}{24 (q-1)
   \left((q-1)^2-p^2\right)} \left(\frac{N_c
   \left(n_f-N_c\right)}{\epsilon}+q
   N_c \left(N_c-n_f\right)+\frac{N_c}{3}
    \left(11 n_f-8
   N_c\right)\right)\,.
   \end{split}
\end{align}


\section{Conclusions}

In this paper we took a step forward in developing the effective theory for
multi-Reggeon calculations~\cite{Caron-Huot:2013fea,Caron-Huot:2017fxr,Caron-Huot:2017zfo,Caron-Huot:2020grv,Caron-Huot:2020vlo,Falcioni:2020lvv,Falcioni:2021buo,Falcioni:2021dgr,Abreu:2024mpk,Buccioni:2024gzo} based on the shock-wave formalism with the aim of applying it to multi-leg amplitudes in the multi-Regge kinematic (MRK) limit. 
In line with previous work on this~\cite{Caron-Huot:2013fea,Caron-Huot:2020vlo,Abreu:2024mpk,Buccioni:2024gzo}, we find that the coupling of a real gluon to the Reggeon field $W$ is not restricted to a single Reggeon transition ${\cal R} g {\cal R}$, but it comes along with a variety of possible multi-Reggeon (MR) transitions. Specifically, besides the single Reggeon transition, the additional ${\cal R} g {\cal R}^2$ and ${\cal R}^2 g {\cal R}^2$ MR transitions are relevant starting from one loop, while 
${\cal R} g {\cal R}^3$, ${\cal R}^2 g {\cal R}^3$ and ${\cal R}^3 g {\cal R}^3$, are relevant starting at two loops. 
One interesting feature is that in sharp contrast to rapidity evolution, which is described by the JIMWLK Hamiltonian and preserves signature, real gluon emission involves transitions between odd signature across one rapidity span and even signature across another.

We applied this formalism to $2\to 3$ scattering, classifying all multi-Reggeon (MR) transitions at both one and two loops. We explicitly computed all one-loop transitions and those contributing to the odd-odd signature component at two loops. This provides us with the essential information for extracting the Lipatov vertex at two loops.

Specializing general-kinematics amplitude computations to the MRK limit, we computed analytically the full set of (non-planar) QCD amplitudes for $gg\to ggg$, $qg\to qgg$ and $qq\to qgq$ scattering at one and two loops. We obtained a closed form analytic result for all components of these amplitudes at leading power in MRK and converted the result to the $t$-channel colour basis. 

Based on the computed amplitudes in MRK and the effective MR computation described above, we isolated the Regge-pole component and used the factorisation formula (\ref{ReggePole}) to extract the Lipatov vertex.
In identifying the Regge-pole component we used the criterion~(\ref{pole-cut}) of ref.~\cite{Falcioni:2021dgr,Falcioni:2021buo}, originally set in the context of $2\to 2$ scattering, namely that the Regge-pole component, i.e.~the part of the octet-octet colour component of the amplitude which factorizes, consists of the sum of the SR
transition, ${\cal R} g {\cal R}$, plus the planar part of the MR
transitions. In turn, the Regge-cut part, which does not factorize, is entirely non-planar.

The extraction of the two-loop Lipatov vertex has been done using the aforementioned three partonic processes, providing a robust check of both the amplitudes and the effective MR computation. Furthermore, 
the universality of the planar part of the MR contribution -- in contrast with the process dependence of its non-planar counterpart -- gives further assurance that we are using the correct separation between the pole and cut components. 
Analyzing the expansion of the factorization formula to three loops, we predict the form of the planar part of the MR contributions to the octet-octet channel at this order. This can be confirmed through explicit computation in the future, to provide further assurance that we understand well the interplay between the MR effective theory and the factorization properties in the MRK limit. We recall that for $2\to2$ scattering, computations have been performed to four loops, providing such evidence~\cite{Falcioni:2021dgr,Falcioni:2021buo}.
We also computed the one-loop amplitudes analytically to ${\cal O}(\epsilon^4)$ (weight 6) and extracted the one-loop Lipatov vertex to this order. These results can be used in the context of future three-loop $2\to 3$ computations.

In section~\ref{sec:Reggeization} we studied the analytic properties of the vertex based on general considerations and clarified its symmetries and the properties of the transcendental functions it contains.
A main ingredient of this analysis has been the comparison between the factorization of eq.~(\ref{REggNLL}) in terms of a complex vertex and eqs.~(\ref{ggNLLsig2}) and (\ref{AmuB}) which involve, before contraction with the polarization vector, real-valued left and right vertices. The reality condition implies that a natural basis of transcendental functions to be used in expressing both the dispersive and absorptive parts of the complex vertex would consist of real-valued functions that are symmetric under a swap of $z$ and $\bar{z}$ as well as imaginary ones, which are antisymmetric, or equivalently, all transcendental functions should be hermitian.   
Another important constraint on the analytic structure of the vertex is that it must satisfy target-projectile symmetry. It implies that the transcendental functions can all be chosen to admit~(\ref{TargetProjectileSymmetryPhi}) and then the same symmetry property must be satisfied by the rational coefficients.

These conclusions have been tested in the context of the one and two loop results for the vertex. They have also been used to derive compact expressions with transparent symmetry properties. A short summary of the properties can be found in the beginning of section~\ref{sec:transcenBases}. Table~\ref{tab:phi_i_properties} summarises the symmetries of the basis of transcendental functions of non-maximal weight appearing in the QCD vertex.

A salient property of the Lipatov vertex is that it is a single-valued function of the kinematic variables $z$ and $\bar{z}$. This is expected on physical grounds, since this complex plane corresponds to a two-dimensional Euclidean space.
We have used this in section~\ref{VertTwoLoopStruct} to write the complete result for the vertex through two loops in terms SVMPLs. 
In agreement with previous work on $2\to 3$ amplitudes in MRK in sYM~\cite{Caron-Huot:2020vlo} and in QCD~\cite{Buccioni:2024gzo}, we find a six-letter alphabet,
$\{z,\bar{z},1-z,1-\bar{z},z-{\bar{z}}, 1-z-{\bar{z}}\}$. The first four letters correspond to physical logarithmic singularities where one of the two $t$ channel momenta vanishes, while the last two correspond respectively to the configurations where all final-state momenta are in a plane, and to the limit where the centrally-emitted gluon is soft. None of these lead to rational kinematic singularities in the vertex, but there are spurious rational poles in the soft limit, which cancel through combinations of transcendental functions. We make this manifest by a basis choice~(\ref{muBasis}) where each function is itself finite. 

We performed a series of checks on our calculations. At one loop we find agreement with ref.~\cite{Fadin:2023aen}, which provides the dispersive and the absorptive parts of the Lipatov vertex through ${\cal O}(\epsilon^2)$. In addition, we checked that all the multi-Reggeon transitions at one loop agree with the effective theory calculation presented in ref.~\cite{Buccioni:2024gzo}. Proceeding to two loops, we compared the multi-Reggeon transitions that contribute to the odd-odd signature components of the amplitude~\cite{Abreu:2024mpk} with the results of~\cite{Buccioni:2024gzo}. Concerning the final ingredient that enter the determination of the two-loop Lipatov vertex, namely the $gg\to ggg$, $qg\to qgg$ and $qq\to qgq$ amplitudes in the MRK limit, we verified that the terms of leading transcendental weight in the $gg\to ggg$ QCD amplitude reproduce the $\mathcal{N}=4$ sYM result of ref.~\cite{Caron-Huot:2020vlo}. Finally, for all the three channels above we checked that the finite remainder of the QCD amplitudes are in agreement with the results of ref.~\cite{Buccioni:2024gzo}.

Let us now provide some outlook regarding the use of the new results. 
The two-loop vertex, along with the already known three-loop Regge trajectory and two-loop impact factors~\cite{Falcioni:2021dgr,Falcioni:2021buo, Caola:2022dfa}, completes the information required to use the MRK Regge-pole factorization formula to predict or resum rapidity logarithms at higher orders with NNLL accuracy.
To achieve complete control of rapidity logarithms at this accuracy in the non-planar theory, the Regge-pole factorization formula needs to be supplemented by Regge cut contributions, corresponding to the non-planar MR components. While these are not yet known at three loops, we have a theory set up to compute these (recall that for $2\to 2$ amplitudes, such MR corrections have already been computed to four loops).

We emphasise that the possibility of using the MRK Regge-pole factorization formula to predict or resum rapidity logarithms at higher orders is not limited to $2\to 3$ scattering. The same information, extracted from $2\to2$ and $2\to 3$ transitions should be applicable to $2\to4$ and higher-point scattering amplitudes. Note though that a new feature arises starting from $2\to 4$ amplitudes, namely that Regge cuts start to contribute also in the planar limit~\cite{Bartels:2008ce,Bartels:2009vkz,Lipatov:2009nt,Dixon:2014voa,DelDuca:2019tur,Bartels:2021thv}. It would be very interesting to push the MR theory further and apply it in this context, where in the planar limit contact can be made with predictions for the remainder functions in sYM.

Finally, the two-loop Lipatov vertex is a key element for determining the BFKL kernel at NNLO. In order to 
compute this kernel (along the lines of the NLO computation in refs.~\cite{Fadin:1998py,Ciafaloni:1998gs,Kotikov:2000pm}) one needs to sum interference diagrams of different multiplicity,
including: 
$i)$~virtual corrections involving the three-loop gluon Regge trajectory~\cite{Falcioni:2021dgr,Falcioni:2021buo, Caola:2022dfa}; 
$ii)$~single-particle production involving the two-loop Lipatov vertex computed here, interfering with the tree-level vertex, as well as the interference of two one-loop Lipatov vertices; 
$iii)$~the central emission of two partons at one loop~\cite{Byrne:2022wzk} interfering with its tree counterpart; 
and finally $iv)$~the central emission of three partons at tree level. Recent progress has been made on determining several of these ingredients~\cite{Byrne:2022wzk,ByrneWIP}, making BFKL theory at NNLL a realistic goal to pursue.


\acknowledgments

We wish to thank Emmet Byrne, Vittorio Del Duca, Claude Duhr, Michael Fucilla, Jenni Smillie, Vasily Sotnikov and Simone Zoia for useful discussions. 
We also wish to thank Federico Buccioni, Fabrizio Caola, Federica Devoto and Giulio Gambuti, who have recently 
published~\cite{Buccioni:2024gzo} an independent extraction of the Lipatov vertex following similar methods, for conducting a careful comparison between our results for the effective MR computation~\cite{Abreu:2024mpk}.
G.D.L.’s work is supported in part by the U.K. Royal Society through Grant URF\textbackslash R1\textbackslash 20109; G.F. is supported by the EU’s Horizon Europe research and innovation programme under the Marie Sk\l{}odowska-Curie grant 101104792, \textit{QCDchallenge};
EG is supported by the STFC Consolidated Grant ``Particle Physics at the Higgs Centre''; 
CM has been partially supported by the Italian Ministry of University and Research (MUR) through grant PRIN20172LNEEZ and by Compagnia di San Paolo through grant TORPS1921EX-POST2101.
GF and EG are grateful to the Galileo Galilei Institute in Florence for hospitality and support during the scientific program {\textit{Mathematical Structures in Scattering Amplitudes}}, where part of this work was done.

\appendix


\section{Tree level impact factors and Lipatov 
vertex for gluon scattering}\label{Tree5g}

In this appendix we calculate explicitly 
the tree level expressions for the impact 
factors and the Lipatov vertex for the 
process 
\begin{align}
\label{5gKinDef}
g\big[(-P_1)^{\oplus},a_1\big] + g\big[(-P_2)^{\oplus},a_2\big] 
\to g\big[P_3^{\oplus},a_3] + g\big[P_4^{\oplus},a_4\big] 
+ g\big[P_5^{\oplus},a_5\big].
\end{align}
To this end we start from the high-energy 
limit of the tree level amplitude, where 
the polarization vectors are written 
explicitly, taking as a starting point 
eq.~(20) of \cite{DelDuca:1995zy}. 
With the notation used in this paper, 
and keeping for now generic helicity 
states, we have 
\bea\label{MTree1} \nn
{\cal M}_{gg \to ggg} &=& 
2 s
\Big[i g_s f^{a_2 a_3 y} \Gamma^{\mu_{2} \mu_{3}}\Big]\,
\varepsilon_{\mu_{2}}^{\lambda_{2} *}(P_2) \,
\varepsilon_{\mu_{3}}^{\lambda_{3}}(P_3) \, \frac{1}{t_2}  \\ 
&&\times 
\Big[i g_s f^{y a_4 x} \,C^{\mu_{4}}(Q_1,Q_2)\Big] \,
\varepsilon_{\mu_{4}}^{\lambda_{4}}(P_4) \,\frac{1}{t_1}  \\ \nn
&&\times 
\Big[i g_s f^{a_1 a_5 x} \Gamma^{\mu_{1} \mu_{5}}\Big] \,
\varepsilon_{\mu_{1}}^{\lambda_{1} *}(P_1) \,
\varepsilon_{\mu_{5}}^{\lambda_{5}}(P_5).
\eea
Comparing this amplitude with the one in 
\eqn{Mtree} specialised for gluons:
\bea\label{MtreeApp} \nn
{\cal M}_{gg \to ggg}^{\rm tree} &=& 
2s \Big[g_s \, \delta_{-\lambda_3, \lambda_2}\, i f^{a_2 a_3 y} 
\, C_{j}^{(0)}(P_2^{\lambda_2},P_3^{\lambda_3}) \Big]\, \frac{1}{t_2} \\ \nn
&&\hspace{0.0cm}\times\, \Big[g_s \, i f^{y a_4 x} \, 
V^{(0)}(Q_1,P_4^{\lambda_4},Q_2) \Big] \, \frac{1}{t_1}  \\ 
&&\hspace{0.0cm}\times\,
\Big[g_s \,\delta_{-\lambda_5, \lambda_1}\, i f^{a_1 a_5 x} \, 
C_{i}^{(0)}(P_1^{\lambda_1},P_5^{\lambda_5}) \Big],
\eea
we immediately identify
\bea \label{eqnImpFactLipVert} \nn
C_{g}^{(0)}(P_2^{\lambda_2},P_3^{\lambda_3}) 
&=& \Gamma^{\mu_{2} \mu_{3}}\,
\varepsilon_{\mu_{2}}^{\lambda_{2} *}(P_2) \,
\varepsilon_{\mu_{3}}^{\lambda_{3}}(P_3), \\ \nn
V^{(0)}(Q_1,P_4^{\lambda_4},Q_2)  &=& C^{\mu_{4}}(Q_1,Q_2) 
\, \varepsilon_{\mu_{4}}^{\lambda_{4}}(P_4), \\ 
C_{g}^{(0)}(P_1^{\lambda_1},P_5^{\lambda_5}) &=&
\Gamma^{\mu_{1} \mu_{5}} \,
\varepsilon_{\mu_{1}}^{\lambda_{1} *}(P_1) \,
\varepsilon_{\mu_{5}}^{\lambda_{5}}(P_5).
\eea
The expression for $\Gamma^{\mu_{2} \mu_{3}}$, 
$\Gamma^{\mu_{1} \mu_{5}}$ and $C^{\mu_4}(Q_1,Q_2)$ 
can be found respectively in eqs.~(21) and~(22) 
of \cite{DelDuca:1995zy}. Taking all momenta 
outgoing we have
\bea \label{ImpactFactorsOur} \nn
\Gamma^{\mu_2 \mu_3} &=& 
g^{\mu_2\mu_3} - \frac{P_2^{\mu_3} P_1^{\mu_2} 
- P_1^{\mu_3} P_3^{\mu_2}}{P_1 \cdot P_2}
- t_2 \frac{P_1^{\mu_3} P_1^{\mu_2}}{2 (P_1 \cdot P_2)^2}, \\ 
\Gamma^{\mu_1 \mu_5} &=& 
g^{\mu_1\mu_5} - \frac{P_2^{\mu_1} P_1^{\mu_5} 
- P_2^{\mu_5} P_5^{\mu_1}}{P_1 \cdot P_2}
- t_1 \frac{P_2^{\mu_5} P_2^{\mu_1}}{2 (P_1 \cdot P_2)^2},
\eea
and
\be \label{LipatovOur1}
C^{\mu}(Q_1,Q_2) = \bigg[ 
(Q_1-Q_2)_{\perp}^{\mu} 
+ \bigg(\frac{P_2 \cdot P_4}{P_1 \cdot P_2} 
+ \frac{Q_1^2}{P_1 \cdot P_4} \bigg) P_1^{\mu} 
- \bigg(\frac{P_1 \cdot P_4}{P_1 \cdot P_2} 
+ \frac{Q_2^2}{P_2 \cdot P_4} \bigg) P_2^{\mu} 
\bigg],
\ee
Notice that e.g.~$Q_{1\perp}$ is still 
a 4-momentum, i.e.~given $Q_1 = (q_1^+ , q_1^- ; \mathbf{q}_1)$
one has $Q_{1\perp} = (0 , 0 ; \mathbf{q}_1)$.

In order to proceed we need to define the 
polarization vectors. To this end we follow 
appendix C of \cite{DelDuca:1995zy}, and define 
the polarization vectors in the left ($L$) or 
right ($R$) gauge, 
following eq.~(50) of \cite{DelDuca:1995zy}:
we have 
\bea 
\label{PolVecLeftOur}
\varepsilon^{\mu}_L(P) &=& 
\varepsilon^{\mu}_{L\perp}(P)
- \frac{P\cdot \varepsilon_{L\perp}(P)}{P \cdot P_2} \, P_2^{\mu}, \\ 
\label{PolVecRightOur}
\varepsilon^{\mu}_R(P) &=& 
\varepsilon^{\mu}_{R\perp}(P)
- \frac{P\cdot \varepsilon_{R\perp}(P)}{P \cdot P_1} \, P_1^{\mu}. 
\eea
In turn, the perpendicular component 
of the polarization vectors, are 
given in eq.~(52), (53), (54) of  
\cite{DelDuca:1995zy}: in our 
notation one has
\bea
\label{PolVecLeftPerpOur} \nn
\varepsilon^{\mu \oplus}_{L \perp}(P) &=& 
\Big( 0,0; \boldsymbol{\varepsilon}^{\oplus}_{L \perp}(P) \Big), 
\qquad {\rm with } \qquad
\boldsymbol{\varepsilon}^{\oplus}_{L \perp}(P) =
\bigg(\frac{1}{\sqrt{2}}, \frac{i}{\sqrt{2}} \bigg),  \\ 
\varepsilon^{\mu \ominus}_{L \perp}(P) &=& 
\Big( 0,0; \boldsymbol{\varepsilon}^{\ominus}_{L \perp}(P) \Big), 
\qquad {\rm with } \qquad
\boldsymbol{\varepsilon}^{\ominus}_{L \perp}(P) =
\bigg(\frac{1}{\sqrt{2}}, - \frac{i}{\sqrt{2}} \bigg),  \\ \nn
\label{PolVecRightPerpOur}
\varepsilon^{\mu \oplus}_{R \perp}(P) &=& 
\Big( 0,0; \boldsymbol{\varepsilon}^{\oplus}_{R \perp}(P) \Big), 
\qquad {\rm with } \qquad
\boldsymbol{\varepsilon}^{\oplus}_{R \perp}(P) =
- \frac{\mathbf{p}}{\mathbf{\bar p}}
\bigg(\frac{1}{\sqrt{2}}, - \frac{i}{\sqrt{2}} \bigg),  \\
\varepsilon^{\mu \ominus}_{R \perp}(P) &=& 
\Big( 0,0; \boldsymbol{\varepsilon}^{\ominus}_{R \perp}(P) \Big), 
\qquad {\rm with } \qquad
\boldsymbol{\varepsilon}^{\ominus}_{R \perp}(P) =
- \frac{\mathbf{p}}{\mathbf{\bar p}}
\bigg(\frac{1}{\sqrt{2}}, \frac{i}{\sqrt{2}} \bigg),  
\eea
with $P \neq P_2$ or $P_1$, and 
\bea
\label{PolVecLeftPerpPBOur} \nn
\varepsilon^{\mu \oplus}_{L \perp}(P_1) &=& 
\Big( 0,0; \boldsymbol{\varepsilon}^{\oplus}_{L \perp}(P_1) \Big), 
\qquad {\rm with } \qquad
\boldsymbol{\varepsilon}^{\oplus}_{L \perp}(P_1) =
\bigg(\frac{1}{\sqrt{2}}, - \frac{i}{\sqrt{2}} \bigg),  \\ 
\varepsilon^{\mu \ominus}_{L \perp}(P_1) &=& 
\Big( 0,0; \boldsymbol{\varepsilon}^{\ominus}_{L \perp}(P_1) \Big), 
\qquad {\rm with } \qquad
\boldsymbol{\varepsilon}^{\ominus}_{L \perp}(P_1) =
\bigg(\frac{1}{\sqrt{2}}, \frac{i}{\sqrt{2}} \bigg),  \\ \nn
\label{PolVecRightPerpPAOur}
\varepsilon^{\mu \oplus}_{R \perp}(P_2) &=& 
\Big( 0,0; \boldsymbol{\varepsilon}^{\oplus}_{R \perp}(P_2) \Big), 
\qquad {\rm with } \qquad
\boldsymbol{\varepsilon}^{\oplus}_{R \perp}(P_2) =
\bigg(\frac{1}{\sqrt{2}}, \frac{i}{\sqrt{2}} \bigg), \\
\varepsilon^{\mu \ominus}_{R \perp}(P_2) &=& 
\Big( 0,0; \boldsymbol{\varepsilon}^{\ominus}_{R \perp}(P_2) \Big), 
\qquad {\rm with } \qquad
\boldsymbol{\varepsilon}^{\ominus}_{R \perp}(P_2) =
\bigg(\frac{1}{\sqrt{2}}, - \frac{i}{\sqrt{2}} \bigg). 
\eea
In order to write explicitly the polarization vectors
in \eqns{PolVecLeftOur}{PolVecRightOur} we need also 
$P\cdot P_1 = \frac{1}{2} p^+ p_{1}^-$, 
$P\cdot P_2 = \frac{1}{2} p^- p_{2}^+$, which follows 
from $P_1 = (0, p_{1}^-; \boldsymbol{0})$ and 
$P_2 = (p_{2}^+, 0; \boldsymbol{0})$. The initial 
state particle polarization vector are fixed to be 
respectively in the $L$ gauge for $P_1$ and in the $R$ gauge, for $P_2$: inserting 
\eqns{PolVecLeftPerpPBOur}{PolVecRightPerpPAOur}
into \eqns{PolVecLeftOur}{PolVecRightOur}, and 
choosing negative helicity, we immediately get
\bea 
\label{PolVecLeftOurP1explicit}
\varepsilon^{\mu \ominus}_L(P_1) &=& 
\varepsilon^{\mu \ominus}_{L\perp}(P_1)
= \bigg( 0,0; \frac{1}{\sqrt{2}}, + \frac{i}{\sqrt{2}}\bigg), \\ 
\label{PolVecRightOurP2explicit}
\varepsilon^{\mu \ominus}_R(P_2) &=& 
\varepsilon^{\mu \ominus}_{R\perp}(P_2)
= \bigg( 0,0; \frac{1}{\sqrt{2}}, - \frac{i}{\sqrt{2}}\bigg), 
\eea
which follows from the fact that $P_1\cdot \varepsilon_{L\perp}(P_1)
=P_2\cdot \varepsilon_{R\perp}(P_2) = 0$. Consider
now the outgoing particles 3, 4 and 5. For particle
3 and 5, whose momenta $P_3$ and $P_5$ are mostly 
along $P_2$ and $P_1$ respectively, we adopt the 
same gauge of particles 1 and 5; furthermore, 
we choose the positive helicity. We obtain 
\bea 
\label{PolVecLeftOurP5}
\varepsilon^{\mu \oplus}_L(P_5) &=& 
\varepsilon^{\oplus}_{L\perp}(P_5)
- \frac{P_5\cdot \varepsilon^{\oplus}_{L\perp}(P_5)}{P_5 \cdot P_2} \, P_2^{\mu} = 
\bigg( \frac{\sqrt{2}\, \mathbf{ p}_5}{p_5^-},0; 
\frac{1}{\sqrt{2}}, \frac{i}{\sqrt{2}}\bigg), \\ 
\label{PolVecRightOurP3}
\varepsilon^{\mu \oplus}_R(P_3) &=& 
\varepsilon^{\mu \oplus}_{R\perp}(P_3)
- \frac{P_3\cdot \varepsilon^{\oplus}_{R\perp}(P_3)}{P_3 \cdot P_1} \, P_1^{\mu} = 
- \frac{\mathbf{p}_3}{\mathbf{\bar p}_3}
\bigg(0,\frac{\sqrt{2}\, \mathbf{\bar p}_3}{p_3^+}; 
\frac{1}{\sqrt{2}}, -\frac{i}{\sqrt{2}}\bigg). 
\eea
Last, we need the polarization vector for 
particle 4, which we choose to have positive 
helicity as well. In this case we are free
to choose between $L$ and $R$ gauge. In the two 
cases we have 
\bea 
\label{PolVecLeftOurPL}
\varepsilon^{\mu \oplus}_L(P_4) &=& 
\varepsilon^{\oplus}_{L\perp}(P_4)
- \frac{P_4\cdot \varepsilon^{\oplus}_{L\perp}(P_4)}{P_4 \cdot P_2} \, P_2^{\mu} = 
\bigg( \frac{\sqrt{2}\, \mathbf{ p}_4}{p_4^-},0; 
\frac{1}{\sqrt{2}}, \frac{i}{\sqrt{2}}\bigg), \\ 
\label{PolVecRightOurPR}
\varepsilon^{\mu \oplus}_R(P_4) &=& 
\varepsilon^{\mu \oplus}_{R\perp}(P_4)
- \frac{P_4\cdot \varepsilon^{\oplus}_{R\perp}(P_4)}{P_4 \cdot P_1} \, P_1^{\mu} =
- \frac{\mathbf{p}_4}{\mathbf{\bar p}_4}
\bigg(0,\frac{\sqrt{2}\, \mathbf{\bar  p}_4}{p_4^+}; 
\frac{1}{\sqrt{2}}, -\frac{i}{\sqrt{2}}\bigg). 
\eea
At this point we still need to remember that 
in the tree-level amplitude of \eqn{MTree1}
the initial state polarization vectors appear 
as complex conjugate, i.e.~we need:
$\varepsilon_{\mu_{2}}^{\ominus *}(P_2)$,
$\varepsilon_{\mu_{1}}^{\ominus *}(P_1)$, 
and $\varepsilon_{\mu_{3}}^{\oplus}(P_3)$,
$\varepsilon_{\mu_{4}}^{\oplus}(P_4)$, 
$\varepsilon_{\mu_{5}}^{\oplus}(P_5)$
for the final-state polarization vectors. 
Summarising, choosing the L gauge for 
$\varepsilon^{\mu \oplus}_L(P_4)$, 
we have 
\bea \label{PolVecLeftOurAmp} \nn
\varepsilon^{\mu \ominus *}_L(P_1) &=& 
\bigg( 0,0; \frac{1}{\sqrt{2}}, -\frac{i}{\sqrt{2}}\bigg), \\ \nn
\varepsilon^{\mu \ominus *}_R(P_2) &=& 
\bigg( 0,0; \frac{1}{\sqrt{2}}, \frac{i}{\sqrt{2}}\bigg), \\ 
\varepsilon^{\mu \oplus}_R(P_3) &=& 
- \frac{\mathbf{p}_3}{\mathbf{\bar p}_3}
\bigg(0,\frac{\sqrt{2}\, \mathbf{\bar p}_3}{p_3^+}; 
\frac{1}{\sqrt{2}}, - \frac{i}{\sqrt{2}}\bigg), \\ \nn 
\varepsilon^{\mu \oplus}_L(P_4) &=& 
\bigg( \frac{\sqrt{2}\, \mathbf{p}_4}{p_4^-},0; 
\frac{1}{\sqrt{2}}, \frac{i}{\sqrt{2}}\bigg), \\ \nn
\varepsilon^{\mu \oplus}_L(P_5) &=& 
\bigg( \frac{\sqrt{2}\, \mathbf{p}_5}{p_5^-},0; 
\frac{1}{\sqrt{2}}, \frac{i}{\sqrt{2}}\bigg). 
\eea
We can now determine the impact factors
by inserting \eqns{ImpactFactorsOur}{PolVecLeftOurAmp} 
into \eqn{eqnImpFactLipVert}: we obtain 
\bea \label{impactfactorscalc} \nn
C_{g}^{(0)}(P_2^{\ominus},P_3^{\oplus}) 
= \Gamma^{\mu_{2} \mu_{3}}\,
\varepsilon_{\mu_{2}}^{\ominus *}(P_2) \,
\varepsilon_{\mu_{3}}^{\oplus}(P_3) 
&=& \frac{\mathbf{p}_3}{\mathbf{\bar p}_3}, \\
C_{g}^{(0)}(P_1^{\ominus},P_5^{\oplus}) 
= \Gamma^{\mu_{1} \mu_{5}}\,
\varepsilon_{\mu_{1}}^{\ominus *}(P_1) \,
\varepsilon_{\mu_{5}}^{\oplus}(P_5)
&=& -1.
\eea
Similarly, using the definitions in 
\eqns{LipatovOur1}{PolVecLeftOurAmp}
we calculate $V^{(0)}(Q_1,P_4^{\oplus},Q_2)$: 
\bea \label{LipatovOur4} \nn
V^{(0)}(Q_1,P_4^{\oplus},Q_2) 
&=& \frac{1}{\sqrt{2}} \bigg[ \mathbf{p}_3
- \mathbf{p}_5 + \bigg(\frac{p_2^+ p_4^-}{p_1^- p_2^+} 
+ \frac{- \mathbf{p}_5 \mathbf{\bar p}_5 }{\frac{1}{2}p_1^- p_4^+} 
\bigg) \frac{\mathbf{p}_4\, p_1^-}{p_4^-} \bigg] \\ 
&=& \sqrt{2} \frac{\mathbf{\bar p}_3\mathbf{p}_5}{\mathbf{\bar p}_4}\,,
\eea
in agreement with eq.~(25) of \cite{DelDuca:1995ki}.
\Eqn{LipatovOur4} is obtained by recalling that 
$P \cdot Q = \frac{1}{2} \big(p^+ q^- + p^- q^+
- \mathbf{p} \mathbf{\bar q}
- \mathbf{\bar p} \mathbf{q}\big)$,
and also that $P_1 \sim p_1^-$, 
$P_2 \sim p_2^+$, and $P_k^2 = p_k^+ p_k^- 
- \mathbf{p}_k \mathbf{\bar p}_k = 0$,
i.e. $p_k^+ p_k^- = \mathbf{p}_k \mathbf{\bar p}_k$
for $k=3$, 4, 5. 

Inserting \eqn{LipatovOur4} 
into \eqn{MtreeApp} we have 
\bea\label{MtreeAppBBB} \nn
{\cal M}_{ij \to i'gj'}^{\rm tree} &=& 
2s \Big[g_s \, {\bf T}_{a_3a_2}^y \,
C_{j}^{(0)}(P_2^{\ominus},P_3^{\oplus}) \Big] \, \frac{1}{t_2} \\ \nn 
&&\hspace{0.0cm}\times\, \Big[g_s \, i f^{y a_4 x} \, 
\sqrt{2} \frac{\mathbf{\bar p}_3\mathbf{p}_5}{\mathbf{\bar p}_4} \Big] 
\, \frac{1}{t_1} \\ 
&&\hspace{0.0cm}\times\,
\Big[g_s \, {\bf T}_{a_5a_1}^x \, 
C_{i}^{(0)}(P_1^{\ominus},P_5^{\oplus}) \Big],
\eea
which can be used to match with 
\eqn{Mtree2} in the main text, and
extract the factors $\mathcal{K}^{(0)}$ 
and $\mathcal{C}_{ij}^{(0)}$ respectively
in \eqns{def:Kij}{def:Cij}.

To conclude, let us express 
the results above in terms of 
the standard helicity basis 
representation of the amplitude.
The relation between the polarization 
vectors in \eqn{PolVecLeftOurAmp}
and the polarization vectors 
expressed in the basis 
\be
\varepsilon^{\pm}_{\mu}(P,K)
= \pm \frac{\langle P \pm |\gamma_{\mu}| 
K \pm\rangle}{\sqrt{2} \langle K \mp | P \pm \rangle},
\ee
where $K$ is a reference vector, 
has been derived in appendix C of 
\cite{DelDuca:1995zy}. In particular, 
one has (eq.~(55) of \cite{DelDuca:1995zy})
\bea \nn
\varepsilon_{\mu}^{\oplus}(P_i,P_2) &=& 
-\frac{\mathbf{\bar p}_i}{\mathbf{p}_i} 
\varepsilon_{L\mu}^{\oplus}(P_i), \\ \nn
\varepsilon_{\mu}^{\oplus}(P_i,P_1) &=& 
-\frac{\mathbf{\bar p}_i}{\mathbf{p}_i} 
\varepsilon_{R\mu}^{\oplus}(P_i), \\ \nn
\varepsilon_{\mu}^{\oplus}(P_1,P_2) &=& 
- \varepsilon_{L\mu}^{\oplus}(P_1), \\ 
\varepsilon_{\mu}^{\oplus}(P_2,P_1) &=& 
\varepsilon_{R\mu}^{\oplus}(P_2).
\eea
Taking into account these relations, we 
determine \eqns{impactfactorscalc}{LipatovOur4} 
in the helicity basis:
\bea \label{impactfactorscalcH} \nn
C_{g}^{(0)}(P_2^{\ominus},P_3^{\oplus}) 
= \Gamma^{\mu_{2} \mu_{3}}\,
\varepsilon_{\mu_{2}}^{\ominus *}(P_2,P_1) \,
\varepsilon_{\mu_{3}}^{\oplus}(P_3,P_1) 
&=& -1, \\
C_{g}^{(0)}(P_1^{\ominus},P_5^{\oplus}) 
= \Gamma^{\mu_{1} \mu_{5}}\,
\varepsilon_{\mu_{1}}^{\ominus *}(P_1,P_2) \,
\varepsilon_{\mu_{5}}^{\oplus}(P_5,P_2)
&=& -\frac{\mathbf{\bar p}_5}{\mathbf{p}_5},
\eea
in agreement with eq.(26) of \cite{DelDuca:1995zy}, 
and
\be \label{LipatovOur4H} 
V^{(0)}(Q_1,P_4^{\oplus},Q_2) 
= - \sqrt{2} \frac{\mathbf{\bar p}_3\mathbf{p}_5}{\mathbf{p}_4}\,,
\ee
in agreement with eq.(27) of \cite{DelDuca:1995zy}.


\section{Colour basis definition}\label{sec:colourBasis}

We label the element of the colour basis as 
$[c_1,c_2]$, where $c_i$ labels the representation 
in the $t_i$-channel in terms of its multiplicity 
number for the group SU(3). We label indices 
in the fundamental representation of SU(3)
as lower indices $i_x$, while indices 
in the adjoint representation are labelled as 
upper indices $a_x$ for particle $x$. Summed indices in 
the adjoint representation are labelled as $w$,
$x$, $y$, $z$. 


\subsection{\texorpdfstring{\boldmath $qq \to qgq$}{qq -> qgq}}

In case of $qq \to qgq$ the colour basis is 
labelled by the indices $(i_1,i_2,i_3,a_4,i_5)$. 
One has 
\begin{align}\label{basisqgq} \nn
c^{[1,8]} &=
\frac{\sqrt{2} }{\sqrt{N_c} \sqrt{N_c^2-1}}\,
\delta_{i_5i_1}\,(t^{a_4})_{i_3i_2}\,, \\ 
c^{[8,1]} &=
\frac{\sqrt{2}}{\sqrt{N_c} \sqrt{N_c^2-1}}\,
\delta_{i_3i_2}\,(t^{a_4})_{i_5i_1}\,, \\ \nn
c^{[8,8]_s} &=
\frac{2 \sqrt{N_c}}{\sqrt{N_c^2-4}\sqrt{N_c^2-1}}\,
(t^{x})_{i_5i_1}\,(t^{y})_{i_3i_2}\,d^{ya_4x}\,, \\ \nn
c^{[8,8]_a} &=
\frac{2}{\sqrt{N_c} \sqrt{N_c^2-1}}\, 
(t^x)_{i_5i_1}\,(t^y)_{i_3i_2}\,if^{ya_4x}\,.
\end{align}
We have for the $\T_{t_i}^2$ matrices
\begin{align}
\T_{t_1}^2 =
\left(
\begin{array}{cccc}
 0 & 0 & 0 & 0 \\
 0 & N_c & 0 & 0 \\
 0 & 0 & N_c & 0 \\
 0 & 0 & 0 & N_c \\
\end{array}
\right), \qquad 
\T_{t_2}^2 =
\left(
\begin{array}{cccc}
 N_c & 0 & 0 & 0 \\
 0 & 0 & 0 & 0 \\
 0 & 0 & N_c & 0 \\
 0 & 0 & 0 & N_c \\
\end{array}
\right).
\end{align}


\subsection{\texorpdfstring{\boldmath $qg \to qgg$}{qg -> qgg}}

It is useful to introduce projectors onto the 
$\mathbf{10}\pm\mathbf{\overline{10}}$, $\mathbf{27}$, 
and $\mathbf{0}$ representations, defined respectively 
as
\begin{subequations}\label{projectors}
\begin{align}
P_{\mathbf{10}+\mathbf{\overline{10}}}^{abcd} =&\,
\frac{1}{2} \left(\delta ^{a c} \delta ^{b d}
-\delta ^{a d} \delta ^{b c}\right)
-\frac{f^{a b e} f^{c d e}}{N_c}\,,
\\
P_{\mathbf{10}-\mathbf{\overline{10}}}^{abcd} =&\,
\frac{1}{2} i d^{a c e} f^{b e d}
-\frac{1}{2} i f^{a c e} d^{b e d}\,
\\
P_{\mathbf{27}}^{abcd} =&\,
\frac{N_c d^{a b e} d^{c d e}}{4 (N_c+2)}
+\frac{1}{2} f^{a d e} f^{c b e}
-\frac{1}{4} f^{a b e} f^{c d e}
+\frac{\delta ^{a b} \delta ^{c d}}{2 (N_c+1)}
+\frac{1}{4} \delta ^{a d} \delta ^{b c}
+\frac{1}{4} \delta ^{a c} \delta ^{b d}\,,
\\
P_{\mathbf{0}}^{abcd} =&\,
-\frac{N_c d^{a b e} d^{c d e}}{4 (N_c-2)}
-\frac{1}{2} f^{a d e} f^{c b e}
+\frac{1}{4} f^{a b e} f^{c d e}
-\frac{\delta ^{a b} \delta ^{c d}}{2 (N_c-1)}
+\frac{1}{4} \delta ^{a d} \delta ^{b c}
+\frac{1}{4} \delta ^{a c} \delta ^{b d}\,.
\end{align}
\end{subequations}
Note that $P_{\mathbf{0}}^{abce}P_{\mathbf{0}}^{abce}
=\frac14(N_c-3)N_c^2(N_c+1)$ which vanishes for $N_c=3$. 

In case of $qg \to qgg$ scattering the quark $q$ is associated
to the line $i \to i'$, while the gluon $g$ is associated 
to the line $j \to j'$. The colour basis elements are labelled 
by the indices $(i_1,a_2,a_3,a_4,i_5)$, and reads 
\begin{align} \label{basisqgg} \nn
c^{[1,8_s]} &= 
\frac{1}{\sqrt{(N_c^2-4)(N_c^2-1)}}\,
\delta_{i_5 i_1} \, d^{a_3 a_4 a_2}\,, \\ \nn
c^{[1,8_a]} &=
\frac{1}{N_c\sqrt{(N_c^2-1)}}\,  
\delta_{i_5 i_1}\,i f^{a_3 a_4 a_2}\,,  \\ \nn 
c^{[8,1]} &=
\frac{\sqrt{2}}{N_c^2-1} 
\,(t^{a_4})_{i_5 i_1} \,\delta^{a_3 a_2} \,, \\ \nn
c^{[8,8_s]_s} &=
\frac{\sqrt{2} N_c}{(N_c^2-4)
\sqrt{N_c^2-1}}\, (t^{x})_{i_5 i_1} \, 
d^{a_3 y a_2} \, d^{y a_4 x}  \\ \nn
c^{[8,8_s]_a} &=
\frac{\sqrt{2}}{\sqrt{\left(N_c^2-4\right) \left(N_c^2-1\right)}} \,
(t^{x})_{i_5 i_1} \, d^{a_3 y a_2} \, i f^{y a_4 x} \,, \\
c^{[8,8_a]_s} &=
\frac{\sqrt{2}}{\sqrt{\left(N_c^2-4\right) \left(N_c^2-1\right)}}\, 
(t^{x})_{i_5 i_1} \, i f^{a_3 y a_2} \, d^{y a_4 x}\,, \\ \nn
c^{[8,8_a]_a} &=
\frac{\sqrt{2}}{N_c \sqrt{\left(N_c^2-1\right)}}\,
(t^{x})_{i_5 i_1}, i f^{a_3 y a_2} \, i f^{y a_4 x}\, , \\ \nn
c^{[8,10+\overline{10}]} &=
\frac{2}{\sqrt{(N_c^2-4)(N_c^2-1)}}\, 
(t^{x})_{i_5 i_1} \, 
P_{\mathbf{10}+\mathbf{\overline{10}}}^{a_3a_2xa_4}\,,  \\ \nn
c^{[8,10-\overline{10}]} &= 
\frac{2}{\sqrt{(N_c^2-4)(N_c^2-1)}}\,
(t^{x})_{i_5 i_1} \, 
P_{\mathbf{10}-\mathbf{\overline{10}}}^{a_3a_2xa_4}\,, \\ \nn
c^{[8,27]} &=
\frac{2 \sqrt{2}}{N_c \sqrt{(N_c+3) (N_c-1)}} \, 
(t^{x})_{i_5 i_1} \, 
P_{\mathbf{27}}^{a_3a_2xa_4}\, , \\ \nn
c^{[8,0]} &=
\frac{2 \sqrt{2}}{N_c \sqrt{(N_c-3) (N_c+1)}}\,
(t^{x})_{i_5 i_1} \,
P_{\mathbf{0}}^{a_3a_2xa_4}\,.
\end{align}

In this basis, the quadratic operators measuring 
the colour flow in the $t$ channel reads 
\begin{align}
\T_{t_1}^2
=&\, \text{diag}
\left[0, 0, N_c, N_c, N_c, N_c, N_c, N_c, N_c, N_c, N_c\right]\, ,
\end{align}
and 
\begin{align}
\T_{t_2}^2
=&\, \text{diag}
\left[N_c, N_c, 0, N_c, N_c, N_c, N_c, 2 N_c, 2 N_c, 2 (N_c+1), 2 (N_c-1)\right]\, .
\end{align}


\subsection{\texorpdfstring{\boldmath $gq \to ggq$}{qg -> qgg}}

In case of $gq \to ggq$ scattering the gluon $g$ is associated
to the line $i \to i'$, while the quark $q$ is associated 
to the line $j \to j'$. The colour basis elements are labelled 
by the indices $(a_1,i_2,i_3,a_4,a_5)$, and reads 
\begin{align} \label{basisggq} \nn
c^{[8_s,1]} &= 
\frac{1}{\sqrt{(N_c^2-4)(N_c^2-1)}}\,
\delta_{i_3 i_2} \, d^{a_5 a_4 a_1}\,, \\ \nn
c^{[8_a,1]} &=
\frac{1}{N_c\sqrt{(N_c^2-1)}}\,  
\delta_{i_3 i_2}\,i f^{a_5 a_4 a_1}\,,  \\ \nn 
c^{[1,8]} &=
\frac{\sqrt{2}}{N_c^2-1} 
\,(t^{a_4})_{i_3 i_2} \,\delta^{a_5 a_1} \,, \\ \nn
c^{[8_s,8]_s} &=
\frac{\sqrt{2} N_c}{(N_c^2-4)
\sqrt{N_c^2-1}}\, (t^{x})_{i_3 i_2} \, 
d^{a_5 y a_1} \, d^{y a_4 x}  \\ \nn
c^{[8_s,8]_a} &=
\frac{\sqrt{2}}{\sqrt{\left(N_c^2-4\right) \left(N_c^2-1\right)}} \,
(t^{x})_{i_3 i_2} \, d^{a_5 y a_1} \, i f^{y a_4 x} \,, \\
c^{[8_a,8]_s} &=
\frac{\sqrt{2}}{\sqrt{\left(N_c^2-4\right) \left(N_c^2-1\right)}}\, 
(t^{x})_{i_3 i_2} \, i f^{a_5 y a_1} \, d^{y a_4 x}\,, \\ \nn
c^{[8_a,8]_a} &=
\frac{\sqrt{2}}{N_c \sqrt{\left(N_c^2-1\right)}}\,
(t^{x})_{i_3 i_2}, i f^{a_5 y a_1} \, i f^{y a_4 x}\, , \\ \nn
c^{[10+\overline{10},8]} &=
\frac{2}{\sqrt{(N_c^2-4)(N_c^2-1)}}\, 
(t^{x})_{i_3 i_2} \, 
P_{\mathbf{10}+\mathbf{\overline{10}}}^{a_5a_1xa_4}\,,  \\ \nn
c^{[10-\overline{10},8]} &= 
\frac{2}{\sqrt{(N_c^2-4)(N_c^2-1)}}\,
(t^{x})_{i_3 i_2} \, 
P_{\mathbf{10}-\mathbf{\overline{10}}}^{a_5a_1xa_4}\,, \\ \nn
c^{[27,8]} &=
\frac{2 \sqrt{2}}{N_c \sqrt{(N_c+3) (N_c-1)}} \, 
(t^{x})_{i_3 i_2} \, 
P_{\mathbf{27}}^{a_5a_1xa_4}\, , \\ \nn
c^{[0,8]} &=
\frac{2 \sqrt{2}}{N_c \sqrt{(N_c-3) (N_c+1)}}\,
(t^{x})_{i_3 i_2} \,
P_{\mathbf{0}}^{a_5a_1xa_4}\,.
\end{align}
With this basis choice, it is easy to check that 
the value of $\T_{t_1}^2$ and $\T_{t_2}^2$ are 
exchanged compared to the case $qg \to qgg$, i.e.
for $gq \to ggq$ one has 
\begin{align}
\T_{t_1}^2
=&\, \text{diag}
\left[N_c, N_c, 0, N_c, N_c, N_c, N_c, 2 N_c, 2 N_c, 2 (N_c+1), 2 (N_c-1)\right]\, ,
\end{align}
and 
\begin{align}
\T_{t_2}^2
=&\, \text{diag}
\left[0, 0, N_c, N_c, N_c, N_c, N_c, N_c, N_c, N_c, N_c\right]\, .
\end{align}


\subsection{\texorpdfstring{\boldmath $gg \to ggg$}{gg -> ggg}}
\label{app:ColorBasis5g}

Using the projectors defined in \eqn{projectors}
we get 
\begin{align}\label{basisggg}  \nn
c^{[1,8_a]} &= 
\frac{1}{(N_c^2-1)\sqrt{N_c}}\, 
\delta^{a_5 a_1} \, i f^{a_3 a_4 a_2}\, , 
\\ \nn
c^{[8_a,1]} &=
\frac{1}{(N_c^2-1)\sqrt{N_c}}\,
\delta^{a_3 a_2} \, i f^{a_1 a_5 a_4}\, , 
\\ \nn
c^{[8_s,8_s]} &=
\frac{\sqrt{N_c}}{(N_c^2-4)\sqrt{N_c^2-1}}\,
d^{a_1 a_5 x} \, i f^{y a_4 x}\, d^{a_2 a_3 y}\, , 
\\ \nn
c^{[8_s,8_a]} &=
\frac{\sqrt{N_c}}{(N_c^2-4)\sqrt{N_c^2-1}}\,
d^{a_1 a_5 x} \, d^{y a_4 x}\, i f^{a_2 a_3 y}\, , 
\\ \nn
c^{[8_a,8_s]} &=
\frac{\sqrt{N_c}}{(N_c^2-4)\sqrt{N_c^2-1}}\,
i f^{a_1 a_5 x} \, d^{y a_4 x}\, d^{a_2 a_3 y}\, , 
\\ \nn
c^{[8_a,8_a]} &=
\frac{1}{\sqrt{N_c^3(N_c^2-1)}}\,
i f^{a_1 a_5 x} \, i f^{y a_4 x}\,  i f^{a_2 a_3 y}\, , 
\\ \nn
c^{[8_s,10-\overline{10}]} &=
\frac{\sqrt{2N_c}}{(N_c^2-4)\sqrt{N_c^2-1}}\,
d^{a_1 a_5 x} \,
P_{\mathbf{10}-\mathbf{\overline{10}}}^{a_3 a_2 x a_4}\,,
\\ \nn
c^{[10-\overline{10},8_s]} &=
\frac{\sqrt{2N_c}}{(N_c^2-4)\sqrt{N_c^2-1}}\,
d^{a_2 a_3 x} \,
P_{\mathbf{10}-\mathbf{\overline{10}}}^{a_1 a_5 a_4 x}\,,
\\ \nn
c^{[8_a,10+\overline{10}]} &=
\frac{\sqrt{2}}{\sqrt{N_c(N_c^2-4)(N_c^2-1)}}\,
i f^{a_1 a_5 x} \,
P_{\mathbf{10}+\mathbf{\overline{10}}}^{a_3 a_2 x a_4}\,,
\\ \nn
c^{[10+\overline{10},8_a]} &=
\frac{\sqrt{2}}{\sqrt{N_c(N_c^2-4)(N_c^2-1)}}\,
i f^{a_2 a_3 x} \,
P_{\mathbf{10}+\mathbf{\overline{10}}}^{a_1 a_5 a_4 x}\,,
\\
c^{[8_a,27]} &=
\frac{2}{\sqrt{N_c^3(N_c+3)(N_c-1)}}\,
i f^{a_1 a_5 x} \,
P_{\mathbf{27}}^{a_3a_2xa_4}\,,
\\ \nn
c^{[27,8_a]} &=
\frac{2}{\sqrt{N_c^3(N_c+3)(N_c-1)}}\,
i f^{a_2 a_3 x} \,
P_{\mathbf{27}}^{a_1a_5a_4x}\,,
\\ \nn
c^{[8_a,0]} &=
\frac{2}{\sqrt{N_c^3(N_c-3)(N_c+1)}}\,
i f^{a_1 a_5 x} \,
P_{\mathbf{0}}^{a_3a_2xa_4}\,,
\\ \nn
c^{[0,8_a]} &=
\frac{2}{\sqrt{N_c^3(N_c-3)(N_c+1)}}\,
i f^{a_2 a_3 x} \,
P_{\mathbf{0}}^{a_1a_5a_4x}\,,
\\ \nn
c^{[10,\overline{10}]_1} &=
\frac{2}{\sqrt{N_c(N_c^2-4)(N_c^2-1)}}\,
P_{\mathbf{10}+\mathbf{\overline{10}}}^{a_1a_5 z x}\,
i f^{y a_4 x} \,
P_{\mathbf{10}+\mathbf{\overline{10}}}^{a_3a_2 y z}\,,
\\ \nn
c^{[10,\overline{10}]_2} &=
\frac{2\sqrt{N_c}}{\sqrt{(N_c^2-4)(N_c^2-1)(N_c+3)(N_c-3)}}\,
\bigg( 
P_{\mathbf{10}-\mathbf{\overline{10}}}^{a_1a_5 z x}\,
d^{y a_4 x} \,
P_{\mathbf{10}+\mathbf{\overline{10}}}^{a_3a_2 y z} \\ \nn
&\hspace{7.0cm}+\,\frac{1}{N_c} 
P_{\mathbf{10}+\mathbf{\overline{10}}}^{a_1a_5 z x}\,
i f^{y a_4 x} \,
P_{\mathbf{10}+\mathbf{\overline{10}}}^{a_3a_2 y z}
\bigg)\,,
\\ \nn
c^{[10+\overline{10},27]} &=
\frac{2\sqrt{2}}{\sqrt{N_c(N_c^2-1)(N_c-2)(N_c+3)}}\,
P_{\mathbf{10}+\mathbf{\overline{10}}}^{a_1a_5 z x}\,
i f^{y a_4 x} \,
P_{\mathbf{27}}^{a_3a_2 y z}\,,
\\ \nn
c^{[27,10+\overline{10}]} &=
\frac{2\sqrt{2}}{\sqrt{N_c(N_c^2-1)(N_c-2)(N_c+3)}}\,
P_{\mathbf{27}}^{a_1a_5 z x}\,
i f^{y a_4 x} \,
P_{\mathbf{10}+\mathbf{\overline{10}}}^{a_3a_2 y z}\,,
\\ \nn
c^{[10+\overline{10},0]} &=
\frac{2\sqrt{2}}{\sqrt{N_c(N_c^2-1)(N_c+2)(N_c-3)}}\,
P_{\mathbf{10}+\mathbf{\overline{10}}}^{a_1a_5 z x}\,
i f^{y a_4 x} \,
P_{\mathbf{0}}^{a_3a_2 y z}\,,
\\ \nn
c^{[0,10+\overline{10}]} &=
\frac{2\sqrt{2}}{\sqrt{N_c(N_c^2-1)(N_c+2)(N_c-3)}}\,
P_{\mathbf{0}}^{a_1a_5 z x}\,
i f^{y a_4 x} \,
P_{\mathbf{10}+\mathbf{\overline{10}}}^{a_3a_2 y z}\,,
\\ \nn
c^{[27,27]} &=
\frac{2\sqrt{2}}{N_c\sqrt{(N_c^2-1)(N_c+3)}}\,
P_{\mathbf{27}}^{a_1a_5 z x}\,
i f^{y a_4 x} \,
P_{\mathbf{27}}^{a_3a_2 y z}\,,
\\ \nn
c^{[0,0]} &=
\frac{2\sqrt{2}}{N_c\sqrt{(N_c^2-1)(N_c-3)}}\,
P_{\mathbf{0}}^{a_1a_5 z x}\,
i f^{y a_4 x} \,
P_{\mathbf{0}}^{a_3a_2 y z}\,.
\end{align}

In this basis, the quadratic operators 
measuring the colour flow in the $t$ 
channel reads 
\begin{align} \nn
\T_{t_1}^2
=&\, \text{diag}
\big[0, N_c, N_c, N_c, N_c, N_c, N_c, 
2N_c, N_c, 2N_c, N_c, 
2(N_c+1),N_c,2(N_c-1), \\
&\hspace{2.0cm}
2N_c,2N_c,2N_c,
2(N_c+1),2N_c,2(N_c-1),
2(N_c+1),2(N_c-1)\big]\, ,
\end{align}
and 
\begin{align} \nn
\T_{t_2}^2
=&\, \text{diag}
\big[N_c, 0, N_c, N_c, N_c, N_c, 
2N_c, N_c, 2 N_c, N_c, 
2 (N_c+1), N_c, 2 (N_c-1), N_c, \\  
&\hspace{2.0cm}
2N_c, 2N_c, 2 (N_c+1), 2N_c,
2 (N_c-1), 2N_c,
2(N_c+1),2(N_c-1) \big]\, .
\end{align}


\subsection{Color space formalism}
\label{Colorflow}

Given the color basis defined above, 
an amplitude is interpreted as a vector 
in color space. This is best expressed
by writing the amplitude in vector form
${\cal M}_{ij \to igj} = 
|{\cal M}_{ij \to igj} \rangle$, 
\cite{Bassetto:1983mvz,Catani:1996jh,Catani:1996vz}, 
then one has
\be\label{ColorAmplitudeVector}
{\cal M}_{ij \to igj} =
\sum_{i} |c^{[i]} \rangle \langle c^{[i]} |  
{\cal M}_{ij \to igj} \rangle 
= \sum_{i} c^{[i]} {\cal M}^{[i]}_{ij \to igj},
\ee
where in the second equality we have implicitly 
defined the vector of color basis element 
$|c^{[i]} \rangle \equiv c^{[i]}$, and the 
corresponding coefficients 
${\cal M}^{[i]}_{ij \to igj} \equiv 
\langle c^{[i]} | {\cal M}_{ij \to igj} \rangle$, 
as used in the main text, for instance in 
\eqn{coloramplitudes}. Within this formalism, 
the action of a color operator $\T_X$ on the 
amplitude reads 
\be\label{ColorOperatorMatrix}
\T_X \, {\cal M}_{ij \to igj} =
\sum_{ij} 
|c^{[j]} \rangle \underbrace{\langle c^{[j]}| \T_X
|c^{[i]} \rangle \langle c^{[i]}|  
{\cal M}_{ij \to igj} \rangle}_{\langle c^{[j]}| 
{\cal M}_{ij \to igj} \rangle} 
= \sum_{ji} c^{[j]} \underbrace{\T_X^{[j][i]}
{\cal M}^{[i]}_{ij \to igj}}_{{\cal M}^{[j]}_{ij \to igj}},
\ee
where $\T_X^{[j][i]} \equiv 
\langle c^{[j]}| \T_X |c^{[i]} \rangle$ 
represents the matrix in color space 
associated with the color operator $\T_X$. 
In the main text we have used the notation 
in the second equality in \eqn{ColorOperatorMatrix}, 
for instance in \eqn{colorOpMatrix}. 

The tree level amplitudes for the 
processes $qq \to qgq$, $qg \to qgg$, 
$gg \to ggg$ have a single non-zero 
element, corresponding to the propagation 
of a gluon from the projectile $i$ to the 
target $j$, with emission of a gluon $g$, 
as indicated in \eqn{TreeAmpColor}:
\bea \label{TreeAmpColorB} \nn
{\cal M}_{qq \to qgq}^{\rm tree} &=& c^{[8,8]_a}\, 
\big({\cal M}_{qq \to qgq}^{\rm tree}\big)^{[8,8]_a}, \\ 
{\cal M}_{qg \to qgg}^{\rm tree} &=& c^{[8,8_a]_a}\, 
\big({\cal M}_{qg \to qgg}^{\rm tree}\big)^{[8,8_a]_a}, \\ \nn 
{\cal M}_{gg \to ggg}^{\rm tree} &=& c^{[8_a,8_a]}\, 
\big({\cal M}_{gg \to ggg}^{\rm tree}\big)^{[8_a,8_a]}. 
\eea
Comparison between eqs.~(\ref{Mtree2}),~(\ref{def:Cij})
and the color bases in eqs.~(\ref{basisqgq}),~(\ref{basisqgg}),
(\ref{basisggq}) and~(\ref{basisggg}) immediately gives 
\bea \label{TreeAmpColorExplicit} \nn
\big({\cal M}_{qq \to qgq}^{\rm tree}\big)^{[8,8]_a} 
&=& \frac{\sqrt{N_c} \sqrt{N_c^2-1}}{2}\, 
C_{i}^{(0)} \, \mathcal{K}^{(0)}\, C_{j}^{(0)} , \\ 
\big({\cal M}_{qg \to qgg}^{\rm tree}\big)^{[8,8_a]_a}
&=& \frac{N_c \sqrt{\left(N_c^2-1\right)}}{\sqrt{2}}\,
C_{i}^{(0)} \, \mathcal{K}^{(0)}\, C_{j}^{(0)} , \\ \nn 
\big({\cal M}_{gg \to ggg}^{\rm tree}\big)^{[8_a,8_a]}
&=& \sqrt{N_c^3(N_c^2-1)} \, 
C_{i}^{(0)} \, \mathcal{K}^{(0)}\, C_{j}^{(0)} . 
\eea

 
\section{Variables conversion in the MRK power expansion}\label{app:vars_conversion}

We provide here details about the conversion between the minimal set of variables \linebreak $(w, \bar{w}, X_{34}, X_{45}, |\mathbf{q}_2|^2)$ used in section \ref{sec:padicreconstruction} and the one used elsewhere, $(z, \bar z, s_{1}, s_{2}, r)$. Since $X_{34}$ and $X_{45}$ are dimensionless, while $s_{1}$, $s_{2}$, and $r$ carry mass dimension squared, we need one more variable to normalize by in the conversion, which we pick to be $|\mathbf{q}_2|^2$ (see eq.~\eqref{q1q2defB}). Nevertheless, since we consider dimensionless functions, $r_i / \mathcal{M}^{(0)}$, this additional variable $|\mathbf{q}_2|^2$ never appears in the reconstruction. The relations read,
\begin{align}
X_{34} &= \frac{1}{x} \frac{s_2 w \bar w}{|\mathbf{q}_2|^2} - w - \bar w -x \frac{|\mathbf{q}_2|^2}{s_2} \\
X_{45} &= 
\frac{1}{x}\frac{s_1 w \bar w}{|\mathbf{q}_2|^2 (1-w) (1-\bar w)}+\frac{-2+w+\bar w}{(1-w) (1-\bar w)}-x \frac{|\mathbf{q}_2|^2}{s_1 w \bar w}\\
w &= z + x \, \frac{r z(1-z)}{s_1 s_2 (z-\bar z)} [s_1 (1-z) + s_2 z] - x^2 \, \frac{r^2 z (1-z)}{s_1^2 s_2^2 (z-\bar z)^3}  \times \nonumber \\ 
&\qquad \Big[
 s_1^2 (1-z)^2 (z^2+\bar z-z \bar z-\bar z^2)+s_2^2 z^2 (z^2+\bar z (2-\bar z) -z (1+\bar z))
\,\, + \nonumber \\
&\qquad 
+s_1 s_2 z (1-z) (-z(1-z)+3 \bar z(1-\bar z) ) 
\Big] \label{w-z-relation}
\end{align}
where $r=s_1 s_2 / s$. Similarly, for $\bar w$, through a $z\leftrightarrow \bar z$ swap of eq.~(\ref{w-z-relation}).


\section{Gluon Regge trajectory, impact factors and vertex}\label{sec:ReggeCoefficients}

In this appendix we provide the impact factors 
at one loop and the Regge trajectory up to two 
loops. Furthermore, we discuss their factorization 
and renormalization scale dependence, for ease of 
comparison with appendix A of \cite{Falcioni:2021buo}, 
where impact factors and Regge trajectory are given
for $\mu^2=\tau=-t$. 

Les us start by recalling the perturbative 
expansions introduced in eqs. (\ref{Regge_traj}), 
(\ref{ImpactPertDef1}), (\ref{ImpactPertDef2}),
(\ref{LipatovPertDef1B}),
indicating the full scale dependence: we have
\be \label{eq:eq:ReggeTrajAllScales}    
\alpha_g(-t,\mu^2) = \frac{g(\mu^2)^2}{4\pi^2}
\alpha_g^{(1)}(-t,\mu^2) +
\left(\frac{g(\mu^2)^2}{4\pi^2}\right)^2
\alpha_g^{(2)}(-t,\mu^2)+ \dots\,,
\ee
for the Regge trajectory, then 
\be \label{eq:ImpFactAllScales}
C_i(-t,\tau,\mu^2) = g(\mu^2)\, C_i^{(0)}(-t,\tau,\mu^2)
\left[1 + \frac{g(\mu^2)^2}{4\pi^2} c_i^{(1)}(-t,\tau,\mu^2) 
+ \dots\right]\,,
\ee
for the impact factors, and 
\bea \label{eq:LipVertAllScales} \nn
V(t_{1},t_{2},|\mathbf{p}_4|^2,\tau,\mu^2)
&=& g(\mu^2)\, V^{(0)}(t_{1},t_{2},|\mathbf{p}_4|^2,\tau,\mu^2) \\
&&\times\, \left[1 + \frac{g(\mu^2)^2}{4\pi^2} 
v^{(1)}(t_{1},t_{2},|\mathbf{p}_4|^2,\tau,\mu^2) + \dots\right]\,,
\eea
for the Lipatov vertex. The perturbative 
coefficients reads respectively
\begin{subequations}
\label{eq:alg}
\begin{align}
\alpha_g^{(1)}(-t,\mu^2) &= 
\bigg(\frac{\mu^2}{-t}\bigg)^{\eps} \frac{r_{\Gamma}}{2\eps} =
\left(\frac{\mu^2}{-t}\right)^{\eps}
\frac{e^{\eps\gamma_E}\Gamma^2(1-\eps)
\Gamma(1+\eps)}{2\eps\,\Gamma(1-2\eps)},  \\
\alpha_g^{(2)}(-t,\mu^2) &= \left(\frac{\mu^2}{-t}\right)^{2\eps}
\bigg\{-\frac{b_0}{16\eps^2} +\frac{1}{8\eps}
\left[\left(\frac{67}{18}-\zeta_2\right)\CA 
-\frac{10T_R\nf}{9}\right] \nn \\
&\hspace{2.5cm}+\,\CA\left(\frac{101}{108}-\frac{\zeta_3}{8}\right) 
- \frac{7\, T_R\nf}{27} + {\cal O}(\eps)\bigg\} \,,
\end{align}
\end{subequations}
for the Regge trajectory, and~\cite{DelDuca:1998kx,DelDuca:1998cx} 
\begin{align} \label{eq:C1Vittorio}
c_q^{(1)} &= \frac{r_\Gamma}{4}
\left(\frac{\mu^2}{-t}\right)^\epsilon
\frac{1}{\eps(1-2\eps)}
\bigg\{C_A\bigg[(1-2\eps) \left(
\log\frac{\tau}{-t} + \psi(1+\epsilon)
-2\psi(-\epsilon)+\psi(1)\right) \nonumber \\
&\hspace{2.0cm}+\,\frac{1}{4(3-2\epsilon)} + \frac{1}{\eps}
-\frac{7}{4} \bigg] +\frac{1}{N_c} 
\bigg(\frac{1}{\eps} - \frac{1-2\eps}{2} \bigg)
-n_f\frac{1-\epsilon}{3-2\epsilon}\bigg\}
-\frac{b_0}{8\epsilon}\,, \\
c_g^{(1)} &=\frac{r_\Gamma}{4}
\left(\frac{\mu^2}{-t}\right)^\epsilon
\bigg\{C_A\bigg[-\frac{2}{\epsilon^2}
+\frac{1}{\epsilon}\bigg(\log\frac{\tau}{-t}
+\psi(1+\epsilon)-2\psi(-\epsilon)
+\psi(1)\bigg) \nonumber\\
&\hspace{2.0cm}+\,\frac{1}{\epsilon(1-2\epsilon)}
\left(\frac{1-\epsilon}{2(3-2\epsilon)}-2\right)\bigg]
+n_f\frac{1-\epsilon}{\epsilon(1-2\epsilon)(3-2\epsilon)}
\bigg\} -\frac{b_0}{8\epsilon}\,,   
\end{align}
respectively for the quark and gluon impact factor. 
Note that in these equations, the parameter $\delta_R$
related to the regularization scheme has been set to 1, 
corresponding to the HV or CDR scheme, 
see~\cite{DelDuca:1998kx,DelDuca:1998cx} for further 
details. In these equations $b_0$ represents the one 
loop coefficient of the beta function: defining 
\begin{equation} \label{BetaQCD} 
\mu^2\frac{d}{d\mu^2}a_s = -\epsilon\, a_s 
- b_0 \, a_s^2 - b_1 \, a_s^3 + \dots \,
\end{equation}
with $a_s = g_s^2/(4\pi)^2$, we have 
\begin{subequations}
\label{eq:betacoeffs}
\begin{align}
b_0 &= \frac{11C_A-2n_f}{3},  \\
b_1 &= \frac{34}{3}C_A^2 - \frac{10}{3}C_A n_f -2 C_F n_f \,.
\end{align}
\end{subequations}

In order to relate the coefficients 
in \eqns{eq:alg}{eq:C1Vittorio} to the one 
in appendix A of~\cite{Falcioni:2021buo},
given at $\mu^2=\tau=-t$, we consider 
the factorised expression for the 
$2\to2$ amplitude
\begin{align}
\text{Disp.}\Big\{\mathcal{M}_{ij\to ij} \Big\} 
&= 2\frac{s}{t}\,C_i(-t,\tau,\mu^2)\,
\left(\frac{s}{\tau}\right)^{C_A \alpha_g(-t,\mu^2)}
\,C_j(-t,\tau,\mu^2).
\end{align}
Expanding to one-loop order one finds
\begin{align} \nn
\label{eq:HEFact1LoopAllScales}
\text{Disp.}\Big\{\mathcal{M}_{ij\to ij}\Big\} 
&= 2\frac{s}{t}\, g(\mu^2)^2\,
C_i^{(0)}(-t,\tau,\mu^2)\, 
C_j^{(0)}(-t,\tau,\mu^2) \Bigg[1 + \\
&\hspace{-2.0cm}+\,\frac{g(\mu^2)^2}{4\pi^2}
\left(c_i^{(1)}(-t,\tau,\mu^2)+c_j^{(1)}(-t,\tau,\mu^2)
+C_A \alpha_g^{(1)}(-t,\mu^2)\log\frac{s}{\tau}\right)
+\dots\Bigg],
\end{align}
Since the amplitude itself is renormalisation group 
invariant, equating the perturbative expansion above
for generic $\mu^2$ and for $\mu^2 = -t$:
\begin{align}
\label{eq:compareRenScale}
g(-t)C_i^{(0)}(-t,\tau,-t) &+ \frac{g(-t)^3}{4\pi^2} 
C_i^{(0)}(-t,\tau,-t)\, c_i^{(1)}(-t,\tau,-t) + \dots 
\nonumber \\
&\hspace{-2.0cm} = \, 
g(\mu^2)C_i^{(0)}(-t,\tau,\mu^2)+\frac{g(\mu^2)^3}{4\pi^2}  
C_i^{(0)}(-t,\tau,\mu^2)\,c_i^{(1)}(-t,\tau,\mu^2)
+ \dots
\end{align}
we are able to relate $C_i(-t,\tau,-t)$ and $C_i(-t,\tau,\mu^2)$.  
To this end we need to solve \eqn{BetaQCD} perturbatively:
up to ${\cal{O}}\left[\alpha(\mu_0^2)^3\right]$ we have
\begin{align}
\label{eq:couplingEvo}
a(\mu^2)&=\frac{a(\mu_0^2)
\left(\frac{\mu_0^2}{\mu^2}\right)^\epsilon}
{1+\frac{a(\mu_0^2)}{\epsilon}\left[1-
\left(\frac{\mu_0^2}{\mu^2}\right)^\epsilon\right]
\left[b_0 + \frac{a(\mu_0^2)}{2}  \, b_1 \left(1 + 
\left(\frac{\mu_0^2}{\mu^2}\right)^\epsilon\right)\right]} 
+ {\cal{O}}\left[a(\mu_0^2)^4\right].
\end{align}
We use the equation above to write the coupling 
$g(-t)=4\pi\sqrt{a(-t)}$, on the LHS of 
\eqn{eq:compareRenScale}, in terms of 
$g(\mu^2)$, obtaining
\begin{subequations}
\label{eq:ImpactFactorsScale1}
\begin{align}
C_g^{(0)}(-t,\tau,\mu^2)&=
\left(\frac{\mu^2}{-t}\right)^{\frac{\epsilon}{2}} 
C_g^{(0)}(-t,\tau,-t),\\
c_g^{(1)}(-t,\tau,\mu^2)&=
\left(\frac{\mu^2}{-t}\right)^\epsilon 
c_g^{(1)}(-t,\tau,-t) - \frac{b_0}{8\epsilon}
\left[1-\left(\frac{\mu^2}{-t}\right)^\epsilon\right],\\
c_g^{(2)}(-t,\tau,\mu^2)&=
\left(\frac{\mu^2}{-t}\right)^{2\epsilon} 
c_g^{(2)}(-t,\tau,-t) + \frac{3b_0^2}{128\epsilon^2}
\left[1-\left(\frac{\mu^2}{-t}\right)^\epsilon\right]^2 \nonumber \\
&\hspace{-1.0cm}-\,\frac{1}{64\epsilon}
\left[1-\left(\frac{\mu^2}{-t}\right)^\epsilon\right]
\bigg\{b_1\left[1+\left(\frac{\mu^2}{-t}\right)^\epsilon\right]
+24b_0\left(\frac{\mu^2}{-t}\right)^\epsilon 
c_g^{(1)}(-t,\tau,-t)\bigg\}.
\end{align}
\end{subequations}

At this point we still need to 
relate the impact factors $C_i(-t,\tau,-t)$,
where the factorization scale is generic, to
the impact factors evaluated at the scale 
$\tau=-t$. Once again, this can be easily 
done by exploiting the scale invariance
of the amplitude. We obtain
\begin{equation}
C_g(\-t,\tau,-t) = C_g(-t,-t,-t)\,
\left(\frac{-t}{\tau}\right)^{-C_A\frac{\alpha_g(-t,-t)}{2}},
\end{equation}
which expanding to second 
order in $\alpha_s$ gives
\begin{subequations}
\label{eq:ImpactFactorsScale2}
\begin{align}
C_g^{(0)}(-t,\tau,-t) &= C_g^{(0)}(-t,-t,-t), \\
c_g^{(1)}(-t,\tau,-t) &= c_g^{(1)}
-\frac{C_A\alpha_g^{(1)}}{2}\,\log\left(\frac{-t}{\tau}\right),\\
c_g^{(2)}(-t,\tau,-t) &= c_g^{(2)}
+\frac{1}{2}\left(\frac{C_A\alpha_g^{(1)}}{2}\log\frac{-t}{\tau}\right)^2
-\frac{C_A}{2}\log\frac{-t}{\tau}\Big[ c_g^{(1)}\,\alpha_g^{(1)} 
+ \alpha_g^{(2)}\Big].
\end{align}
\end{subequations}
where $c_g^{(1)}\equiv c_g^{(1)}(-t,-t,-t)$, 
$\alpha_g^{(1)}\equiv\alpha_g^{(1)}(-t,-t)$, 
and similarly $c_g^{(2)}\equiv c_g^{(2)}(-t,-t,-t)$, 
$\alpha_g^{(2)}\equiv\alpha_g^{(2)}(-t,-t)$, 
are given in the ancillary file of 
\cite{Falcioni:2021buo}. Combining 
\eqn{eq:ImpactFactorsScale1} with 
\eqn{eq:ImpactFactorsScale2} we obtain
\begin{subequations}
\label{eq:ImpactFactorsScale3}
\begin{align}
C_g^{(0)}(-t,\tau,\mu^2) &=
\left(\frac{\mu^2}{-t}\right)^{\frac{\epsilon}{2}} 
C_g^{(0)}(-t,-t,-t), \\
c_g^{(1)}(-t,\tau,\mu^2) &= 
\left(\frac{\mu^2}{-t}\right)^\epsilon 
\left[c_g^{(1)}-\frac{C_A\,\alpha_g^{(1)}}{2}
\log\left(\frac{-t}{\tau}\right)\right] 
- \frac{b_0}{8\epsilon}
\left[1-\left(\frac{\mu^2}{-t}\right)^\epsilon\right], \\
c_g^{(2)}(-t,\tau,\mu^2) &= 
\left(\frac{\mu^2}{-t}\right)^{2\epsilon}
\bigg\{c_g^{(2)}+\frac{C_A^2}{8}
\big(\alpha_g^{(1)}\big)^2
\log^2\left(\frac{-t}{\tau}\right) \nonumber \\
&\hspace{2.0cm}-\, \frac{C_A}{2}
\log\left(\frac{-t}{\tau}\right)
\Big( c_g^{(1)}\,\alpha_g^{(1)} 
+ \alpha_g^{(2)}\Big)\bigg\} \nonumber \\
&\hspace{0.0cm}-\, \frac{1}{64\epsilon}
\left[1-\left(\frac{\mu^2}{-t}\right)^\epsilon\right]
\bigg\{b_1\left[1+\left(\frac{\mu^2}{-t}\right)^\epsilon
\right] \nonumber \\
&\hspace{2.0cm}+\,24b_0\left(\frac{\mu^2}{-t}\right)^\epsilon 
\bigg[c_g^{(1)}-\frac{C_A\,\alpha_g^{(1)}}{2}\,
\log\left(\frac{-t}{\tau}\right)\bigg] \bigg\} \nonumber\\
&\hspace{0.0cm}+\, \frac{3}{128\epsilon^2}
b_0^2\left[1-\left(\frac{\mu^2}{-t}\right)^\epsilon\right]^2.
\end{align}
\end{subequations}

Repeating the same procedure for the 
Regge trajectory we find 
\begin{subequations}
\label{eq:ReggeTrajectoryScale}
\begin{align}
\alpha_g^{(1)}(-t,\mu^2) &= 
\alpha_g^{(1)}(-t,-t)\left(\frac{\mu^2}{-t}\right)^\epsilon, \\
\alpha_g^{(2)}(-t,\mu^2) &= 
\alpha_g^{(2)}(-t,-t)\left(\frac{\mu^2}{-t}\right)^{2\epsilon} 
- \alpha_g^{(1)}(-t,-t)\left(\frac{\mu^2}{-t}\right)^\epsilon\,
\frac{b_0}{4\epsilon} \left[1-\left(\frac{\mu^2}{-t}\right)^\epsilon\right].
\end{align}
\end{subequations}

Once the scale dependence for the impact 
factors and Regge trajectory has been obtained, 
it can be used in the factorization formula for 
the $2\to 3$ scattering amplitudes (\eqn{REggNLL}
in the main text) to derive the scale dependence 
of the Lipatov vertex. After some work we find 
that the Lipatov vertex evaluated at a generic 
renormalization scale $\mu^2$ is related to the 
vertex at the scale $\mu^2 = |\mathbf{p}_4|^2$ 
as follows:
\begin{subequations}
\label{eq:LipatovVertexScale1}
\begin{align}
V_0(t_1,t_2,|\mathbf{p}_4|^2,\tau,\mu^2) 
&= V_0(t_1,t_2,|\mathbf{p}_4|^2,\tau,|\mathbf{p}_4|^2)
\left(\frac{\mu^2}{|\mathbf{p}_4|^2}\right)^\frac{\epsilon}{2}, \\
v^{(1)}(t_1,t_2,|\mathbf{p}_4|^2,\tau,\mu^2) 
&= v^{(1)}(t_1,t_2,|\mathbf{p}_4|^2,\tau,|\mathbf{p}_4|^2)
\left(\frac{\mu^2}{|\mathbf{p}_4|^2}\right)^\epsilon
-\frac{b_0}{8\epsilon}\left[1
-\left(\frac{\mu^2}{|\mathbf{p}_4|^2}\right)^\epsilon\right], \\
v^{(2)}(t_1,t_2,|\mathbf{p}_4|^2,\tau,\mu^2) 
&= v^{(2)}(t_1,t_2,|\mathbf{p}_4|^2,\tau,|\mathbf{p}_4|^2)
\left(\frac{\mu^2}{|\mathbf{p}_4|^2}\right)^{2\epsilon}
+\frac{3\,b_0^2}{128\epsilon^2}\left[\left(\frac{\mu^2}{|\mathbf{p}_4|^2}\right)^\epsilon-1\right]^2 \nonumber \\
&\hspace{-3.5cm}-\,\frac{1}{64\epsilon}
\left[1-\left(\frac{\mu^2}{|\mathbf{p}_4|^2}\right)^\epsilon\right]
\left\{b_1\left[1+\left(\frac{\mu^2}{|\mathbf{p}_4|^2}
\right)^\epsilon\right]
+24b_0\left(\frac{\mu^2}{|\mathbf{p}_4|^2}\right)^\epsilon 
v^{(1)}(t_1,t_2,|\mathbf{p}_4|^2,\tau,|\mathbf{p}_4|^2)\right\}.
\end{align}
\end{subequations}
Furthermore, the relation between
the Lipatov vertex evaluated at a 
generic factorization scale $\tau$
and the vertex at $\tau = |\mathbf{p}_4|^2$
is given by
\begin{align}\label{vSamescale}
V(t_1,t_2,|\mathbf{p}_4|^2,\tau,\mu^2) 
= V(t_1,t_2,|\mathbf{p}_4|^2,|\mathbf{p}_4|^2,\mu^2)
\left(\frac{|\mathbf{p}_4|^2}{\tau}
\right)^{-\frac{1}{2}\big[\alpha_g(t_1,\mu^2)
+\alpha_g(t_2,\mu^2)\big]},
\end{align}
and expanding to second order 
in $\alpha_s$ we obtain
\begin{subequations}
\label{eq:LipatovVertexScale2}
\begin{align}
V_0(t_1,t_2,|\mathbf{p}_4|^2,\tau,|\mathbf{p}_4|^2) 
&= V_0(t_1,t_2,|\mathbf{p}_4|^2,|\mathbf{p}_4|^2,|\mathbf{p}_4|^2), \\
v^{(1)}(t_1,t_2,|\mathbf{p}_4|^2,\tau,|\mathbf{p}_4|^2) &= 
v^{(1)}(t_1,t_2,|\mathbf{p}_4|^2) \nonumber \\
&\hspace{-2.0cm}-\,\frac{\alpha_g^{(1)}C_A}{2}
\log\frac{|\mathbf{p}_4|^2}{\tau}
\left[\left(\frac{|\mathbf{p}_4|^2}{-t_1}\right)^\epsilon
+\left(\frac{|\mathbf{p}_4|^2}{-t_2}\right)^\epsilon \right], \\
v^{(2)}(t_1,t_2,|\mathbf{p}_4|^2,\tau,|\mathbf{p}_4|^2) 
&= v^{(2)}(t_1,t_2,|\mathbf{p}_4|^2) \nonumber \\
&\hspace{-2.0cm}-\, v^{(1)}(t_1,t_2,|\mathbf{p}_4|^2)
\frac{\alpha_g^{(1)}C_A}{2}\log\frac{|\mathbf{p}_4|^2}{\tau} 
\left[\left(\frac{|\mathbf{p}_4|^2}{-t_1}\right)^\epsilon
+\left(\frac{|\mathbf{p}_4|^2}{-t_2}\right)^\epsilon\right] 
\nonumber \\ 
&\hspace{-2.0cm}-\,
\frac{\alpha_g^{(1)}C_Ab_0}{8\epsilon}
\log\frac{|\mathbf{p}_4|^2}{\tau}
\left[\left(\frac{|\mathbf{p}_4|^2}{-t_1}\right)^{2\epsilon}
+\left(\frac{|\mathbf{p}_4|^2}{-t_2}\right)^{2\epsilon}
-\left(\frac{|\mathbf{p}_4|^2}{-t_1}\right)^\epsilon
-\left(\frac{|\mathbf{p}_4|^2}{-t_2}\right)^\epsilon\right]
\nonumber \\
&\hspace{-2.0cm}+\, \frac{C_A}{8}
\log\frac{|\mathbf{p}_4|^2}{\tau}
\Bigg\{(\alpha_g^{(1)})^2C_A\log\frac{|\mathbf{p}_4|^2}{\tau}
\bigg[\left(\frac{|\mathbf{p}_4|^2}{-t_1}\right)^\epsilon
+\left(\frac{|\mathbf{p}_4|^2}{-t_2}\right)^\epsilon\bigg]^2
\nonumber \\
&\hspace{0.0cm}-\,4\alpha_g^{(2)}
\bigg[\left(\frac{|\mathbf{p}_4|^2}{-t_1}\right)^{2\epsilon}
+\left(\frac{|\mathbf{p}_4|^2}{-t_2}\right)^{2\epsilon}\bigg]\Bigg\}
\end{align}
\end{subequations}
Combining \eqn{eq:LipatovVertexScale1}
with \eqn{eq:LipatovVertexScale2} we have
\begin{subequations}
\label{eq:LipatovVertexScale3}
\begin{align}
V_0(t_1,t_2,|\mathbf{p}_4|^2,\tau,\mu^2) 
&= V_0(t_1,t_2,|\mathbf{p}_4|^2,|\mathbf{p}_4|^2,|\mathbf{p}_4|^2)
\left(\frac{\mu^2}{|\mathbf{p}_4|^2}\right)^\frac{\epsilon}{2} , \\
v^{(1)}(t_1,t_2,|\mathbf{p}_4|^2,\tau,\mu^2) 
&= v^{(1)}(t_1,t_2,|\mathbf{p}_4|^2)\left(\frac{\mu^2}{|\mathbf{p}_4|^2}\right)^\epsilon
+\frac{b_0}{8\epsilon}
\left[\left(\frac{\mu^2}{|\mathbf{p}_4|^2}\right)^\epsilon-1\right]
\nonumber \\ 
&\hspace{1.0cm}-\,\frac{\alpha_g^{(1)}C_A}{2}
\log\frac{|\mathbf{p}_4|^2}{\tau}
\left[\left(\frac{|\mathbf{p}_4|^2}{-t_1}\right)^\epsilon
+\left(\frac{|\mathbf{p}_4|^2}{-t_2}\right)^\epsilon \right], \\
v^{(2)}(t_1,t_2,|\mathbf{p}_4|^2,\tau,\mu^2) 
&= v^{(2)}(t_1,t_2,|\mathbf{p}_4|^2)
\left(\frac{\mu^2}{|\mathbf{p}_4|^2}\right)^{2\epsilon}
\nonumber \\
&\hspace{-2.0cm}+\,v^{(1)}(t_1,t_2,|\mathbf{p}_4|^2)
\Bigg\{\frac{3b_0}{8\epsilon}
\left(\frac{\mu^2}{|\mathbf{p}_4|^2}\right)^\epsilon
\left[\left(\frac{\mu^2}{|\mathbf{p}_4|^2}\right)^\epsilon-1\right]
\nonumber \\
&\hspace{0.0cm}-\,\frac{\alpha_g^{(1)}C_A}{2}
\log\frac{|\mathbf{p}_4|^2}{\tau}
\left(\frac{\mu^2}{|\mathbf{p}_4|^2}\right)^{2\epsilon}
\bigg[\left(\frac{|\mathbf{p}_4|^2}{-t_1}\right)^\epsilon
+\left(\frac{|\mathbf{p}_4|^2}{-t_2}\right)^\epsilon\bigg]
\Bigg\} \nonumber \\
&\hspace{-2.0cm}+\,
\frac{3b_0^2}{128\epsilon^2}\Bigg[\left(\frac{\mu^2}{|\mathbf{p}_4|^2}\right)^\epsilon-1\Bigg]^2
+\frac{b_1}{64\epsilon}\Bigg[\left(\frac{\mu^2}{|\mathbf{p}_4|^2}\right)^{2\epsilon}-1\Bigg] 
\nonumber \\
&\hspace{-2.0cm}-\,\frac{\alpha_g^{(1)}C_Ab_0}{16\epsilon}
\left(\frac{\mu^2}{|\mathbf{p}_4|^2}\right)^\epsilon
\log\frac{|\mathbf{p}_4|^2}{\tau}
\Bigg\{\left(\frac{\mu^2}{|\mathbf{p}_4|^2}\right)^\epsilon\bigg[\left(\frac{|\mathbf{p}_4|^2}{-t_1}\right)^\epsilon
+\left(\frac{|\mathbf{p}_4|^2}{-t_2}\right)^\epsilon 
\nonumber \\
&\hspace{0.0cm}+\,2
\left(\frac{|\mathbf{p}_4|^2}{-t_1}\right)^{2\epsilon}
+\,2\left(\frac{|\mathbf{p}_4|^2}{-t_2}\right)^{2\epsilon}\bigg]
-3\bigg[\left(\frac{|\mathbf{p}_4|^2}{-t_1}\right)^{\epsilon} 
+\left(\frac{|\mathbf{p}_4|^2}{-t_2}\right)^{\epsilon}\bigg]
\Bigg\} \nonumber \\
&\hspace{-2.0cm}+\,\frac{C_A}{8}\left(\frac{\mu^2}{|\mathbf{p}_4|^2}\right)^{2\epsilon}
\log\frac{|\mathbf{p}_4|^2}{\tau}
\Bigg\{C_A \, \big(\alpha_g^{(1)}\big)^2 
\log\frac{|\mathbf{p}_4|^2}{\tau}
\bigg[\left(\frac{|\mathbf{p}_4|^2}{-t_1}\right)^{\epsilon}
+\left(\frac{|\mathbf{p}_4|^2}{-t_2}\right)^{\epsilon} \bigg]^2 
\nonumber \\
&\hspace{0.0cm}-4\alpha_g^{(2)}\bigg[
\left(\frac{|\mathbf{p}_4|^2}{-t_1}\right)^{2\epsilon}
+\left(\frac{|\mathbf{p}_4|^2}{-t_2}\right)^{2\epsilon}
\bigg]\Bigg\}.
\end{align}
\end{subequations}

 
\section{The dipole formula in multi-Regge kinematics}
\label{sec:DipoleFormula}

According to the infrared factorization theorem
an $n$-particle scattering amplitude has the 
following multiplicative structure~\cite{Catani:1998bh,Sterman:2002qn,Aybat:2006mz,Aybat:2006wq,Gardi:2009qi,Gardi:2009zv,Dixon:2009ur,Becher:2009cu,Becher:2019avh,Almelid:2015jia,Almelid:2017qju,Magnea:2021fvy,Falcioni:2021buo} 
\be \label{IRfacteq}
{\cal M}_n \left(\{p_i\},\mu, \as (\mu^2) \right) 
= {\bf Z}_n \left(\{p_i\},\mu, \as (\mu^2) \right)
{\cal H}_n \left(\{p_i\},\mu, \as (\mu^2) \right),
\ee
where on the left-hand side one has the 
infrared divergent $n$-particle amplitude
in dimensional regularization ${\cal M}_n$, 
while on the right-hand side ${\cal H}_n$ is 
the finite reminder, and infrared divergences
are factorized in the renormalization factor 
${\bf Z}_n$, which has the exponential structure
\be \label{RGsol}
{\bf Z}_n \left(\{p_i\},\mu, \as (\mu^2) \right) =   
{\cal P} \exp \left\{ -\frac{1}{2}\int_0^{\mu^2} 
\frac{d \lambda^2}{\lambda^2}\, {\bf \Gamma}_n 
\left(\{p_i\},\lambda, \as(\lambda^2) \right) \right\}\,,
\ee
and in turn ${\bf \Gamma}_n$ represents the 
soft anomalous dimension for scattering of 
massless partons ($p_i^2=0$). Through three 
loops, it is given by
\be \label{gammaSoft1}
{\bf \Gamma}_{n}\left(\{p_i\},\lambda, \as(\lambda^2) \right) =
{\bf \Gamma}_{n}^{\rm dip.}\left(\{p_i\},\lambda, \as(\lambda^2) \right)
+{\bf \Delta}_{n}\left(\{\rho_{ijkl}\}\right), 
\ee
where 
\be \label{gammadip1}
{\bf \Gamma}_{n}^{\rm dip.}\left(\{p_i\},\lambda, \as(\lambda^2) \right)
= -\frac{\gamma_{K}  (\as)}{2} \sum_{i<j} 
\log \left(\frac{-s_{ij}}{\lambda^2}\right) 
\, \T_i \cdot \T_j \,+\, \sum_i \gamma_i (\as) \,,
\ee
and the argument of the functions indicates that 
the dependence on the scale is both explicit and 
via the $4-2\eps$ dimensional coupling, which 
obeys the renormalization group equation
\beq\label{betagamma}
\beta(\as,\eps) \equiv \frac{d\as}{d\ln \mu}= 
-2\eps \,\as - \frac{\as^2}{2\pi} \sum_{n = 0}^{\infty} 
b_n \, \left(\frac{\as}{\pi}\right)^n\,,
\eeq 
with $b_0=\frac{11}{3}C_A-\frac{2}{3} n_f$. 

Aim of this appendix is to specialise \eqn{gammaSoft1}
to the $2 \to 3$ parton amplitudes considered in this paper. 
We focus in particular on the dipole component, \eqn{gammadip1}, 
and neglect the quadrupole term ${\bf \Delta}_{n}\left(\{\rho_{ijkl}\}\right)$.
The latter starts at three loops and therefore it is not relevant 
for the analysis of the high-energy limit considered in the main 
text, which concerns scattering amplitudes up to two loops. For 
the purpose of this section we follow closely 
\cite{DelDuca:1995hf,DelDuca:2011ae} and parameterize the 
external momenta $P_1,\dots,P_5$ in the lab frame as follows:    
\be
P_1 = (0, x_1 \sqrt{S}, \mathbf{0}), \qquad \qquad 
P_2 = (x_2 \sqrt{S}, 0, \mathbf{0}), 
\ee
where $S$ is the hadronic centre of mass energy, 
such that $s_{12} = x_1 x_2 S$, and 
\be
P_i = (|\mathbf{p}_i| \,e^{y_i}, 
|\mathbf{p}_i| \, e^{-y_i}, \mathbf{p}_i),
\qquad {\rm with} \qquad i = 3,4,5.
\ee
In turn, the transverse momenta $\mathbf{p}_i$
are parameterized as $\mathbf{p}_i = 
|\mathbf{p}_i| (\cos \phi_i, \sin\phi_i)$, where
$\phi_i$ is the azimuthal angle between the vector 
$\mathbf{p}_i$ and an arbitrary vector in the 
transverse plane. Momentum conservation requires
\bea\label{momentum_scaling1} \nn
\mathbf{0} &=& \mathbf{p}_3 +
\mathbf{p}_4 + \mathbf{p}_5, \\[0.1cm]
x_1 &=& \frac{1}{\sqrt{S}}
\Big(|\mathbf{p}_3| e^{-y_3} + |\mathbf{p}_4| e^{-y_4} 
+ |\mathbf{p}_5| e^{-y_5}\Big), \\ \nn
x_2 &=& \frac{1}{\sqrt{S}}
\Big(|\mathbf{p}_3| e^{y_3} + |\mathbf{p}_4| e^{y_4} 
+ |\mathbf{p}_5| e^{y_5}\Big),
\eea 
and the kinematic invariants of \eqn{kinhel1} reads
\bea\label{MandelstamRegge2} \nn
s_{12} = x_1 x_2 S
&=& \sum_{i,j=3}^5 |\mathbf{p}_i| |\mathbf{p}_j| 
\, e^{y_i - y_j}, \\ \nn 
s_{1i} = -2 P_1 \cdot P_i 
&=& -\sum_{j=3}^5 |\mathbf{p}_i| |\mathbf{p}_j| \, e^{y_i - y_j}, 
\qquad i = 3,4,5, \\ 
s_{2i} = -2 P_2 \cdot P_i 
&=& -\sum_{j=3}^5 |\mathbf{p}_i| |\mathbf{p}_j| \, e^{-(y_i - y_j)}, 
\qquad i = 3,4,5 \\[0.2cm] \nn
s_{ij} = 2 P_i \cdot P_j
&=& |\mathbf{p}_i| |\mathbf{p}_j|  \big[\cosh(y_i - y_j) 
- \cos(\phi_i - \phi_j)\big], 
\qquad i,j = 3,4,5.
\eea
In terms of the rapidity variables $y_i$, the 
multi-Regge limit of \eqn{hel_def0} reads 
\be\label{hel_def1} 
y_3 \gg y_4 \gg y_5.
\ee
In this limit $x_1 \simeq |\mathbf{p}_5|/\sqrt{S}\,e^{-y_5}$, 
$x_2 \simeq |\mathbf{p}_3|/\sqrt{S}\, e^{y_3}$, and the 
invariants of \eqn{MandelstamRegge2} can be approximated as 
\bea\label{MandelstamRegge3} \nn
s_{12} &\simeq& |\mathbf{p}_3| |\mathbf{p}_5| \,e^{y_3 - y_5}, \\ \nn 
s_{1i} &\simeq& - |\mathbf{p}_i| |\mathbf{p}_5| \,e^{y_i - y_5}, 
\qquad i = 3,4,5 \\ 
s_{2i} &\simeq& - |\mathbf{p}_3| |\mathbf{p}_i| \, e^{y_3 - y_i}, 
\qquad i = 3,4,5 \\ \nn
s_{ij} &\simeq& |\mathbf{p}_i| |\mathbf{p}_j| \, e^{|y_i - y_j|},
\qquad i,j = 3,4,5.
\eea
Specialising \eqn{gammadip1} to the case $n = 5$ 
and taking the high-energy limit according to 
\eqn{MandelstamRegge3} one has 
\bea \label{gammadip3} \nn
{\bf \Gamma}_5^{\rm dip.}\left(\{p_i\},\lambda, \as(\lambda^2) \right)
&=& -\frac{\gamma_{K}  (\as)}{2} \bigg\{ 
\bigg[\log \frac{|\mathbf{p}_3|}{\lambda} 
+\log \frac{|\mathbf{p}_5|}{\lambda} -i \pi +y_3 - y_5\bigg] 
\, \T_1 \cdot \T_2 \\ \nn
&&\hspace{0.5cm}
+\,\bigg[\log \frac{|\mathbf{p}_3|}{\lambda} 
+\log \frac{|\mathbf{p}_5|}{\lambda} +y_3 - y_5\bigg]
\, \T_1 \cdot \T_3 \\ \nn
&&\hspace{0.5cm}
+\,\bigg[\log \frac{|\mathbf{p}_4|}{\lambda} 
+\log \frac{|\mathbf{p}_5|}{\lambda} +y_4 - y_5\bigg] 
\, \T_1 \cdot \T_4 \\ \nn
&&\hspace{0.5cm}
+\,2 \log \frac{|\mathbf{p}_5|}{\lambda} \, \T_1 \cdot \T_5 
+ 2\log \frac{|\mathbf{p}_3|}{\lambda} \, \T_2 \cdot \T_3 \\ 
&&\hspace{0.5cm}
+\,\bigg[\log \frac{|\mathbf{p}_3|}{\lambda} 
+\log \frac{|\mathbf{p}_4|}{\lambda} +y_3 - y_4\bigg] 
\, \T_2 \cdot \T_4 \\ \nn
&&\hspace{0.5cm}
+\,\bigg[\log \frac{|\mathbf{p}_3|}{\lambda} 
+\log \frac{|\mathbf{p}_3|}{\lambda} +y_3 - y_5\bigg] 
\, \T_2 \cdot \T_5 \\ \nn
&&\hspace{0.5cm} 
+\,\bigg[\log \frac{|\mathbf{p}_3|}{\lambda} 
+\log \frac{|\mathbf{p}_4|}{\lambda} -i \pi +y_3 - y_4\bigg] 
\, \T_3 \cdot \T_4 \\ \nn
&&\hspace{0.5cm}
+\,\bigg[\log \frac{|\mathbf{p}_3|}{\lambda} 
+\log \frac{|\mathbf{p}_5|}{\lambda} -i \pi +y_3 - y_5\bigg] 
\, \T_3 \cdot \T_5 \\ \nn
&&\hspace{0.5cm}
+\,\bigg[\log \frac{|\mathbf{p}_4|}{\lambda} 
+\log \frac{|\mathbf{p}_5|}{\lambda} -i \pi +y_4 - y_5\bigg]
\, \T_4 \cdot \T_5 \Bigg\}+ \sum_{i=1}^{5} \gamma_i (\as) \,.
\eea
The expression in \eqn{gammadip3} 
can be expressed in terms of the colour operators 
in eqs.~(\ref{def:Tt1}),~(\ref{def:Tt2}) 
and~(\ref{def:Ts}) by means of the identities 
\be\label{colcons}
\sum_{i = 1}^5 {\bf T}_i = 0, 
\qquad \qquad 
\Big(\sum_{i = 1}^5 {\bf T}_i\Big)^2 
= \sum_{i = 1}^5 C_i + 2 \sum_{j>i=1}^{5}
{\bf T}_i \cdot {\bf T}_j.
\ee
After some elaboration one has  
\bea \label{gammadip4} \nn
{\bf \Gamma}_5^{\rm dip.}\left(\{p_i\},\lambda, \as(\lambda^2) \right)
&=& -\frac{\gamma_{K}  (\as)}{2} \bigg[ 
- \big(y_4 - y_5\big) \T_{t_1}^2 
- \big(y_3 - y_4\big) \T_{t_2}^2
- i \pi \, \T_{s}^2  \\ \nn
&&\hspace{1.0cm}
-\frac{1}{2}\bigg(\log \frac{|\mathbf{p}_5|^2}{\lambda^2} - i\pi\bigg)
\big( C_1 + C_5\big)
-\frac{1}{2}\bigg(\log \frac{|\mathbf{p}_4|^2}{\lambda^2} - i\pi\bigg) C_4 \\
&&\hspace{1.0cm}
-\frac{1}{2}\bigg(\log \frac{|\mathbf{p}_3|^2}{\lambda^2} - i\pi\bigg)
\big(C_2 + C_3\big) \bigg] + \sum_{i=1}^{5} \gamma_i (\as) \,,
\eea
which is consistent with eq. (6.8) of 
\cite{DelDuca:2011ae}. Using \eqn{Ts2ToTpm}
it is possible to write the dipole formula 
in terms of the operators $\T_{(\pm,\pm)}$
introduced in \eqn{TpmDef}, in such a way 
to make symmetries under the exchanges 
$2 \leftrightarrow 3$ and $1 \leftrightarrow 5$ 
manifest. Taking into account that in the 
high-energy limit $C_1 = C_5 \equiv C_i$, 
$C_2 = C_3 \equiv C_j$, and identifying 
$C_4 \equiv C_v = C_A$, one obtains
\bea \label{gammadip5} \nn
{\bf \Gamma}_5^{\rm dip.}\left(\{p_i\},\lambda, \as(\lambda^2) \right)
&=& -\frac{\gamma_{K}  (\as)}{2} \bigg[ 
- \big(y_4 - y_5\big) \T_{t_1}^2 
- \big(y_3 - y_4\big) \T_{t_2}^2 \\ 
&&\hspace{-1.0cm}-\,\frac{i \pi}{2} \bigg(
\T_{(++)} + \T_{(+-)} + \T_{(-+)} + \T_{(--)} \bigg) \\ \nn
&&\hspace{-1.0cm}
- C_i \log \frac{|\mathbf{p}_5|^2}{\lambda^2}  
- C_j \log \frac{|\mathbf{p}_3|^2}{\lambda^2} 
-\frac{C_v}{2}\bigg(\log \frac{|\mathbf{p}_4|^2}{\lambda^2} - i\pi \bigg) 
\bigg] + \sum_{i=1}^{5} \gamma_i (\as) \,.
\eea
The operator $\T_{(++)}$ commutes 
with $\T^2_{t_1}$ and $\T^2_{t_2}$, 
furthermore one has
\be\label{TppToTt12}
\T_{(++)} = - \frac{1}{2}
\big(\T^2_{t_1}+\T^2_{t_2}-C_4\big).
\ee
This relation can be used in 
\eqn{gammadip5} to obtain
\bea \label{gammadip6} \nn
{\bf \Gamma}_5^{\rm dip.}\left(\{p_i\},\lambda, \as(\lambda^2) \right)
&=& -\frac{\gamma_{K}  (\as)}{2} \bigg[ 
-\bigg(y_4 - y_5 - \frac{i \pi}{4}\bigg) \T_{t_1}^2 
-\bigg(y_3 - y_4 - \frac{i \pi}{4}\bigg) \T_{t_2}^2 \\ 
&&\hspace{-1.0cm}-\,\frac{i \pi}{2} \bigg(
\T_{(+-)} + \T_{(-+)} + \T_{(--)} \bigg) \\ \nn
&&\hspace{-1.0cm}
- C_i \log \frac{|\mathbf{p}_5|^2}{\lambda^2}  
- C_j \log \frac{|\mathbf{p}_3|^2}{\lambda^2} 
-\frac{C_v}{2}\bigg(\log \frac{|\mathbf{p}_4|^2}{\lambda^2} 
- \frac{i\pi}{2}\bigg) \bigg] 
+ \sum_{i=1}^{5} \gamma_i (\as) \,.
\eea
Using this result it appears natural, in the high-energy 
limit, to factorize the dipole renormalization factor 
${\bf Z}_5^{\rm dip.}$ in \eqn{RGsol} as follows:
\bea \label{ZtildeZijZgfact} \nn
{\bf Z}_5^{\rm dip.} \left(\{p_i\},\mu, \as (\mu^2) \right)
&=& Z_i\bigg(\frac{|\mathbf{p}_5|}{\mu}, \as(\mu^2), \eps \bigg)
\,Z_j\bigg(\frac{|\mathbf{p}_3|}{\mu}, \as(\mu^2), \eps \bigg) \\
&&\hspace{0.0cm} \times \, 
Z_v\bigg(\frac{p_{4\perp}}{\mu}, \as(\mu^2), \eps \bigg)
\, {\bf \hat Z}\Big(\{y_i - y_j\}, \as(\mu^2), \eps \Big).
\eea
In this equation the last 
factor ${\bf \hat Z}$ reads 
\bea \label{ZetaHatDef}  \nn
{\bf \hat Z}\Big(\{y_i - y_j\}, \as(\mu^2), \eps \Big) 
&=& \exp \Bigg\{ K \Big(\as(\mu^2), \eps \Big)
\bigg[ \tilde\eta_1 \,\T^2_{t_1}+\tilde\eta_2 \,\T^2_{t_2} \\ 
&&\hspace{1.0cm}+\,\frac{i \pi}{2} \Big(
\T_{(+-)} + \T_{(-+)} + \T_{(--)} \Big) \bigg]\Bigg\} \,,
\eea
where 
\be\label{eta12Def} 
\tilde\eta_1 = y_4-y_5  - \frac{i \pi}{4},
\qquad \qquad  
\tilde\eta_2 = y_3-y_4  - \frac{i \pi}{4},
\ee
and the factor $K(\as(\mu^2), \eps)$
is the well-known integral over the 
scale of the cusp anomalous dimension:
\beq \label{intK}
K\Big(\as(\mu^2), \eps \Big) =
- \frac{1}{4} \int_0^{\mu^2} \frac{d \lambda^2}{\lambda^2} 
\, \hat{\gamma}_K \left( \alpha_s (\lambda^2) \right) \,.
\eeq
The rapidity differences $\tilde \eta_1$, $\tilde \eta_2$
in \eqn{eta12Def} constitutes the natural 
generalization of the signature-even combination 
of logarithms $L = \log (s/|t|) - i \pi/2$, 
introduced in case of $2\to 2$ parton scattering 
(cf. with eq.~(2.9) of \cite{Falcioni:2021buo}). 
The relation with the rapidity factors in 
\eqn{complexLogsB} can be easily found to be
\bea\label{eta12DefPrime} \nn 
\eta_1 &=& \frac{1}{2}\big[\log(1-z) + \log(1-\bar z) \big]
+ \log \frac{|\mathbf{p}_4|^2}{\tau} + \tilde \eta_1,  \\
\eta_2 &=& \frac{1}{2}\big[\log(z) + \log(\bar z) \big]
+ \log \frac{|\mathbf{p}_4|^2}{\tau} + \tilde \eta_2.
\eea
Let us also point out that in this section 
we keep explicit dependence on the IR renormalisation 
scale $\mu$, and fix the Regge factorization scale 
to $\tau = |\mathbf{p}_4|^2$. Next, we define 
\be
\gamma_{i}(\al_s) \equiv \gamma_{1}(\al_s) = \gamma_{5}(\al_s),
\qquad 
\gamma_{j}(\al_s) \equiv \gamma_{2}(\al_s) = \gamma_{3}(\al_s), 
\ee 
such that the factors $Z_{i/j}$ in 
\eqn{ZtildeZijZgfact} reads 
\be \label{ZetaijDef}
Z_{i/j}\bigg(\frac{|\mathbf{p}_{5/3}|}{\mu}, \as(\mu^2), \eps \bigg) 
= \exp\left\{-\frac{1}{2}\int_0^{\mu^2}\frac{d\lambda^2}{\lambda^2}
\left[ C_{i/j} \frac{\gamma_K(\al_s)}{2}   
\log\frac{|\mathbf{p}_{5/3}|^2}{\lambda^2}+2\gamma_{i/j}(\al_s)\right] \right\}\,,
\ee
and correspond to the collinear 
singularities associated to the 
impact factors $C_{i/j}$, according 
to \eqns{ImpFactIRsubtract}{ImpFactIRsubtract2}.
Last, the factor $Z_v$ in 
\eqn{ZtildeZijZgfact} reads
\be \label{ZetagDef}
Z_v\bigg(\frac{|\mathbf{p}_{4}|}{\mu}, \as(\mu^2), \eps \bigg) 
= \exp\Bigg\{
-\frac{1}{2}\int_0^{\mu^2}\frac{d\lambda^2}{\lambda^2}
\bigg[ C_{v} \frac{\gamma_K(\al_s)}{4} \bigg(  
 \log\frac{|\mathbf{p}_{4}|^2}{\lambda^2} - \frac{i \pi}{2} \bigg)
+\gamma_{v}(\al_s) \bigg] \Bigg\}\,,
\ee
where we identified 
\be
\gamma_{v}(\al_s) \equiv \gamma_{4}(\al_s).
\ee 
\Eqns{ZetaijDef}{ZetagDef} can be 
expressed in terms of the factors 
\beqa \label{intKD} 
K_{D}\Big(\as(\mu^2), \eps \Big) &=&
- \frac{1}{4} \int_0^{\mu^2} \frac{d \lambda^2}{\lambda^2} 
\, \hat{\gamma}_K \left( \alpha_s (\lambda^2) \right) 
\log\frac{\mu^2}{\lambda^2}  \,, \\
\label{intKB}
K_{B_i}\Big(\as(\mu^2), \eps \Big) &=&
- \frac{1}{2} \int_0^{\mu^2} \frac{d \lambda^2}{\lambda^2} 
\, \hat{\gamma}_i \left( \alpha_s (\lambda^2) \right) \,.
\eeqa
Then we have 
\bea \label{ZetaijDefB} \nn
Z_{i/j}\bigg(\frac{|\mathbf{p}_{5/3}|}{\mu}, \as(\mu^2), \eps \bigg) 
&=& \exp\Bigg\{ \bigg[ K \Big(\as(\mu^2), \eps \Big) 
\log\frac{|\mathbf{p}_{5/3}|^2}{\mu^2} \\  
&&\hspace{1.0cm}+\, K_{D}\Big(\as(\mu^2), \eps \Big)\bigg] C_{i/j} 
+ 2 K_{B_i}\Big(\as(\mu^2), \eps \Big) \Bigg\}\,,
\eea
and 
\bea \label{ZetagDefB} \nn
Z_v\bigg(\frac{|\mathbf{p}_{4}|}{\mu}, \as(\mu^2), \eps \bigg) 
&=& \exp\Bigg\{ \bigg[ K \Big(\as(\mu^2), \eps \Big) 
\bigg( \log\frac{|\mathbf{p}_{4}|^2}{\lambda^2} - \frac{i \pi}{2} \bigg) \\  
&&\hspace{1.0cm}+\, K_{D}\Big(\as(\mu^2), \eps \Big) \bigg] \frac{C_{v}}{2}
+ K_{B_v}\Big(\as(\mu^2), \eps \Big) \Bigg\}\,.
\eea


\subsection*{Perturbative expansion}
 
Given the non-commuting nature of the color 
operators in \eqn{ZetaHatDef}, the expansion 
in powers of $\as$ of \eqn{ZtildeZijZgfact}
requires the repeated application of the 
Zassenhaus formula: 
\beq
e^{k(X+Y)} = e^{k X} \, e^{k Y} \, 
e^{-\frac{k^2}{2} [X,Y]} \, e^{\ord(k^3)},
\eeq
where $X$, $Y$ represent two non-commuting
color operators, and $k$ a c-number. 
In what follows we define the 
perturbative expansion of the factors $K_m(\as)$
in eqs.~(\ref{intK}),~(\ref{intKD}) and~(\ref{intKB})
as follows:
\beq \label{intKexpansion}
K_m\Big(\as(\mu^2), \eps \Big) =
\sum_{n} \bigg( \frac{\as(\mu^2)}{\pi}\bigg)^n 
K^{(n)}_m\Big(\mu^2, \eps \Big),
\eeq
then we have 
\bea \nn
K^{(1)} &=& \frac{\gamma_K^{(1)}}{4 \eps}, \\
K^{(2)} &=& -\frac{b_0 \gamma_K^{(1)}}{32 \eps^2} 
+ \frac{\gamma_K^{(2)}}{8 \eps},
\eea
\bea \nn
K_D^{(1)} &=& -\frac{\gamma_K^{(1)}}{4 \eps^2}, \\
K_D^{(2)} &=& \frac{3 b_0 \gamma_K^{(1)}}{64 \eps^3} 
- \frac{\gamma_K^{(2)}}{16 \eps^2},
\eea
\bea \nn
K_{B_q}^{(1)} &=& \frac{\gamma_q^{(1)}}{2 \eps}, \\
K_{B_q}^{(2)} &=& -\frac{b_0 \gamma_q^{(1)}}{16 \eps^2} 
+ \frac{\gamma_q^{(2)}}{4 \eps},
\eea
\bea \nn
K_{B_g}^{(1)} &=& \frac{\gamma_g^{(1)}}{2 \eps}, \\
K_{B_g}^{(2)} &=& -\frac{b_0 \gamma_g^{(1)}}{16 \eps^2} 
+ \frac{\gamma_g^{(2)}}{4 \eps},
\eea
where in turn 
\bea \nn
\gamma_K^{(1)} &=& 2, \\
\gamma_K^{(2)} &=& \bigg(\frac{67}{18} - \frac{\pi^2}{6}\bigg) C_A 
- \frac{5}{9} n_f,
\eea
\bea \nn
\gamma_q^{(1)} &=& -\frac{3}{4} C_F, \\ \nn
\gamma_q^{(2)} &=& 
 \frac{C_F^2}{16} \bigg(-\frac{3}{2} + 2 \pi^2 - 24 \zeta_3\bigg) 
 + \frac{C_A C_F}{16} \bigg(-\frac{961}{54} - \frac{11}{6} \pi^2 
 26 \zeta_3\bigg) \\
&& +\, \frac{C_F T_R n_f}{16} \bigg(\frac{130}{27} + \frac{2}{3} \pi^2\bigg),
\eea
\bea \nn
\gamma_g^{(1)} &=& -\frac{b_0}{4}, \\
\gamma_g^{(2)} &=& 
 \frac{C_A^2}{16} \bigg(-\frac{692}{27} + \frac{11}{18} \pi^2 
 + 2 \zeta_3\bigg) 
 + \frac{C_A n_f}{32} \bigg(\frac{256}{27} - \frac{2}{9} \pi^2\bigg) 
 + \frac{C_F n_f}{8},
\eea
and in turn
\be
b_0 = \frac{1}{3} \big(11 C_A - 2 n_f\big), \qquad \qquad
b_1 = \frac{1}{6} \Big[17 C_A^2 - \big(10 C_A + 6 C_F\big) T_R  n_f\Big].
\ee
Then, at tree level and first order
in perturbation theory one has 
\bea \label{ZtildeZijZgfact-01} \nn
{\cal M}^{(0)} &=& {\cal H}^{(0)}, \\ \nn
{\cal M}^{(1)} &=& \bigg\{ 
K^{(1)}\bigg[\Big( \eta_1 \T_{t_1}^2 + \eta_2 \T_{t_2}^2\Big)
+\frac{i\pi}{2}\Big( \T_{(--)} + \T_{(+-)} + \T_{(+-)} \Big) \\ \nn 
&&\hspace{0.5cm}+\, C_i \log\frac{|\mathbf{p}_{5}|^2}{\mu^2}
+\frac{C_v}{2} \bigg(\log\frac{|\mathbf{p}_{4}|^2}{\mu^2} - \frac{i \pi}{2}\bigg)
+C_j  \log\frac{|\mathbf{p}_{3}|^2}{\lambda^2} \bigg] \\ 
&&\hspace{0.5cm}+\, 
K_D^{(1)} \bigg(C_i +\frac{C_v}{2} + C_j \bigg) 
+ 2K_{B_i}^{(1)}+ K_{B_v}^{(1)}+ 2K_{B_j}^{(1)}
\bigg\} {\cal H}^{(0)} + {\cal H}^{(1)}, 
\eea
and second order the 
factorization formula gives 
\bea \label{ZtildeZijZgfact-2} \nn
{\cal M}^{(2)} &=& \Bigg\{
\frac{1}{2}\Big(K^{(1)}\Big)^2 
\Bigg[\eta_1^2  \Big(\T_{t_1}^2\Big)^2 
+ \eta_1 \eta_2 \Big\{\T_{t_1}^2, \T_{t_2}^2\Big\} 
+ \eta_2^2  \Big(\T_{t_2}^2\Big)^2 \\ \nn
&& +\, \frac{i\pi}{2} 
\bigg(\eta_1 \Big\{\T_{(--)} + \T_{(-+)} + \T_{(+-)}, 
\T_{t_1}^2 \Big\} 
+ \eta_2 \Big\{\T_{(--)} + \T_{(-+)} + \T_{(+-)}, 
\T_{t_2}^2\Big\} \\ \nn
&& -\, \frac{\pi^2}{4} 
\Big(\T_{(--)} + \T_{(-+)} + \T_{(+-)}\Big)^2 \Bigg] 
+ K^{(2)} \Bigg[\eta_1 \T_{t_1}^2 + \eta_2 \T_{t_2}^2 \\ \nn
&&+\,\frac{i\pi}{2} \Big(\T_{(--)} + \T_{(-+)} + \T_{(+-)}\Big)\Bigg]
+ \frac{C_i^2}{2} \Bigg[\Big(K_D^{(1)}\Big)^2 
+ 2 K_D^{(1)} K^{(1)} 
\log\bigg(\frac{|\mathbf{p}_{5}|^2}{\mu^2}\bigg) \\ \nn
&&+\, \Big(K^{(1)}\Big)^2 
\log^2\bigg(\frac{|\mathbf{p}_{5}|^2}{\mu^2}\bigg) \Bigg] 
+ \frac{C_v^2}{4} \Bigg[ \frac{1}{2}\Big(K_D^{(1)}\Big)^2 
+ K^{(1)} K_{D}^{(1)} 
\bigg(\log\bigg(\frac{|\mathbf{p}_{4}|^2}{\mu^2}\bigg) 
- \frac{i \pi}{2} \bigg) \\ \nn
&& +\, \frac{1}{2}\Big(K^{(1)}\Big)^2 
\bigg(\log^2\bigg(\frac{|\mathbf{p}_{4}|^2}{\mu^2}\bigg) 
- i \pi \log\bigg(\frac{|\mathbf{p}_{4}|^2}{\mu^2}\bigg)
- \frac{\pi^2}{4}\bigg) \Bigg] \\ \nn
&&+\, \frac{C_j^2}{2} \Bigg[\Big(K_{D}^{(1)}\Big)^2 
+ 2 K_{D}^{(1)} K^{(1)} 
\log\bigg(\frac{|\mathbf{p}_{3}|^2}{\mu^2}\bigg) 
+ \Big(K^{(1)}\Big)^2  
\log^2\bigg(\frac{|\mathbf{p}_{3}|^2}{\mu^2}\bigg) \Bigg] \\ \nn
&&+\, C_i \Bigg[\bigg(K^{(1)} K_{D}^{(1)} 
+ \Big(K^{(1)}\Big)^2 
\log\bigg(\frac{|\mathbf{p}_{5}|^2}{\mu^2}\bigg)\bigg) 
\bigg(\eta_1  \T_{t1}^2 + \eta_2  \T_{t_2}^2 \\ \nn
&& +\, \frac{i\pi}{2} \Big(\T_{(--)} + \T_{(-+)} + \T_{(+-)}\Big)
\bigg) + K^{(2)} 
\log\bigg(\frac{|\mathbf{p}_{5}|^2}{\mu^2}\bigg) \\ \nn
&&+\, \bigg(K^{(1)} 
\log\bigg(\frac{|\mathbf{p}_{5}|^2}{\mu^2}\bigg) 
+K_{D}^{(1)}\bigg)\Big(2 B_i^{(1)} + 2 B_j^{(1)} + B_v^{(1)}\Big) 
 + K_{D}^{(2)} \\ \nn
&&+\, \frac{C_v}{2} \bigg[\Big(K_{D}^{(1)}\Big)^2 
+ K^{(1)} K_{D}^{(1)} 
\bigg(\log\bigg(\frac{|\mathbf{p}_{4}|^2}{\mu^2}\bigg) 
+ \log\bigg(\frac{|\mathbf{p}_{5}|^2}{\mu^2}\bigg) 
- \frac{i \pi}{2}\bigg) \\ \nn
&&+\, \Big(K^{(1)}\Big)^2
\log\bigg(\frac{|\mathbf{p}_{5}|^2}{\mu^2}\bigg) 
\bigg(\log\bigg(\frac{|\mathbf{p}_{4}|^2}{\mu^2}\bigg) 
- \frac{i \pi}{2}\bigg) \bigg] \Bigg] \\ \nn
&&+\, C_v \Bigg[ \frac{1}{2}\Big(K^{(1)}\Big)^2  
\bigg(\Big(\eta_1 \T_{t1}^2 + \eta_2 \T_{t_2}^2\Big)
\log\bigg(\frac{|\mathbf{p}_{4}|^2}{\mu^2}\bigg) 
- \frac{i \pi}{2} \bigg(\eta_1 \T_{t_1}^2 + \eta_2 \T_{t_2}^2 \\ \nn
&& -\, \big(\T_{(--)} + \T_{(-+)} + \T_{(+-)}\Big) 
\log\bigg(\frac{|\mathbf{p}_{4}|^2}{\mu^2}\bigg)\bigg) 
+ \frac{\pi^2}{4} 
\Big(\T_{(--)} + \T_{(-+)} + \T_{(+-)}\Big)\bigg) \\ \nn
&& +\, K^{(1)} \bigg(\frac{1}{2}K_{D}^{(1)} 
\Big(\eta_1 \T_{t_1}^2 + \eta_2 \T_{t_2}^2\Big) 
- \frac{i\pi}{2} \bigg(B_i^{(1)} + B_j^{(1)} 
+ \frac{1}{2}B_v^{(1)} \\ \nn
&& -\, \frac{1}{2}K_{D}^{(1)} 
\Big(\T_{(--)} + \T_{(-+)} + \T_{(+-)}\Big)\bigg) 
+ \bigg(B_i^{(1)} + B_j^{(1)} + \frac{1}{2} B_v^{(1)}\bigg)
\log\bigg(\frac{|\mathbf{p}_{4}|^2}{\mu^2}\bigg) \\ \nn
&&+\, \frac{K^{(2)}}{2} 
\bigg(\log\bigg(\frac{|\mathbf{p}_{4}|^2}{\mu^2}\bigg) 
- \frac{i\pi}{2}\bigg) + K_{D}^{(1)}
\Big(B_i^{(1)} + B_j^{(1)} + \frac{1}{2}B_v^{(1)}\Big)  
+ \frac{1}{2}K_{D}^{(2)}\bigg) \Bigg] \\ \nn
&&+\, C_j \Bigg[\bigg(K^{(1)} K_{D}^{(1)} + \Big(K^{(1)}\Big)^2
\log\bigg(\frac{|\mathbf{p}_{3}|^2}{\mu^2}\bigg) 
\bigg(\eta_1 \T_{t_1}^2 + \eta_2 \T_{t_2}^2 \\ \nn
&& +\, \frac{i\pi}{2} \Big(\T_{(--)} + \T_{(-+)} + \T_{(+-)}\Big)
\bigg) + K^{(2)} 
\log\bigg(\frac{|\mathbf{p}_{3}|^2}{\mu^2}\bigg) \\ \nn
&&+\, \bigg( K^{(1)} \log\bigg(\frac{|\mathbf{p}_{3}|^2}{\mu^2}\bigg) 
+K_{D}^{(1)}\bigg)\Big(2 B_i^{(1)} + 2 B_j^{(1)} + B_v^{(1)}\Big) 
+ K_{D}^{(2)} \\ \nn
&&+\, \frac{C_v}{2} \bigg[
\Big(K_{D}^{(1)}\Big)^2 + K^{(1)} K_{D}^{(1)} 
\bigg(\log\bigg(\frac{|\mathbf{p}_{3}|^2}{\mu^2}\bigg) 
+ \log\bigg(\frac{|\mathbf{p}_{4}|^2}{\mu^2}\bigg) 
- \frac{i\pi}{2}\bigg) \\ \nn
&& +\, \Big(K^{(1)}\Big)^2
\log\bigg(\frac{|\mathbf{p}_{3}|^2}{\mu^2}\bigg) 
\bigg(\log\bigg(\frac{|\mathbf{p}_{4}|^2}{\mu^2}\bigg) 
- \frac{i \pi}{2}\bigg)\bigg] \Bigg] \\ \nn
&& +\, C_i  C_j \Bigg[ \Big(K_{D}^{(1)}\Big)^2 
+ K_{D}^{(1)}  K^{(1)} \bigg(
\log\bigg(\frac{|\mathbf{p}_{3}|^2}{\mu^2}\bigg) 
+ \log\bigg(\frac{|\mathbf{p}_{5}|^2}{\mu^2}\bigg)
\bigg) \\ \nn
&&+\, \Big(K^{(1)}\Big)^2  
\log\bigg(\frac{|\mathbf{p}_{3}|^2}{\mu^2}\bigg)
\log\bigg(\frac{|\mathbf{p}_{5}|^2}{\mu^2}\bigg) 
+ 2 \Big(B_i^{(1)}\Big)^2 + 4 B_i^{(1)} B_j^{(1)} 
+ 2 \Big(B_j^{(1)}\Big)^2  \\ \nn
&& +\,  2 B_i^{(2)} + 2 B_j^{(2)} + 2 B_i^{(1)} B_v^{(1)} 
+ 2 B_j^{(1)} B_v^{(1)} + \frac{1}{2}\Big(B_v^{(1)}\Big)^2 
+ B_v^{(2)} \\ \nn
&& +\, K^{(1)} \bigg(\eta_1 \T_{t_1}^2 
+ \eta_2 \T_{t_2}^2 + \frac{i \pi}{2} \Big(\T_{(--)} + \T_{(-+)} + \T_{(+-)}\Big) \bigg) \\ \nn
&& \times\, 
\Big(2 B_i^{(1)} + 2 B_j^{(1)} + B_v^{(1)}\Big) \Bigg]
\Bigg\} {\cal H}^{(0)} \\ \nn
&&+ \,\bigg\{ 
K^{(1)}\bigg[\Big( \eta_1 \T_{t_1}^2 + \eta_2 \T_{t_2}^2\Big)
+\frac{i\pi}{2}\Big( \T_{(--)} + \T_{(-+)} + \T_{(+-)} \Big)
\bigg] \\ \nn 
&&+\, C_i \bigg[K_D^{(1)} +
K^{(1)} \log\frac{|\mathbf{p}_{5}|^2}{\mu^2} \bigg]
+ \frac{C_v}{2} \bigg[K_D^{(1)} +
K^{(1)}\bigg(\log\frac{|\mathbf{p}_{4}|^2}{\mu^2} 
- \frac{i \pi}{2}\bigg) \bigg] \\ 
&&\hspace{0.5cm}+\, C_j \bigg[ K_D^{(1)} + 
K^{(1)} \log\frac{|\mathbf{p}_{3}|^2}{\lambda^2}
\bigg]+ 2K_{B_i}^{(1)}+ K_{B_v}^{(1)}+ 2K_{B_j}^{(1)}
\bigg\} {\cal H}^{(1)} + {\cal H}^{(2)},
\eea
where we have defined 
\be
\big\{{\bf \cal O}_1, {\bf \cal O}_2\big\}
= {\bf \cal O}_1 {\bf \cal O}_2 
+ {\bf \cal O}_2 {\bf \cal O}_1,
\ee
with ${\bf \cal O}_1$, ${\bf \cal O}_2$
two non-commuting color operators. 

To conclude this section, let us recall that
in section \ref{sec:oddOdd2} we have evaluated 
only the odd-odd amplitude at two loops, thus 
we need to extract the ${\cal M}^{(-,-)}$ 
component from \eqn{ZtildeZijZgfact-2}. 
In general, at all orders in perturbation 
theory this is obtained expanding systematically
${\bf Z}$ and ${\cal H}$ into their $(\pm,pm)$
components. Then one has 
\be
{\cal M}^{(-,-)} = {\bf Z}^{(-,-)} {\cal H}^{(+,+)}
+ {\bf Z}^{(-,+)} {\cal H}^{(+,-)} + {\bf Z}^{(+,-)} {\cal H}^{(-,+)}
+ {\bf Z}^{(+,+)} {\cal H}^{(-,-)}.
\ee
In practice, at two loops we have
\bea \label{ZtildeZijZgfact-2MinusMinus} \nn
{\cal M}^{(-,-,2)} &=& \Bigg\{
\frac{1}{2}\Big(K^{(1)}\Big)^2 
\Bigg[\eta_1^2  \Big(\T_{t_1}^2\Big)^2 
+ \eta_1 \eta_2 \Big\{\T_{t_1}^2, \T_{t_2}^2\Big\} 
+ \eta_2^2  \Big(\T_{t_2}^2\Big)^2 \\ \nn
&& -\, \frac{\pi^2}{4} 
\Big(\T^2_{(--)} + \T^2_{(-+)} + \T^2_{(+-)}\Big) \Bigg] 
+ K^{(2)} \Big[\eta_1 \T_{t_1}^2 + \eta_2 \T_{t_2}^2 \Big] \\ \nn
&&+\, \frac{C_i^2}{2} \Bigg[\Big(K_D^{(1)}\Big)^2 
+ 2 K_D^{(1)} K^{(1)} 
\log\bigg(\frac{|\mathbf{p}_{5}|^2}{\mu^2}\bigg) 
+ \Big(K^{(1)}\Big)^2 
\log^2\bigg(\frac{|\mathbf{p}_{5}|^2}{\mu^2}\bigg) \Bigg] \\ \nn
&&+\, \frac{C_v^2}{4} \Bigg[ \frac{1}{2}\Big(K_D^{(1)}\Big)^2 
+ K^{(1)} K_{D}^{(1)} 
\bigg(\log\bigg(\frac{|\mathbf{p}_{4}|^2}{\mu^2}\bigg) 
- \frac{i \pi}{2} \bigg) \\ \nn
&& +\, \frac{1}{2}\Big(K^{(1)}\Big)^2 
\bigg(\log^2\bigg(\frac{|\mathbf{p}_{4}|^2}{\mu^2}\bigg) 
- i \pi \log\bigg(\frac{|\mathbf{p}_{4}|^2}{\mu^2}\bigg)
- \frac{\pi^2}{4}\bigg) \Bigg] \\ \nn
&&+\, \frac{C_j^2}{2} \Bigg[\Big(K_{D}^{(1)}\Big)^2 
+ 2 K_{D}^{(1)} K^{(1)} 
\log\bigg(\frac{|\mathbf{p}_{3}|^2}{\mu^2}\bigg) 
+ \Big(K^{(1)}\Big)^2  
\log^2\bigg(\frac{|\mathbf{p}_{3}|^2}{\mu^2}\bigg) \Bigg] \\ \nn
&&+\, C_i \Bigg[\bigg(K^{(1)} K_{D}^{(1)} 
+ \Big(K^{(1)}\Big)^2 
\log\bigg(\frac{|\mathbf{p}_{5}|^2}{\mu^2}\bigg)\bigg) 
\bigg(\eta_1  \T_{t1}^2 + \eta_2  \T_{t_2}^2\bigg) \\ \nn
&& + K^{(2)} 
\log\bigg(\frac{|\mathbf{p}_{5}|^2}{\mu^2}\bigg) 
+ \bigg(K^{(1)} 
\log\bigg(\frac{|\mathbf{p}_{5}|^2}{\mu^2}\bigg) 
+K_{D}^{(1)}\bigg)\Big(2 B_i^{(1)} + 2 B_j^{(1)} + B_v^{(1)}\Big) \\ \nn
&&+\, K_{D}^{(2)} + \frac{C_v}{2} \bigg[\Big(K_{D}^{(1)}\Big)^2 
+ K^{(1)} K_{D}^{(1)} 
\bigg(\log\bigg(\frac{|\mathbf{p}_{4}|^2}{\mu^2}\bigg) 
+ \log\bigg(\frac{|\mathbf{p}_{5}|^2}{\mu^2}\bigg) 
- \frac{i \pi}{2}\bigg) \\ \nn
&&+\, \Big(K^{(1)}\Big)^2
\log\bigg(\frac{|\mathbf{p}_{5}|^2}{\mu^2}\bigg) 
\bigg(\log\bigg(\frac{|\mathbf{p}_{4}|^2}{\mu^2}\bigg) 
- \frac{i \pi}{2}\bigg) \bigg] \Bigg] \\ \nn
&&+\, C_v \Bigg[ \frac{1}{2}\Big(K^{(1)}\Big)^2  
\Big(\eta_1 \T_{t1}^2 + \eta_2 \T_{t_2}^2\Big)
\bigg(\log\bigg(\frac{|\mathbf{p}_{4}|^2}{\mu^2}\bigg) 
- \frac{i \pi}{2} \bigg)\\ \nn
&& +\, K^{(1)} \bigg(\frac{1}{2}K_{D}^{(1)} 
\Big(\eta_1 \T_{t_1}^2 + \eta_2 \T_{t_2}^2\Big) 
- \frac{i\pi}{2} \bigg(B_i^{(1)} + B_j^{(1)} 
+ \frac{1}{2}B_v^{(1)}\bigg) \\ \nn
&& +\, \bigg(B_i^{(1)} + B_j^{(1)} + \frac{1}{2} B_v^{(1)}\bigg)
\log\bigg(\frac{|\mathbf{p}_{4}|^2}{\mu^2}\bigg)
+ \frac{K^{(2)}}{2} 
\bigg(\log\bigg(\frac{|\mathbf{p}_{4}|^2}{\mu^2}\bigg) 
- \frac{i\pi}{2}\bigg)\\ \nn
&&+\, K_{D}^{(1)}
\Big(B_i^{(1)} + B_j^{(1)} + \frac{1}{2}B_v^{(1)}\Big)  
+ \frac{1}{2}K_{D}^{(2)}\bigg) \Bigg] \\ \nn
&&+\, C_j \Bigg[\bigg(K^{(1)} K_{D}^{(1)} + \Big(K^{(1)}\Big)^2
\log\bigg(\frac{|\mathbf{p}_{3}|^2}{\mu^2}\bigg) 
\bigg(\eta_1 \T_{t_1}^2 + \eta_2 \T_{t_2}^2 \bigg)\\ \nn
&&+\, K^{(2)} 
\log\bigg(\frac{|\mathbf{p}_{3}|^2}{\mu^2}\bigg) 
+ \bigg( K^{(1)} \log\bigg(\frac{|\mathbf{p}_{3}|^2}{\mu^2}\bigg) 
+K_{D}^{(1)}\bigg)\Big(2 B_i^{(1)} + 2 B_j^{(1)} + B_v^{(1)}\Big) \\ \nn
&&+\, K_{D}^{(2)} + \frac{C_v}{2} \bigg[
\Big(K_{D}^{(1)}\Big)^2 + K^{(1)} K_{D}^{(1)} 
\bigg(\log\bigg(\frac{|\mathbf{p}_{3}|^2}{\mu^2}\bigg) 
+ \log\bigg(\frac{|\mathbf{p}_{4}|^2}{\mu^2}\bigg) 
- \frac{i\pi}{2}\bigg) \\ \nn
&& +\, \Big(K^{(1)}\Big)^2
\log\bigg(\frac{|\mathbf{p}_{3}|^2}{\mu^2}\bigg) 
\bigg(\log\bigg(\frac{|\mathbf{p}_{4}|^2}{\mu^2}\bigg) 
- \frac{i \pi}{2}\bigg)\bigg] \Bigg] \\ \nn
&& +\, C_i  C_j \Bigg[ \Big(K_{D}^{(1)}\Big)^2 
+ K_{D}^{(1)}  K^{(1)} \bigg(
\log\bigg(\frac{|\mathbf{p}_{3}|^2}{\mu^2}\bigg) 
+ \log\bigg(\frac{|\mathbf{p}_{5}|^2}{\mu^2}\bigg)
\bigg) \\ \nn
&&+\, \Big(K^{(1)}\Big)^2  
\log\bigg(\frac{|\mathbf{p}_{3}|^2}{\mu^2}\bigg)
\log\bigg(\frac{|\mathbf{p}_{5}|^2}{\mu^2}\bigg) 
+ 2 \Big(B_i^{(1)}\Big)^2 + 4 B_i^{(1)} B_j^{(1)} 
+ 2 \Big(B_j^{(1)}\Big)^2  \\ \nn
&& +\,  2 B_i^{(2)} + 2 B_j^{(2)} + 2 B_i^{(1)} B_v^{(1)} 
+ 2 B_j^{(1)} B_v^{(1)} + \frac{1}{2}\Big(B_v^{(1)}\Big)^2 
+ B_v^{(2)} \\ \nn
&& +\, K^{(1)} \bigg(\eta_1 \T_{t_1}^2 
+ \eta_2 \T_{t_2}^2 \bigg) 
\Big(2 B_i^{(1)} + 2 B_j^{(1)} + B_v^{(1)}\Big) \Bigg]
\Bigg\} {\cal H}^{(0)} \\ \nn
&&+ \,\bigg\{ 
K^{(1)}\bigg[\Big( \eta_1 \T_{t_1}^2 
+ \eta_2 \T_{t_2}^2\Big) \bigg] 
+ C_i \bigg[K_D^{(1)} +
K^{(1)} \log\frac{|\mathbf{p}_{5}|^2}{\mu^2} \bigg] \\ \nn 
&&+\, \frac{C_v}{2} \bigg[K_D^{(1)} +
K^{(1)}\bigg(\log\frac{|\mathbf{p}_{4}|^2}{\mu^2} 
- \frac{i \pi}{2}\bigg) \bigg] 
+ C_j \bigg[ K_D^{(1)} + 
K^{(1)} \log\frac{|\mathbf{p}_{3}|^2}{\lambda^2}
\bigg] \\ \nn
&&\hspace{0.5cm}+\, 2K_{B_i}^{(1)}+ K_{B_v}^{(1)}+ 2K_{B_j}^{(1)}
\bigg\} {\cal H}^{(-,-,1)} 
+ K^{(1)} \frac{i\pi}{2}\Big[\T_{(+-)} {\cal H}^{(-,+,1)} \\ 
&&+\, \T_{(-+)} {\cal H}^{(+,-,1)} 
+ \T_{(--)} {\cal H}^{(+,+,1)} \Big]
+ {\cal H}^{(-,-,2)}.
\eea


\bibliographystyle{JHEP}
\bibliography{biblio.bib}

\end{document}